\documentclass[12pt,lmodern,numbered,hyperref,print,index]{report}

\newcommand{\La}{\mathcal{L}}

\newcommand{\Z}{\mathcal{Z}}
\newcommand{\adq}{\overline{\Psi}}
\newcommand{\q}{\Psi}
\newcommand{\hs}{\hspace{.5mm}}

\newcommand{\LA}{\left \langle}
\newcommand{\RA}{\right \rangle}
\newcommand{\mB}{\hat{\mu}_B}

\newcommand{\Ns}{N_{\sigma}}
\newcommand{\Nt}{N_{\tau}}
\newcommand{\Tpc}{T_{pc}}
\newcommand{\ord}{\mathcal{O}}
\newcommand{\M}{\mathcal{M}}
\newcommand{\mT}{\left(\frac{\mu_I}{T}\right)}

\newcommand{\NR}{N_R}
\newcommand{\dP}{\Delta P}
\newcommand{\D}{\bar{D}}
\newcommand{\Ka}{\mathcal{K}}
\newcommand{\Ob}{\mathcal{O}}
\newcommand{\muB}{\mu_B}
\newcommand{\muI}{\mu_I}

\newcommand{\Wc}{\mathcal{W}}
\newcommand{\dN}{\mathcal{N}}

\newcommand{\be}{\begin{equation}}
\newcommand{\ee}{\end{equation}}
\newcommand{\nsum}[1][1.4]{
    \mathop{%
        \raisebox
            {-#1\depthofsumsign+1\depthofsumsign}
            {\scalebox
                {#1}
                {$\displaystyle\sum$}%
            }
    }
}
\usepackage[a4paper, tmargin=35mm,bmargin=32mm, inner=25mm, outer=25mm]{geometry}

\input{Preamble/packages.tex}
\usepackage{lmodern}
\usepackage[T1]{fontenc}
\usepackage{calligra}

\numberwithin{equation}{section}
\titleformat{\chapter}[display]{\normalfont \huge \filright \bfseries\color{black}}{\chaptertitlename\ \thechapter}{5mm}{\Huge}
\titlespacing{\chapter}{5pt}{5pt}{5pt}
\graphicspath{{./Figures/}}

\DeclareMathOperator{\Tr}{Tr}

\newcommand\thetitle{EXPONENTIAL RESUMMATION OF QCD AT FINITE CHEMICAL POTENTIAL}
\newcommand\thescholar{Sabarnya Mitra}

\newcommand\theyear{2023}
\newcommand\thedepartment{Centre for High Energy Physics} 
\newcommand\thecollege{Indian Institute of Science Bengaluru} 

\newcolumntype{P}[1]{>{\centering\arraybackslash}p{#1}}

\begin{document}

\newlength{\depthofsumsign}
\setlength{\depthofsumsign}{\depthof{$\sum$}}
\newlength{\totalheightofsumsign}
\newlength{\heightanddepthofargument}


\pagestyle{fancy}
\renewcommand{\chaptermark}[1]{\markboth{#1}{#1}}
\fancyhead[R]{\leftmark}
\fancyhead[L]{\textbf{\thepage}}
\cfoot{}

	\pagenumbering{roman}
	
    \thispagestyle{empty}
\graphicspath{{Figures/JPEG/}{Figures/}}
\doublespacing
\begin{center}
\vspace{8mm}   
\huge
\textbf{\thetitle} \\
\vspace{18mm}  

\begin{figure}[h!]
\centering
\includegraphics[width=0.53\textwidth]{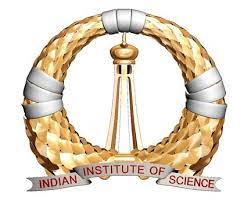}  
\end{figure}

\vspace{.5cm}
 
\Large \textbf{\thescholar}

\vspace{.5cm}



\par\vspace{4mm}

\large
{ \thedepartment\\
\thecollege\\
\theyear
}

\vspace{4mm}

This dissertation is submitted for the degree of \\
\emph{Doctor of Philosophy}

\end{center}
\thispagestyle{empty}
\onehalfspacing                   
	\cleardoublepage                         


\begin{spacing}{1.3}
    
    \phantomsection 
    \addcontentsline{toc}{chapter}{Dedication} 
\thispagestyle{empty}
{\color{white}.}
\vspace{95mm}
\begin{center}

\Large \textit{To \\Maa, Baba \\ \& \\ Dida}
\end{center}
\newpage  
	\cleardoublepage

    \phantomsection 
    \addcontentsline{toc}{chapter}{Declaration} 
\graphicspath{{Figures/PNG/}{Figures/}}
\vspace{-4.5\baselineskip}

\begin{center}
{\bf\large Declaration }\\
\par\vspace{5mm}
\end{center}
\noindent I declare that this written submission represents my ideas in my own words and where others' ideas or words have been included, I have adequately cited and referenced the
original sources. I also declare that I have adhered to all principles of academic honesty and integrity and have not misrepresented or fabricated or falsified any idea/ data/ fact/ source in my submission. I understand that any violation of the above will be cause for disciplinary action by the University and can also evoke penal action from the sources which have thus not been properly cited or from whom proper permissions have not been taken when needed.

\begin{flushright}
\vspace{25mm}
\begin{tabular}{r}
\thescholar \\
\end{tabular}
\end{flushright}

\vspace{25mm}
\begin{tabular}{@{}p{1cm}p{5cm}@{}}
Date: & February 2023\\
\end{tabular}
  
	\cleardoublepage
    \phantomsection 

    \phantomsection 
    \addcontentsline{toc}{chapter}{Acknowledgement}
    \vspace{-6.0\baselineskip}
\textbf{\begin{center}
		{\large \textbf{ACKNOWLEDGMENTS}}
\end{center}}
\vspace{0.1\baselineskip}
It is a great pleasure for me to express my heartfelt thanks and sincere gratitude to my Ph.D. supervisor \textcolor{blue}{Prof. Prasad Hegde} of Centre for High Energy Physics, Indian Institute of Science, for his far-flung wisdom, excellent vision and expertise, enthusiastic and prompt involvement, persistent academic encouragement as well as unprecedented mental support and guidance during the planning and development of this research work and making it a success. I also gratefully acknowledge his painstaking efforts in thoroughly going through and improving the manuscripts without which this work could never have been completed. 
\par I am highly obliged to Centre for High Energy Physics department and all the department members, including  chairman Prof. Justin David, faculties, students as well as \textcolor{blue}{Indian Institute of Science} and all institute members for providing all the facilities, help and gifting me a congenial, conducive environment for carrying out the research work uninterruptedly.
\par I am out of superlatives for my parents and my grandmother. I am endlessly obliged to them for their moral support, boundless love, encouragement and blessings and also for being there always beside me tolerating my fluctuating and unjustified cynical moods time and again all throughout this tenure of my doctoral studies. Special mention to my guardian Bibekananda Goswami who has been my go-to guide and teacher besides my family and giving me unconditional support in my times of desolation and loneliness.  
\par I wish to express my sincere appreciation and thanks to my dear friends Sourabh, Abirlal, Avinaba, Debottam, Samrat for their help and support in these times. My endless appreciation and heartfelt gratitude also to my phd batch mates and friends in the institute Rhitaja, Souvik, Samriddhi, Kartick, Pabitra, Prabhat, Adithi, Ahmadullah, Rishabh, Lokesh as well as 
my beloved juniors Budhaditya, Samudra, Sudeepan, Tanay, Shreya, Camellia, Arindam, Mainak. I also cannot but mention
my senior cum singing companions Pratik, Ratan, as well as Gobinda, Aranya, Parthiv, Alam and Dibyendu for their valuable advice and helping me glide through all thick and thin.
 I also would like to express my deep and sincere thanks to all other persons whose names do not appear here, for helping me either directly or indirectly in all even and odd times.
\par I am also thankful to the anonymous reviewers of my research publications. Their comments and suggestions were very helpful in shaping my research work and also motivated me towards more fundamental and deeper understanding. I also express my sincere acknowledgement to the anonymous referees of this thesis for their probing comments, insightful suggestions and corrections which have been very instrumental in improving the quality of this thesis. 

I also express my sincere gratitude to Prof. Frithjof Karsch, Prof. Christian Schmidt, Prof. Swagato Mukherjee and other members of the HotQCD collaboration for their inputs and valuable discussions and for the permission to use their data and use the GPU cluster at Bielefeld University, Bielefeld, Germany.

\par Finally, I am indebted and grateful to the Almighty for bestowing on me, the ability of perseverance in this strenuous yet enjoyable and fascinating endeavor.

\vspace{1.2cm}
\begin{flushright}
	\textbf{\thescholar}
\end{flushright}

	\cleardoublepage

    \phantomsection 
    \addcontentsline{toc}{chapter}{Publications based on this thesis}
    \vspace{-4.0\baselineskip}
\textbf{\begin{center}
		{\huge \textbf{Publications based on this thesis}}
\end{center}}

\vspace{1cm}
\begin{enumerate}
    \item $\textbf{S.~Mitra}$, P.~Hegde and C.~Schmidt , $\textbf{Phys. Rev. D 106, 034504}$, \\arXiv: $\boldsymbol{2205.08517}$ [$\textbf{hep-lat}$]

    \vspace{.5cm}
    
    \item $\textbf{S.~Mitra}$, P.~Hegde and C.~Schmidt , $\textbf{PoS LATTICE 2022 (2023) 153}$, \\
    arXiv:$\boldsymbol{2209.07241}$ [$\textbf{hep-lat}$]
    
    \vspace{.5cm}
    
    \item $\textbf{S.~Mitra}$ and P.~Hegde , arXiv:$\boldsymbol{2209.11937}$ [$\textbf{hep-lat}$]

    \vspace{.5cm}
    
    \item $\textbf{S.~Mitra}$ and P.~Hegde , arXiv:$\boldsymbol{2302.02360}$ [$\textbf{hep-lat}$]
\end{enumerate}

\vspace{1cm}

\textbf{
\begin{center}
{\huge \textbf{Publications not based on this thesis}}
\end{center}
}

\vspace{1cm}

\begin{enumerate}
  \item $\textbf{S.~Mitra}$ , arXiv:$\boldsymbol{2303.12063}$ [$\textbf{hep-lat}$]
\end{enumerate}
 
	\cleardoublepage
 
    \phantomsection 
    \addcontentsline{toc}{chapter}{Abstract}
	\vspace{-4.0\baselineskip}
\textbf{\begin{center}
		{\huge \textbf{Abstract}}
\end{center}}

\vspace{1cm}
 A comprehensive study of the QCD phase diagram is one of the challenging and open problems in high energy physics. Having significant astrophysical implications, this is also important in constructing the chronological evolution of the universe. With this aim, this thesis describes the behaviour of thermodynamic observables like pressure and number density with changing chemical potential $\mu$, through the method of an unbiased exponential resummation of lower order Taylor series of these observables at a finite $\mu$. We address the problem of biased estimates, which manifest uncontrollably in exponential resummation and which become severe in the domain of large values, higher orders of $\mu$ and also in observables which are higher $\mu$ derivatives of the thermodynamic potential. We show that our new formalism of unbiased exponential resummation can eliminate these biased estimates exactly upto a given order of $\mu$, and can capture important contributions of higher order Taylor series for all our working temperatures starting from hadronic phase to the plasma phase, including the crossover region. We also demonstrate that this new formalism is highly efficient in saving appreciable computational time and storage space for computations. 
	\clearpage

 \end{spacing}
	\tableofcontents
	\cleardoublepage
 
    \phantomsection 
	\listoffigures
	\cleardoublepage
    \phantomsection 
	\listoftables
	\cleardoublepage
    \phantomsection 
     \phantomsection 
 
	\pagenumbering{arabic}

 \begin{spacing}{1.3}
 \chapter{Motivation and overview of thesis}
\label{Chapter 1}

\graphicspath{{Figures/Chapter-1figs/PDF/}{Figures/Chapter-1figs/}}
\vspace{1cm}

The indomitable desire of knowing the unknown, the unflinching curiosity of exploring the unexplored remains an age-long passion of us. And it is this very inquisitiveness of humans along with their close association with Nature, that has led to the birth and advent of science. Although, we have come a long way alongside science, this quest of science and humans to comprehend life and nature in a more deeper way, on a more fundamental scale remains unfazed and unwavering. And it is this endless exploration towards the fundamental building blocks of Nature that promises to understand Nature better and improve quality of lives, in cohesion and harmony with Nature. The reductionist policy of science is key in that endeavor that leads our curiosity towards the most fundamental constituent elements and particles and encourages us to explore their properties and behavior.

To quell this curiosity about what is it that constitutes and makes everything up as we see them around us, we delve into this highly mysterious world of particle
physics. It is well known that fundamental particles exist in two species marked by their inherent intrinsic
spins.

\begin{itemize}
\item Fermions with half-integer spins~\cite{Pathria.2011,Reif.2010,Blundell.2010}. These are particles that usually interact with each other in a process of interaction. These satisfy Pauli exclusion principle, meaning that no two fermions with same spin can exist together in a single energy state. For example, electron ($e^{-}$), positron ($e^{+}$), proton ($p$), quarks ($q$) etc.
\item Bosons with integral spins~\cite{Pathria.2011,Reif.2010,Blundell.2010}. These are mediators of interactions in an interaction process. They are regarded as the "force carriers" travelling between the particles involved in the interaction. These are not exclusive, in fact they like to crowd a single energy state as much as possible, as opposed to fermions. For example, photon ($\gamma$), gluon ($g$), Higgs boson ($H$) etc.
\end{itemize}

\vspace{1cm}

 It is the interaction between these fundamental particles by virtue of their individual charges at
various energy scales which is responsible for all the myriad phenomena we see in our universe (except
gravity) ranging from cosmological galactic scales like within the stars to quantum microscopic scales within the nucleus of atoms.

Speaking about fundamental interactions, till date, all forms of interactions existent among particles in all scales, ranging from quantum to cosmological, manifest in four forms of fundamental interactions as shown~\cite{Griffiths.2008}:

\renewcommand{\arraystretch}{1.5}
\begin{table}[ht]
\centering 
\begin{tabular}{|P{2.9cm}|P{2.3cm}|P{1.5cm}|P{3cm}|P{3cm}|} 
\hline

Interaction & Relative magnitude & Charge & Particles & Mediators    \\ \hline
Strong & $1$ & Color & quark ($q$) & gluon ($g$)  \\ \hline
Electromagnetic & $10^{-3}$ & Electric & quark ($q$) and lepton ($l$) & photon ($\gamma$)  \\ \hline
Weak & $10^{-15}$ & Flavor & flavored $q$ and $l$ & $W^{+},W^{-},Z^{0}$  \\ \hline
Gravitational & $10^{-39}$ & Mass & massive $q$ and $l$ & graviton  \\ \hline
\end{tabular}

\vspace{6mm}

\caption{\hspace{1mm}All about the four fundamental interactions till date}
\label{table:four fundamental int} 
\end{table}
\renewcommand{\arraystretch}{1}

The above \autoref{table:four fundamental int} depicts the four fundamental interactions along with their relative magnitudes and also the particles as well as the mediators involved in the respective process of interactions. The usual naive notion in particle physics is that these particles experience the interactions or the forces via these mediators. Although this argument seems adequate and is easy to understand on preliminary grounds, a more fundamental reason of all these interactions is the respective charges of the particles. These particles by virtue of their charges, interact with each other and the corresponding mediators "mediate" the exchange process of these corresponding charges and manifest their effects in form of the corresponding interaction. For example, two particles with zero electric charge in any one of them cannot interact via electromagnetic forces. Similar reasoning holds true for other forms of interactions as well. 

In all these interaction processes, all the individual charges as enlisted in the above \autoref{table:four fundamental int} remain conserved apart from the $4$-momentum conservation (conservation of mass and energy). The Feynman diagram in particle physics is a very  useful pictorial representation, often used for illustrating these interactions and for exploring their underlying dynamics. As proposed by Feynman, it beautifully articulates these interactions

\begin{figure}[H]
     \centering
     \includegraphics[width=0.70\textwidth]{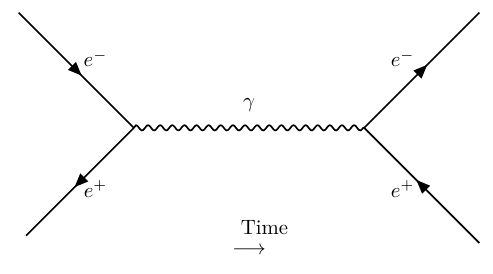}
     \caption{Feynman Diagram illustrating electromagnetic interaction between electron ($e$) and positron ($e^+$) via the exchange of the mediator photon ($\gamma$)}
     \label{fig:electromagnetic interaction}
 \end{figure}
 
\vspace{2cm}

\begin{figure}[H] 
\centering
\includegraphics[width = 0.30\textwidth]{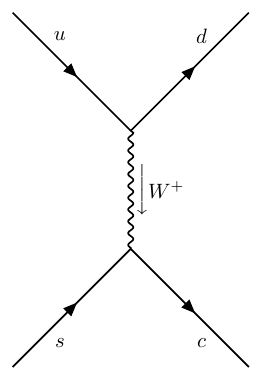} 
\hspace{.3cm} 
\includegraphics[width = 0.30\textwidth]{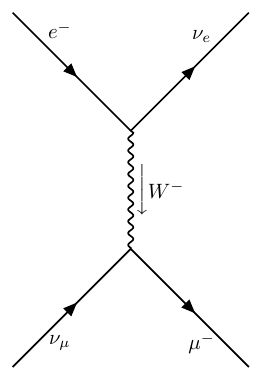} 
\hspace{.3cm} 
\includegraphics[width = 0.30\textwidth]{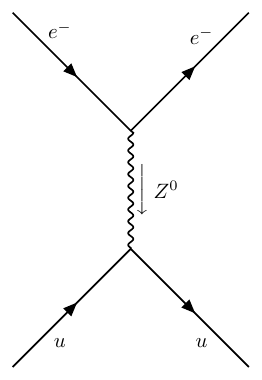} 
\caption{Feynman Diagrams showing weak interaction among quark-quark (left), lepton-lepton (centre) and quark-lepton (right) via $W^+$, $W^-$ and $Z^0$ bosons} 
\label{fig:weak interaction} 
\end{figure}

\begin{figure}[H]
     \centering
     \includegraphics[width=0.55\textwidth]{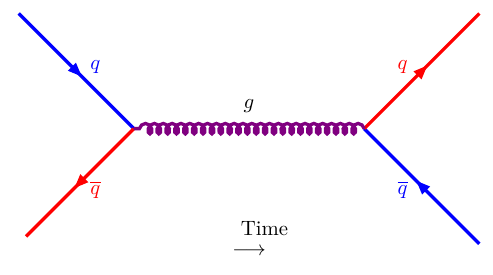}
     \caption{Feynman Diagram illustrating strong interaction among quark antiquark pair through exchange of color charges of the particles mediated by gluon}
     \label{fig:strong interaction}
 \end{figure}
 
\hspace{-5mm}and also vividly portrays the charge conservation of every interaction. The \autoref{fig:electromagnetic interaction} illustrates the Feynman diagram for the electromagnetic interaction and describes electron-positron scattering $e^- + e^+ \rightarrow e^- + e^+$, also called Bhabha scattering. \autoref{fig:weak interaction} depicts the weak interactions among the quarks and leptons of different flavors which are mediated by the massive $W^+$, $W^-$ and $Z^0$ bosons. Since, $W^+$ and $W^-$ possess an electric charge, the weak interaction mediated by them are often referred as charged weak interaction. Similarly, the interaction due to exchange of $Z^0$ boson are called neutral weak interactions. \autoref{fig:strong interaction} signifies the strong interaction mediated by the gluon. The quarks and antiquarks have their own color charges and these gluons bear the signature of these exchanged color charges, thereby mediating the strong interaction between a quark-antiquark ($q\overline{q}$) pair. However, there is still enough ambiguity regarding the correct way of including the gravitational interaction in the present framework of the Standard Model of particle physics for which it still remains beyond the scope of incorporation into the Standard Model. That is why, we have abstained from including a possible Feynman diagram for massive particles mediating via graviton, which itself remains fairly unexplored till date.

At this point, it is important to motivate the Standard Model of Particle physics. There are twelve fundamental particles comprising six quarks and six leptons. These six quarks are up $(u)$, down $(d)$, charm $(c)$, strange $(s)$, top $(t)$ and bottom $(b)$ quark. The six leptons comprise three particles and their corresponding neutrinos. These leptonic particles are electron $(e)$, muon $(\mu)$, tau $(\tau)$ and the corresponding neutrinos are electron neutrino $(\nu_e)$, muon neutrino $(\nu_{\mu})$ and tau neutrino $(\nu_{\tau})$. They are tabulated vertically in increasing order of the three generations namely first, second and third generations along with their corresponding anti-particles. In mediator sector, there are five mediators by considering the $W^+$ and $W^-$ bosons as a single boson (due to almost similar masses) and including the famous Higgs boson ($H$). Apart from these two massive mediator bosons, there is a massive $Z^0$ boson and the massless gluon $(g)$ and photon $(\gamma)$. Unfortunately till now, we do not know of any way to incorporate graviton ($G$) in the Standard Model. 

\begin{figure}
    \centering
    \includegraphics[width = 0.82\textwidth]{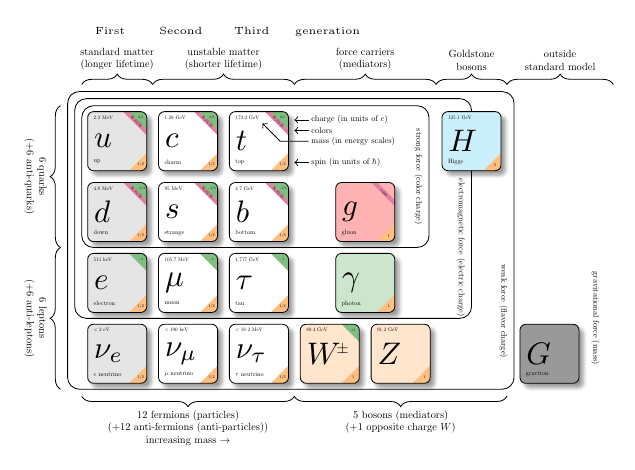}
    \caption{The Standard Model of Particle Physics }
    \label{fig:Standard Model of Particle Physics}
\end{figure}

The Standard Model is by far the most elaborate well-established theoretical framework of fundamental interactions existing till date. The underlying hierarchy and subsequent generation-wise classification of these particles are constructed in such a manner, so that particles get placed in higher generations in the order of increasing masses and decreasing lifetimes. This is because lifetime of a particle $t \sim 1/m$, where $m$ is the mass of the particle. The above \autoref{fig:Standard Model of Particle Physics} also very beautifully illustrates that the relative magnitude of the interactions (forces) decrease from strong force to gravitational, as one traverses from inner region towards the outer region of the Standard model. All the fermions have spin $\frac{1}{2}$ and the mediator bosons $(g, \gamma, W^{\pm}, Z^0)$ are vector bosons having spin $1$, except Higgs boson $(H)$ which, being a spin $0$ boson is a scalar boson. Graviton is postulated to be a spin-$2$ particle, which at present continues to remains outside the boundaries of Standard Model. All the lepton particles have electric charge $-e$ and all the lepton neutrinos have zero electric charge\footnote{Saying "chargeless" will be ambiguous, because for a clear understanding, one needs to specify which of the four charges of interactions have magnitude zero} irrespective of generations. Similarly in the quark family the up, charm and top quarks all bear the same electric charge $+\frac{2}{3} e$ whereas the down, strange and bottom quarks have electric charge $-\frac{1}{3} e$. While quarks participate in all the four fundamental interactions, leptons do not
 participate in strong interactions. While leptons are fundamental and can exist freely in nature, these quarks due to the confinement property by virtue of their color charges, cannot exist freely in nature and by virtue of strong interactions among the color charges possessed by them, they bind together to form hadrons. Strictly speaking, leptons like $\tau$ belonging to higher generations decay very quickly to lower generation leptons or bound state of quarks, obeying the energy momentum conservation principle and other laws of conservation for the respective interaction process. But this decay often happens via weak interaction, as opposed to free quarks which forms hadrons through strong interaction and which is almost $10^{15}$ times stronger than weak forces. Hence, one can safely claim that the chances of finding a free lepton in nature are fairly much higher than that of a free quark. Depending on their existence in nature therefore, particles are classified as leptons and hadrons. Hadrons are of two types as follows:
 
 \begin{itemize}
     \item Baryons : These are bound states comprising three quarks. For example, proton and neutron have quark contents $\boldsymbol{uud}$ and $\boldsymbol{udd}$ respectively. Proton is the lightest baryon with a mass $\sim 1$ GeV.
     \item Mesons : These are two quark bound states consisting of a quark-antiquark pair. Pion with a mass of about $140$ MeV is the lightest pion. There are three pions 
     \begin{enumerate}
     \item $\pi^+ \longrightarrow u\overline{d}$, 
     \item $\pi^- \longrightarrow d\,\overline{u}$ and 
     \item $\pi^0$ with a quantum superposition of $u\overline{u}$ and $d\overline{d}$.
     \end{enumerate}
 \end{itemize}

\renewcommand{\arraystretch}{2}
\begin{table}
\centering 
\begin{tabular}{|P{2.5cm}|P{3cm}|P{3cm}|}
\hline
Generation & Leptons & Quarks    \\ \hline 
First & $e, \nu_e$ & $u , d$  \\ \hline
Second & $\mu, \nu_{\mu}$ & $c , s$  \\ \hline
Third & $\tau, \nu_{\tau}$ & $t , b$  \\ \hline
\end{tabular}

\vspace{6mm}

\caption{Generation-wise arrangement of twelve fundamental particles}
\label{table:fundamental particles} 
\end{table}
\renewcommand{\arraystretch}{1}

Of many areas of study that are still in exploration in particle physics, one is that of the widely successful theory of sub-nuclear particles called Quantum Chromodynamics (QCD)~\cite{Gross:2022QCD}. It is a theory that helps us explain the strong force felt by quarks and gluons. Gluons are the force carriers bosons for the strong force. They interact among themselves and also with quarks. Gluons are of eight different types marked by their color combination and quarks come in three color charges. It is this color charge and the resulting dynamics from the exchange of these color charges that lends the name “chromodynamics” to the theory. This terminology became prevalent after drawing an analogy between the three primary colors and the three “colors” of quarks and it is entirely unrelated to the regular meaning of colors.

Despite the appreciable electromagnetic repulsive interactions among the protons, it is this interaction between the constituent quarks of proton and neutron that keep them strongly bound within the nucleus of an atom which constitutes the fundamental building block of matter and universe. Proton is the lightest baryon having quark content $(uud)$ with a mass of $938$ MeV. Neutron is slightly heavier than proton with a mass of about $940$ MeV and quark content $(udd)$. 

\begin{figure}[H]
     \centering
     \includegraphics[width=0.6\textwidth]{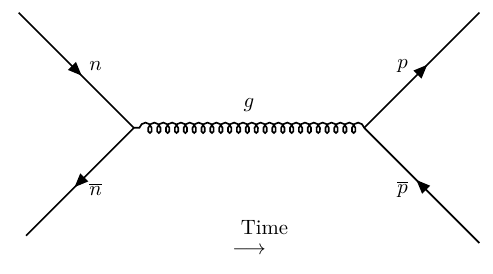}
     \caption{Strong interaction between neutron and proton via gluon}
     \label{fig:neutron proton}
 \end{figure}
 
The features of QCD are quite peculiar and two such peculiar properties of QCD are color confinement and asymptotic freedom. 

The color-charged particles although hypothesized quite
successfully, cannot be isolated and thus observed independently. They combine to form color-neutral
hadrons. This property that makes them exist in only color-neutral states is called color confinement~\cite{Wilson.1974,Chaichian:1999colorconfinement}.
When tried to pull apart from within the hadrons, the interaction between them increases with
increasing distance, which requires large energy. This excess energy ends up generating pairs of
quark-antiquark which then ends up as multiple color-neutral hadrons. On the other hand, at short
distances, they behave like free particles independent of interaction. This behavior of quarks to run
free of interactions at short distances is termed asymptotic freedom~\cite{Gross.2005,Wilczek.1993}. 

\begin{figure}
     \centering
     \includegraphics[width=0.85\textwidth]{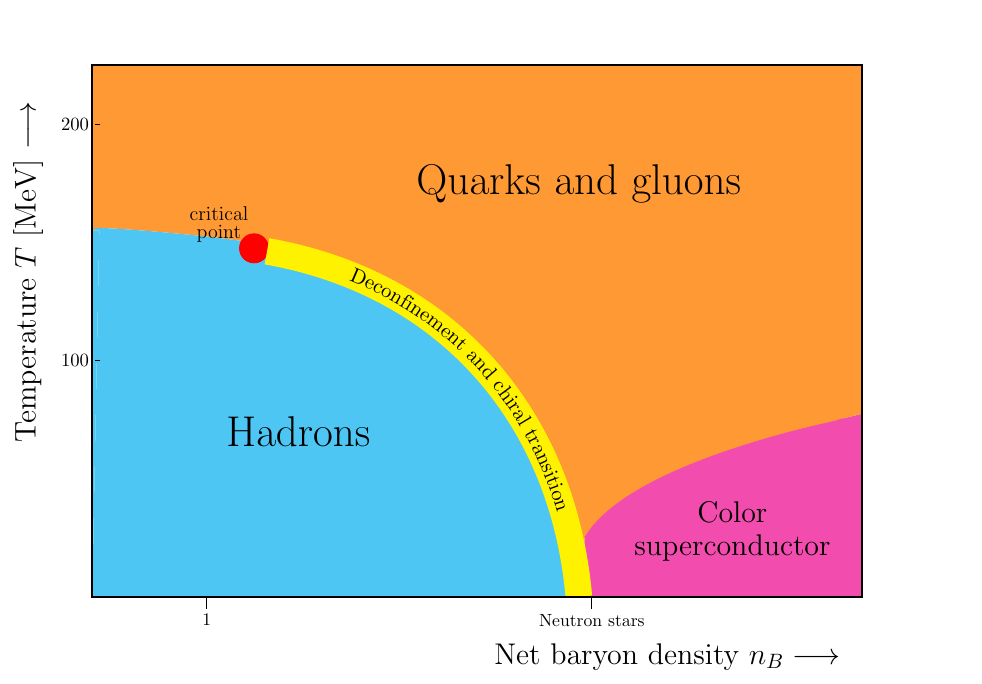}
     \caption{QCD Phase Diagram in the Temperature Baryon density plane}
     \label{fig:Phase diagram}
 \end{figure}

This small interaction at a short
distance allows us to do a perturbative calculation at this scale but fails at larger length scales. The
scale of interaction is given by $\Lambda_{QCD} \sim 260$ MeV~\cite{Irving.2006} which is where the coupling coefficient $\alpha_s \sim 1$. Hence, we have to remain at an energy scale greater that $\Lambda_{QCD}$ or conversely at a length scale smaller than that corresponding to $\Lambda_{QCD}$ to justify our perturbative calculations.
We need other methods to probe larger length scales and smaller energy scales, where the regime is strictly non-perturbative and usual perturbative techniques fail to explain the physics in this domain. And this is precisely the content of this thesis, where one will get an idea about the pertinent problems of analysis in this regime and the different probing approaches or techniques to somewhat prolong these problems and explain the physics of this important non-perturbative large length scale regions. In fact, this regime forms a significant part of the QCD phase diagram as shown in \autoref{fig:Phase diagram}.

One of the primary goals of QCD is to explain and map the different phases of the QCD phase diagram~\cite{Halasz:1998phasediagram, Rajagopal:1999phasediagram,Stephanov:2006phasediagram,Fukushima:2011phasediagram}. This is also immensely important for uncovering the physics and constructing the equation of state of early universe~\cite{McGuigan:2008earlyuniverse,Florkowski:2010earlyuniverse,Castorina:2015earlyuniverse} which is believed to exist after the occurrence of Big Bang for a time scale of the order of some microseconds. In this phase diagram, QCD predicts that with increasing thermal energy in the form of higher temperatures, there occurs a phase transition from hadronic phase to a Quark-gluon plasma phase. Such a change in
state was confirmed in RHIC (Relativistic Heavy Ion Collider)~\cite{Luo:2017RHIC} and LHC (Large Hadron Collider)~\cite{Beech:2010LHC} experiments where this new state of Quark-Gluon
Plasma (QGP) was observed~\cite{Gavai.2005,Mueller.1993,Braun.2007,Harris.1996}, only for a fraction of a second after the collisions. In these
experiments, heavy element atoms, like Gold atoms, are collided at extremely high energies thus
allowing us to observe the system at very small length scales. In the QGP phase~\cite{Bratkovskaya:2012QGP}, the hadronic states
dissolve into individual quarks and gluons making them the independent degrees of freedom. They
are free from interaction thereby removing the restriction of confinement.
Experimental verification of this state gives further credibility to the QCD theory. Despite all these developments, a lot remains to be explored and found out regarding the behavior of this phase transition, as portrayed in the QCD phase diagram.

Let us look at the phase diagram in temperature ($T$) - baryon chemical potential ($\mu_B$) plane to understand it better (\autoref{fig:Phase diagram}). The diagram depicts the scenario with quark masses set to their physical
values. There are three different phases, namely - QGP, hadronic gas~\cite{Hatsuda:1995hadron}, and the conjectured color
superconductor~\cite{Rajagopal:2001colorsuper}. We see that at finite non-zero chemical potential, there is a first-order phase transition strictly demarcating the hadronic and the QGP phases. This line culminates in at a critical point which has a
second-order phase transition at a specific point in the $T-\mu_B$ plane. This is because, chiral symmetry arguments (leading order tree-level chiral perturbation theory) including model calculations suggest that the phase transition between the hadronic and QGP phases is of $1^{st}$ order starting from some finite value of $\mu_B$. And a first order phase transition line can only end at a $2^{nd}$ order critical point, after which crossover~\cite{Steinbrecher:2018QCDcrossover,Borsanyi:2020QCDcrossover,Bazavov.2018,Li:2020QCDcrossover,Guenther:2021QCDcrossover} starts. This is what exactly happens if one proceeds along the phase transition demarcation towards the crossover near $\mu_B=0$. This phase transition is often called the chiral phase transition joining the crossover at a chiral critical point because, this transition is believed to demarcate hadronic and QGP phases, which respectively are believed to break and preserve chiral symmetry, by virtue of finite and almost vanishing hadron masses. 

\begin{figure}[ht]
     \centering
     \includegraphics[width=0.68\textwidth]{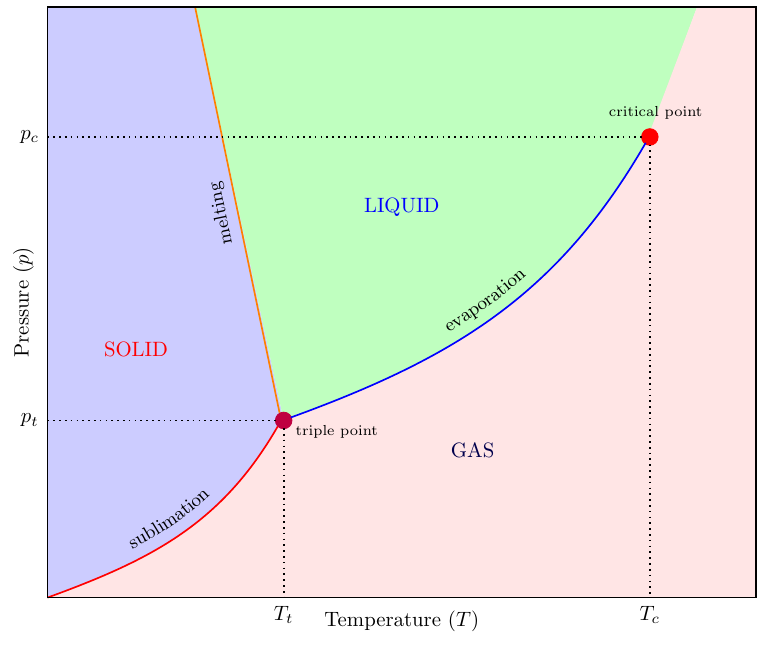}
     \caption{Solid liquid Phase diagram ending at a critical point}
     \label{fig:Solid liquid Phase diagram}
 \end{figure}

It is very much like the liquid-gas
phase transition where we have a first-order phase transition ending at a critical point beyond which
we can continuously change from one phase to another without having to go through a phase transition as shown in \autoref{fig:Solid liquid Phase diagram}. Upon further decreasing the chemical potential for the QGP diagram, even though
the change is continuous, the change in the properties is quite rapid marking a crossover transition. This rapidity is measured by the range of temperature, or the thermal width around the crossover temperature. More precisely, this is determined by the change of the observable value with respect to temperature, or mathematically, the slope calculated at the crossover temperature.

This rapid crossover at $\mu_B = 0$ is estimated around $T = 157$ MeV (about $10^{12}$ K), with an error of around $1.5$ MeV~\cite{Bazavov.2018,Tanmoy.2014}. There is another phase change
predicted at high chemical potential and low temperatures of the order of $0.1$ eV (about $10^3$ K), where we expect
to obtain a color superconductor phase~\cite{Alford.2001,Alford.2008} of strongly correlated quarks and existing in Cooper
pairs. 

Understanding the nature of this hadron QGP phase transition and the order of chiral phase
transition is a very important puzzle to enhance our understanding regarding strong interactions and its underlying dynamics. The conditions
at which QGP exists also mimic the conditions of the early Universe and thus understanding the
properties of QGP is imperative for understanding the state of the early Universe and thereby, comprehend its evolution into the present form in the long run. One of the key
questions still looming around is the location of the critical point on the diagram. As the crossover
region is near $157$ MeV, which is less than $\Lambda_{QCD}$ scale, analytically analyzing that region using
perturbation theory is not feasible. Hence, we need other non-perturbative methods to understand
the properties of that regime. 

Lattice QCD~\cite{Lombardo:2002latticeQCD, Allton:2005latticeQCD, Karsch:2009latticeQCD, Philipsen:2010latticeQCD, Karsch:2022latticeQCD} is one such method. It uses the computational method
that simulates the system using Monte Carlo techniques after the discretisation of the continuum spacetime.
One trouble upon discretising fermions on a lattice is the emergence of the doublers. The number
of simulated fermions goes as $2^{d+1}$, where $d+1$ is the dimension of the spacetime being discretised on lattice. One resorts to different discretisation schemes to get rid of these doublers~\cite{Gattringer.2009,Degrand.2006,Stacey.1982}. But for understanding the
chiral symmetry restoration, our lattice action should necessarily possess that symmetry. However,
on the lattice, one cannot have a system free of doublers and with chiral symmetry simultaneously~\cite{Nielsen.1981}.
Thus, we need a discretisation scheme where the chiral symmetry is obeyed as well as where the
doublers get removed in the chiral limit.

There is another caveat involved with employing lattice QCD. The current techniques available limit
the simulation to only zero quark chemical potential and enable reliable extrapolations only up to very small values of finite chemical potential. This is because, while trying to add finite chemical potential in the QCD Lagrangian and investigate finite density QCD regime which forms the crux of the phase diagram, it
runs into a problem called the sign problem making the prediction in this region computationally difficult and extremely unreliable with current
techniques. There have been attempts to formulate different approaches by which one can bypass this difficulty and it is a topic of current research. One of them is exponential resummation approach which forms the central part of discussion in this thesis. Thus, if we are to explore the phase diagram we need to carefully tread around the above
two challenges with lattice QCD.

 In this thesis, we have dealt more with the second caveat and have thoroughly investigated the behaviour of important thermodynamic observables like excess pressure and number density as a function of chemical potential. In the lines of the QCD phase diagram, the detailed exploration is carried out in this thesis in temperatures which are almost equally below and above the chiral crossover temperature. This is roughly about $157$ MeV, duly set by the quark masses, couplings and other necessary parameters of the lattice considered in this thesis. The study includes the crossover region too between the hadronic and the QGP phases. As mentioned in the abstract before, the new novel approach of unbiased exponential resummation is central to this study and more features will hopefully unfold gradually to the readers in the subsequent chapters of this thesis.

	\cleardoublepage
\chapter{Introduction to QCD and the structure of thesis}
\label{Chapter 2}

\graphicspath{{Figures/Chapter-2figs/PDF/}{Figures/Chapter-2figs/}}

While addressing anything, be it a problem or a topic, it is imperative to mention the following three things beforehand. This is to ensure that the idea of the working domain remains vivid and unambiguously clear to all, thereby leading to smooth motivated working towards the problem. 

\begin{itemize}
    \item The relevant scale of study, like length scale, time scale, scales of momentum and energy etc.
    \item The relevant observer, or the frame of reference
    \item The relevant dimensionality
\end{itemize}

\section{Domain of QCD}

What we are going to do is to introduce QCD briefly and try specifying its domain in the spectrum of physics in this section. Keeping these three above mentioned points in mind, we can say that regarding the discussion of QCD, we will restrict ourselves in a domain characterised by 

\begin{itemize}
\item Quantum microscopic length scales, relativistic energy and momentum scales described by the usual Einstein dispersion relation $p^2 = m^2c^2$, where $p^2 = p^{\mu}p_{\mu}$. $p_{\mu}$ and $m$ are the usual covariant four momentum and mass of the particle and $c$ is the speed of light, as we know from our knowledge of relativity. As usual, $p_{\mu} = \left(p_0,p_1,p_2,p_3\right)$ defined in a Minkowski spacetime which in the realm of special relativity, is characterised by a mostly negative or a mostly positive metric with elements $g_{\mu\nu}$ where $\mu,\nu=0,1,2,3$ and $g_{\mu\nu}=0$ for $\mu \neq \nu$. The former corresponds to $diag\left(g\right) = \left(+1,-1,-1,-1\right)$ and the latter signifies $diag\left(g\right) = \left(-1,+1,+1,+1\right)$, with $diag\left(g\right)$ depicting the diagonal elements of the metric $g$. 

\item We ourselves are the frame of reference here, because, it is us who are observing and enumerating results at the end of the day.

\item Since, the energy scales are relativistic and length scales are microscopic and well within quantum regime, it is evident that QCD is a relativistic theory and its dimensionality must therefore be a $\left(d+1\right)$ spacetime, instead of just $d$ space and time separately.
\end{itemize}

In relativistic regime, we express temperature not in the usual units of Kelvin but in units of electron-volt (eV). This conversion is trivial with Boltzmann constant $k_B \sim 10^{-4} eV/K$. So, the normal room temperature of $300$K corresponds to $\sim 10^{-8}$ MeV, which can easily offer the idea that where we are positioned at present along the vertical temperature axis of the QCD phase diagram in \autoref{fig:Phase diagram}. In fact, we not only remain within the deep hadronic regime, we reside very close to the origin ($T=0,\,\muB=0$ in MeV scales) of the phase diagram. The crossover temperature $T=157$ MeV corresponds to $\sim 10^{12}$ K, which is unimaginably hot with respect to our present day-to-day temperature. 

With this idea about the relativistic temperature scales and the aforementioned domain, let us proceed introducing QCD step by step.

\section{Quantum field theories : A naive overview}

As we found that QCD is a relativistic theory, hence, it must provide scope for particle creation and annihilation processes, which is implied by Relativistic quantum mechanics. Hence, unlike the usual non-relativistic quantum mechanics, the probability of finding a particle in space is not conserved. In fact, this probability now becomes a function of time. This suggests that particles are no more fundamental entities, since it is possible to find a state of vacuum with no particles. It is from this, emerged the concept of fields, which became the fundamental entities in place of particles. 

Particles are replaced by the term "particle fields", like electron fields, proton fields, because particles are considered to be the excited form of fields. These fields are mathematical functions of spacetime in relativistic realm and so are the momentum fields. Then arose the canonical quantisation relations between the fields and their momentum counterparts and thus, the concept of quantum field theories (QFT) arose. These theories describe interaction among the relevant fields and hence, do describe many particle systems~\citep{Das.2008,Peskin.2018,Schwartz.2014,Weinberg.1995,Ryder.1996}. As a result, the action and the Lagrangian forms the core starting point of these theories and the corresponding Euler-Lagrange equations provide the equation(s) of motion of these fields encoding their inherent dynamics as described the corresponding QFT. Both the action and Lagrangian density are Lorentz invariant, that is, they remain invariant under Lorentz transformation of spacetime. Similar arguments hold true for QCD also, which is a valid legitimate QFT. The usual relations follow : 

\begin{equation}
    S = \int L \hspace{1mm} dt = \int \mathcal{L} \hspace{1mm}d^4x \hspace{3mm},\hspace{3mm} L = \int \mathcal{L} \hspace{1mm} d^3x
    \label{eq:action}
\end{equation}

where $S$ is the action, $L$ is the Lagrangian, $\mathcal{L}$ is the Lagrangian density and $d^4x=d^3x\hspace{1mm}dt$ is the differential four volume. It must be noted that Lagrangian density $\mathcal{L}$ is Lorentz invariant, whereas the true Lagrangian $L$ is not, since the 3 volume is not Lorentz invariant, as clear from the above \autoref{eq:action}. In the following section of this chapter, we refer Lorentz invariant Lagrangian density $\La$ as the Lagrangian.

\section{QCD from QED : Towards the gauge theory}
The strong interaction is a short range interaction apart from the weak interaction and remains existent only within the nuclear length scales $\sim$ of the order of femtometer $\left(1\,fm = 10^{-15} \,m\right)$. Unlike gravitational and electromagnetic forces which are long range forces, having an $1/r$ dependence (Newton's law of gravitation and Coulomb's law of electromagnetism) in classical length scales, the strong and weak interactions are described by pure quantum theories. The physics of strong interactions is described by the theory of Quantum Chromodynamics (QCD)
to capture the interaction between quarks and gluons, where the former are fermions with spin $1/2$ and the latter are spinless bosons as mentioned in \autoref{table:fundamental particles}.

\subsection{QED}

The theory of QCD differs from Quantum Electrodynamics (QED) in terms of the gauge symmetries followed by the Lagrangian of the theory. These gauge symmetries are a common and typical feature of gauge theories, in which the Lagrangian remains invariant under Lie groups of continuous internal transformations called gauge transformations. 
\begin{itemize}
    \item Internal, because they are related only to field transformations. More mathematically, they transform the functional forms of spacetime, not the spacetime like the spacetime transformation.
    \item Continuous, because, these transformations are characterised by a transformation parameter, which is free to assume continuous values in its range.
\end{itemize}

The chronology of steps for formulating QED and establishing it as an interacting $U\left(1\right)$ gauge theory are as follows:

\begin{itemize}
    \item The fermion Dirac Lagrangian is given by 
    
    \begin{equation*}
        \La_{Dirac} = \sum_f \overline{\Psi}_f\left(x\right)\left[i\gamma^{\mu}\partial_{\mu} - m_f\right]\Psi_f\left(x\right),
    \end{equation*} 
    
    where $\Psi_f\left(x\right)$, $\overline{\Psi}_f\left(x\right)$ are the fermion and adjoint fermion fields of flavor $f$ as a function of four spacetime point $x$. The Lagrangian is obtained as the sum over all the available flavors of the fermion fields in the theory. $m_f$ is the mass of the fermion field.
    \item The above Lagrangian had a global $U\left(1\right)$ symmetry, characterised by 
    \begin{equation*}
     \Psi_f\left(x\right) \to e^{-i\theta} \hs \Psi_f\left(x\right)   , \quad \overline{\Psi}_f\left(x\right) \to    \overline{\Psi}_f\left(x\right) \hs e^{i\theta}
    \end{equation*}

    \item We demand that the interacting Lagrangian must have a local $U\left(1\right)$ symmetry which is effectively global $U\left(1\right)$ symmetry with transformation parameter $\theta\left(x\right)$ now a function of every spacetime point. 
    
    \item Under this transformation, we found that although the mass term for every flavor remains invariant, the kinetic term for every flavor $f$ i.e. $\overline{\Psi}_f\left(x\right)\left(i\gamma^{\mu}\partial_{\mu}\right)\Psi_f\left(x\right)$ does not remain invariant as there is a symmetry breaking term $\sim \gamma^{\mu}\partial_{\mu} \theta\left(x\right)$.
    
    \item To cancel this gauge symmetry breaking terms, we introduce new local fields $A_{\mu}\left(x\right)$ having dependence on spacetime points $x$, which must transform under $U\left(1\right)$ local symmetry as 
    \begin{equation*}
     A_{\mu} \to A_{\mu} + \frac{1}{g} \hs \partial_{\mu} \hs \theta\left(x\right)   
    \end{equation*}
    where $g$ is a dimensionless parameter in $(3+1)$ dimensional spacetime, identified with particle gauge coupling.
    \item The usual derivative $\partial_{\mu}$ now transforms to a covariant derivative $D_{\mu}$ defined as 
    \begin{equation*}
     \partial_{\mu} \to  D_{\mu} = \partial_{\mu} + i g A_{\mu}   
    \end{equation*}
    
    \item These covariant derivatives transform under $U\left(1\right)$ local symmetry as 
    \begin{equation*}
    D_{\mu}\Psi\left(x\right) = e^{-i\theta} D_{\mu} \Psi\left(x\right), \overline{\Psi}\left(x\right) D_{\mu} = \overline{\Psi}\left(x\right) D_{\mu} \hs e^{i\theta}  
    \end{equation*}
     
    \item The commutation relation among these covariant derivatives give the gauge field anti-symmetric tensor given by
    \begin{equation*}
    F_{\mu\nu} = -\frac{i}{g} \hs \left[D_{\mu},D_{\nu}\right]    
    \end{equation*}
    \item Since, $U\left(1\right)$ is an Abelian group, hence, $[A_{\mu},A_{\nu}]=0$. Hence, $F_{\mu\nu} = \partial_{\mu}A_{\nu} - \partial_{\nu}A_{\mu}$
    \item We find $F_{\mu\nu}$ as gauge invariant. $F^2$ is therefore a gauge invariant Lorentz scalar. Hence, $F^2$ is a legitimate term for the Lagrangian.
    \item Hence, the final QED Lagrangian is given by 
    \begin{equation}
    \La_{QED} = \sum_f \overline{\Psi}^{\left(f\right)}\left(i\gamma^{\mu}D_{\mu} - m^{\left(f\right)}\right)\Psi^{\left(f\right)} - \frac{1}{4}F^2    
    \label{eq:QED Lagrangian}
    \end{equation}
    \item The three dynamical equations of motion for QED are therefore as follows:
    \begin{align}
     (i\gamma^{\mu}D_{\mu} - m) \q &= 0 \notag \\    
     \adq(i\gamma^{\mu}D_{\mu} + m) &= 0  \notag \\
     \partial_{\mu}F^{\mu\nu} = J^{\nu} &= g \adq \gamma^{\nu} \q 
    \end{align}
     where $J^{\mu} = (J^0, \Vec{\boldsymbol{J}})$ is the $4$-current density.
    \item Since, we know total charge $Q = \int J^0 \hs d^3x$, hence here 
    \begin{equation}
     Q = \int g \hs \adq \hs \gamma^{0} \hs \q \hs d^3x  = g \int |\q|^2 \hs d^3x   
    \end{equation}
    
     This identifies coupling parameter $g$ to some factor of charge in QED.
    
\end{itemize}

\subsection{QCD : Comparison with QED}

\begin{figure}
    \centering
    \includegraphics[width=0.47\textwidth]{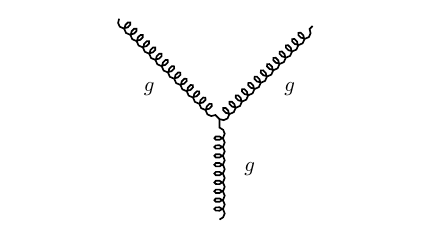}
    \includegraphics[width=0.47\textwidth]{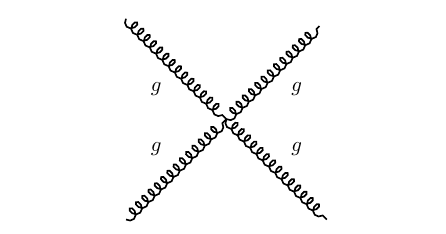}
    \caption{Cubic (left) and quartic (right) interaction among gluons}
    \label{fig:Gluons cubic and quartic}
\end{figure}

  The exact procedure is followed in case of QCD also, in order to establish it as a gauge theory describing interactions. The difference with QED, is that unlike QED, the fermion fields here possess three color degrees of freedom in addition to usual Dirac indices. Hence, for a gauge theory, the Lagrangian of QCD must invariant under the rotation among $3$ color degrees of freedom and in this case, the gauge group is $SU\left(3\right)$. As opposed to QED Lagrangian mentioned in \autoref{eq:QED Lagrangian}, the QCD Lagrangian is given by

    \begin{equation}
    \La_{QCD} = \sum_f \overline{\Psi}_a^{\left(f\right)}\left(i\gamma^{\mu}D_{\mu}^{ab} - m^{\left(f\right)}\right)\Psi_b^{\left(f\right)} - \frac{1}{4} \Tr[F^2]    
    \label{eq:QCD Lagrangian}
    \end{equation}
    where $a,b=1,2,3$ are the three color indices identified with the three color charges $\left(\textcolor{red}{\textbf{red}}, \textcolor{green}{\textbf{green}}, \textcolor{blue}{\textbf{blue}}\right)$ of quarks. This imply that the particle fields in QCD have four Dirac indices and three color indices and the $4$ dimensional Dirac space and the $3$ dimensional color space are decoupled from one another. A $12 \times 12$ matrix can be looked upon as a $4 \times 4$ matrix in Dirac space with each component being a $3\times 3$ matrix in color space or in the other way, a $3 \times 3$ matrix in color space with each matrix element effectively being a $4 \times 4$ matrix in Dirac space.

Here, in \autoref{eq:QCD Lagrangian}, the sum is over all the flavours of quarks. Unlike in QED where we do not see self-interaction
of photons, we have terms cubic and quartic in the gauge fields in QCD which gives rise to self-interaction of gluons along with their interaction with the quarks. On a mathematical note, the non-Abelian nature of $SU(3)$ group is responsible for this self-interaction among the mediators and one therefore can intuitively understand that this behaviour will be preserved for any $SU(N)$ Yang-Mills theory. And it is this self-interaction among gluons (mediators), apart from the usual quark (particle) gluon interaction, that makes QCD very unique quantum field theory (QFT) with peculiar properties like confinement and asymptotic freedom. In QCD, the quarks form the fundamental representation, whereas the gluons form the adjoint representation of $SU(3)$. Similarly, for a general $SU(N)$ Yang Mills theory, the particle fields form the fundamental representation, whereas the $N^2-1$ mediator fields as the group generators form the adjoint representation of the gauge theory. A naive comparative discussion between QED and QCD are tabulated as follows:

  \renewcommand{\arraystretch}{1.5}
\begin{table}[H]
\centering 
\begin{tabular}{|P{3cm}|P{4.9cm}|P{4.9cm}|} 
\hline
QFT & QED & QCD \\ \hline
Theory & Abelian gauge theory & Non-Abelian gauge theory \\ \hline 
Properties & No such properties & Color Confinement and Asymptotic freedom \\ \hline
Gauge group & Abelian $U\left(1\right)$ & Non-abelian $SU\left(3\right)$ \\ \hline
Group generators & Identity & $\neq I_{3}$ \\ \hline
Structure constants & $f_{abc}=0$ & $f_{abc} \neq 0$ \\ \hline
Mediators & No photon-photon coupling & Cubic and quartic coupling among gluons \\ \hline 
Gauge field transformations & $A_{\mu} \to A_{\mu} + \frac{1}{g} \partial_{\mu} \theta$ & $A_{\mu}^a \to A_{\mu}^a + \frac{1}{g} \partial_{\mu} \theta^a + f_{abc}A_{\mu}^b \theta^c$ \\ \hline
Gauge field tensor & $F_{\mu\nu} = \partial_{\mu}A_{\nu} - \partial_{\nu} A_{\mu}$ & $F_{\mu\nu}^a = \partial_{\mu}A_{\nu}^a - \partial_{\nu} A_{\mu}^a + gf_{abc} A_{\mu}^b A_{\nu}^c$ \\ \hline
Gauge field term in $\mathcal{L}$ & $\frac{1}{4} F_{\mu\nu}F^{\mu\nu}$ & $\frac{1}{4} \Tr [F_{\mu\nu}F^{\mu\nu}]$ \\ \hline 

\end{tabular}

\vspace{6mm}

\caption{\hspace{1mm} A naive overview of the differences among QED and QCD}
\label{table:QED and QCD} 
\end{table}
\renewcommand{\arraystretch}{1}

\section{SU(3) group : A quick overview}

The Lie Algebra for a group with generators $T$ satisfies

\begin{equation}
    \left[T^a,T^b\right] = if^{abc}\hspace{1mm}T_c
    \label{eq:Lie}
\end{equation}
where $f_{abc}$ are the structure constants of the group and bear the signature of the group and its generators. For an $SU(N)$ group, the number of generators is given by $N^2-1$, due to the constraint of unitarity and unit determinant. It can be easily shown that these generators are Hermitian and traceless matrices of order $N \times N$.

 Being a continuous group, all the elements or matrices in $SU(N)$ must be continually connected to identity. With this property along with the usual properties of a group, we find that an element of $SU(N)$ is given by $M(\theta) = \exp{\left(i \hspace{1mm} \theta \hspace{1mm} T_a\right)}$, with generators $T_a$ and transformation parameter $\theta \in \mathbb{R}$. We find with $\theta=0$, $M(\theta)= I_n$. 
 \begin{itemize}
\item Now, being unitary, $\left[M(\theta)\right]^{\dagger} M(\theta) = I_n$ implying that $\left(T_a\right)^{\dagger} = T_a$ , suggesting $T_a$ is Hermitian.
\item With determinant $1$ and using $\det \left(e^P\right) = e^{\Tr \hs \left(P\right)}$, for any complex non-singular matrix $P$, we find $\Tr \left(T_a\right) = 0$, implying tracelessness of generators $T_a$.
\end{itemize}

The above arguments even follow for $N=3$. For $SU(3)$, the $8$ generators are $T_a = \lambda_a/2$, where the $\lambda$'s in standard Gell-Mann basis is given by~\cite{Costa.2012} 
\vspace{4mm}

$\lambda_1 = \begin{pmatrix}
 0 & 1 & 0\\
 1 & 0 & 0\\
0 & 0 & 0
\end{pmatrix} \hspace{26mm} \lambda_2 = \begin{pmatrix}
 0 & -i & 0\\
 i & 0 & 0\\
0 & 0 & 0
\end{pmatrix} \hspace{26mm} \lambda_3 = \begin{pmatrix}
 1 & 0 & 0\\
 0 & -1 & 0\\
0 & 0 & 0
\end{pmatrix}$ 

 \vspace{4mm}

$\lambda_4 = \begin{pmatrix}
 0 & 0 & 1\\
 0 & 0 & 0\\
1 & 0 & 0
\end{pmatrix} \hspace{26mm} \lambda_5 = \begin{pmatrix}
 0 & 0 & -i\\
 0 & 0 & 0\\
i & 0 & 0
\end{pmatrix} \hspace{26mm} \lambda_6 = \begin{pmatrix}
 0 & 0 & 0\\
 0 & 0 & 1\\
0 & 1 & 0
\end{pmatrix}$ 

\vspace{4mm}

$\lambda_7 = \begin{pmatrix}
 0 & 1 & 0\\
 1 & 0 & 0\\
0 & 0 & 0
\end{pmatrix}  \hspace{26mm} \lambda_8 = \dfrac{1}{\sqrt{3}}\begin{pmatrix}
 1 & 0 & 0\\
 0 & 1 & 0\\
0 & 0 & -2
\end{pmatrix}$

\vspace{7mm}

These set of $\lambda$ satisfy $\left[\lambda_a, \lambda_b\right] = 2i f_{abc} \lambda_c$. In case of $SU(2)$ where the generators are the usual Pauli matrices $\Vec{\boldsymbol{\sigma}} = \{ \sigma^a, \sigma^b, \sigma^c\}$, the structure constants $f^{abc}$ are the usual Levi-Civita symbols $\epsilon^{abc}$.
These $\lambda$ also form a complete set of Hermitian matrices.

\section{Structure of thesis}

 In this thesis, we have dealt more with the second caveat as mentioned above and have thoroughly investigated the behaviour of important thermodynamic observables like excess pressure and number density as a function of chemical potential. In the lines of the QCD phase diagram, this exploration is conducted in this thesis in temperatures which are equally below and above the chiral crossover temperature $\sim 157$ MeV, including the crossover region itself.

     In Chapter $3$, we introduce the subject of lattice QCD and comprehensively argue for its essence in the study of QCD phase diagram. We also discuss the different important aspects of lattice QCD, besides outlining the notorious sign problem, which obscures numerical analysis and computations in the finite density or equivalently finite chemical potential regime of the QCD phase diagram. 
    
     Chapter $4$ discusses the different methodologies adopted to circumvent this sign problem. We present reweighting of complex measure to real at finite baryon chemical potential $\mu_B$ or other similar class of chemical potentials, where the analysis does suffer from a sign problem. We also highlight the method of analytic continuation to real from imaginary chemical potentials, where there is no sign problem to curtail calculations. We briefly mention some newer methods like contour deformation, Lifschitz thimbles and complex Langevin method, all of which were developed to avoid the sign problem. We enlighten Taylor expansion of observables in terms of $\mu$ in a bit more detail, where we stress upon the slow convergence and non-monotonic behaviour of Taylor series, requiring Taylor calculations to sufficiently high order in $\mu_B$.
    
    In Chapter $5$, we motivate resummation approaches as solution to the setbacks of Taylor series. We briefly touch upon Pad{\'e} resummation, before moving onto the discussion of exponential resummation at finite baryon chemical potential $\mu_B$ in detail, narrating about its benefits over the usual Taylor series in the form of capturing contributions to all orders in $\mu_B$, as well as its drawbacks in form of uncontrollably emerging biased estimates of $D_n^B$, which are $n$ point correlation functions for baryon chemical potential $\muB$.

    In Chapter $6$, we focus on the origin of these biased estimates in exponential resummation and enlighten the schematic structure of random volume sources nested inside every gauge field configuration constituting the gauge ensemble. While highlighting the essence of estimating $n$-point baryon correlation functions $D_n^B$ and using random volume sources for this purpose, we also discuss the two kinds of stochastic bias which we have encountered in the process, in great detail.
    
    In Chapter $7$, we present the method of cumulant expansion, which allow us to replace biased with unbiased estimates by truncating the resummed series order-by-order in $\mu$. We perform this for isospin chemical potential $\muI$, where there is no sign problem and so that the method of calculation becomes clear and less tedious. We also validate and substantiate our arguments with necessary figures. But at the end, we do gain all these at the expense of the valuable reweighting factor and phasefactor, which constitute the partition function.
    
    Plugging in the loophole and taking a leaf out of the cumulant expansion approach, our new work on unbiased exponential resummation is vividly portrayed in Chapter $8$, discussing all the important aspects and features. We establish that this new formalism can exactly reproduce Taylor series upto a finite order in $\mu$, irrespective of whether computations at finite $\mu$ suffer from sign problem or not. We demonstrate this for all the working temperatures which lie in both the hadronic and QGP phases, equidistant from the crossover temperature. We also study the crossover region and validate the theoretical aspect of our new formalism. We also present the significant computational benefits of this formalism, and compare the phasefactor results apart from plotting the roots of partition function and comparing between them obtained in a biased and unbiased manner respectively.

    We finally give a brief summary of the entire thesis and conclude along with describing the future scope and outlook in the final Chapter $9$.

	\cleardoublepage
 	
\chapter{Lattice QCD and sign problem}
\label{Chapter 3}

\graphicspath{{Figures/Chapter-3figs/PDF/}{Figures/Chapter-3figs/}}

The energy scale for QCD is set by the $\Lambda_{QCD} \sim 260$ MeV. The coupling constant $\alpha_{s}$ which is
a dimensionless constant, upon quantization, is a function of the energy scale of the system. For
energies greater than $\Lambda_{QCD}$, the coupling constant is $\alpha_{s} < 1$, thus allowing a perturbative calculation and QCD exhibits property of asymptotic freedom.
However, for a system with lesser energy than $\Lambda_{QCD}$, the coupling constant is $\alpha_{s} > 1$. It is in this
region that the perturbation theory breaks down, the interactions start becoming stronger and the theory of QCD starts showing signs of confinement.
It is this non-perturbative regime that lattice QCD is very instrumental as traditional methods are
not effective.
Next, we introduce this statistical system on a lattice. The way that works is by converting the
continuous system of fields to the fields residing on a lattice. For this, we first quantize the fields
using Euclidean path integrals and then discretizes the spacetime converting it into a lattice. Let us
pursue it briefly.

\section{Path integral on lattice}
Given the classical QCD Lagrangian in \autoref{eq:QCD Lagrangian} in Minkowski spacetime, the key to quantizing fields on the lattice is by using the Euclidean path integral. We convert the Minkowski spacetime of our field theory to the Euclidean
spacetime using the Wick rotation~\cite{Das.2008,Peskin.2018,Schwartz.2014,Weinberg.1995,Ryder.1996}, which is, by analytically continuing time to an imaginary value using
the transformation $t \to -i\tau$, where $t$ is the Minkowski time and $\tau$ is the Euclidean time. Upon doing this, we obtain the Euclidean Lagrangian as follows 

\begin{equation}
    \La_{E} = \frac{1}{4} \Tr \left[F^2\right] + \sum_f \overline{\Psi}_f\left(x\right)\left[\gamma^{\mu}\left(\partial_{\mu} + igT_aA_{\mu}^a\right) + m_f\right]\Psi_f\left(x\right)
    \label{eq:Euclidean Lagra}
\end{equation}

where $\La_E$ is the QCD Lagrangian defined in Euclidean spacetime. $A_{\mu}^a$ are the gluon fields with Dirac index $\mu$ and color index $a$. This gives us the action as follows 

\begin{equation}
S_E\left[\overline{\Psi}, \Psi, A\right] = \int d^4 x_E \hspace{1mm} \bigg[ \frac{1}{4} \Tr \left[F^2\right] + \sum_f \overline{\Psi}_f\left(x\right)\left[\gamma^{\mu}\left(\partial_{\mu} + igT_aA_{\mu}^a\right) + m_f\right]\Psi_f\left(x\right)\bigg]
\label{eq:Euclidean action}
\end{equation}

where $S_E\left[\overline{\Psi}, \Psi, A\right]$ is the Euclidean action and $d^4 x_E = d \tau \hspace{1mm} d^3 \Vec{\boldsymbol{x}}$ is the Euclidean version of the four volume differential. $T_a$ are the $SU\left(3\right)$ generators of QCD. 

Now using the Feynman path integral, $\left(D+1\right)$ dimensional Euclidean quantum field theory system
can be converted into to a $\left(D\right)$ dimensional quantum statistical system. We use this similarity to
formulate our system of quantum fields in QCD to that of a statistical problem. This can be seen
using the transition amplitude. A transition amplitude sums the probability of all the possible paths
taken by the system from initial state to final state with the probability for the path given by an exponentially decaying Boltzmann
weight factor $e^{-S_E}$. Now to relate it with a statistical system, we trace over the states of transition
amplitude. This gives us a quantity

\begin{equation}
\Z = \int \mathcal{D} A \hspace{1mm} \mathcal{D}\overline{\Psi} \hspace{1mm} \mathcal{D} \Psi \hspace{1mm} e^{-S_E\left[\overline{\Psi}, \Psi, A\right]} 
\label{eq:path partition}
\end{equation}

Evaluation of trace over fields require that we have a periodic boundary
conditions for bosonic fields and anti-periodic boundary conditions for the fermionic field. This anti-periodic boundary condition is a result of their anti-commuting behaviour [36, 37]. Now in usual statistical mechanics, partition function is like 

\begin{equation}
    \Z \sim \exp{\left(-\beta E\right)}
    \label{eq:stat part func}
\end{equation}

where $\beta = 1/T$ is the inverse temperature of the system. Hence, $\Z$ in \autoref{eq:path partition} resembles the partition function of \autoref{eq:stat part func} when we identify
\begin{equation}
    \beta = \frac{1}{T} = \tau = it
    \label{eq:QCD to stat}
\end{equation}

with the symbols having conventional meanings. Thus, with the above procedure like in \autoref{eq:QCD to stat}, we can transform our quantum system into a statistical one at
a finite temperature and obtain the partition function from the transition amplitude~\cite{Feynman.2010}. With the knowledge of the partition function at our disposal, we can use usual statistical techniques to obtain expectation values of any observable, like as follows

\begin{equation}
    \LA \Ob \RA = \frac{1}{\Z} \int \mathcal{D} A \hspace{1mm} \mathcal{D}\overline{\Psi} \hspace{1mm} \mathcal{D} \Psi \hspace{1mm} e^{-S_E\left[\overline{\Psi}, \Psi, A\right]} \hs \hs \Ob
    \label{eq:observable}
\end{equation}

\section{Action discretisation and gauge invariance}

We introduce an $N_{\sigma}^3 \times N_{\tau}$ lattice $\textbf{L}$ in Euclidean spacetime , which is a set of Euclidean spacetime points as follows : 
\begin{equation}
\textbf{L} = \{\left(n_1,n_2,n_3,n_4\right) \hspace{1mm}|\hspace{1mm} n_i \in \mathbb{N} \cup \{0\},  \hspace{1mm} 0 \leq n_1,n_2,n_3 < N_{\sigma}, \hspace{1mm} 0 \leq n_4 < N_{\tau} \}    
\label{eq:lattice}
\end{equation}

where $N_{\sigma}$ and $N_{\tau}$ define spatial and temporal extent of the lattice with the lattice regularized spacetime points. $a$ is the lattice spacing which is the distance between adjoining
lattice sites. The above representation of points in \autoref{eq:lattice} are in units of $a$. This is the case for an isotropic lattice. In an an-isotropic lattice, one can have different lattice spacings in spatial and temporal direction, like $a_{\sigma}$ and $a_{\tau}$ respectively. In our analysis we have used an isotropic lattice with $a_{\sigma} = a_{\tau} = a$.

Since we only have a finite extent of the lattice, we need to impose proper boundary conditions. Of the 
different boundary conditions available [38], the most common boundary condition used is a periodic
boundary condition requiring the condition $x_i + Na = x_i$, where $N=N_{\sigma}$ for $i=1,2,3$ and $N=N_{\tau}$ for $i=4$. We have used this periodic boundary condition on the lattice for our work and purpose. Hence, in this chapter and all throughout, by lattice, we only mean isotropic lattice, if not mentioned otherwise. This is a generalization of the toroidal boundary conditions specified in \autoref{Appendix 6}. The discretization
of the path integral in \autoref{eq:path partition} sets the temperature scale. On lattice, we get
\begin{equation}
    \beta = \frac{1}{T} = aN_{\tau}
    \label{eq:temp scale}
\end{equation}

We can decompose the Euclidean action $S_E\left[\overline{\Psi}, \Psi, A\right]$ into fermionic action \\$S_F\left[\overline{\Psi}, \Psi, A\right]$ and gluonic action $S_G\left[A\right]$. 

\subsection{Fermion action}

In continuum spacetime, the free fermionic action $S_F^0$ (with zero gauge fields), obtained by imposing $A=0$ in the fermionic part of \autoref{eq:QCD Lagrangian}, can be written as

\begin{equation}
    S_F^0\left[\overline{\Psi}, \Psi \right] = \int d^4x \hspace{1mm} \overline{\Psi}\left(x\right) \left[\gamma_{\mu}\partial^{\mu} + m\right] \Psi \left(x\right)
    \label{eq:Free fermions}
\end{equation}

While casting this action as in \autoref{eq:Free fermions} on a lattice, the partial derivative gets replaced with finite differences owing to the spacetime discretisation. Hence, considering a single flavored fermion, the lattice version of the above action looks as follows :  

\begin{equation}
    S_F^0\left[\overline{\Psi}, \Psi \right] = a^4 \sum_{n \in \textbf{L}} \overline{\Psi}\left(n\right) \Bigg[\sum_{\mu=1}^4 \gamma_{\mu} \hspace{1mm}\frac{\Psi\left(n+a\hat{\mu}\right) - \Psi\left(n-a\hat{\mu}\right)}{2a} + m\hspace{1mm}\Psi\left(n\right)\Bigg]
    \label{eq:Free fermions on lattice}
\end{equation}

where $m$ is the mass of the fermion and $\hat{\mu}$ being the unit vector in $\mu$ direction, where $\mu=1,2,3,4$. Now, as per the usual norms of QFT and gauge theory, we demand invariance of the lattice discretised action in \autoref{eq:Free fermions on lattice} under the local $SU\left(3\right)$ transformation, equivalently rotation in the $3$ color space. On lattice $\textbf{L}$,  we can impose the same for the quark and adjoint quark fields, $\Psi$ and $\overline{\Psi}$ as follows [10, 39] : 

\begin{equation}
\Psi\left(n\right) \to \Psi^{'}\left(n\right) = \Omega\left(n\right)  \Psi\left(n\right) , \hs \hs \hs \hs \overline{\Psi}\left(n\right) \to \overline{\Psi}^{'}(n) = \overline{\Psi}\left(n\right) \Omega^{\dagger}\left(n\right)    
\label{eq:gauge tran of fermions on lattice}
\end{equation}

where $\Omega\left(n\right)$ are elements of $SU\left(3\right)$ group, defined on lattice site $n$. We refer every spacetime point of lattice $\textbf{L}$, or every element of set $\textbf{L}$ in \autoref{eq:lattice} as lattice site. Now, the free fermionic action in \autoref{eq:Free fermions on lattice} has terms of the form $\overline{\Psi}\left(n\right) \Psi\left(n \pm a\hat{\mu}\right)$ which represent product of fermionic fields on adjacent lattice sites. Clearly these terms are not invariant under gauge transformations given in \autoref{eq:gauge tran of fermions on lattice}, because 

\begin{align}
    \adq (n) \q (n \pm a\hat{\mu}) &\to \adq^{'} (n) \q^{'} (n \pm a\hat{\mu}) \notag \\
    &= \adq \left(n\right) \Omega^{\dagger} \left(n\right) \Omega \left(n \pm a\hat{\mu}\right) \q \left(n \pm a\hat{\mu}\right)
\end{align}

and there is no gauge invariance since, $\Omega^{\dagger} \left(n\right) \Omega \left(n \pm a\hat{\mu}\right) \neq I_3$. Hence to preserve gauge invariance of lattice free fermion action, we need to introduce new fields $U_{\mu}\left(n\right)$, which will transform as follows
\begin{equation}
    U_{\mu}\left(n\right) \to U_{\mu}^{'}\left(n\right) = \Omega\left(n\right) U_{\mu}\left(n\right) \Omega^{\dagger}\left(n \pm a\hat{\mu}\right)
    \label{eq:gauge tran of bosons on lattice}
\end{equation}

under the gauge transformation of quark fields depicted in \autoref{eq:gauge tran of fermions on lattice}.

\begin{figure}[H]
    \hspace{2mm}
    \includegraphics[width=0.80\textwidth]{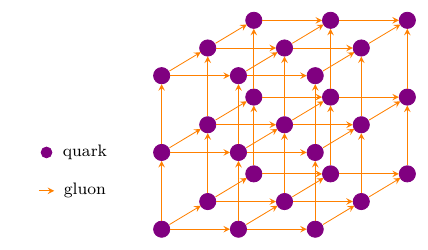}
    \caption{Naive description of quarks and gluons of QCD on a lattice}
    \label{fig:quarks and gluons on lattice}
\end{figure}

These new fields are the gauge fields which turn out to be the lattice version of gluon fields in QCD. On lattice, the fermionic fields are placed on the lattice sites, whereas these gauge fields are placed on the links conjoining successive adjoining lattice sites as shown in the above \autoref{fig:quarks and gluons on lattice}. According to the placement of quarks, these gauge fields can be directed and link adjoining quarks in $\hat{\mu}$ direction, with $\mu=1,2,3,4$. Hence, these gauge fields $U_{\pm \mu}(n)$ are often referred as link variables. 

With the fermionic and gauge fields as link variables on lattice and their respective gauge transformations in \autoref{eq:gauge tran of fermions on lattice} and \autoref{eq:gauge tran of bosons on lattice}, the gauge invariant fermion action is given by 

\begin{align}
    S_F\left[\overline{\Psi}, \Psi \right] &=  \notag\\
    a^4 \sum_{n \in \textbf{L}} &\overline{\Psi}\left(n\right) \Bigg[\sum_{\mu=1}^4 \gamma_{\mu} \hspace{1mm}\frac{U_{\mu}\left(n\right)\Psi\left(n+a\hat{\mu}\right) - U_{-\mu}\left(n\right)\Psi\left(n-a\hat{\mu}\right)}{2a} + m\hspace{1mm}\Psi\left(n\right)\Bigg]
    \label{eq:fermionaction}
\end{align}

\vspace{3mm}

where $U_{\mu}(n) = U_{-\mu}^{\dagger}(n+a\hat{\mu})$ and $U_{-\mu}(n) = U_{\mu}^{\dagger}(n-a\hat{\mu})$ and $\gamma_{-\mu} = -\gamma_{\mu}$. Hence, we find that similar to the continuum picture, the lattice version of QCD also demands that the interaction between the fermionic and gauge fields is required to impose gauge invariance on fermionic action, a free fermionic action does not preserve gauge invariance or invariance under gauge transformations of \autoref{eq:gauge tran of fermions on lattice}.

Based on the similarities in gauge transformation properties of the gauge transporter $G(x,y)$, where $x,y$ are two spacetime points in continuum picture 
\begin{equation*}
    G(x,y) = P \exp{\left(i \int_{C_{xy}} A \cdot ds\right)}
\end{equation*}

where $C_{xy}$ is an arbitrary path going from point $x$ to point $y$~\cite{Wilson.1974,Gattringer.2009}.
The gauge fields on lattice or the gauge links $U_{\mu}\left(n\right)$, is related to the gauge field $A_{\mu}\left(n\right)$ defined in continuum spacetime by the relation [39]
\begin{equation}
U_{\mu}\left(n\right) = \exp{\left[iaA_{\mu}(n)\right]}    
\label{eq:links to fields}
\end{equation}

which rightly corresponds to $U_{\mu}(n) = I_3$ for $a=0$, implying the case when there is no lattice and both $\adq$ and $\q$ are defined on site $n$, corresponding to mass term $\propto \adq(n) \q(n)$. Since, the matrices $U_{\mu}\left(n\right)$ are $SU(3)$ matrices and matrix operations happen in $3$-dimensional color space, hence the identity is $I_3$. In the continuum limit $a \to 0$, we expand above \autoref{eq:links to fields} and keeping terms upto linear powers in $a$, we find 

\begin{align}
    U_{\mu}\left(n\right) &= I_3 + iaA_{\mu}\left(n\right) + \Ob(a^2) \notag \\
    U_{-\mu}\left(n\right) &= I_3 - iaA_{\mu}\left(n-a\hat{\mu}\right) + \Ob(a^2)
    \label{eq:relations}
\end{align}

Using these equations of \autoref{eq:relations}, we find that the interaction part of the fermion action is given by 

\begin{align}
    S_F^I &= ia^4 \sum_{n \in \textbf{L}} \sum_{\mu=1}^4 \adq\left(n\right)  \frac{\gamma_{\mu}}{2} \bigg[A_{\mu}(n) \q(n+a\hat{\mu}) + A_{\mu}\left(n-a\hat{\mu}\right) \q\left(n-a\hat{\mu}\right)\bigg] \notag \\
    &= ia^4 \sum_{n \in \textbf{L}} \sum_{\mu=1}^4 \adq\left(n\right) \gamma_{\mu} A_{\mu}\left(n\right) \q\left(n\right) + \Ob\left(a\right)
    \label{eq:inteaction term}
\end{align}

where we have used $\q(n \pm a\hat{\mu}) = \q(n) + \Ob(a)$ and $A_{\mu}(n-a\hat{\mu}) = A_{\mu}(n) + \Ob(a)$. \autoref{eq:inteaction term} therefore establishes that we recover the continuum form of interaction term while expanding the lattice version in terms of $a$. As mentioned before, all $n$ are presented in units of lattice spacing $a$.
\subsection{Wilson's Gauge action and plaquette}

\begin{figure}[ht]
    \centering
    \includegraphics[width=1\textwidth]{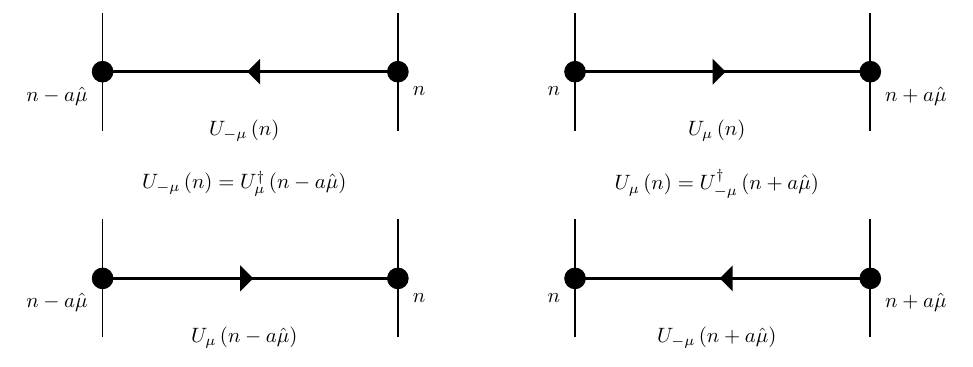}
    \caption{The geometric interpretation of link variables on lattice}
    \label{fig:Wilson lines}
\end{figure}

In the following \autoref{fig:Wilson lines}, we illustrate the geometric setting of the gauge fields which are the link variables on lattice, where the black blobs represent (quark) fermionic fields positioned on lattice sites $n$ and  $n \pm a\hat{\mu}$.

\vspace{3mm}

Given a trajectory $T$ on lattice constituting $k$ gauge fields constructed from $k$ arbitrary points on lattice in $k$ arbitrary directions, the matrix $M_T$ can be constructed as follows:

\begin{equation}
    M_T\left[U\right] =  U_{\mu_0}(n_0) \hs U_{\mu_1}(n_1) \hs \cdots \hs U_{\mu_k}(n_k) 
    \label{eq:Trajectory}
\end{equation}

where $n_l=n_{l-1} + a\mu_{l-1}$ for $1 \leq l \leq k$,  

we find that the gauge transformation of $M_T\left[U\right]$ of \autoref{eq:Trajectory} reads 

\begin{equation}
    M_T[U] \to M_T[U^{'}] = \Omega\left(n_0\right) M_T\left[U\right] \Omega^{\dagger} \left(n_k\right)
    \label{eq:traj gauge tran}
\end{equation}

So, hence, as per the gauge transformation of $T\left[U\right]$ as given in the above \autoref{eq:traj gauge tran}, a gauge invariant term or object will be $\adq(n_0) M_T\left[U\right] \q (n_k)$ which is familiar to us, because while imposing gauge invariance on Lagrangian $\mathcal{L}$, we convert $\partial_{\mu} \to D_{\mu}$ and obtain a $\adq \gamma^{\mu}A_{\mu} \q$ containing term. 

\begin{figure}[ht]
    \centering
    \includegraphics[width=0.3\textwidth]{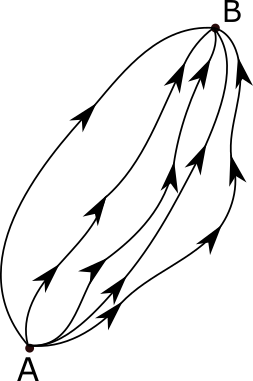}
    \caption{Paths between points $A$ and $B$}
    \label{fig:many paths}
\end{figure}

Another important thing to note here is that the imposition of gauge invariance on lattice depends only on the end points of the path constructed by the gauge links, it is independent of the nature of trajectory (similar to familiar "conservative and non-conservative forces"). A convenient diagram of five paths between two fixed points is given in the following \autoref{fig:many paths}. For all these paths, the term $\adq(A) P_n \q (B)$ is gauge invariant, where $P_n$ for $n=1,2,3,4,5$ is $n^{th}$ path joining points $A$ and $B$.

Following the above philosophy, a gauge invariant object constructed from pure gauge fields on lattice is $\Tr \left(M_L\left[U\right]\right)$, where $L[U]$ is a closed loop constructed by these gauge links or variables. The plaquette is the shortest, non-trivial closed loop constructed on a lattice and the plaquette variable $U_{\mu\nu}\left(n\right)$ is defined as 

\begin{align}
    U_{\mu\nu}\left(n\right) &= U_{\mu}\left(n\right) U_{\nu} \left(n+a\hat{\mu}\right) U_{-\mu} \left(n+a\hat{\mu} + a\hat{\nu}\right) U_{-\nu} \left(n+a\hat{\nu}\right) \notag \\
    &= U_{\mu}\left(n\right) U_{\nu} \left(n+a\hat{\mu}\right) U_{\mu}^{\dagger} \left(n+a\hat{\nu}\right) U_{\nu}^{\dagger} \left(n\right)
\end{align}

It is trivial to observe that $U_{\nu\mu} = U_{\mu\nu}^{\dagger}$. The Wilson's form of gauge action is a sum over all the possible plaquettes with every plaquette traversed with only one orientation. This sum is carried over all possible lattice sites, where the plaquettes are located along with a sum over all possible Lorentz indices $1 \leq \mu < \nu \leq 4$, which amount to ${}^4C_2=6$ possibilities. 
\begin{figure}[ht]
    \centering
    \includegraphics[width=1\textwidth]{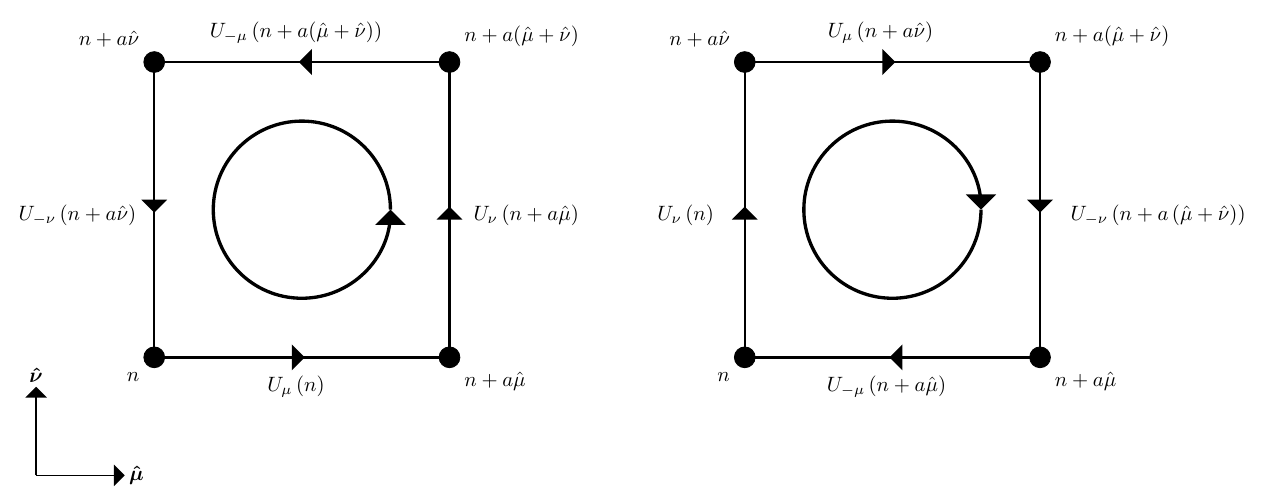}
    \caption{The four links variables constituting the plaquette. The circle indicates the order and sense in which the links are considered through the plaquette. Anti-clockwise (left) and clockwise (right)}
    \label{fig:Wilson loop}
\end{figure}

The gauge action on lattice, generalised or $N$ colors, is given as follows 

\begin{equation}
    S_G\left[U\right] = \frac{2N}{g^2} \hs \mathlarger{\sum}_{n \in \textbf{L}} \hs \hs \mathlarger{\sum}_{\mu < \nu} \hs \text{Re}\Big[\Tr \big(I_3 - U_{\mu\nu}\left(n\right)\big)\Big]
    \label{eq:Wilson gauge action}
\end{equation}

In continuum limit, for the simplest case $N=1$,  expanding the above \autoref{eq:Wilson gauge action} in terms of $a$, we find 

\begin{equation}
    S_G\left[U\right] = \frac{2}{g^2} \hs \mathlarger{\sum}_{n \in \textbf{L}} \hs \hs \mathlarger{\sum}_{\mu < \nu} \hs \text{Re}\Big[\Tr \big(I_3 - U_{\mu\nu}\left(n\right)\big)\Big] = \frac{a^4}{2g^2} \hs \mathlarger{\sum}_{n \in \textbf{L}} \hs \hs \mathlarger{\sum}_{\mu < \nu} \hs \Tr \left[F_{\mu\nu}(n)^2\right] + \Ob\left(a^2\right)
    \label{eq:continuum gauge action}
\end{equation}

Thus the Wilson's gauge action of \autoref{eq:Wilson gauge action} is equal to the continuum form upto $\Ob(a^2)$. The factor $a^4$ along with the sum over the lattice sites in the above \autoref{eq:continuum gauge action} is the outcome of the discretisation of spacetime integral. 

\section{Monte Carlo method}

Having obtained the expressions for the expectation value of an operator on a lattice as in \autoref{eq:observable}, we now have to calculate it computationally since, it is not possible to solve the path integral by analytical means. We wish to average the value of the operator over the whole phase space of gauge field configurations. However, it is not very feasible to span the entire phase space computationally. So instead we use a sampling technique that gives us a good representation of the entire phase
space, without having to span the entire phase space, thereby saving a great deal of computational time. This is called the importance sampling. It gives us a subset of all the possible states. We choose this subset with a Boltzmann weight $\propto \exp{\left(-S_E\left[U\right]\right)}$, with gauge field configuration $U$ and $S_E$, being the Euclidean action. So that it gives an appropriate representation of all possible states, or configurations, over which the system can traverse. The process we employ for it is called Markov chain Monte Carlo simulation~\cite{Gattringer.2009,Newman.1999}.

\subsection{Importance sampling and Markov Chain}
\label{subsec:Markov}
We generate the sample space of configurations based on the model of a Markov chain where the probability of manifestation of an immediate next configuration depends only upon the present configuration of the system and independent of the previous record or history of configurations. This is the basic principle of the Markov chain which makes the transition probability going from one configuration to another, a function of only these two configurations. We denote this transition probability from $U$ to $U^{'}$ as $T(U^{'}|U)$. This constitutes product of the selection probability $P_S(U^{'}|U)$, which is the probability that the algorithm will generate configuration $U^{'}$ starting from configuration $U$ and acceptance probability $P_A(U^{'}|U)$, which is the probability that the system accepts configuration $U^{'}$, given that the algorithm generated configuration $U^{'}$ from $U$. This is given by

\begin{equation}
    T(U^{'}|U) = P_S(U^{'}|U) P_A(U^{'}|U)
\end{equation}

where $T(U^{'}|U) = P(U_n=U^{'}|U_{n-1}=U)$. These transition probabilities obey

\begin{equation}
    0 \leq T\left(U^{'}|U\right)\leq 1 \quad , \quad \sum_{U^{'}} T\left(U^{'}|U\right) = 1
    \label{eq:prob}
\end{equation}

In this context, the other two necessary conditions that need to be followed and hence, imposed during the generation of algorithm are as follows : 

\begin{itemize}
    \item The condition of $\textbf{Ergodicity}$ which implies that the system must be able to attain any possible configuration in the sample of possible configurations from any arbitrary starting state (configuration) in a finite number of Markov steps. This is possible only when the transition probability is positive for any pair of $U$, $U^{'}$.
    
    \item  The condition of $\textbf{Detailed Balance}$ which guarantees that for any pair of configurations say $U$, $U^{'}$, the probability of the system attaining $U^{'}$ from $U$ is equal to the system attaining $U$ from $U^{'}$, ensuring similar preference of configurations. In mathematical terms, this balance equation is given by
    \begin{align}
       &\sum_U T(U^{'}|U) P(U) = \sum_U T(U|U^{'}) P(U^{'}) \notag \\
       &\Rightarrow \sum_U T(U^{'}|U) P(U) = P(U^{'})  
       \label{eq:detailed balance}
    \end{align}
\end{itemize}

The above final form of detailed balance condition in \autoref{eq:detailed balance} is obtained using \autoref{eq:prob}. In an actual calculation, the observables are calculated only after the system has reached equilibrium, or in the language of Monte Carlo simulation, has traversed a sufficient number of equilibrating Monte Carlo steps. This state of equilibrium is understood from the near-uniformity in the observable with evolving time or Monte-Carlo steps.

\begin{figure}[H]
    \centering
    \includegraphics[width=0.80\textwidth]{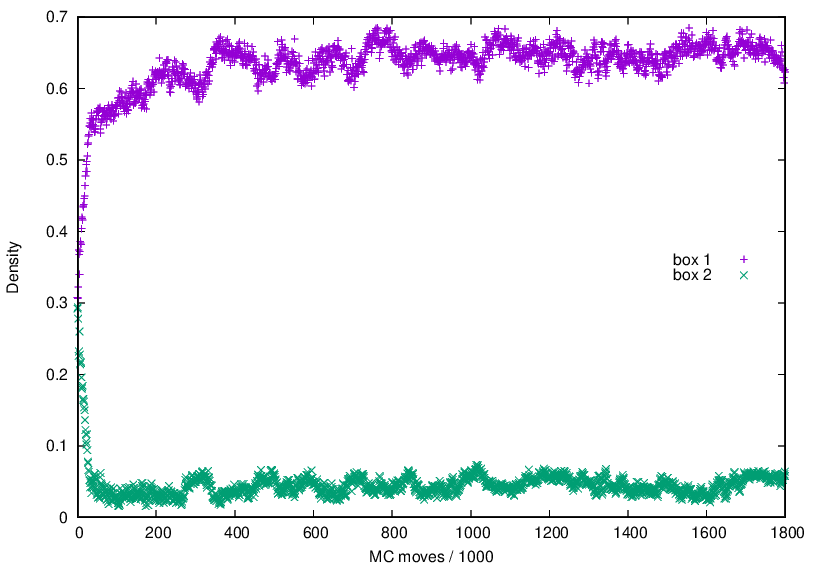}
    \caption{The density of particles in two boxes $1$ and $2$ forming a Gibbs ensemble as the system approaches the state of equilibrium. From~\cite{MonteCarlopicture}}
    \label{fig:Monte-Carlo equilibrium}
\end{figure}

The above \autoref{fig:Monte-Carlo equilibrium} offers a naive idea about Monte-Carlo equilibration which forms one of the chief ingredients of our work. 
This figure describes Monte-Carlo simulation for Gibbs ensemble of liquid and gas attaining liquid-gas equilibrium in its coexistence region~\cite{1988.Gibbs}. We know that as the density of a single-component fluid is increased at a given temperature, it eventually transits into liquid state from a gaseous phase. However, for intermediate densities below a critical value of temperature, there exists a phase in which both the liquid and gaseous phases coexist, with equilibrium values of respective densities $\rho_G$ and $\rho_L$. This is precisely the coexistence region where both the phases attain equilibrium. The box 1 and box 2 mentioned in the above figure describes the liquid and gaseous phases respectively, in which molecules or components of both the phases are allowed to mix keeping fixed the total number of molecules $N$, pressure $p$ and temperature $T$ ($NpT$ ensemble) for the combined liquid-gas system. The combined system equilibrate with time and this time evolution of the ensemble is measured with the time, which is numerically computed in the units of of Monte-Carlo (MC) steps. 
 As seen from the figure, the equilibrium condition can be estimated to set in box $1$ after around $1000 \times 10^3$ Monte-Carlo steps and around $200 \times 10^3$ steps in box $2$ respectively. This is naively how the equilibrium of a liquid gas ensemble can be captured through Monte-Carlo methods.

Two popular choices for the choice of initial configurations to start the Markov chain are as follows :

\vspace{2mm}

\begin{itemize}
\item  $\textbf{Ordered start}$ : All the gauge link matrices are set to identity. This corresponds to trivial plaquette variables and minimal gauge action which is approximately expected for small gauge coupling $g$ (large $\beta$). In fact strictly speaking, this corresponds to no gauge coupling at all, since the fermions are free in this case. In QCD, this corresponds to the infinite temperature limit where the degrees of thermal agitation are very much dominant and very nearly suppress the coupling effects resulting to very low or no coupling between fermions and gluons.

\item  $\textbf{Random start}$ : The gauge link matrices are chosen randomly such that they are still members of the $SU(3)$ group and follow the respective group properties. This condition is consistent with finite temperature QCD, where the coupling effects continue to contribute beside the thermal degrees of freedom.
\end{itemize}

Even though the system reaches equilibrium from any supplied configuration by virtue of ergodicity, the choice of the initial configuration should be such that it takes less time. The time it takes to reach equilibrium is called the equilibration time and it depends upon the proximity of the initial configuration to the equilibrium region. Thus, it is better to initialize the Markov chain for a particular temperature with an equilibrated configuration having a temperature, in its close thermal vicinity.

Now, the simplicity of the Markov chain to generate new configurations has certain shortcomings. An important one of them is that the local updates of the configurations cause auto correlation between the subsequent configurations. This correlation can cause systematic errors for various observables and overlook the error of the observables. One way to solve this issue is by disregarding intermediate configurations, such that we are left with only uncorrelated configurations in the sample. However, this is computationally wasteful
and reduces the number of available configurations. We do not keep all the configurations but instead of rejecting the intermediate ones,
we choose a step size where correlations are still present and use statistical tools like jackknife sampling and bootstrap sampling to get a correct error estimate. We have used bootstrap sampling of configurations to calculate observables in our work.

A quantitative measure of auto correlation among different configurations is given by the auto correlation time which we can obtain
from the auto correlation function. The auto correlation function for an observable $\Ob$ is given by
\begin{equation*}
A(t) = \LA \Ob_i \Ob_{i+t} \RA - \LA \Ob_i \RA \LA \Ob_{i+t} \RA 
\end{equation*}

with $\Ob_i$ being the observable at the $i^{th}$ time step and $t$ is the temporal separation between trajectory steps for which
we are evaluating the auto correlation function. The auto correlation function has an exponential decay characterised by the auto correlation time $\tau$ given by 
\begin{equation*}
A(t) \sim \exp{\left(-\frac{t}{\tau}\right)}
\end{equation*}

This auto correlation time $\tau$ provides a measure for the separation between trajectories which can be safely considered independent for the analysis.

\section{Rational Hybrid Monte Carlo Method}

The gauge field configurations used in our work have been generated using the Rational Hybrid Monte Carlo (RHMC) algorithm in order to stimulate the staggered action. This algorithm uses the rational approximation technique to get the fractional power of the fermionic determinant using pseudofermionic fields~\cite{Gattringer.2009}. The stochastic process used to propose the next viable configuration is the Hybrid Monte Carlo (HMC) which by itself is a combination of a molecular dynamics evolution and Metropolis acceptance test. Let us have a brief overview of all these constituent steps.

\subsubsection{Pseudofermions}

Till now, we have seen that the vacuum expectation value of an observable $\Ob$ can be expressed as 

\begin{align*}
    \LA \Ob \RA &= \frac{1}{\Z} \int \mathcal{D}U \hs \Ob \hs \det \M \left[U\right] \hs e^{-S_G\left[U\right]} \\
    \Z &= \int \mathcal{D}U \det \M \left[U\right] \hs e^{-S_G\left[U\right]}
\end{align*}

Calculating the determinant of the fermion matrix $\M$ numerically is an expensive job. To go around this difficulty, we introduce the concept of pseudofermions~\cite{Fucito:1985pseudofermions}. Implementing this pseudofermionic approach  facilitates the numerical evaluation of fermionic determinant. This is because, in this method it becomes possible to replace the Grassmann fermionic fields with a bosonic field and employ Gaussian integrals. We introduce $N$ number of complex pseudofermionic field $\phi_p = \phi_p^R + i\phi_p^I$ on each lattice site. We write the determinant of an arbitrary non-singular matrix $A$ as

\begin{equation}
    \det A \sim C\left(N\right) \int \mathcal{D}[\phi_p^R] \mathcal{D}[\phi_p^I] \exp{\left(-\phi_p^{\dagger} A^{-1} \phi_p\right)}
    \label{eq:pseudofer}
\end{equation}

In the above \autoref{eq:pseudofer}, the constant $C(N)$ is a non-zero constant dependent on number of pseudofermions $N$. One of the important properties of the fermion matrix $\M$ is that both the continuum and lattice versions of the matrix satisfies the $\gamma^5$ hermiticity given by~\cite{Hamber.1983} 

\begin{equation}
    \M^{\dagger} = \gamma^5 \M \hs \gamma^5
    \label{eq:gamma hermiticity}
\end{equation}

and this immediately implies $\det \M^{\dagger} = \det \M$. Hence, we can write $\det \M = \det (\M \M^{\dagger})^{1/2}$. Using this, we find 

\begin{equation}
    \det \M \sim C\left(N\right) \int \mathcal{D}[\phi_p^R] \mathcal{D}[\phi_p^I] \exp{\left(-S_{\phi_p}\right)}
    \label{eq:det M}
\end{equation}

where in the above \autoref{eq:det M}, $S_{\phi_p} = \phi_p^{\dagger} (\M \M^{\dagger})^{-1/2} \phi_p$.  Also, note that $(\M \M^{\dagger})^{-1/2}$ does not couple even and odd lattice sites and therefore diagonalises the full lattice into lattices containing only odd and even sites. A site $(n_1,n_2,n_3,n_4)$ is even if $\sum_{l=1}^4 n_l$ is even, otherwise odd. Keeping this in mind, we get 

\begin{align}
    \det (\M \M^{\dagger}) = \det (\M \M^{\dagger})_{odd} \hs \det (\M \M^{\dagger})_{even} \notag \\
    \det (\M \M^{\dagger})_{odd} = \det (\M \M^{\dagger})_{even} = \det \M
    \label{eq:pseudofermion apply}
\end{align}

Hence, we find from the above \autoref{eq:pseudofermion apply}, the computation of $\det \M$ requires the knowledge of only odd or even regime of lattice, which is basically just one half of the full lattice. Hence, the degree of computational expensiveness or time also gets reduced by half. Also, in \autoref{eq:pseudofermion apply}, where $(\M \M^{\dagger})_{even}$ is the $\M$ operator in the lattice with only even sites and $(\M \M^{\dagger})_{odd}$ is the same for lattice with odd sites only. Thus by introducing the pseudofermions, we effectively replace the Grassmann field variable by simple field variable and by taking advantage of the Gaussian integral, we make evaluating the determinant tractable. Thus, our expression reduces to

\begin{align*}
    \LA \Ob \RA &= \frac{1}{\Z} \int \mathcal{D}\left[U,\phi_p\right] \hs \Ob \hs e^{-S_{\phi_p}} \hs e^{-S_G\left[U\right]} \\
    \Z &= \int \mathcal{D}\left[U,\phi_p\right]  \hs e^{-S_{\phi_p}} \hs e^{-S_G\left[U\right]}
\end{align*}

\subsubsection{Rational Approximation}

The fourth rooting of the fermion determinant helps us get rid of the additional three degenerate components of the staggered fermions, as we will discuss in \autoref{sec:staggered fermions}.
To calculate it, we use a rational approximation. 

This is based on the idea of approximation theory in which one uses simpler functions to approximate an otherwise complicated function $f(x)$ and mimic its behaviour. This is typically done using polynomial or rational functions, in which these polynomials are called optimal polynomials. The goal is to choose a degree and the coefficients of the polynomial $P(x)$ and also select a domain $x \in (a,b)$ such that the maximum value of $|f(x) - P(x)|$ is minimised over the chosen domain $(a,b)$. 

For a general exponent $n_q = N_q/8$ where $N_q$ is the number of quarks with the same quark mass $m_q$, we use the following ansatz for the rational approximate of $\M \M^{\dagger}$:

\begin{equation}
    R(\M \M^{\dagger}) = (\M \M^{\dagger})^{-n_q} = \lambda_0 I_4 + \sum_k \frac{\lambda_k I_4}{\M \M^{\dagger} + \zeta_k I_4}
\end{equation}
where $\lambda_0, \lambda_i$ and $\zeta_i$ are the coefficients for the rational approximation. $R(\M \M^{\dagger})$ is the rational approximate form of $\M \M^{\dagger}$. These coefficients are obtained using the Remez algorithm which is optimized for the spectrum of fermion matrix $\M$ having the lower bound eigenvalue $m_q^2$. We stop our discussion on rational approximation here, for further details regarding rational approximation, refer to~\cite{Clark.2006}.

\subsection{Hybrid Monte Carlo}

While simulating the Markov Chain Monte Carlo sampling, we want our gauge configurations to accommodate as many changes as possible
so as to be less correlated (less auto correlation) and at the same time have high acceptance rates (see discussion on Markov Chain in \autoref{subsec:Markov}). Moreover, the non-local action due to the fermion determinant means that one needs to calculate the determinant again for each Markov step, even if only a single gauge link $U_{\mu}(n)$ is altered. Thus it would be more practically efficient if one can update as many gauge link variables in a single step. Doing this in a naive manner, however, would lead to a small
acceptance probability as it would result in a large change of the action making it inefficient. This is exactly where the algorithm of Hybrid Monte Carlo (HMC) comes in handy which can update multiple gauge links and offer higher acceptance rates simultaneously~\cite{Duane.1987}.

In order to implement HMC, we introduce another field $\pi_{\mu}(n) = \pi_{\mu}^{a}(n) T_a$ conjugate to the gauge field link variable $U_{\mu}(n)$ with $T_a$ with $1 \leq a \leq 8$ being the eight $SU(3)$ generators. Clearly, these $\pi_{\mu}(n)$ are traceless hermitian matrices. Then, we create a Hamiltonian-like operator using pseudofermionic fields $\phi_p$ and conjugate gauge fields $\pi_{\mu}$ which becomes the statistical weight for the generation of configurations [10]. The modified equations look like 

\begin{align}
    \LA \Ob \RA &= \frac{1}{\Z} \int \mathcal{D}\left[U,\phi_p, \pi \right] \hs \Ob \hs \exp{\big(-H\left[U,\phi_p,\pi\right]\big)} \notag \\
    \Z &= \int \mathcal{D}\left[U,\phi_p, \pi \right] \hs \exp{\big(-H\left[U,\phi_p,\pi\right]\big)}
\end{align}

with the Hamiltonian given by

\begin{equation}
    H[U,\phi_p, \pi] = \frac{1}{2} \sum_{n,\mu} \Tr \left[\pi_{\mu}^2\left(n\right)\right] + S_{\phi_p} + S_G
\end{equation}

where $S_G$ is the gauge action and $S_{\phi_p}$ is the action of pseudofermions, described by pseudofermionic fields $\phi_p$. Using Hamiltonian dynamics, one can traverse on a constant energy curve for the system described by the Hamiltonian $H$. Any configuration
on a constant energy curve guarantees that the final configuration attained after traversing the curve will be equally likely as the configuration the system started from, since the Boltzmann weights $\propto \exp{(-H)}$ are the same for both the initial and final configurations. Thus, it enables us to explore a greater range or subset of configuration space with lesser correlation and at the same time promises a larger acceptance ratio. As mentioned earlier, we implement this process using Molecular
dynamics followed by the Metropolis Acceptance test.

\subsubsection{Evolution of Molecular dynamics}

Given the Hamiltonian $H(q(t),p(t),t)$ of a system, we can construct the dynamics of the system from the Hamilton's canonical equations of motion 

\begin{equation}
    \dot{q} = \frac{\partial q}{\partial t} = \frac{\partial H}{\partial p} \quad,\quad \dot{p} = \frac{\partial p}{\partial t} = -\frac{\partial H}{\partial q}
    \label{eq:canonical equations}
\end{equation}
where $q$ and $p$ are the generalised coordinate and generalised conjugate momentum as a function of time $t$. The evolution with time is such that the system preserves temporal symmetry and from our knowledge of symmetry and mechanics, we know that time symmetry implies conservation of energy which is a constant of motion for the Hamiltonian.

To implement it numerically, we evolve the system in discrete time steps, each step generating a trajectory in the phase space described by the Hamiltonian $H$. We take the step size to be $\epsilon$. We follow the trajectory to generate a new configuration.
However, there will be many numerical errors due to the evolution occurring in discretised temporal steps, which is why we resort to a Metropolis step, which we have discussed subsequently.

To make sure the necessary condition of the detailed balance is followed on each step, we employ
the technique of leapfrog integration scheme to evolve the system. It guarantees us the reversibility
of the trajectory, $T(U \to U^{'})=T(U^{'} \to U)$, which implies the transition probability is equal for both the configurations $U$ and $U^{'}$. In this process, the generalised coordinate $q$ is evolved in $n$ steps of length $\epsilon$, and the conjugate momentum $p$
is evolved first with a half step $\epsilon/2$, followed by $n - 1$ full steps and then by a half step again.

Applying this process to our system containing link variables $U$ as the primary fields and conjugate fields $\pi$ , we
evolve them accordingly. But we also do have the pseudofermions that need to be included. For
that, we first generate the pseudofermionic fields $\phi_p = \M \sigma$, by randomly generating $\sigma$ which are
distributed according to a probability weight distribution $\exp{(-\sigma^{\dagger}\sigma)}$. This directly gives us the appropriate
distribution of pseudofermionic field. Next, given a gauge configuration U, the $\pi$ fields are generated according to
the distribution $\exp{(-\Tr[\pi_{\mu}^2])}$. This requires generating $8$ real numbers for each lattice site $n$ and direction $\mu, 1 \leq \mu \leq 4$, each real number corresponding to each of the $8$ generators of $SU(3)$ group. Following these steps, we reach a configuration which we then need to accept or reject according to the Metropolis acceptance rule, which we discuss next.

\subsubsection{Metropolis acceptance}

Having fixed the selection probability for the new configuration, we now need to fix our acceptance
probability such that it fulfills the detailed balance condition~\cite{Gattringer.2009,Newman.1999}. The reversibility of the leapfrog integrator as mentioned before, ensures that the transition probabilities $T(U \to U^{'})=T(U^{'} \to U)$, which is exactly the detailed balance condition. We have therefore

\begin{align}
    P_S(U) P_A(U \to U^{'}) = P_S(U^{'}) P_A(U^{'} \to U) \notag \\
    \Rightarrow \frac{P_A(U^{'} \to U)}{P_A(U \to U^{'})} = \frac{P_S(U)}{P_S(U^{'})} = \exp{(S[U^{'}]-S[U])}
    \label{eq:Metro}
\end{align}

since, $P_S(U) \propto \exp{(-S[U])}$. Here $P_S[U]$ is the selection probability that the system selects configuration $U$ and $P_A(U \to U^{'})$ is the acceptance probability that the system accepts the transition and moves from configuration $U$ to new configuration $U^{'}$. The above \autoref{eq:Metro} fixes only the ratio of the probabilities. As we would like to have a high acceptance probability for aforementioned reasons, we set the larger acceptance ratio between the two transition directions to $1$, which is the maximum
possible value of a probability measure. Thus, we get our acceptance rate as

\begin{equation}
    P(U \to U^{'}) = \text{min}\Big[1, \hs \exp{\big(S\left[U\right] - S[U^{'}]\big)}\Big]
    \label{eq:acceptance}
\end{equation}
If the acceptance probability is 1, then as per the above \autoref{eq:acceptance}, $S[U] > S[U^{'}]$. This implies the new configuration $U^{'}$ is more energetically favorable, and hence, the new configuration is always selected, which is usually expected from a thermodynamic system. If the acceptance probability is less than one, then a random number $r$ is generated from a uniform distribution
in the interval $[0,1)$ using a random number generator, and the new configuration $U^{'}$ is only accepted if $r < P(U \to U^{'})$. Otherwise, we take the same configuration and start a new molecular dynamics trajectory~\cite{Gattringer.2009,Newman.1999}.

It is this Monte Carlo algorithm with implementation of pseudofermions, rational approximation, Hybrid Monte Carlo algorithm, the molecular dynamics along with  Metropolis acceptance that gives us a set of configurations of $SU(3)$ gauge links on the lattice. We can calculate our observables on these generated configurations and obtain their estimation values.

\section{Symanzik improvement}
\label{sec:Symanzik improvement}
While introducing the QCD action on the lattice we had to discretise the derivative terms that show up in the continuum action, apart from the integrals. It is found that this discretisation give rise to symmetric diﬀerences for the ﬁrst derivatives in the fermion action, leading to discretization eﬀects. Typically the discretization eﬀects are $\sim \Ob \left(a\right)$ for fermions and of $\sim \Ob \left(a^2\right)$ for the gauge ﬁelds. They vanish only in the continuum limit $a \to 0$. Performing the continuum limit is, however, a nontrivial task.
As one decreases a, the number of lattice points has to increase, such that the physical volume remains constant. In exact continuum, the number of physical points must be infinity, which is impossible to consider from a pragmatic point of view. Hence, in a numerical
simulation one always works with ﬁnite lattice spacing $a$ and the discretization errors have to be considered, by including them in the extrapolation to vanishing $a$.

An elegant way of approaching this problem is a systematic reduction
of the discretization errors to different orders in $a$. We have already mentioned that the discretisation scheme one chooses is not unique. Also other discretisation approach can converge to the same formal continuum limit. In particular one may combine diﬀerent terms to obtain a lattice action with reduced discretization eﬀects. For example, adding an extra term to the Wilson fermion action and matching its coeﬃcient appropriately, one can reduce the discretization error from $\Ob\left(a\right)$ to $\Ob\left(a^2\right)$. A systematic implementation of these ideas is the Symanzik improvement which we will discuss it very naively.

\subsubsection{A naive example}

For the discussion of improvement, we consider a toy example which already
contains most of the steps that will be taken when improving lattice QCD.
We consider the symmetric discretization of the derivative $f^{'}\left(x\right)$ for some
function $f\left(x\right)$ of a single real variable x:

\begin{equation}
    \frac{f(x+a) - f(x-a)}{2a} = f^{(1)}(x) + \frac{a^2}{3!} f^{(3)}(x) + \frac{a^4}{5!} f^{(5)}(x) + \Ob(a^6)
    \label{eq:start equation}
\end{equation}

where $f^{(k)}(x)$ is the $k^{th}$ derivative of $f(x)$. 

The strategy for improvement is to add a
discretised expression, which is an expression constituting $f(x), f(x \pm a), f (x \pm 2a)$ to the left-hand side of \autoref{eq:start equation} such that the correction terms on the right-hand side are canceled up to the required order in $a$. By $\Ob(a^k)$ improvement, we mean that there are no terms upto $\Ob(a^k)$, the discretisation effects start from $\Ob(a^{k+1})$ on-wards. For improvement of $\Ob(a^2)$,  we therefore make the following correction

\begin{equation}
    \frac{f(x+a) - f(x-a)}{2a} +c \hs a^2 \hs D^{(3)}[f](x) = f^{(1)}(x) + \Ob(a^4)
    \label{eq:improvement}
\end{equation}

where $D^{(3)}[f]$ is a discretised expression obeying $D ^{(3)} [f] \approx f^{(3)} + \Ob(a^2)$ and $c \neq 0$ is some constant.
Using the following values 

\begin{equation}
D^{(3)}[f](x) = \frac{f(x+2a) - 2f (x+a) + 2f(x-a) - f(x-2a)}{2a^3} \quad, c =  -\frac{1}{6}    
\label{eq:choice}
\end{equation}

we find that $\Ob(a^2)$ improvement is achieved. We remark, however, that
the choice in \autoref{eq:choice} is not unique, and, terms including $f(x \pm 3a)$ could
have been used.

Let us summarize the steps taken in our toy example, which already outline
the approach for improving lattice QCD:

\begin{itemize}
\item We start from a simple discretised expression for the quantity of interest
like the ﬁrst derivative $f^{(1)}$ in our example.
\item Correction terms are identiﬁed using continuum language and higher derivatives in the above example considered.
\item The correction terms have certain symmetries like having only odd derivatives in the given 
example and are ordered according to their mass dimensions.
\item In order to achieve improvement, discretised versions of the correction
terms are added with suitable coeﬃcients, such that corrections up to the
desired order in $a$ vanish.
\item The choice of the discretised correction terms is not unique.
\end{itemize}

Exactly the same steps and features do appear in the improvement of lattice
QCD. The main diﬀerence is the determination of the coeﬃcients of different orders of $a$. In the above
example the coeﬃcient c followed from simple algebraic considerations. Due
to the nonlinear nature of QCD and the necessary renormalization schemes considered, the determination of the corresponding coeﬃcients in QCD is much more involved
and must be done using suitable and appropriate perturbative or non-perturbative matching procedure. The approach to improvement outlined here is known as Symanzik
improvement program~\cite{Symanzik.1983.187,Symanzik.1983.205,Weisz.1985,Karsch.2001}.





\section{Staggered fermions}
\label{sec:staggered fermions}

Before going into staggered fermion prescription, it is important to analyse the fermion doubling problem, because, the whole idea of staggered fermionic action emerges as a solution to this problem. We also observe how the addition of an extra term, introduced by Wilson, called the Wilson term, removes fermion doubling problem.

\subsection{Fermion doubling problem}
\label{subsec:Fermion doubling problem}
The naive non-interacting free fermion action formulated on an isotropic lattice $\textbf{L}$ of lattice spacing $a$ is given by 

\begin{equation}
    S_F\left[\adq,\q\right] = a^4 \hs \mathlarger{\sum}_{n \in \textbf{L}} \hs \hs \hs \adq\left(n\right) \left[\sum_{\mu=1}^4 \frac{\gamma_{\mu}}{2a} \Big(\q\left(n+a\hat{\mu}\right) - \q\left(n-a\hat{\mu}\right)\Big) + m \hs \q\left(n\right)\right]
    \label{eq:action in appen}
\end{equation}

where $m$ is the mass of the fermion and $\adq, \q$ are the adjoint fermion and fermion fields respectively. Here, we consider only a single flavor of fermion field and hence, the otherwise sum over flavor indices are suppressed.

Since, the above action in \autoref{eq:action in appen} is bilinear in $\adq$ and $\q$, we can write the action in the following form

\begin{equation}
    S_F\left[\adq,\q\right] = a^4 \mathlarger{\sum}_{n,m \in \textbf{L}} \hs \hs \hs \sum_{a,b,\alpha,\beta} \adq \left(n\right)_{\alpha}^a \hs D\left(n|m\right)_{\alpha\beta}^{ab} \hs \q\left(m\right)_{\beta}^b 
    \label{eq:bilinear}
\end{equation}

The Dirac operator $D$ on lattice with Dirac indices $\alpha,\beta$ and color indices $a,b$, for interacting fermion is given by

\begin{equation}
    D\left(n|m\right)_{\alpha\beta}^{ab} = \mathlarger{\sum}_{\mu=1}^4 \frac{(\gamma_{\mu})_{\alpha\beta}}{2a} \bigg[U_{\mu}^{ab}\left(n\right) \hs \delta_{n+a\hat{\mu},m} - U_{-\mu}^{ab}\left(n\right) \hs \delta_{n-a\hat{\mu},m}\bigg] + m \hs \delta_{\alpha\beta} \hs \delta_{nm} \hs \delta^{ab}
    \label{eq:interacting}
\end{equation}

For free fermions, there is no gauge fields $U_{\pm \mu}(n)$ and hence, there is no color exchange, preserving color charge conservation, following which, the color indices in the above \autoref{eq:interacting} are dropped. Hence, here $U_{\pm \mu}(n) = I_3$ and hence, the free Dirac operator would look like

\begin{equation}
    D\left(n|m\right)_{\alpha\beta}^{ab} = \mathlarger{\sum}_{\mu=1}^4 \frac{(\gamma_{\mu})_{\alpha\beta}}{2a} \bigg[\delta_{n+a\hat{\mu},m} - \delta_{n-a\hat{\mu},m}\bigg] + m \hs \delta_{\alpha\beta} \hs \delta_{nm}     \label{eq:position_lat_Dirac}
\end{equation}

The Fourier transform (see \autoref{Appendix 6}) of lattice Dirac operator as given in above \autoref{eq:position_lat_Dirac} is given by~\cite{Gattringer.2009} 

\begin{equation}
    \Tilde{D}(p) = m \hs I_4 + \frac{i}{a} \sum_{\mu=1}^4 \gamma_{\mu} \sin{\left(p_{\mu}a\right)}
\end{equation}

The fermion propagator dictates the behaviour of $n$-point correlation functions and hence, it is important to analyse it. The fermion propagator is the inverse of the lattice Dirac operator, given by $D^{-1}\left(n|m\right)$ and we obtain it from the following inverse Fourier transformation:

\begin{equation}
    D^{-1}(n|m) = \frac{1}{\left|\Tilde{\textbf{L}}\right|} \sum_{p \hs \in \hs \Tilde{\textbf{L}}} \Tilde{D}^{-1}(p) e^{ip \cdot (n-m)a}
\end{equation}

where $|\textbf{L}|$ and $\left|\Tilde{\textbf{L}}\right|$ is the total number of lattice sites on lattice $\textbf{L}$ and $\Tilde{\textbf{L}}$ respectively. The latter is the lattice in the conjugate $4$-momentum space. The fermion propagator in the momentum space is given by

\begin{equation}
    \Tilde{D}^{-1}(p) = \frac{m \hs I_4 - i \hs a^{-1}\sum_{\mu} \gamma_{\mu} \sin{(p_{\mu}a)}}{m^2+a^{-2}\sum_{\mu}\sin^2{(p_{\mu}a)}}
    \label{eq:mom sp prop}
\end{equation}

where we have the anti-commutator $\{ \gamma_{\mu}, \gamma_{\nu}\} = 2 \delta_{\mu\nu} I_4$.
For free fermions, this analysis is best performed in the momentum space, which draws our attention to the momentum space propagator as given in \autoref{eq:mom sp prop}. 

\subsubsection{Massive case}

We look for the singularities of the fermion propagator, which may manifest in form of poles, branch cuts etc. For this, we need to carefully analyse the denominator of the fermion propagator given in \autoref{eq:mom sp prop}.

In continuum limit $a \to 0$, this becomes $m^2+p^2$, where $p^2 = \sum_{k=1}^4 p_k^2$, since the spacetime being Euclidean, the metric is an Euclidean metric. Hence, for real $p_k$, there is no pole since $m > 0$. In fact, all the poles lie on the surface of a $4$-hypersphere with radius $im$ in a complex $4$-momentum space, for example, $(im,0,0,0)$ is a pole. 

In lattice version, $\sin^2{(x)}$ eliminates the possibility of having a negative quantity. Hence, $m^2+a^{-2} \sum_{\mu} \sin^2{(p_{\mu}a)}$ is always greater than $0$, and hence the above fermion propagator in \autoref{eq:mom sp prop} has only imaginary poles, for massive fermions. 

\subsubsection{Massless case}

In the case of massless fermions, for a fixed momentum $p$, the propagator looks like:

\begin{equation*}
    \Tilde{D}^{-1}(p) \Big|_{m=0} = \frac{-ia^{-1}\sum_{\mu}\gamma_{\mu}\sin{\left(p_{\mu}a\right)}}{a^{-2}\sum_{\mu} \sin^2{\left(p_{\mu}a\right)}}
\end{equation*}

which in continuum limit $a \to 0$, becomes $-i a \sum_{\mu}\gamma \cdot p/p^2$, where $\gamma \cdot p = \sum_{\mu} \gamma_{\mu}p_{\mu}$ and $p^2 = \sum_{\mu} p_{\mu}p_{\mu}$. We do not use the contravariant and covariant indices here, since we are working in a Wick rotated Euclidean spacetime.

In this case in continuum limit, the propagator has a pole at $(0,0,0,0)$ which corresponds to the true physical pole for the continuum Dirac operator. However, in lattice version of the propagator, we have many other poles, since $\sum_{\mu} \sin^2{(p_{\mu}a)}=0$, implying that for each $\mu$, $\sin{(p_{\mu}a)}=0$. Since, $-\pi/a < p_{\mu} \leq \pi/a$, following the boundary conditions given in \autoref{Appendix 6}, hence, the spectrum of poles looks like 

\begin{equation}
    p = (0,0,0,0), (\pi/a,0,0,0), \cdots, (\pi/a, \pi/a, \pi/a, \pi/a)
    \label{eq:all fermion doublers}
\end{equation}

Thus, in above \autoref{eq:all fermion doublers}, we find $15$ other poles containing at least one $\pi/a$, apart from the physical pole $(0,0,0,0)$. This is the fermion doubling problem and these $15$ poles arising due to lattice discretisation effects are called fermion doublers, or simply doublers. 
doublers.

\subsection{Wilson's correction to fermion doublers}

One of the first approaches to remove fermion doublers was introduced by Wilson. The fermions having action including the Wilson term and with no doublers are called Wilson fermions. The Wilson corrected Dirac operator in $4$-momentum space is given by 

\begin{equation}
    \Tilde{D}(p) = m \hs I_4 + \frac{i}{a} \hs \sum_{\mu=1}^4 \gamma_{\mu} \sin{\left(p_{\mu}a\right)} + \boldsymbol {I_4} \hs \boldsymbol{\frac{1}{a}} \hs \sum_{\mu=1}^4 \boldsymbol{\left(1-\cos{\left(p_{\mu}a\right)}\right)}
    \label{eq:Wilson fermion}
\end{equation}

The bold extra term in \autoref{eq:Wilson fermion} is the Wilson term, which vanishes for the true pole $p=(0,0,0,0)$ and adds an extra contribution $2/a$ for all the aforementioned doublers. This term acts like an additional mass term and the mass of the doublers is given by $m + 2n/a$, where $n=15$ is the number of doublers. In continuum limit $a \to 0$, the doublers become highly massive and decouple from the theory. Hence, we find that there is only the true physical pole $(0,0,0,0)$ remaining with no doublers.

\section{Staggered prescription}

Staggered fermions, often called Kogut–Susskind fermions~\cite{Kogut.1975}, are fermions described by a staggered fermionic action on lattice. This action is obtained from implementing a staggered formulation on the usual fermionic action on lattice. In this formulation, the $16$-fold degeneracy of the naive fermion discretization, is reduced to only four quark flavors, while at the same time a remnant chiral symmetry is maintained. This $16$-fold degeneracy owes its origin to the fermion doubling problem, which gives rise to $15$ unphysical poles in addition to the true physical pole, found in the continuum limit $a \to 0$ of lattice with lattice spacing $a$. All these $16$ quark degrees of freedom are mass degenerate i.e. they all have the same mass, irrespective of whether they are massive or massless. The different sets of $4$-momenta components give rise to $16$ poles and hence, $16$ flavors. 

As known from the familiar Nielsen-Ninomya theorem, it is not possible to implement chiral symmetry on lattice without running into fermion doubling problem. In any case, either one is not possible to achieve, a naive proof of which is presented above in \autoref{subsec:Fermion doubling problem}. The staggered formulation is certainly an improvement, in the sense that it reduces the number of doublers by preserving a remnant chiral symmetry on the lattice. This staggered behaviour is achieved by a transformation which mixes Dirac and lattice indices, distributing the $16$ quark degrees of freedom on a $4$ hypercube form of the lattice, having $2^4=16$ corners or vertices, on which these $16$ quarks can be placed. Each of the $16$ poles or doublers (since, each of the four momentum components are allowed to become $0$ or $\pi/a$) correspond to $16$ mass degenerate flavors of quark field. The naive free fermion action given in \autoref{eq:Free fermions on lattice} contains a symmetry which allows one to reduce the number of doublers, therefore the severity of the doubling problem, without compromising the chiral symmetry. This symmetry is therefore confined within the kinetic term, since, we know the kinetic term of fermionic action preserves the chiral symmetry, whereas the mass term is the chiral symmetry breaking term. 

To implement this symmetry, we perform a staggered transformation of $\adq(n)$ and $\q(n)$ mixing the lattice and Dirac indices in order to eliminate the matrices $\gamma_{\mu}$ and these are given by~\cite{Gattringer.2009}

\begin{equation}
    \adq^{'}(n) = \adq\left(n\right) \hs \gamma_1^{n_1} \hs \gamma_2^{n_2} \hs \gamma_3^{n_3} \hs \gamma_4^{n_4}  \quad \quad \q^{'}(n) =  \gamma_4^{n_4} \hs \gamma_3^{n_3} \hs \gamma_2^{n_2} \hs \gamma_1^{n_1} \hs \q\left(n\right)
    \label{eq:stag transformation}
\end{equation}

In the above \autoref{eq:stag transformation}, all the lattice indices are in the units of lattice spacing $a$, since, strictly speaking $n=(a\hs n_1,a\hs n_2,a\hs n_3,a\hs n_4)$. The mass term remains invariant under staggered transformation (\autoref{eq:stag transformation}), since $\gamma_{\mu}^2 = I_4$, for each $\mu$ in Euclidean 
spacetime. The main game happens in the kinetic term which are of the form $\sim \adq(n) \gamma_{\mu} \q(n+a\hat{\mu})$. Through some mathematical algebra, it can be explicitly shown that the free fermion action in \autoref{eq:Free fermions on lattice} reduces to the following form 

\begin{equation}
S_F^{(di)}\left[\overline{\Psi}, \Psi \right] = a^4 \sum_{n \in \textbf{L}} \overline{\Psi}\left(n\right) \boldsymbol{I_4} \Bigg[\sum_{\mu=1}^4 \eta_{\mu}(n) \hspace{1mm}\frac{\Psi\left(n+a\hat{\mu}\right) - \Psi\left(n-a\hat{\mu}\right)}{2a} + m\hspace{1mm}\Psi\left(n\right)\Bigg]
\label{eq:diag action}
\end{equation}

where in the above \autoref{eq:diag action}, we have introduced the staggered sign functions $\eta_{\mu} = \left(-1\right)^{P_{\mu}}$ where $P_{\mu} = \sum_{l < \mu} n_l$ for $\mu \geq 2$ and $\eta_1=1$ . This therefore implies

\begin{equation}
    \eta_1\left(n\right) = 1, \quad \eta_2\left(n\right) = \left(-1\right)^{n_1}, \quad \eta_3\left(n\right) = \left(-1\right)^{n_1+n_2}, \quad \eta_4\left(n\right) = \left(-1\right)^{n_1+n_2+n_3} 
\end{equation}

The new action $S_F^{(di)}$ in \autoref{eq:diag action} therefore, is diagonal in Dirac space, since the gamma matrices $\gamma_{\mu}$ are now replaced by identity $\boldsymbol{I_4}$ and hence has the same form for all the four Dirac components. The staggered transformation in \autoref{eq:stag transformation} basically enables a diagonalisation of the Dirac operator in the Dirac space. 

The staggered fermion action is obtained by keeping only one of the four identical components, which means that in the staggered action, there will be no Dirac indices.  and there will only be $16/4=4$ quark degrees of freedom. Coupling with gauge fields $U_{\mu}(n)$, the staggered fermion action $S_F^{(ST)}$ is given by~\cite{Kilcup.1987}

\begin{align}
    &S_F^{(ST)}\left[\overline{\xi}, \xi \right] = a^4 \sum_{n \in \textbf{L}} \overline{\xi}\left(n\right) \hs \boldsymbol{I_4} \hs H(n,\mu) \quad \text{where}\notag \\
    &H(n,\mu) = \Bigg[\sum_{\mu=1}^4 \eta_{\mu}(n) \hspace{1mm}\frac{U_{\mu}(n) \hs \xi\left(n+a\hat{\mu}\right) - U_{\mu}^{\dagger}(n-a\hat{\mu}) \hs \xi\left(n-a\hat{\mu}\right)}{2a} + m\hspace{1mm}\xi\left(n\right)\Bigg]
    \label{eq:stag ferm action}
\end{align}

where $\xi(n)$ and $\overline{\xi}(n)$ are redefined fermion fields having only color indices, but no Dirac structure with no Dirac indices.

\subsection{Tastes of staggered fermions: A naive overview} 
  As known from the basic QFT, the Feynman rules are formulated using the Green's functions and this is essentially done for free theory, when there are no gauge fields. The vertices, propagator all come along with the formulation of these Feynman rules. Following the lines, we evaluate the propagator of lattice Dirac operator in the conjugate $4$-momentum space which is obtained as a Fourier transform of the $4$-lattice. The poles or singularities of the fermion propagator give the co-ordinates of the fermions or fermion fields of the theory. Essentially, these are all free fermions, since everything is evaluated in the free case, as per the norms and rules of usual QFT. 

In fact, it is not possible to evaluate the propagator for interacting theory in presence of gauge fields, since the Fourier transform does not become possible. This is because the gauge fields of the gauge theory as they are named, have a local symmetry and it is this spacetime dependence of fields in continuum or the lattice site dependence in lattice which is the local nature of fields that inhibits Fourier transform whereas taking global symmetry and non-gauge fields therefore, does not pose a problem.

Anyway, we find that the momentum space fermion propagator has $16$ poles, including the true physical continuum pole $(a\to 0)$. indicating that the naive free fermion action on lattice describes $16$ mass-degenerate fermions or fermion fields. In QCD, these are all quarks or quark fields, retaining their usual spinor structure with Dirac and color indices.  This is also attributed to the on-shell nature of the poles, irrespective of massive or massless fermionic action. 

The staggered transformation performs a diagonalisation of the fermionic action, containing information about $16$ doublers in Dirac space. All these poles are called doublers, since each of the components can have only two values, either $0$ or $\pi/a$. Through this diagonalisation, the new action is diagonal in Dirac space, in fact all the four components are equal, since the action is written completely in terms of identity matrix $I_4$, proving that diagonalisation happened in Dirac space which is a $4$-dimensional space.

We discard three of the four identical components and the staggered action, peeled off from the diagonal fermionic action contains only this Dirac component. This means effectively, the fermion field in staggered action has no Dirac indices with no Dirac structure. It therefore only contains color degrees of freedom. But now, the staggered transformation has made a drastic difference over the non-staggered situation; it has made the transformation $\gamma_{\mu} \to \eta_{\mu}(n) \hs I_4$. Thus the staggered transformation not only carried out diagonalisation, it also ensures that the fermionic action mimics an effective interacting theory, it is just that the gauge fields are substituted with $\eta_{\mu}(n)$, both having a local signature. Gauge fields in QCD are $SU(3)$ matrices, whereas $\eta_{\mu}(n)$ are just signed numbers. So, the staggered transformation is free from the fermion doubling problem, having a global chiral symmetry. 

Now, this transformation started from a free action and ended with an effective interacting action and this is the trick. Had the action described only one fermion, it could have never described interaction, because self-interaction of particles in free case is never allowed in QFT. Somewhere or the other, the effect of having $16$ fermions start making sense, and it may eventually be that something is happening among these $16$ mass degenerate fermions. 

The fact is that the free action on lattice effectively described a $16$ spinor structure, each of which corresponds to the $16$ fermions or doublers. Fermion doubling effect being a characteristic of the lattice effect, it does not change under staggered transformation. The latter, doing a diagonalisation in Dirac space, each of the four identical components has the $16 \times 16$ structure. Since the four components are identical, these hidden $16 \times 16$ structures must also be identical. They cannot be non-singular and hence, can be expressed as block diagonal form. 

On expressing one of these four $16 \times 16$ matrix (in staggered space now, where we are dealing with only one out of four $16 \times 16$ matrices) in a block diagonal form, we get four matrices as effective diagonal elements, each of which is $4 \times 4$ dimensional. We interpret them as $4$ tastes of staggered fermions, each of which has the familiar $4$-spinor structure. We consider grouping together the $16$ sites of a hypercube and place these $16$ doubler fermions on the sites of hypercube. We consider non-intersecting hypercubes in $\hat{\mu}$ direction with labels $h_{\mu}$ with origins separated by $2a$~\cite{Gattringer.2009,Degrand.2006}, where $a$ is the lattice spacing. The fermionic action in taste space with the taste fields is given by

\begin{align}
    S_F[\adq,\q] &= b^4 \nsum[1.8]_h \Bigg[\nsum[1.6]_{t=1}^4 \left( m \hs \adq^{\left(t\right)}\left(h\right) \q^{\left(t\right)}\left(h\right) + \nsum[1.4]_{\mu=1}^4 \hs \adq^{\left(t\right)}\left(h\right) \hs \gamma_{\mu} \hs \nabla_{\mu} \hs q^{\left(t\right)}\left(h\right)\right) \notag \\
    &-\frac{b}{2} \nsum[1.4]_{t,t^{'}=1}^4 \nsum[1.4]_{\mu=1}^4 \adq^{\left(t\right)}\left(h\right) \gamma_5 \left(\tau_5\tau_{\mu}\right)_{ts} \hs \Delta_{\mu} \hs \q^{\left(s\right)} \left(h\right)\bigg]
    \label{eq:Fermionic action in taste space}
\end{align}

where $b=2a$ and $\tau_{\mu} = \gamma_{\mu}^T$, which are gamma matrices defined in taste space. Also, we have the following definitions of the symbols:

\begin{align}
    \nabla_{\mu}f\left(h\right) &= \frac{f\left(h + b\hs \hat{\mu}\right)-f\left(h - b\hs \hat{\mu}\right)}{2b} \notag \\
    \Delta_{\mu}f\left(h\right) &= \frac{f\left(h + b\hs \hat{\mu}\right)+f\left(h - b\hs \hat{\mu}\right)-2f\left(h\right)}{b^2}
\end{align}

The ﬁrst two terms in \autoref{eq:Fermionic action in taste space} are diagonal in taste space and represent the mass and kinetic terms for the four tastes of fermions expected to be described by \autoref{eq:stag ferm action}. The third term looks similar to a Wilson term, but mixes the diﬀerent tastes. This taste symmetry-breaking term reduces the symmetry of the kinetic term which is invariant under independent vector and axial rotations for each of the four tastes. The taste-breaking term is only invariant under the remaining symmetry $U\left(1\right) \times U\left(1\right)$ given by the rotations

\begin{align}
  \q \to e^{-i\theta} \hs \q  , \quad  \adq \to \adq \hs e^{i\theta}, \quad
  \q \to e^{i\theta \Gamma_5} \hs \q, \quad \adq \to \adq \hs e^{i\theta \Gamma_5}
\end{align}

where we have deﬁned the taste-mixing generator $\Gamma_5 = \gamma_5 \otimes \tau_5$. For a detailed derivation of the above ~\autoref{eq:Fermionic action in taste space}, please refer to~\cite{Gattringer.2009}. 

\subsection{Taste breaking effect}

The second term in \autoref{eq:Fermionic action in taste space} 
leads to mixing between the different tastes and breaks the degeneracy among the different tastes. Additionally, in the interacting theory, we find fermions of different tastes within a hypercube interact by exchanging gluons with momenta of the order of the cutoff scale. A large fluctuation in link variables gives rise to larger taste-breaking effects. Hence, reducing these unphysical ultraviolet fluctuations or the order of the cutoff (upper bound), will cause in reducing and suppressing the interaction among them. This can be done by smoothing each gauge link with a weighted sum of the neighboring paths keeping the endpoints fixed, referred to as smearing.

\section{Highly Improved Staggered Quarks (HISQ)}

The smearing algorithm that we use for our analysis gives us an improved action called Highly
Improved Staggered Quarks (HISQ) action~\cite{Follana.2007,Bazavov.2010,Bazavov:2011nk,Bazavov:2014pvz}. In HISQ action, the taste symmetry breaking effects are eliminated up to $\Ob(a^4)$ and this is achieved through smearing techniques. 

Smearing is a process used to improve the signal-to-noise ratio of observables calculated on lattice. This is very useful and important for providing reliable, accurate results and also for obtaining correct physical information from lattice QCD, where discretisation errors and statistical fluctuations can become pronounced. In smearing, the gauge field configurations are redistributed and are averaged over shorter lattice distances by considering and mixing the contributions from different lattice sites. This mixing among different lattice sites augments low-energy lying modes and illuminates low-energy physics by revealing information about physical quantities like hadron masses and decay constants. The redistribution of fields and shorter distance averaging reduces the effect of noise and improves the quality of signal, involving results of observables.

The HISQ action 
consists of two stages of smearing, namely fat link smearing and thin link smearing. A detailed discussion of this however, remains beyond the scope of this thesis. Naively in fat link smearing, the original gauge links are replaced by smeared gauge links that are constructed by averaging out the neighbouring gauge links, keeping fixed the endpoints that remain connected by these links. This form of smearing redistributes the field configurations extensively and also smoothens the high-frequency fluctuations, thereby reducing the impact of short distance lattice artifacts. In thin link smearing, the number of links considered for averaging is less and    

\begin{figure}[H]
    
    \includegraphics[width=0.99\textwidth]{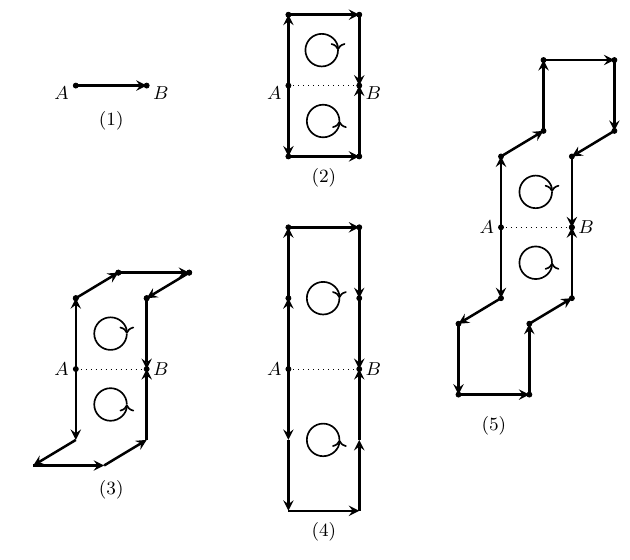}
    \caption{The kinds of paths between two fixed lattice points $A$ and $B$ for smearing gauge links in HISQ action. ($1$)  1-link path $L_{\mu}^1$, ($2$) 3-link path $L_{\mu}^3$, ($3$) 5-link path $L_{\mu}^5$, ($4$) Lepage term or coplanar 5-link path $L_{\mu}^{5f}$ and ($5$) 7-link path $L_{\mu}^7$. All of them start from $A$ and end at $B$ in two possible sense of direction shown by circular arcs.}
    \label{fig:All links}
\end{figure}

In the fat link smearing stage of HISQ action, the gauge links are replaced by averaging over the contributions of the adjoining links of the lattice, we smooth over $1$-link, $3$-link (fat3), 5-link (fat5),
and 7-link (fat7) with appropriate coefficients. The diagrammatic depiction of the paths is done in
3.3. This smoothing is represented as $\mathcal{S}_7^{L}$ and it is given by

\begin{equation}
    U_{\mu}(n) \to \mathcal{S}_7^{L} \hs U_{\mu} (n) = d_1 L_{\mu}^1(n) + d_3 L_{\mu}^3(n) + d_5 L_{\mu}^5(n) + d_7 L_{\mu}^7(n)
    \label{eq:1st stage smearing}
\end{equation}

Summing over these links takes it away from the $U(3)$ elements of the gauge variable and so we
project the obtained smeared link to a $U(3)$ element. In the second stage of smearing, the 5-link Lepage loop~\cite{Lepage.1999} is added along with all other terms of $\mathcal{S}_7^L$ given in \autoref{eq:1st stage smearing} and therefore is given by

\begin{equation}
    \mathcal{S}_7^{L}\hs U_{\mu}(n) \to \mathcal{S}_7^{(\mathcal{L})}\hs U_{\mu}(n) = \mathcal{S}_7^{L}\hs U_{\mu}(n) + d_{5f}\hs L_{\mu}^{5f}(n) 
\end{equation}

\renewcommand{\arraystretch}{1.5}

\begin{table}[ht]
\centering 
\begin{tabular}{|P{2.5cm}|P{2cm}|P{2cm}|} 
\hline
Coefficient & $\mathcal{S}_7^L$ & $\mathcal{S}_7^{(L)}$    \\ \hline 
$d_1$ & $1/8$ & $1$   \\ \hline
$d_3$ & $1/16$ & $1/16$    \\ \hline
$d_5$ & $1/64$ & $1/64$    \\ \hline
$d_7$ & $1/384$ & $1/384$    \\ \hline
$d_{5f}$ & $0$ & $-1/8$    \\ \hline
\end{tabular}

\vspace{6mm}

\caption{The smearing coefficients for the first $\mathcal{S}_7^{L}$ and second stage $\mathcal{S}_7^{(\mathcal{L})}$}
\label{table:smearing coefficients} 
\end{table}

The coefficients for both the stages are given in above \autoref{table:smearing coefficients}. They are obtained from perturbation theory results to minimise taste breaking and their effects~\cite{Bazavov.2010}. 

\begin{figure}[H]
    
    \includegraphics[width=0.99\textwidth]{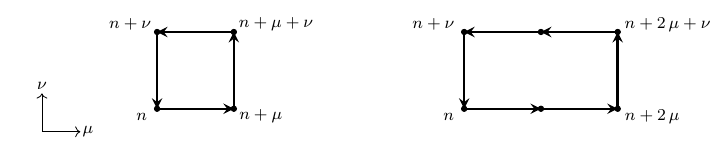}
    \caption{$1 \times 1$ (left) and $2 \times 1$ loops (right) used for Symanzik improvement on a lattice with unit lattice spacing.}
    \label{fig:1x1 and 2x1 loops}
\end{figure}

The Symanzik improvement for fermions that we
use here is by adding the Naik term~\cite{Naik.1989}, which is added to the naive term in the staggered action along with the choice of appropriate coefficients such that $\Ob(a^2)$ contribution in the kinetic terms get cancelled out at the tree level. This Naik term is a $3$-link term and in its presence, the Symanzik improved staggered fermion action is given by 

\begin{align}
    S_{sk} &= a^4 \sum_{n \in \mathbb{L}} \overline{\xi}(n) 
    \left(\sum_{\mu=1}^4 \eta_{\mu}(n) \,\frac{9}{16}\left[\frac{U_{\mu}(n) \,\xi(n+a\hat{\mu}) - U_{\mu}^{\dagger}(n-a\hat{\mu})\,\xi(n-a\hat{\mu}))}{2a} \right] + m\,\xi(n)\right)  \notag \\
    &- a^4 \sum_{n \in \mathbb{L}} \overline{\xi}(n) 
    \left(\sum_{\mu=1}^4 \eta_{\mu}(n)\,\frac{1}{48} \left[\frac{N_{\mu}(n)\,\xi(n+3a\hat{\mu}) - N_{\mu}^{\dagger}(n-a\hat{\mu})\,\xi(n-3a\hat{\mu})}{6a}\right]\right)
\end{align}
The Naik term contributes the term $N_{\mu}(n)$ which is given by the product of three adjoining links $U_{\mu}(n)\,U_{\mu}(n+a\,\hat{\mu})\,U_{\mu}(n+2a\,\hat{\mu})$. The relative coefficients between the terms have been tuned accordingly so that they eliminate discretisation errors from $\Ob(a^2)$ at the tree level. As illustrated in the above \autoref{fig:1x1 and 2x1 loops}, we use only $1 \times 1$ and $2 \times 1$ loops~\cite{Beinlinch.1999} in the gluonic action. On improvement, the gluonic action is given as 

\begin{equation}
    S_g^{(im)}[U] = \sum_{n \in \mathbb{L}} \sum_{\mu<\nu} 
    \alpha_1 \,\left(1 - \frac{1}{3}\,\text{Re} \,\text{Tr} \,U_{\mu\nu}^{1\times 1}(n)\right) +\alpha_2 \,\left(1 - \frac{1}{6}\,\text{Re}\, \text{Tr} \, U_{\mu\nu}^{2\times 1}(n)\right) 
    \label{eq:gluon corrections}
\end{equation}
In the above \autoref{eq:gluon corrections}, the $U_{\mu\nu}^{1\times 1}(n)$ and $U_{\mu\nu}^{2\times 1}(n)$ represent the usual plaquette expressions for the $1 \times 1$ and $2 \times 1$ loops in \autoref{fig:1x1 and 2x1 loops}. The coefficients $\alpha_1$ and $\alpha_2$ in the above Symanzik-improved gluon action is tuned so that $\Ob(a^2)$ discretisation effects are nullified at the tree level. The factor $1/6$ is because this method of improvement employs six such $2 \times 1$ rectangles in the lattice, and one out of six identical contributions are considered. These Symanzik improvements (see \autoref{sec:Symanzik improvement}) reduce discretisation effects due to lattice, and cancel out errors appearing upto $\Ob(a^2)$.

It has been shown that the HISQ action
greatly reduces the pion mass splitting for finite lattice spacing~\cite{Follana.2008} with taste exchange interactions
at least about $4$ times smaller than other popular staggered actions.

\section{Utility of staggered formulation}

Staggered fermions are widely used for dynamical simulations. The reason is that due to reduced number of degrees of freedom (no Dirac structure), staggered fermions are numerically cheaper to simulate and less expensive. This staggered formulation also causes explicit breaking of the $U(4) \times U(4)$ symmetry group, which is the symmetry group of the usual massless fermionic action in continuum spacetime, to $U(1)_V \times U(1)_{\epsilon}$. The $U(1)_V$ corresponds to the usual conservation of baryon number. The $U(1)_{\epsilon}$ which is a subgroup of 
 $SU(4)_A$, is the remnant chiral symmetry preserved by the staggered fermions.
 However, a problem is that the number of doublers is reduced from sixteen to four only. These four doublers are called tastes. Hence the staggered action describes four tastes of quarks, while in a realistic $2+1$ flavor QCD simulation, one usually have two light mass degenerate up ($u$) and down ($d$) quarks and one heavier strange quark ($s$), with the ratio $m_s/m_l$ being $27$ physically. In order to suitably reduce the number of degrees of freedom to match with the number of flavors in real QCD simulation, it has been proposed to express the QCD partition function in form of a path integral with an eﬀective action $S_{eff}$ in terms of a decaying Boltzmann weight given by

\begin{equation}
    \exp{(-S_{eff})} = \exp{(-S_G)} [\det \M_{st}(m_{ud})]^{1/2} [\det \M_{st}(m_{s})]^{1/4} 
    \label{eq:stag corr}
\end{equation}

where we have 

\begin{equation}
    \left[\det \M_{st}(m_{ud})\right]^{1/2} = \left[\det \M_{st}(m_{u})\right]^{1/4} \left[\det \M_{st}(m_{d})\right]^{1/4}
\end{equation}

Here,in \autoref{eq:stag corr}, $m_{ud}$ is the average of $u$ and $d$ quark masses and $m_s$ is the strange quark mass. From a mathematical point of view, taking the quartic roots of the individual determinants is not problematic, since as we have shown it is real and positive. This arises from the fact that the partition function obtained as the path integral spanning over all possible gauge configurations contains a probability density function as its integrand, which is $e^{-S_G} \det \M$, with fermion matrix $\M$. Since, a probability density function is positive definite, hence $\det \M$ is positive, since $e^{-S_G}$ is positive, without loss of generality. 

However, this procedure is quite nontrivial from a conceptual perspective and we must ask ourselves if the universality class remains the same in this approach. Probably even more important is the question whether the eﬀective action can be expressed in the form of a local lattice ﬁeld theory. For a snapshot of the ongoing debate about these issues, see references~\cite{Adler.1969,Adler.1969.1517,Bell.1969,Hooft.1976}. Although the conceptual problems are not all resolved, simulations with staggered fermions have found good agreement with experimental results. Examples are found in~\cite{Goldstone.1961,Nicola.2016,Nicola.2018,Sandmeyer.2019}.

\section{Chemical potential on lattice : Sign problem}

From our thermodynamic knowledge, we know that $\mu N$ has the dimensions of energy, since free energy contains a term $\sim \mu N$ where $\mu$ is the chemical potential and $N$ is the particle number. Hence, from a dimensional perspective, a term containing $\mu N$ is legitimate to add to the Lagrangian of the theory, having mass dimension $1$, similar to energy. Following this dimensional trait, we observe the chemical potential enters in the expression of Dirac operator in lattice getting coupled with the quark number density $N_q$, which is the spatial volume integral over the temporal component $\adq(x) \hs \gamma_4 \hs \q(x)$ of the conserved Euclidean $4$-current $\adq \hs \gamma_{\mu} \hs \q$ in continuum $3+1$ dimensional spacetime with spacetime points $x$ as shown below. 

\begin{equation}
    N_q = \int d^3x \adq(x) \gamma_4 \q(x)
\end{equation}

The Lagrangian density $\mathcal{L}$ therefore contains $\mu \hs \adq(x) \hs \gamma_4 \hs \q(x)$. With this introduction of $\mu$ in the lattice gauge theory, the Dirac operator in momentum space is given by~\cite{Gattringer.2009}

\begin{equation}
    \Tilde{D}_{\mu}(p) = \Tilde{D}_{0}(p) + a\mu \gamma_4
    \label{eq:mu Dirac op}
\end{equation}

where 

\begin{equation}
    \Tilde{D}_0(p) = m I_4 + \frac{i}{a} \sum_{\mu=1}^4 \sin{\left(p_{\mu}a\right)}
    \label{eq: mu zero Dirac op}
\end{equation}

$\Tilde{D}_{\mu}(p)$ is the Dirac operator with finite chemical potential $\mu$ in \autoref{eq:mu Dirac op} and $\Tilde{D}_{0}(p)$ is the same for zero chemical potential in \autoref{eq: mu zero Dirac op}. However, the introduction of $\mu$ in this linear way incorporates ultraviolet divergences in the continuum limit ($a \to 0$). These UV divergences remain upto $\mathcal{O}(a^{-4})$. An exponential term of the form $\exp{(a\mu)}$ is commonly used\,\cite{Gattringer.2009} to override these divergences, using which we have calculated the different correlation functions in our work. Although this term works well beyond fourth power of momentum $\sim a^{-1}$, the linear term is mostly used in this domain following the linear $\mu$ formalism\,\cite{Gavai.2012linear,GAVAI.2015}, as the former is computationally expensive and almost gives similar agreeable results to the latter. This is also, because these ultraviolet divergences vanish beyond $\mathcal{O}(a^{-4})$. and this is the same we have used in the work followed in this thesis. 

The introduction of $\mu$ also poses a serious technical
drawback, which we have described here. For non-zero finite chemical potential, the Dirac operator is no longer $\gamma^5$ hermitian, and the modified hermiticity equation for Dirac operator becomes 

\begin{equation}
    D^{\dagger}(-\mu) = \gamma^5 \hs D(\mu) \hs \gamma^5
\end{equation}

Consequently for non-vanishing real
$\mu$, the determinant of the Dirac operator becomes complex. See \autoref{Appendix 9} for a quick proof. A
non-vanishing real $\mu$ creates a particle–antiparticle asymmetry which obscures
the determinant, being real and a straightforward application of importance
sampling. This is the complex measure problem~\cite{Alford.2001.61}. The staggered formulation also fails to work, since taking the fourth root of the determinant would now give branch cuts and multi-valued functions. 

The reweighting procedure, as we will observe and discuss in the next chapter, re-enables Monte-Carlo techniques by reweighting the measure at a zero $\mu$, but then the observable becomes complex and the complex phases appearing in the observable gives rise to the notorious sign problem. Thus, the complex measure problem assumes the form of a sign problem~\cite{Fodor.2002,deForcrand.2009,Nagata.2022}, the problem which has remained a stern hurdle in our exploration of finite density QCD. And it is this very problem that has compelled us, every now and then to look for different methods to circumvent it and explore finite density QCD, which is important for understanding the phase diagram conclusively. 

 	\cleardoublepage
\newcommand{\benn}{\nonumber\begin{equation}}
\newcommand{\eenn}{\nonumber\end{equation}}
\def\bea{\begin{eqnarray}} \def\eea{\end{eqnarray}}
\def\beann{\begin{eqnarray*}} \def\eeann{\end{eqnarray*}}

\chapter{Different approaches to the sign problem in QCD}
\label{Chapter 4}
\vspace{2cm}

\graphicspath{{Figures/Chapter-4figs/PDF/}{Figures/Chapter-4figs/}}
\graphicspath{{Figures/Chapter-4figs/EPS/}{Figures/Chapter-4figs/}}

\section{A brief Introduction}

As pointed out in the last chapter, the complex measure problem for real finite $\muB$ in the path integral of partition function restricts standard Monte Carlo sampling and related Monte-Carlo techniques to a very small value of non-vanishing finite real $\muB$. In fact, the quest for the optimal simulation strategy for non-vanishing chemical potential is far from being settled. In this chapter, we discuss some of the oft-used prevailing approaches for a study of finite-density QCD. Almost all the different approaches so far have analytically continued and extrapolated results obtained from measurements for real determinants to the actual parameter or observable values, one is interested to calculate or is of central interest in the study. Mostly, there have been two groups of such extrapolations:

\begin{itemize} 
\item Using results determined for purely imaginary $\mu$ with $(\mu^2 < 0)$ and subsequent analytic continuation to real $\mu$ with $\mu^2 > 0$  via the fit of a power series ansatz in $\mu$ and Pad{\'e} rational expansion or reconstruction of the fugacity expansion coeﬃcients via Fourier transformation.
\item Using Taylor expansion of observables in terms of $\hat{\mu} \equiv \mu/T$, for finite temperature analysis.
\item Using measurements at $\mu=0$ and extrapolating with the help of reweighting.
\end{itemize}

Quenched simulations~\cite{Barbour.1986,Kogut.1995,Barbour.1998} at baryon chemical potential ﬁrst led to confusing results. It was found at finite temperature $T$, that the critical value of $\mu_B$ decreased with the pion mass $m_{\pi}$ following $\mu_{B,c} \propto m_{\pi}/2$, which is expected to vanish in the chiral limit (limit of zero quark masses). On the other hand at $T=0$, one expects that the transition is near $\mu_B \approx m_p/3$, with proton mass $m_p$ since the proton is the lightest
baryon, in the limit of mass degenerate up and down quarks in $2$ flavor QCD. Both this observation and argument led to confusion regarding the correctness of transition. For zero $\mu_B$, the quenched theory is a theory with $N_f$ quarks and $N_f$ conjugate quarks~\cite{Stephanov.1996}, instead of $N_f \to 0$ limit of QCD. This implies that for a simulation, we need to attend carefully to the phase of the quark determinant and dynamical fermions are necessary to obtain a $\mu_{B,c} \approx m_p/3$ like situation. Most simulations in that context are
therefore done with dynamical fermions, mainly of the staggered type, similar to the lines of our work with HISQ and a Symanzik improved gauge action, both of which are mentioned in the previous~\autoref{Chapter 3}.
Large temperature $T \gg T_c$ corresponds to very small temporal extension in lattice as $T = 1/aN_{\tau}$, with lattice spacing $a$ and $N_{\tau}$ temporal sites and one expects that the system approaches eﬀectively a non-relativistic 3D gauge theory~\cite{Ginsparg.1980,Thomas.1981,Kajantie.1997,Hart.2000,Hart.2001,Kajantie.2003}. In our work, we have not considered the chiral limit and also, our highest working temperature $T \approx 176$ MeV, which makes $T/T_{pc} \approx 1.12$. $T_{pc}$ is the pseudo-critical temperature which is roughly $157$ MeV, for the values of couplings and quark masses, we have used (Refer to LCP in \autoref{Appendix 2}). Hence, it is safe to restrict our discussion of methods for the full $4$ dimensional system.

\section{Approach of analytical continuation}
\label{sec:Analytic continuation}

The idea of analytic continuation from imaginary to real chemical potentials~\cite{Borsanyi.2018,Ratti.2018,Gunther.2016} gets motivated from the fact that Monte-Carlo simulations of thermodynamic observables at purely imaginary $\mu$ do not suffer from a sign problem. This is  explicitly proven in \autoref{Appendix 4} of this thesis. Hence, one adopts the following strategy in this approach : 
perform independent simulations at different values of the imaginary chemical potential $\mu = i \mu_I$, where $\mu_I$ is the imaginary part of $\mu$, fit the results with an ansatz, and analytically continue the ansatz to real $\mu$.

If the ansatz is a polynomial or a power series in $\mu$, then the fit parameters are the usual Taylor coefficients. The power series is expanded about values of $\mu$, say $\mu_E$ which lie within the radius of convergence $R$ of the series, such that $\left|\mu - \mu_E\right| \hs < R$, where $R$ is the distance from the origin ($\mu=0$) to the closest singularity of the observable calculated. A
standard method in the theory of analytic functions is to perform a sequence of expansions around points located suitably in the convergence domain of the preceding series. Other methods involve optimal mappings or Pad{\'e} expansions. The latter is a systematic method to replace the power series by
a rational function which has identical expansion coeﬃcients(see \autoref{sec:Pade resummation}).
Although this approach has been used mostly to determine the pseudo-critical temperature $T_{pc}$ as a function of the chemical potential $\mu$, it has
also been applied to the pressure and other observables.

At low temperature, the pressure is best described by a hadron resonance gas ansatz. For $T \geq 0.95 \hs T_{pc}$,
this ansatz becomes poor, and a better description is obtained by a Taylor expansion, which is sensitive to sixth order Taylor coefficient $c_6$, appearing alongside $(\mu_B/T)^6$ in a Taylor series. Similar observations have been made in Ref.~\cite{Takaishi.2010} on a smaller lattice with small lattice four volume, where all derivatives
in up $(\mu_u)$ and down $(\mu_d)$ quark chemical potentials up to $4^{th}$ order have been calculated as a function of quark or baryon chemical potential. Similar study has been done implementing this approach in Ref.~\cite{D'Elia.2009} where simulations are performed only with the quark number density, that is, the first derivative of the pressure, as a function of imaginary baryon and isospin chemical potentials.

 It turns out that convergence is rapid for several observables studied, like the chiral condensate and screening masses~\cite{Hart.2000}, as well as the position of the crossover (the
pseudo-critical temperature)~\cite{deForcrand.2007,DEFORCRAND.2002,D'Elia.2003,D'Elia.2004}. Padé approximants (refer to \autoref{sec:Pade resummation}) may allow
extension of the extrapolation range beyond the convergence circle of the
power series. 

A important technical issue should be addressed: how to choose the simulated values of imaginary chemical 
potential and the statistics for each value, so as to maximize the accuracy on a given set of Taylor coefficients.
Larger values of $\mu_I$ increase the sensitivity to the desired higher-order terms, but also enhance the truncation error in the fitted Taylor polynomial.

\section{Taylor Expansion}
\label{sec:Taylor Expansion}

The previous \autoref{sec:Analytic continuation} on analytic continuation approach paves the way automatically for the study of Taylor Expansion of thermodynamic observables like excess pressure, number density in terms of chemical potential $\mu$~\cite{Allton.2005,Bazavov.2017,Bollweg.2021}. The $CP$ invariance of QCD and time reversal of gauge configurations provide the following symmetry for partition function $\Z$

\begin{equation}
    \Z(-\mu) = \Z(\mu)
    \label{eq:CP symmetry and time reversal}
\end{equation}

Thus $\Z$ and observables that follow $CP$ symmetry of QCD and time reversal invariance of gauge configurations are even functions in $\mu$. We know, the thermodynamic pressure in a homogeneous thermodynamic system of volume $V$ and temperature $T$ is given by 

\begin{equation}
P = \frac{T}{V} \ln \Z \quad  \Rightarrow \quad \frac{P}{T^4} = \frac{1}{VT^3} \ln \Z
\label{eq:Basic thermodynamics pressure equation}
\end{equation}

Equating \autoref{eq:CP symmetry and time reversal} and \autoref{eq:Basic thermodynamics pressure equation}, we find the corresponding Taylor series for dimensionless excess pressure $\Delta P/T^4$ is even in $\mu/T$ and dimensionless number density $\mathcal{N}/T^3$ odd in $\mu/T$. Up to $\Ob(\mu^N)$, they are given as follows:

\begin{equation}
    \frac{\Delta P}{T^4} = \frac{P(\mu)}{T^4} - \frac{P(0)}{T^4} = \nsum[1.5]_{n=1}^{N/2} c_{2n} \left(\frac{\mu}{T}\right)^{2n}
    \label{eq:Taylor series of excess pressure}
\end{equation}

\begin{equation}
    \frac{\mathcal{N}}{T^3} = \frac{\partial}{\partial \mu}\bigg[\frac{\Delta P}{T^4}\bigg] = \nsum[1.5]_{n=1}^{N/2} c_{2n} \left(\frac{\mu}{T}\right)^{2n-1}
    \label{eq:Taylor series of number density}
\end{equation}

The coeﬃcient of the quadratic term in \autoref{eq:Taylor series of excess pressure} i.e. $c_2$ is the quark number susceptibility. It is a first order $\mu$ derivative of the quark number density of \autoref{eq:Taylor series of number density} and a second derivative of excess pressure and therefore, $\ln \Z$ with respect to the chemical
potential, as per \autoref{eq:Basic thermodynamics pressure equation}, calculated at $\mu=0$

\begin{equation}
    c_2 = \frac{\partial}{\partial(\mu/T)} \bigg[\frac{\mathcal{N}}{T^3}\bigg]\Bigg|_{\mu=0} = \frac{1}{VT^3} \frac{\partial^2 \ln \Z}{\partial(\mu/T)^2}\Bigg|_{\mu=0}  
    \label{eq:quark number susceptibility}
\end{equation}

Due to the path integral of $\Z$ in the form of fermion determinant as in \autoref{sec:Exp resummation} and also by virtue of the formula $\det \left[\exp{\left(M\right)}\right] = \exp{\left[\Tr \ln \left(M\right)\right]}$ ,such derivatives involve traces like of the following (more about this in \autoref{Appendix 3} and also in \autoref{Chapter 6}):

\begin{equation}
    \frac{\partial \ln \det \M}{\partial \mu}  = \Tr \bigg[\M^{-1} \frac{\partial \M}{\partial \mu}\bigg]
\end{equation}

 These Taylor coefficients $c_n$ are functions of the temperature $T$, since the observables are functions of $T$. A detailed study of Taylor expansion for different orders in $\mu$ enables one to understand the intrinsic thermodynamic behaviour of the observables and other relevant properties in terms of the convergence of the series~\cite{Allton.2005,Choe.2002}. Although formally the radius of convergence $\rho$ of the above series in \autoref{eq:Taylor series of excess pressure} is given by

 \begin{equation}
     \rho = \lim_{n \to \infty} \sqrt{\frac{c_n}{c_{n+2}}}
     \label{eq:Radius of convergence}
 \end{equation}
 for practical purposes, one usually computes estimates of the radius of convergence at a given temperature $T$ using different ordered Taylor coefficients for $n=2,4,6,...$ and obtains a measure of how rapid or slow the convergence is, for the given $T$. This determines the efficacy of the Taylor series in the study of observables, as compared to other approaches. The square root in \autoref{eq:Radius of convergence} is because the Taylor series of \autoref{eq:Taylor series of excess pressure} is even in $\mu/T$, due to the aforementioned $CP$ symmetry of QCD and time reversal invariance of gauge field configurations.


\begin{figure}[H]
    \centering
    \hspace{-0.1cm}
    \includegraphics[width=0.32\textwidth]{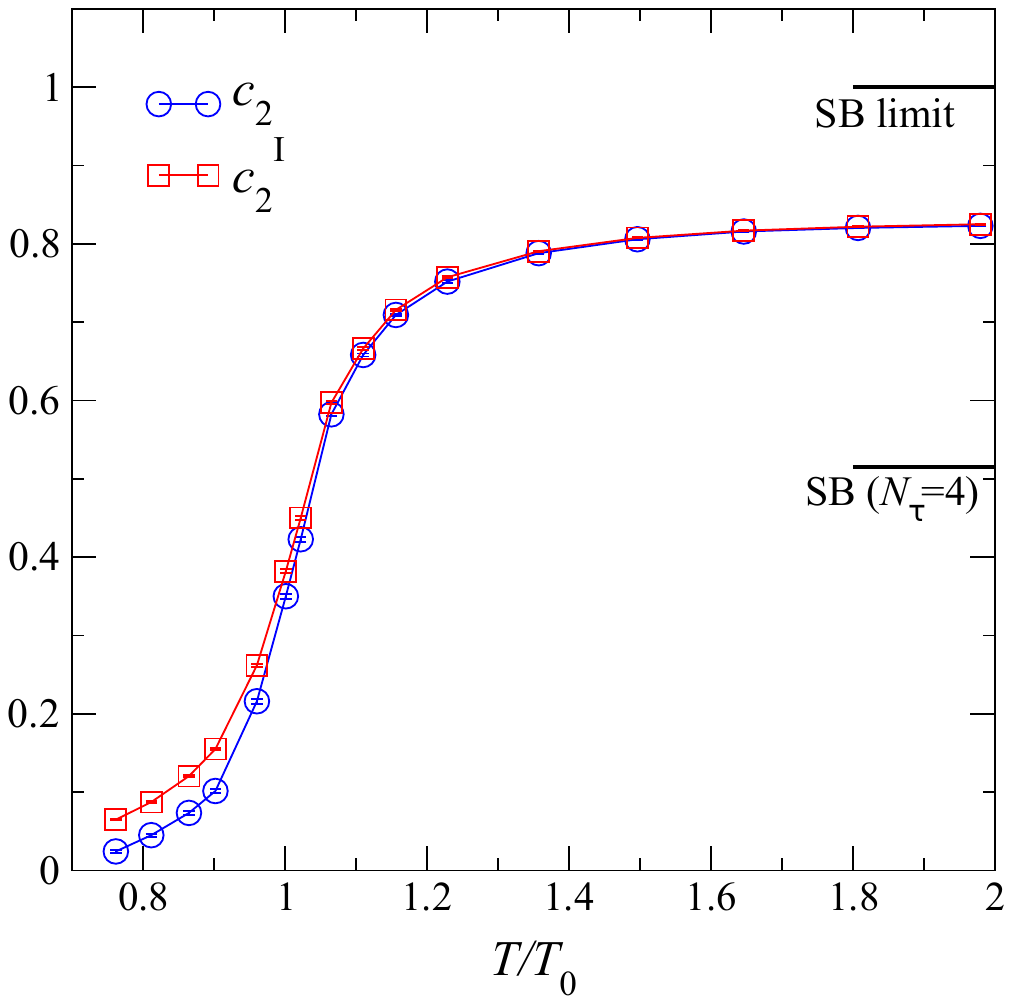}
    \includegraphics[width=0.32\textwidth]{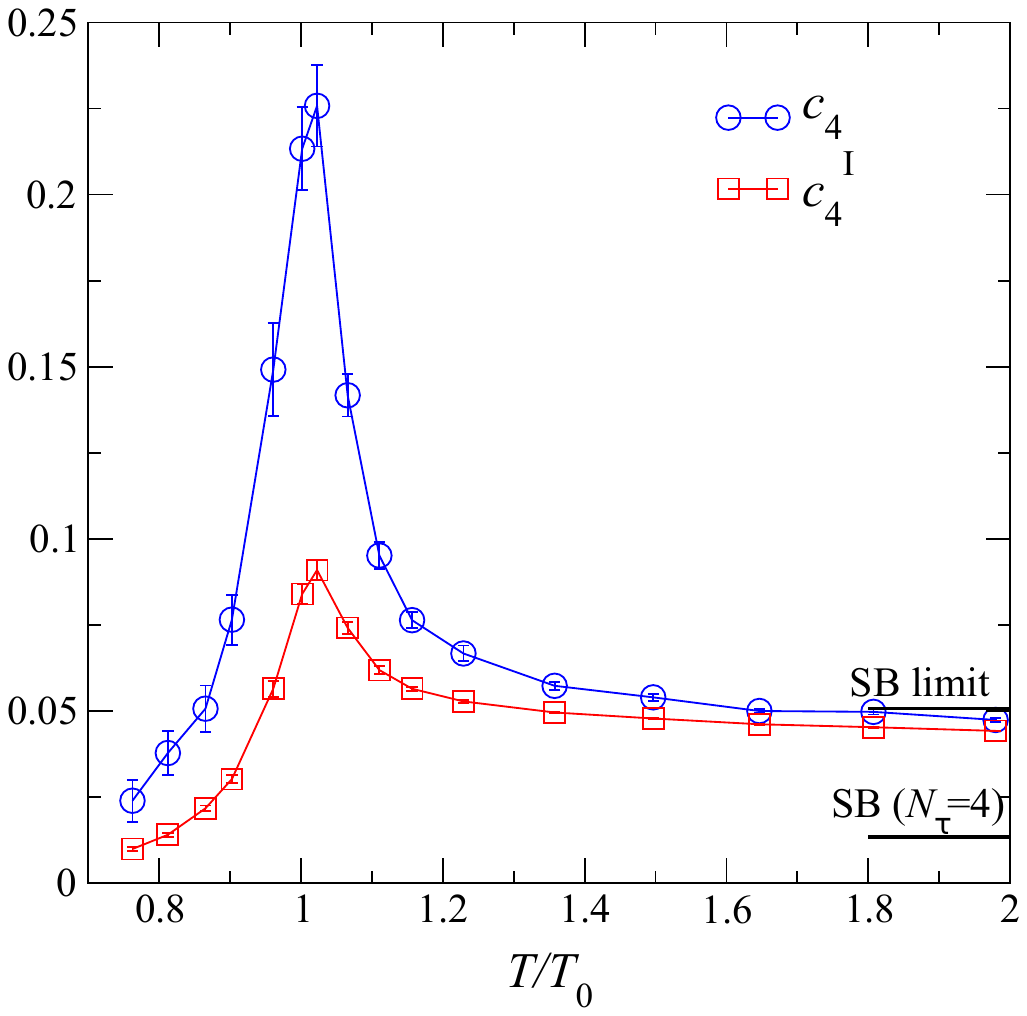}
    \includegraphics[width=0.32\textwidth]{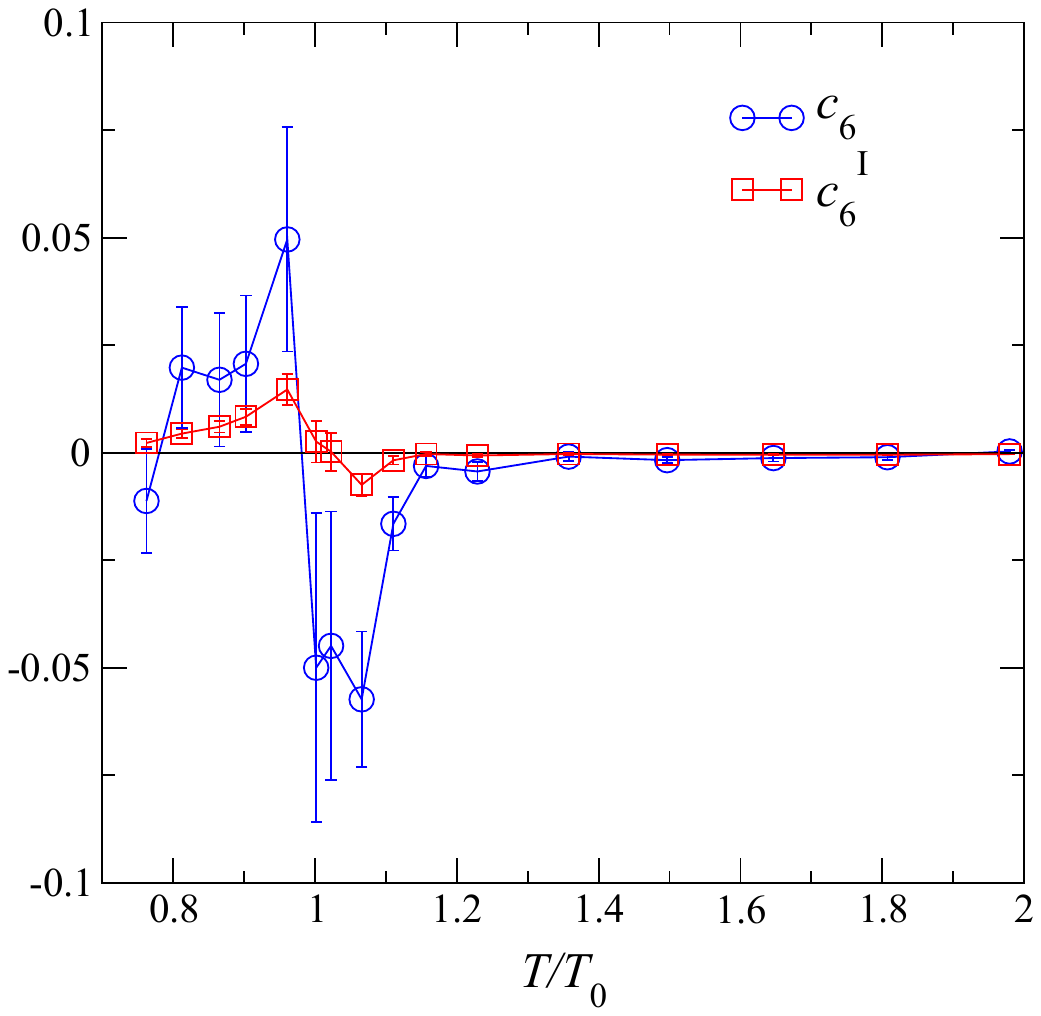}
    \caption{The quark number Taylor expansion coefficients $c_n^q$ and isospin Taylor expansion coefficients $c_n^I$ for $n = 2, 4$ and $6$ as function of $T/T_{0}$, where $T_0$ is the transition temperature. This figure is borrowed from~\cite{Allton.2003}.}
    \label{fig: The Taylor coefficients Allton}
\end{figure}

 In the thermodynamic limit phase transitions will be associated with singularities in some observables. The coeﬃcients of the expansion can be used to estimate the position of the closest singularity, as described before. As seen in \autoref{fig: The Taylor coefficients Allton}, the quark number and the isospin susceptibilities exhibits a pronounced peak at the transition~\cite{Allton.2003}, where they have worked on a $16^3 \times 4$ lattice with the transition temperature $T_0 \approx 170$ MeV. The isospin susceptibilities are second order derivatives of thermodynamic potential $\Omega \sim \ln \Z$ with respect to dimensionless isospin chemical potential $\mu_I/T$. Finite size scaling considerations then may give hints on the type of phase transition or crossover. A whole lot of works~\cite{Choe.2002,Gavai.2003,Allton.2003,Gavai..2005,Allton.2002} have considered analyzing the quark number susceptibility and other expansion coeﬃcients in this way. However, the computation for $c_n, n \geq 6$ is extremely tedious and highly expensive and also for thermodynamic pressure, we find the Taylor series has a very slow convergence (more in \autoref{sec:Exp resummation}),i.e. $\rho \approx 1$, as per \autoref{eq:Radius of convergence}. This ultimately directs one towards different resummation techniques, like Pad{\'e} resummation and Exponential resummation, discussed in \autoref{sec:Pade resummation} and \autoref{sec:Exp resummation} respectively of the next \autoref{Chapter 5}.

\section{Reweighting method}
\label{sec:Reweighting}

An alternative to  Taylor series expansion is reweighting approach. Reweighting is a standard method in Monte Carlo approaches for statistical spin systems and has been successfully used to improve interpolation between Monte Carlo results at diﬀerent couplings and also for analytical continuation from
real to complex couplings. It has been particularly useful for determining the singularities or partition function zeroes for complex couplings (Lee–Yang zeroes~\cite{Peng.2015,Basar.2021} and Fisher zeroes~\cite{Liu.2019}).

In QCD thermodynamics, these couplings are related to temperature $T$ of the system. The principal idea is to simulate gauge ensembles generated at $\mu=0$ at a given $T$ and by reweighting path integral integrands, evaluate observables at some finite value of $\mu$ for the same $T$. This extrapolation is done to make the integral measure real and Monte-Carlo averaging effective. This is surely a special case for the more general multi-parameter reweighting, where the extrapolations are done for both the temperature $T$ and the chemical potential $\mu$. The efficacy of this extrapolation depends on the extent of overlap between the distribution of ensembles for the target parameters and the parameters of the simulated theories. Let's discuss this overlap issue briefly.

\subsection{Overlap problem}

A generic reweighting method aims to construct expectation values in a desired target theory $t$, with parameters $U$, the path-integral measure $W_t(U)$, and partition function of the target theory $\Z_t = \int \mathcal{D}U \hs W_t(U)$, using simulations from a theory $s$ with positive definite real weights $W_s(U)$ and partition function $\Z_s = \int \mathcal{D}U \hs W_s(U)$ via the formula as following:

\begin{align}
\big \langle \Ob(U) \big \rangle_t &= \frac{1}{\Z_t} \int \mathcal{D}U \hs W_t(U) \hs \Ob(U) \notag \\
&= \frac{\int \mathcal{D}U \hs W_s(U) \hs (W_t/W_s) \hs \Ob(U)}{\int \mathcal{D}U \hs W_s(U) \hs (W_t/W_s)}
= \frac{\left \langle \frac{W_t}{W_s} \hs \Ob \right \rangle_s}{\left \langle \frac{W_t}{W_s} \right \rangle_s}
\label{eq:reweighting equation model}
\end{align}

where $\left \langle \Ob \right \rangle_t$ and $\left \langle \Ob \right \rangle_s$ are the expectation values of $\Ob$ in the target and the simulated theories respectively.

In the language of the above, when the target theory is lattice QCD at a finite chemical potential, the target weights $W_t(U)$ have
wildly fluctuating phases and this is precisely the infamous sign problem. In addition to this problem, generic
reweighting methods also suffer from an overlap problem. This happens when the probability distribution of the
reweighting factor $W_t/W_s$ obtained from different configurations in the ensemble sample has a long tail, which cannot be sampled efficiently in standard
Monte Carlo simulations. The overlap problem is present even in cases and situations when the target theory does not suffer from a sign problem, such as reweighting to a theory with a different bare inverse gauge coupling $\beta$.

\begin{figure}[H]
    \centering
    \includegraphics[width=0.6\textwidth]{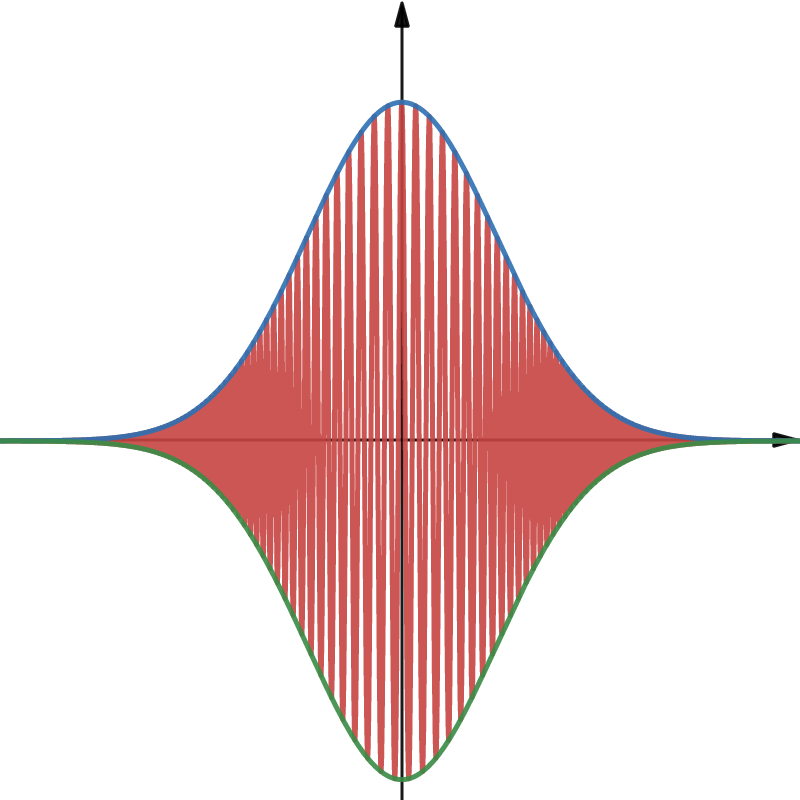}
    \caption{An example of a highly oscillating integrand $W_t/W_s$ providing a naive illustration of the sign problem as well as the overlap problem.}
    \label{fig: Reweighting integrand}
\end{figure}

The above \autoref{fig: Reweighting integrand} is a culmination of both sign problem and overlap problem. The former causes highly fluctuating behaviour of the integrand forming the measure of a path integral, whereas the latter is responsible for the long tail of the distribution. This suggests qualitatively at least, that majority of the gauge configurations of the sample considered do not provide sufficient relative contribution of the reweighting factor values. Although strictly speaking not true, we will find that the blue and green envelopes of the above \autoref{fig: Reweighting integrand} give a naive pictorial representation and idea about the unsigned and signed phase quenched reweighting factors. These quantities being free of the phasefactor, do not exhibit rapid oscillations with increasing value of the chemical potential. On the other hand, the rapid oscillations shown in red demonstrates the complete reweighting factor naively, which is the product of phase quenched reweighting factor and phasefactor. Besides phase quenched part, it also captures the contribution of the phasefactor. These high magnitude of oscillations reflect the oscillatory nature and behaviour of the phasefactor, and we will eventually find that this phasefactor does have the functional form of cosine in chemical potential. All this requires extensive study regarding numerical integration of highly oscillatory functions~\cite{Milovanovic:2013oscillatoryintegration}. In computational terms, these oscillations cause lower values of Metropolis acceptance discussed in \autoref{Chapter 3} and hence, a lot of important Monte-Carlo time goes by without capturing good number of effective configuration from the sample with appreciable values of reweighting factors. As a result, the Monte-Carlo sampling method is rendered highly inefficient in such situations with a long tail of the distribution of reweighting factors. 

In case of $\mu_B$, this overlap problem gets suppressed by the dominant sign problem and one mostly deals with the sign problem rather than the former in this case.  The overlap problem comes to forefront predominantly for isospin chemical potential $\mu_I$, where there is no sign problem. One also resorts to sign quenched reweighting to avoid sign problem in $\mu_B$ and focus on the impending pertinent overlap problem. 

\subsection{Sign problem}

The method of reweighting enables one to find a numerical estimate of the path integral of $\Z$ using Monte-Carlo simulations. This is because, this method scales the otherwise complex measure of the integral with a real-valued function and includes the complex ratio of the measures inside the observable part of the integral by keeping the measure real. This happens because, the fermion determinant is real at $\mu=0$ and complex for real $\mu \neq 0$, except for isospin chemical potential $\mu_I$, where the fermion determinant remains real for all real values of $\mu_I$. For a quick proof, refer to \autoref{Appendix 9}. In general, for reweighting a target theory with $(\beta, \mu)$ by simulating a theory with $(\beta_0,\mu_0)$, the general mathematical expression for the partition function $\Z(\beta,\mu)$ of the target theory looks as follows : 

\begin{align}
    \Z(\beta,\mu) &= \int \mathcal{D}U e^{-S_G(\beta,U)} \det \M(\mu,U) \notag \\
    &= \int \mathcal{D}U e^{-S_G(\beta_0,U)} \det \M(\mu_0,U) \left[e^{\Delta S_G(\beta,U)} \hs \frac{\det \M(\mu,U)}{\det \M(\mu_0,U)}\right] \notag \\
    &= \left \langle e^{\Delta S_G(\beta,U)} \hs \frac{\det \M(\mu,U)}{\det \M(\mu_0,U)} \right \rangle_{(\beta_0,\mu_0)}
    \label{eq:reweighting equation in our work}
\end{align}

where $\Delta S_G(\beta,U) = S_G(\beta,U) - S_G(\beta_0,U)$ and the ratio of the two fermion determinants at $\mu$ and $\mu_0$, $\det \M(\mu)/\det \M(\mu_0)$ is the reweighting factor in the above \autoref{eq:reweighting equation in our work}. This reweighting factor $\det \M(\mu)/\det \M(0)$, being a complex number can be written in form of $Re^{i\theta}$, where $R = f(\beta,\mu\hs |\hs \beta_0,\mu_0)$ and $\theta = g(\beta,\mu\hs |\hs \beta_0,\mu_0)$ given as  

\begin{align}
   R= f(\beta,\mu\hs |\hs \beta_0,\mu_0) = \left|\frac{\det \M(\beta,\mu)}{\det \M(\beta_0,\mu_0)}\right|, \quad 
   \theta =  g(\beta,\mu\hs |\hs \beta_0,\mu_0) = \tan^{-1}\left[\frac{\text{Im}(\M/\M_0)}{\text{Re}(\M/\M_0)}\right]
    \label{eq;phase quenched reweighting factor and phasefactor}
\end{align}

where $\M \equiv \M(\beta,\mu)$ and $\M_0 \equiv \M(\beta_0,\mu_0)$. $R$ is called the phase-quenched reweighting factor and $\theta$ is the phaseangle and $\left \langle \cos{(\theta)} \right \rangle$ is the gauge ensemble averaged phasefactor. As discussed above, calculation of this reweighting factor goes through sign problem and overlap problem. In case of baryon chemical potential $\mu_B$, the sign problem becomes highly severe with increasing real $\mu_B$ and more dominant over overlap problem. The average phasefactor with the phaseangle in \autoref{eq;phase quenched reweighting factor and phasefactor}, reduces and becomes zero with increasing value of $\muB$, meaning the sign fluctuations among the reweighting factor values from different configurations are severe and needs to be taken care of. 

In our work, the parameters are chemical potential $\mu_X$, where $X \in (B,S,I)$ and temperature $T$ which dictates the inverse gauge coupling $\beta$ of the theory. This is guided by the Line of Constant Physics $(LCP)$ of the theory, which has been vividly discussed in \autoref{Appendix 2}. The basis transformation for $\mu$ from $(u,d,s) \to (B,S,I)$ is presented in \autoref{Appendix 7}. The set $B,S,I$ represent the baryon, strangeness and isospin chemical potentials respectively. For our work, we perform reweighting in $\mu$ only, which means in our case, the simulated theory is at $(\beta_0, \mu_X=0)$ and the target theory is at $(\beta_0, \mu_X \neq 0)$ where $\beta_0$ is the value of inverse gauge coupling. The data we have worked on, provides gauge ensembles generated at $\beta_0$ and $\mu=0$. We have worked with $\mu_B$ and $\mu_I$ only, meaning $X=B,I$. 

Recent developments have been made along the lines of contour deformations of path integral~\cite{Detmold.2020,Detmold.2021}, complex Langevin dynamics~\cite{Aarts.2009,Gert.2011,Gert.2013,Aarts.2013,Sexty.2013,Kogut.2019} and Lefschitz thimbles~\cite{Cristoforetti.2012,Schmidt.2017,Fukuma.2019} to go around this sign problem. Despite their proven abilities to circumvent the sign problem, application of these methods however continues to remain very limited in explicit QCD. A detailed discussion of these new approaches along with their application in the paradigm of QCD, is however beyond the scope of this thesis. Instead, we try motivating resummation techniques in the next chapter.

 	\cleardoublepage
 	\newcommand{\todo}[1]{\textcolor{red}{#1}}
\newcommand{\Dn}[1]{\bar{D}_{#1}}
\newcommand{\Cn}[1]{\bar{C}_{#1}}
\newcommand{\Gn}[1]{\bar{\cal G}_{#1}}
\newcommand{\cb}{\chi^B}
\newcommand{\av}[1]{\left\langle{#1}\right\rangle}
\newcommand{\pt}{\Delta P^T}
\newcommand{\pr}{\Delta P^R}
\newcommand{\ndt}{\mathcal{N}^T}
\newcommand{\ndr}{\mathcal{N}^R}
\newcommand{\zr}{\mathcal{Z}^R}
\newcommand{\thr}{\Theta^R}
\newcommand{\tht}{\Theta^E}
\newcommand{\nth}[1]{{#1}^\text{th}}
\newcommand{\detrt}[2]{\frac{\det M\left({#1}\right)}{\det M\left({#2}\right)}}
\newcommand{\dbeta}{\Delta\beta}
\newcommand{\dT}{\Delta T}
\newcommand{\dmaa}{\Delta m}
\newcommand{\hm}{\hat{\mu}_B}
\newcommand{\hms}{\hat{m}}
\newcommand{\hdm}{\Delta\hat{m}}
\newcommand{\tc}[1]{\textcolor{red}{#1}}

\chapter{Resummation methods}
\label{Chapter 5}

\graphicspath{{Figures/Chapter-5figs/PDF/}{Figures/Chapter-5figs/}}

\section{Motivation and Introduction}

 In this chapter, we focus on the different resummation techniques, namely the Pad{\'e} resummation and the exponential resummation, where the latter is being discussed in more detail, leaving the former on a brief and naive introductory note. This is because, the latter forms an instrumental part for the new work and discussion that has been subsequently presented from \autoref{Chapter 6} to \autoref{Chapter 8} of this thesis.
 
 However, both the resummation approaches stem from the viewpoint that the conventional Taylor series expansion possesses a slow rate of convergence. Mathematically, this means that the estimates of the successive ratios of the subsequent Taylor coefficients, as the per \autoref{eq:Radius of convergence} are always close to $1$. This implies that higher order corrections can never be ignored and it is not correct to consider the series convergent. Consequently, one should not never rely and work with a truncated version of the series. Therefore, despite the pertinent reluctance, one must evaluate the Taylor series up to sufficiently high order in $\mu$, and obtain sufficient number of estimates for the radius of convergence of the series. 

 However, there is another problem that lurks in this venture. The computation of higher order Taylor coefficients, specially from $c_6$ onwards gets extremely expensive computationally and requires lots of lattice resources and substantial computational time{~\cite{Allton.2005,Bazavov.2017,Bollweg.2021,Choe.2002,Gavai.2003,Allton.2003,Gavai..2005,Allton.2002}}. One therefore is motivated to ask whether something can be done with the knowledge of lower order Taylor series, possibly knowing Taylor coefficients $c_2$ and $c_4$ only. Resummation of lower Taylor series comes to our rescue in this respect, by virtue of which, one can capture contribution to all orders in $\mu$. Although, different resummation methods come with their own subtleties and nuances, the objective of capturing contributions to all orders in $\mu$ is the common objective for all the available resummation techniques.

\section{Pad{\'e} Resummation}
\label{sec:Pade resummation}

 We discuss Pad{\'e} resummation very briefly, primarily giving a naive mathematical notion without delving into physics in detail. Nevertheless, this resummation is a very important resummation method which is often used for the study of finite density equation of state, finite baryon chemical potential $\muB$ crossover and conserved charge cumulants for finite density in lattice QCD~\cite{Jurk.1978, Pasztor.2021, Goswami.2022, Jishnu.2022}.

  Let us discuss the mathematics and try to motivate Pad{\'e} resummation naively. Given a power series expansion of a function $f(x)$, so that

\begin{equation}
    f(x) = \nsum[1.1]_{k=0}^{\infty} c_k \hs x^k
    \label{eq:Pade starting power series}
\end{equation}

This expansion is the fundamental starting point of any analysis using Pad{\'e} approximants. The main idea in Pad{\'e} resummation is to approximate the behaviour of the above function $f(x)$ and its power series (Taylor) form in $x$ in \autoref{eq:Pade starting power series} by using a rational function. These rational functions are termed as Pad{\'e} approximants of different order. A Pad{\'e} approximant of the form $[M/N]$ is a rational function defined as follows:

\begin{equation}
    P\left[M/N\right] = \frac{a_0+a_1\hs x+ \cdots +a_M\hs x^M}{b_0+b_1\hs x+ \cdots +b_N\hs x^N} 
    = \frac{\nsum[1.2]_{l=1}^L a_l\hs x^l}{\nsum[1.2]_{m=1}^M b_m\hs x^m}
    \label{eq:Pade approximant}
\end{equation}

In the above \autoref{eq:Pade approximant}, there are total $M+N+1$ undetermined coefficients, with $b_0=1$ for definiteness. So, we are left with $M+1$ coefficients in the numerator and $N$ in the denominator. We can always choose an arbitrary basis for the denominator such that the leading order term becomes $1$. Note that there are all total $M+N+1$ undetermined coefficients, hence the above Pad{\'e} approximant can approximate Taylor series upto $\Ob(x^{M+N})$. So, in the notation of formal power series, 

\begin{equation}
    f(x) = \nsum[1.1]_{k=0}^{\infty} c_k \hs x^k = P\left[M/N\right] + \Ob\left(x^{M+N+1}\right)
    \label{eq:formal power series}
\end{equation}

where $P\left[M/N\right]$ is the Pad{\'e} approximant of order $\left[M/N\right]$ as given in \autoref{eq:Pade approximant}. There exists extensive formulae~\cite{Baker.2010} for determining the $N$ denominator coefficients $b_n,\hs  1 \leq n \leq N$ and the $M+1$ numerator coefficients $a_m, \hs 0 \leq m \leq M$, from which the Pad{\'e} approximant can be fully constructed in a functional form, and agrees with $f(x)$ up to $\Ob(x^{M+N})$. These are used often to approximate a given Taylor series in actual calculation of observables in lattice QCD.

Every power series expanded around $x=0$ has a circle of convergence $\left|x\right| = R$, with radius $R$. The power series converges for $\left|x\right|<R$,  and diverges for $\left|x\right|>R$. If $R \to \infty$, then the power series represents an analytic function (functions analytic everywhere in the complex $x$ plane) and the series may be summed directly for any value of $x$ to yield the function $f(x)$. If $R = 0$, the power series is undoubtedly formal. It contains information about $f(x)$, but no clear indication about how this information is to be used. However, if a sequence of Pad{\'e} approximants of the formal power series as in \autoref{eq:formal power series} converges to a function $g(x)$ for $x \in \mathbf{D}$, where $\mathbf{D}$ is the domain of validity of $x$, then we may reasonably conclude that $g(x)$ is a function with the given power series. If the given power series converges to the same function for $\left|x\right|<R$ for some finite value of $R$($0 < R < \infty$), then a
sequence of Pad{\'e} approximants may converge for $x \in \mathbf{D}$ where $\mathbf{D}$ is a domain larger than $\left|x\right|<R$. We will then have extended our domain of convergence. This is frequently a practical approach to what amounts to analytic continuation and have largely contributed to the significance of using Pad{\'e} resummation method apart from the conventional Taylor series expansion.

We defer from discussing more about the mathematics and physical significance of Pad{\'e} approximants. For more detailed discussion and comprehensive outlook regarding Pad{\'e} approximants and results of the resummation in the context of lattice QCD thermodynamics, we refer to the following references~\cite{Baker.2010,Brezinski.2011,Jurk.1978,Pasztor.2021,Goswami.2022,Jishnu.2022}.

\section{Exponential Resummation}
\label{sec:Exp resummation}

\subsection{Introduction}
 \hspace{2mm}We now are in a position to elaborate about the second important form of resummation, apart from the aforementioned Pad{\'e} resummation in \autoref{sec:Pade resummation}. And this is the exponential resummation. 
 
 As mentioned before, the Lattice QCD results for the QCD
Equation of state (EoS) are precise to a great extent, providing substantial conclusive information about the dynamical modeling of heavy-ion
collisions, over an extensive range of collision energy~\cite{Bernhard.2016,Everett.2020,Monnai.2019,Parotto.2018}
and, thereby, assist in the experimental explorations of the QCD phase diagram in the
$T$-$\muB$ plane. But, due to the fermion sign problem~\cite{Alford.2001.61,Fodor.2002,deForcrand.2009,Nagata.2022,Barbour.1986,Kogut.1995}, it is difficult to carry out computations in QCD on lattice directly at $\muB\ne0$. Although some recent
progress have been achieved~\cite{Cristoforetti.2012, Sexty.2013, Fukuma.2019, Aarts.2009,
Aarts.2013, Fodor.2015}, direct lattice computations for determining EoS of QCD at finite non-zero $\muB$
with physical quark masses, fine lattice spacings and large lattice volumes continue to remain evasive and arduous. Hence, the present state-of-the-art lattice QCD Equation of state for $\muB>0$ has been
obtained using the approaches of analytic continuation~\cite{Borsanyi.2018,Ratti.2018,Gunther.2016,Takaishi.2010,D'Elia.2009} and Taylor expansion~\cite{Allton.2005,Bazavov.2017,Bollweg.2021,Choe.2002,Gavai.2003,Allton.2003,Gavai..2005,Allton.2002} methods as mentioned in \autoref{sec:Taylor Expansion} and \autoref{sec:Analytic continuation} respectively. Although, the subsequent formalism and calculations hold true for any chemical potential $\mu$, in this chapter, the relevant one will be $\muB$, since it is directly related to the QCD phase diagram. Hence, all the discussions in the subsequent sections of this chapter will be in terms of $\muB$. A quick review of these two methods ensue before going into the exponential resummation and talking about its results and other relevant features.

In the Taylor
expansion method, the excess pressure is expanded in powers of $\muB$ around $\muB=0$ and one directly computes the Taylor coefficients at $\muB=0$. For the analytic continuation,
one performs simulations at purely imaginary values of $\muB$ where there is no fermion sign problem, perform a fitting of these simulation results with a power series in $\muB$ chosen as an ansatz to determine the Taylor
coefficients at $\muB=0$ and then figure out the EoS at real $\muB>0$ based on these Taylor
coefficients. As far as the limitations are concerned, it is well-known that the applicability of the Taylor
expansion as well as the analytic continuation are limited by the zeros or singularities,
nearest to $\muB=0$, of the partition function $\Z$ in the entire complex $\muB$
plane~\citep{Stephanov.2006, Almasi.2019, Mukherjee.2019}. 

Although in principle, the locations of these singularities can be traced by expanding the partition function $\Z$ as a power series in real or imaginary $\muB$ in terms of Pad{\'e} approximants~\cite{Baker.2010,Brezinski.2011,Jurk.1978,Pasztor.2021,Goswami.2022,Jishnu.2022}, 
in reality, with the knowledge of only the first few lowest order Taylor coefficients, this becomes a very difficult and laborious task. In practice,
one restricts the Equation of state calculations to a specific domain in $T$ and $\muB$, that avoids any pathological non-monotonicity in the calculation of relevant observables
in the truncated Taylor series~\cite{Bazavov.2017, Gunther.2016}. In addition to that,
these methods do not prove to be adequate and sufficiently reliable enough to ascertain numerically the radius of convergence of the Taylor series. Also, they do not credibly reveal the severity of the fermion problem,
\textit{i.e.} how rapidly the phase of the partition function fluctuates as $\muB$ is
increased.  It is possible to determine the zeros as well as its average phase by reweighting the fermion determinant to $\muB\ne0$~\citep{Fodor.2001, Fodor.2004, Ejiri.2004, Saito.2013,
Giordano.2020}. However, it is because of the computational expenses associated with the exact determination of the fermion determinant, at present this method is limited within the regime of large lattice spacings and small lattice volumes.

In this chapter, we present an overview about the basic formalism of exponential resummation. We also present some of the results, hopefully convincing about its ability of capturing contributions of finite ordered correlation functions to all orders in $\muB$ and also, indicating the singularities of partition function from the zeroes of phasefactor.

\subsection{The method and formalism}
\label{sec:exp resum formalism}

 \hspace{2mm}The Taylor expansion to $\Ob\left({\muB^N}\right)$ of the excess
pressure, $\Delta P(T,\muB) \equiv P(T,\muB)-P(T,0)$, is given by
\begin{equation}
  \frac{\pt_N}{T^4} = \nsum[1.4]_{n=1}^N \frac{\cb_n}{n!} \left(\frac{\muB}{T}\right)^n ,
  \label{eq:pt}
\end{equation}
where the Taylor coefficients are defined as
\begin{align}
  \chi_n^B(T) = \frac{1}{VT^3} \frac{\partial^n\ln \Z(T,\muB)}{\partial(\muB/T)^n}\Bigg\vert_{\muB=0} 
  \label{eq:chi}
\end{align}
where $V$ and $T$ are the volume and temperature of the thermodynamic system.  Here, the QCD partition function is a grand canonical partition function denoted as $\Z$, which is given by 
\begin{equation}
 \Z(\muB,T) = \int \mathcal{D}U e^{-S_G[T,U]} \det \M(\muB,T,U)    
\end{equation}
 
 where $U$ are the $SU(3)$ gauge fields of QCD,  $S_G$ is
the pure gauge action and $\M$ is the fermion matrix. The system is studied in thermal equilibrium at a temperature $T$. The volume dependence in $\Z$ and $\M$ are suppressed, since the system is already in a thermodynamic limit.  Each $\cb_n$ consists of sum
of terms like $\av{\left(D_i^B\right)^a \left(D_j^B\right)^b \cdots \left(D_k^B\right)^c}$ 
satisfying the relation $i \cdot a + j \cdot b + \cdots +
k \cdot c=n$~\cite{Allton.2005,Gavai.2003}, where
\begin{align}
   D_n^B(T) = \frac{1}{n!}\,\frac{\partial^n\ln \Z(T,\muB)}{\partial(\muB/T)^n}\Bigg\vert_{\muB=0}
  \label{eq:D}
\end{align}
and the $\LA\cdot\RA$ denotes the average over gauge field ensembles generated at $\muB=0$, which implies that 

\begin{equation}
    \LA \Ob(\muB) \RA = \frac{1}{\Z} \int \mathcal{D}U  e^{-S_G[T,U]} \det \M(T,\muB=0,U) \hspace{1mm} \Ob(\muB,U)
    \label{eq:exp_ob}
\end{equation}


\begin{figure}[H]
  \centering
  \includegraphics[width=0.67\textwidth, height=0.30\textheight]{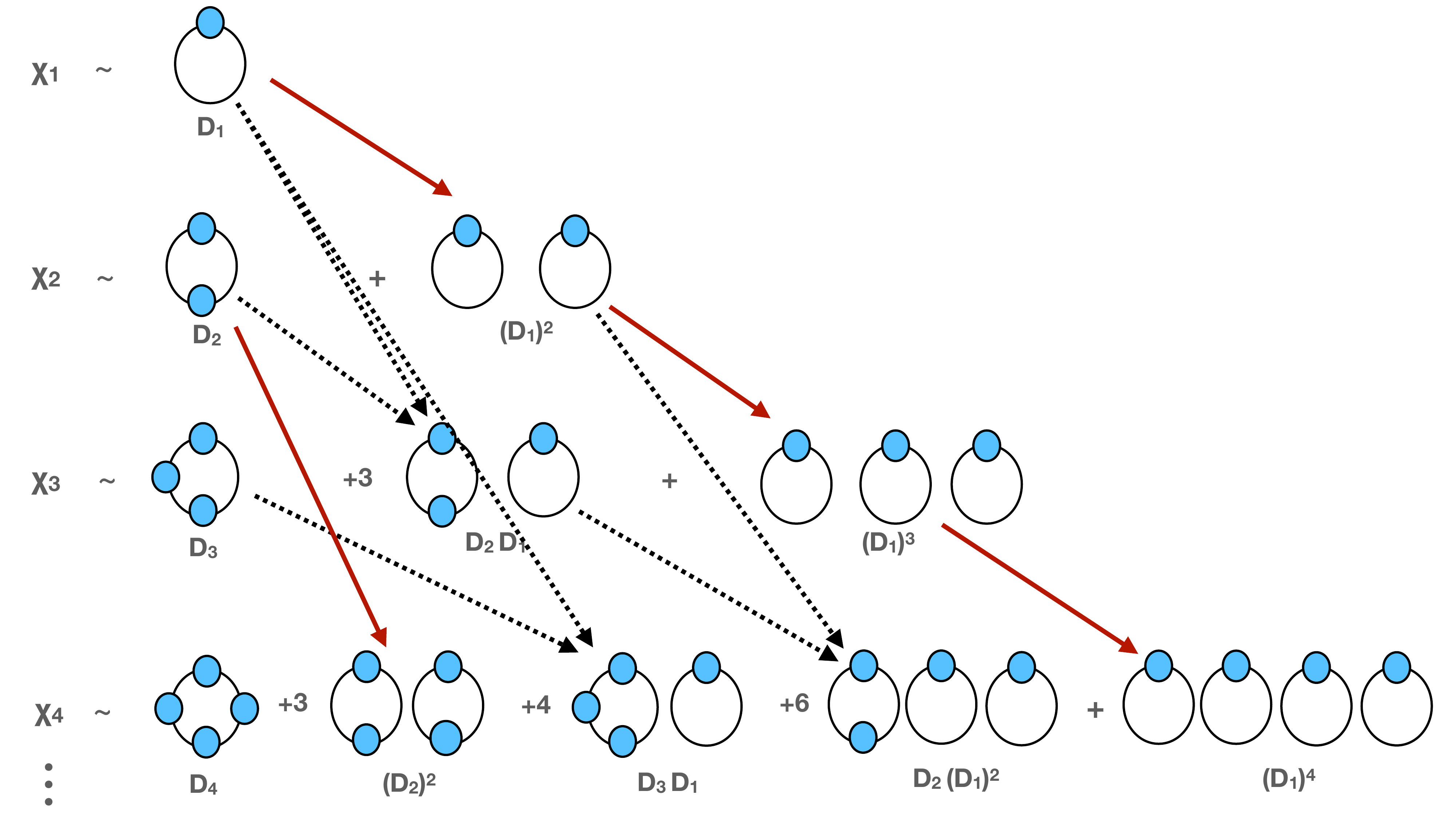}
  \caption{\hspace{2mm}Contributions of different $D_n$ to different baryon number susceptibilities $\chi_n^B$. Each blob represents
  insertion of the $0^{th}$ component of the conserved current. Solid red and
  dotted black lines represent directly exponentiated and cross terms respectively~\cite{Mondal.2022}.}
  \label{fig:D}
\end{figure}

where the symbols have their conventional meanings. \autoref{eq:exp_ob} reveals the evaluation of observable $\Ob$ at a finite $\muB$ from gauge ensembles generated at $\muB=0$. 

In case of the continuum theory, $D_n^Y$ can be physically interpreted as the scaled integrated $n$-point
correlation function of the zeroth component of the conserved 
four current $J_{\alpha}^Y(x)$ at a space-time point $x$, given by  

\begin{equation*}
 D_n=\int dx_1
\cdots dx_n \hs J_0(x_1) \cdots J_0(x_n)  
\end{equation*}

where $J_{\alpha}^Y = \left(J_0^Y, \Vec{\boldsymbol{J}}^Y\right)$ and $x_n \equiv \left(x_{\beta}\right)_n$, with $x_{\beta} = \left(x_0, \Vec{\boldsymbol{x}}\right)$. The label $Y$ denotes the chemical potential considered here. According to \autoref{Appendix 7}, $Y \in (B,S,I)$ and so, accordingly are the currents and the subsequent correlation functions. The label $n$ in $(x_{\beta})_n$ point to site $n$ in the lattice. 

Due to the $CP$ symmetry
of QCD, all $D_n$ are purely imaginary for odd $n$ and purely real for even $n$(see \autoref{Appendix 11}), for all $X$. This is because, the symmetry is independent of chemical potential introduction in the theory. It is only the even $n$ terms that contribute to \autoref{eq:pt}. This is completely similar to the argument of why the Taylor series of thermodynamic observables form an even series in $\mu$. In practice, lattice QCD computations of the
$\cb_N$ involve computations of all $D_n$ for $n\le N$ as intermediate steps, and
$\cb_N$ are obtained from combinations of $D_n$ and their powers. For an extensive and elaborate list, refer to \autoref{Appendix 3}.

Contributions of various combinations of $D_n$ to the few lowest order Taylor
coefficients are sketched in \autoref{fig:D}. If one considers the factorials coupled with the appropriate
powers of $\muB/T$ associated with each $D_n$ in the sum of \autoref{eq:pt}, it is not
difficult to realize that all contributions of each $D_n$ to $\pt$ can be resummed
into exponential forms. For example, contributions of $D_1^n$ from all $\cb_n$ in
\autoref{eq:pt} can be resummed as $\exp\left[\bar{D}_1^B \times (\muB/T)\right]$. Similarly, contributions of all
$D_2^n$ can be resummed as $\exp\left[\bar{D}_2^B \times (\muB/T)^2\right]$, and so on. All of these exponential forms can be found by following the $\textcolor{red}{\textbf{red}}$ arrows in \autoref{fig:D}. It is also easy to see
that the contributions of the mixed terms like $D_1^nD_2^m$ arise from
$\exp\left[\bar{D}_1^B \times (\muB/T)\right] \times \exp\left[\bar{D}_2^B \times (\muB/T)^2\right]$, which can be traced along the dotted lines of \autoref{fig:D}. Hence, it is therefore possible to write down a 
resummed version of the Taylor series of \autoref{eq:pt}, which is given by

\begin{align}
  \frac{\pr_N}{T^4}  = \frac{1}{VT^3} \ln \left \langle \exp{\Bigg[ \nsum[1.4]_{n=1}^N \bar{D}_n^B \left(\frac{\muB}{T}\right)^n \Bigg]} \right \rangle 
  \label{eq:pr}
\end{align}
where the symbols have their conventional meanings. The $\bar{D}_n^B$ gives an estimate of $D_n^B$ for every configuration, by averaging over all the random volume estimates of the correlation function present in the given gauge field configuration (more of this in \autoref{Chapter 6}). For any chemical potential flavor, this is given by 

\begin{equation}
    \Bar{D}_n = \frac{1}{N_R} \nsum[1.3]_{r=1}^{N_R} D_n^{(r)}
\end{equation}

where $N_R$ is the number of random Gaussian volume sources present in the gauge configuration (details in \autoref{Chapter 6}). Eqn.\autoref{eq:pr} provides finite ordered EoS up to infinite orders in $\muB$.  The $\pr_N$ can be considered as a $\muB$-dependent effective action obtained by resumming the first $N$-point correlation
functions of the conserved current. Expansion of $\pr_N$ in powers of $\muB/T$ yields
an infinite series in $\muB/T$, in addition to the truncated Taylor series given by

\begin{equation}
  \frac{\pr_N}{T^4} = \frac{\pt_N}{T^4} + \left[\nsum[1.3]_{n>N}^\infty \left \langle \left(\bar{D}_1^B\right)^i \cdots \left(\bar{D}_N^B\right)^j \right \rangle \right] \left(\frac{\muB}{T}\right)^n   
\end{equation}

 where $i,j$ are integers satisfying $0 \leq i,j \leq N$ satisfying $1 \cdot i + \cdots + N \cdot j = n$. The angular brackets $\left \langle \cdot \right \rangle$, as before represent average over gauge ensembles generated at $\muB=0$.
 The Taylor expanded ($\ndt_N$) and the resummed ($\ndr_N$) net baryon-number
densities can be straightforwardly obtained as a single $\muB$ derivative of
$\pt$ and $\pr$ in \autoref{eq:pt} and \autoref{eq:pr}, respectively.

\subsection{Connection with reweighting factor and phasefactor}

The resummed version in \autoref{eq:pr} also highlights the connection between  the
Taylor expansion and the reweighting method. Because we have seen that, in the reweighting method the following 

\begin{equation*}
 \frac{\Z(T,\muB)}{\Z(T,0)} = \left \langle \frac{\det \M(T,\muB)}{\det \M(T,0)} \right \rangle   
\end{equation*}

can be calculated, if
computationally feasible, by exactly evaluating the ratio of the fermion matrix
determinants on the gauge fields generated at $\muB=0$. In more realistic lattice
calculations with large volumes, exact evaluations of the determinant ratios might not
be computationally feasible. 

This is because, for a lattice with total number of lattice sites $V$, which is equivalent to $4$-volume of lattice given by $V=N_{\sigma}^3 \times N_{\tau}$, the fermion matrix $\M$ have a total of $12V$ components. This factor of $12$ comes from the fact that every matrix element on a given lattice point $\left(p_1,p_2,p_3,p_4\right)$ is given by $\M_{\gamma}^i, \gamma = 1,2,3,4$ (Dirac indices) and $i=a,b,c$ (color indices), since QCD is a $SU(3)$ gauge theory with $3$ colors defined on a $4$-dimensional spacetime. For large lattice volumes, it therefore becomes extremely difficult to analytically construct the structure of $\M$, implying that it is also analytically tedious and expensive to construct the fermion determinant and therefore $\M^{-1}$. 

In this case, one may henceforth consider evaluating $\det\left[\M(T,\muB)\right]$ within
some approximation scheme to obtain approximate partition function
$\zr_N(T,\muB)\approx \Z(T,\muB)$. Following the steps of the Taylor expansion, one such
approximation scheme can be expansion of $\det\left[\M \left(T,\muB\right)\right]$ in powers of $\muB/T$. Keeping
in mind that $\det\left[\M\right] = \exp\left[\Tr \left(\ln \M\right) \right]$ and \autoref{eq:D}, one can immediately recognize

\begin{align}
   \frac{\zr_N(T,\muB)}{\Z(T,0)} = \av{  \exp\left[ \nsum[1.2]_{n=1}^N \bar{D}_n^B \left(\frac{\muB}{T}\right)^n \right] } .
  \label{eq:Z}
\end{align}

As mentioned before, the $CP$ symmetry dictates that the correlation functions $D_n$ are purely real for even $n$ and purely imaginary for odd $n$, which means that the RHS of \autoref{eq:Z} is complex. The $CP$ symmetry also implies that the partition function $\Z$ must be real and every estimate of $\Z$, obtained from different gauge configuration must also be real. This suggests that not only the LHS of \autoref{eq:Z} is real, the observable within the angular brackets in \autoref{eq:Z} is also real. Hence, with all such arguments, \autoref{eq:Z} is modified on computational grounds, which yields the following 

\begin{align}
   \frac{\zr_N(T,\muB)}{\Z(T,0)} = \av{ \text{Re} \left[\exp\left( \nsum[1.2]_{n=1}^N \bar{D}_n^B \left(\frac{\muB}{T}\right)^n \right) \right]} 
  \label{eq:Z corrected}
\end{align}

where, the real part of the complex exponential is extracted to preserve the equality and the $CP$ symmetry. Likewise, in the realm of exponential resummation, the corrected version of excess pressure of \autoref{eq:pr} is given by

\begin{align}
  \frac{\pr_N}{T^4}  = \frac{1}{VT^3} \ln \av{ \text{Re} \left[\exp\left( \nsum[1.2]_{n=1}^N \bar{D}_n^B \left(\frac{\muB}{T}\right)^n \right) \right]} 
  \label{eq:pr corrected}
\end{align}

Following the aforementioned arguments about the properties of $D_n$, we can define the following 

\begin{equation}
    \exp\left[ \nsum[1.2]_{n=1}^N \bar{D}_n^B \left(\frac{\muB}{T}\right)^n \right] = \mathcal{R} = Re^{i\theta}
    \label{eq:R_exp_itheta}
\end{equation}

where $\mathcal{R}$, $R$ are the reweighting factor and the phase-quenched reweighting factor respectively. $\theta$ is the phase-angle.
A measure of the severity of the sign problem is
given by the average phase factor $\av{\cos\theta_N^R}$ for $\zr_N$ (with $\muB$ real) where $\theta_N^R$ is given by
\begin{align}
  \theta_N^R = \nsum[1.2]_{n=1}^{N/2} \text{Im}\left[\bar{D}_{2n-1}\right] \left(\frac{\muB}{T}\right)^{2n-1} 
  \label{eq:theta}
\end{align}

where $\text{Im}\left(\bar{D}_{2n-1}\right)$ denotes the imaginary part of $\bar{D}_{2n-1}$. This happens because, $\text{Re}\left(R \hspace{1mm}e^{i\theta}\right) = R\hspace{1mm}\cos{\theta}$, as per Eqns.~\autoref{eq:Z corrected} and \autoref{eq:pr corrected}. For a more explicit derivation and explanation of these formulae in case of more generic complex chemical potentials, refer to \autoref{Appendix 4}.

An expansion of average phasefactor $\av{\cos\theta_N^R}$ in $\muB/T$ leads to the Taylor expanded measure of the
average phase of the partition function $\Z$~\cite{Ejiri.2004, Allton.2005}. As the sign problem becomes more severe, the average phase
$\av{\cos\thr_N}\approx0$ and resummed results also show signs of breakdown~\cite{Mondal.2022}.
Furthermore, although $\pt_N$ 
can be evaluated for any complex value of $\muB$, $\pr_N$ becomes undefined when $\text{Re}[\zr_N]\le0$ for a given $N$ and
statistics, leading to a natural breakdown of the resummed results.
The location of
the zeros of $\zr_N$ in the complex $\muB$ plane will indicate the $\muB$ region
where such resummation can be applicable. It goes without saying, that for any given $N$ the region of
applicability of $\pt_N$ cannot exceed the same for $\pr_N$, since the latter captures finite ordered contributions of $D_n$ to all orders in $\muB$, as opposed to finite orders in $\muB$ in the former. To illustrate this, we refer to the following diagram cited from Ref.~\cite{Mondal.2022}.

\begin{figure}[H]
  \centering
  \includegraphics[width=0.67\textwidth, height=0.30\textheight]{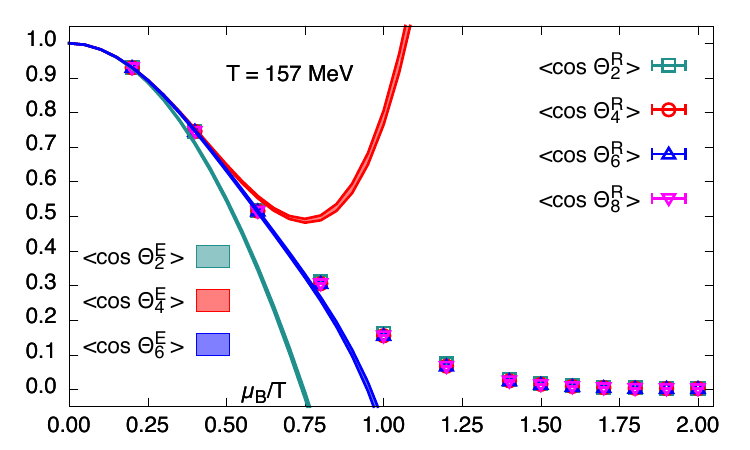}
  \caption{$N^{th}$ order Taylor estimates $\left \langle cos \Theta_N^E \right \rangle$ and resummed estimates $\left \langle cos \Theta_N^R \right \rangle$ of average phasefactor as a function of $\muB/T$ at $T=157$ MeV~\cite{Mondal.2022}.}
  \label{fig:Mondal average phasefactor plot}
\end{figure}

The main takeaway of the above \autoref{fig:Mondal average phasefactor plot} is that the Taylor series, owing to slow convergence and non-monotonic behaviour as well as being finite order in $\muB$ cannot capture the zeroes of the phasefactor properly. On the other hand, the exponential resummation indicate average phasefactor zeroes and hence, breakdown of resummed results for $\muB/T \approx 1.7$, by capturing contributions of $n \leq N$ point correlation functions to all orders in $\muB$. Regarding the data, an extensive description of the relevant lattice calculations and scale setting are mentioned in Ref.~\cite{Mondal.2022}.

 	\clearpage	
	\chapter{Biased and Unbiased Estimates}
\label{Chapter 6}
\graphicspath{{Figures/Chapter-6figs/PDF/}{Figures/Chapter-6figs/}}

\section{Origin : The starting point}
\hspace{2mm}As shown in \autoref{Appendix 3}, the calculation of $n$-point correlation functions requires the evaluation of different traces comprising operator products of $\M^{-1}$ and different ordered $\muB$ derivatives of fermion matrix $\M$. We have already argued in detail in \autoref{Chapter 2} and also in \autoref{Chapter 4} (see \autoref{sec:exp resum formalism}) about the associated difficulty in exact calculation of $\M^{-1}$. Also, the genuine non-simulated finite temperature thermodynamics of Nature being in thermodynamic limit and in continuum spacetime, it is intended to approach these limits even from the lattice point of view. 

It is sufficient at this point to understand that since the latter requires lattice spacing $a \to 0$ and the former requires at least volume $V \to \infty$, hence the number of lattice points in spatial direction $N_{\sigma} \to \infty$. This is because in a $N_{\sigma}^3 \times N_{\tau}$ lattice with lattice spacing $a$, we know $V=\left(N_{\sigma}a\right)^3$ and $T=\left(N_{\tau}a\right)^{-1}$, where $V$ and $T$ are the volume and temperature of the system respectively. As a result, the lattices in closest proximity to the true physical picture are lattices with large volumes and fine lattice spacings. Since, as mentioned before, the number of components or elements of fermion matrix $\M$ is $12V_4$, where $V_4=N_{\sigma}^3 \cdot N_{\tau}$ is the four volume of the lattice, hence on such lattices, it is extremely tedious and difficult to construct even the matrix $\M$, and subsequently the fermion determinant and $\M^{-1}$. 

Owing to this difficulty in computing $\M^{-1}$ exactly, we have to numerically estimate these $D_n^B$. This is because, as mentioned in \autoref{Appendix 3}, all these correlation functions are linear combinations of some traces, each of which contains $\M^{-1}$, along with appropriate $\mu$ derivatives of $\M$. We have also seen that the values of these correlation functions are essential to compute thermodynamic observables like $\dP/T^4$ and $\dN/T^3$ as we have seen in \autoref{eq:pr}. And that is done using random volume sources within every gauge field configuration in the gauge ensemble. 

We explain more about this while talking about biased and unbiased estimates in this section, which are a direct consequence of using finite number of random volume sources, as we will see eventually in this chapter. In the following section, we schematically present the overall structure and arrangement of the random volume sources and how they remain nested inside every gauge field configurations in the data used. All the arguments and discussions presented in this chapter hold good for any chemical potential, despite we have centralised our discussion around baryon chemical potential $\muB$ here only.


\section{Random volume sources : Stochastic averaging}

\begin{figure}
    \centering
    \includegraphics[width=1.\textwidth]{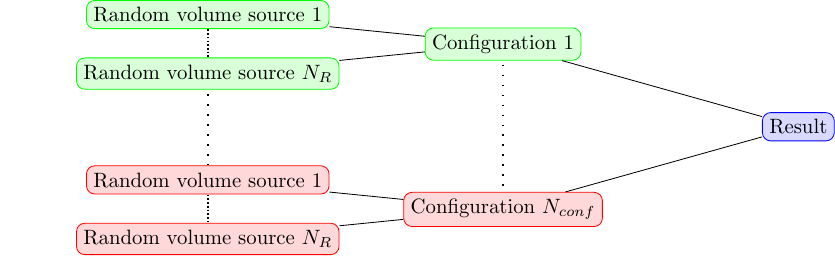}
    \caption{A schematic overview of the nesting structure of different random volume sources inside each of the available gauge field configuration which gives the estimate of $n$-point correlation functions $D_n$ ($D_n^B$ here)}
    \label{fig:flowchart}
\end{figure}

\hspace{2mm}The following \autoref{fig:flowchart} gives a rough schematic overview as to how the different random volume sources are nested inside every gauge field configuration. Hence, as the above figure suggests, there are a total of $N_{conf}$ gauge configurations and every such gauge configuration constitutes a further of $\NR$ number of random volume sources, each of which, shelters an estimate of $D_n^B$ for every $n$.

Each of these random volume estimates of $D_n^B$ are stochastically generated using conjugate gradient algorithm of numerically approximating and evaluating $\M^{-1}$. All these random volume estimates for a given gauge configuration are averaged out to obtain an estimate of $D_n^B$ at the level of individual gauge field configurations. It is then the final observables are calculated using angular brackets $\left \langle \cdot \right \rangle$ like in \autoref{eq:pr} and other equations involving thermodynamic observables, indicating average over all such gauge configurations present in the gauge ensemble. The stochastic averaging at the level of random vectors or volume sources is given as follows : 

 \begin{equation}
        \bar{D}_n^B = \frac{1}{\NR} \sum_{r=1}^{\NR} D_n^{B(r)} 
        \label{eq:stoaver}
      \end{equation}

    where in the above \autoref{eq:stoaver}, $D_n^{B(r)}$ is the estimate of $D_n^B$ found in the $r^{th}$ random volume source. Also, the $\bar{D}_n^B$ in this \autoref{eq:stoaver} is exactly the same as used in \autoref{eq:pr} and \autoref{eq:Z}. We will see in the subsequent sections how this way of evaluating $D_n^B$ gives rise to stochastic bias in form of biased and unbiased estimates.

  \section{Stochastic bias and its kinds}

  \hspace{2mm} Although, we will explicitly use $D_n^B$ notation, the following sets of arguments and discussions hold equally true for other forms of chemical potentials $\mu$.  We find that with the aforementioned way of calculating $D_n^B$ in \autoref{eq:stoaver}, we come across two kinds of stochastic biases as follows :

  \subsection{Estimate bias}

   \hspace{2mm}This form of stochastic bias accompanies every estimate of $D_n^B$ obtained from individual gauge configurations, for every $n$. This bias originates from the use of finite number of random volume sources used per configuration to estimate $D_n^B$ for each $n$. The bias decreases with increasing number of random volume sources and increases with decreasing number of sources. 

   In infinite limit of random vectors, if $\eta_k^{(r)}$ signify $r^{th}$ random source from the set of $\NR$ random volume sources defined at lattice site $k = (k_1,k_2,k_3,k_4)$, then the set of random vectors satisfy the following orthonormal relation
   \begin{equation}
\lim_{\NR\to\infty} \frac{1}{\NR} \sum_{n=1}^{\NR} \eta_i^{(n)\dagger}\eta_j^{(n)} = \delta_{ij}
\label{eq:Orthonormal}
\end{equation}

Then, for operator $\Ob$, the value of $\Tr \Ob$ can be determined as follows:
\begin{align}
\lim_{\NR\to\infty}\frac{1}{\NR}\sum_{n=1}^{\NR}\eta^{(n)\dagger}\Ob\eta^{(n)}
= \lim_{\NR\to\infty}\frac{1}{\NR}\sum_{n=1}^{\NR}\sum_{i,j}\;\eta^{(n)\dagger}_i\Ob_{ij}\;\eta^{(n)}_j,\\
= \sum_{i,j}\Ob_{ij}\left[\lim_{\NR\to\infty}\frac{1}{\NR}\sum_{n=1}^{\NR} \eta_i^{(n)\dagger}\eta_j^{(n)}\right]
= \sum_i\Ob_{ii} \equiv \Tr \Ob
\end{align}

where the trace is obtained by summing over identical lattice sites.

The above \autoref{eq:Orthonormal} holds true strictly for infinite $\NR$. For finite $\NR$, the set of random vectors do not satisfy the orthonormal relation illustrated in \autoref{eq:Orthonormal} and consequently, an estimate of $\Tr \Ob$ is biased, due to some additional corrections on top of $\delta_{ij}$ in \autoref{eq:Orthonormal}. Consequently, since in general, every $n$-point correlation function $D_n$ can be expressed as linear combinations in terms of traces of operators, as mentioned in \autoref{Appendix 3}, the estimates of $D_n$ contain stochastic bias for finite $\NR$ and only unbiased in infinite limit. 

The effect of regulating $\NR$ on the values of thermodynamic observables like excess pressure ($\dP/T^4$) and number density ($\dN/T^3$) are vividly demonstrated in \autoref{Chapter 7}, where we observe an appreciable change in the values of respective observables as one goes from $\NR=500$ to $\NR=250$. We also find this difference to enhance and become drastic for higher orders in isospin chemical potential $\muI$ and also more in $\dN/T^3$ over $\dP/T^4$, establishing that higher order $\mu$ derivatives of thermodynamic potential or free energy are expected to show greater change. We will continue this discussion here in detail in \autoref{Chapter 7}.

\subsection{Formalism bias}

 \hspace{2mm} The central work of this thesis revolves around eliminating this formalism bias precisely upto a finite order in $\mu$ ; in our case, it is baryon ($\muB$) and isospin ($\muI$) chemical potentials. This form of bias arises when any estimate of $D_n^B$ obtained from any arbitrary configuration after stochastic averaging of the corresponding random volume estimates, is raised to any integral positive definite non-linear powers. To put it in simple mathematical terms, using \autoref{eq:stoaver}, we have as follows : 

  \begin{align}  
    \left(\overline{D}_n^B\right)^m = \Bigg[\frac{1}{\NR} \sum_{r=1}^{\NR} D_n^{B(r)}\Bigg]^m &= 
    \Bigg[\left(\frac{1}{\NR}\right)^m \sum_{r_1=1}^{\NR}...\sum_{r_m=1}^{\NR} D_n^{B(r_1)}\hspace{1mm}...\hspace{1mm} D_n^{B(r_m)}\Bigg] \notag \\
    &= \left(\frac{1}{\NR}\right)^m\mathop{\sum^{\NR}...\sum^{\NR}}_{C(r_1\neq \cdots \neq r_m)} D_n^{B(r_1)}\hspace{1mm}...\hspace{1mm}D_n^{B(r_m)} \label{eq:Biased}\\
     &+\left(\frac{1}{\NR}\right)^m\mathop{\sum^{\NR}...\sum^{\NR}}_{\textcolor{red}{\boldsymbol{r_1\neq \cdots \neq r_m}}} D_n^{B(r_1)}\hspace{1mm}...\hspace{1mm}D_n^{B(r_m)} \label{eq:Unbiased} 
    \end{align}

\vspace{2mm}

 In the above \autoref{eq:Unbiased}, beside equation label \autoref{eq:Biased}, the notation \\$C(r_1\neq \cdots \neq r_m)$ represents the complement or negation of the condition \\$r_1 \neq r_2 \neq \cdots \neq r_m$. This has a profound implication as to why the terms in the above \autoref{eq:Biased} and \autoref{eq:Unbiased} give biased and unbiased estimates respectively. Let us understand the implication carefully which will also eventually clarify the etymological justification of these terms.

 As we understood before, we have a sample of stochastically generated $\NR$ estimates inside every gauge field configuration for a given baryon correlation function $D_n^B$. These $\NR$ estimates are stored within $\NR$ random volume sources and we have already seen the detailed schematic structure in \autoref{fig:flowchart}. There are $m$ number of $D_n^B$, with each correlation function bearing the label of a given random volume source, out of a total of $\NR$ sources. The notation $\textcolor{red}{\boldsymbol{r_1\neq ...\neq r_m}}$ implies that all the $m$ random volume sources out of a possible $\NR$ sources are different ; no two random volume sources out of this collection of $m$ sources are same. Then all the $m$ random volume estimates are treated equally in the sense, that each of them are raised to unit powers. Thus these form of calculating random volume estimates gives rise to unbiased estimates~\cite{Voinov.2012}. 
 
 From this argument, one arrives at a very significant conclusion and that is, unbiased $m^{th}$ power of a given $n$-point correlation function for any $n$, is only possible if and only if $m \leq \NR$. Because, if it is not so, then we have $C(m \leq \NR) = m > \NR$. 
 
 In that case, constructing $m$ different random volume sources from a total of $\NR$ sources is not possible. So, we will have at least a pair of sources out of these $m$ sources, whose labels are identical, implying that at least one random volume estimate of $D_n^B$ is repeated at least twice. Then this gives rise to biased estimates, because, unlike unbiased estimates, different random volume estimates are raised to different powers, suggesting that all the estimates are not treated on an equal footing, unlike the $\emph{equal-a-priori}$ policy of unbiased estimates. It is understood that all $m, \NR, n$ are positive definite integers. 
 
 In our work, we have random volume sources $\NR \sim \Ob(500)$ and we have dealt upto $4^{th}$ unbiased powers of $D_n$ i.e. $1 \leq m \leq 4$. Hence, it is always possible to calculate unbiased estimates of $D_n^{\mu}$ in our domain of work, where $D_n^{\mu}$ is the $n$-point correlation function calculated for any generic chemical potential $\mu$.
 
 	\cleardoublepage
 	 \newcommand{\Ca}{\mathcal{C}}

\chapter{Cumulant Expansion}
\label{Chapter 7}

\graphicspath{{Figures/Chapter-7figs/PDF/}{Figures/Chapter-7figs/}}

\vspace{.8cm}

\hspace{3mm}My doctoral work has two parts. In this chapter, I have presented the first part of my doctoral work, which is about analysing biased estimates order-by-order in $\mu$, using the method of cumulant expansion.  I have added necessary plots for adequate explanation and have tried to make my presentation of the chapter more understandable to the readers. I have also mentioned necessary references in chapter. This chapter is a kind of reflection of my recent paper~\cite{Mitra.2022.prd}.
 
\section{Motivation and Introduction}
\label{sec:cumu motiv}
\hspace{5mm}Let us have a quick and brief summary of all that has been covered till now, before going into the main topic. This is to maintain the continuity among different methods.

As outlined before, the Equation of State (EoS) of strongly-interacting matter is an important input and a crucial aspect in the hydro dynamical modeling of heavy-ion collisions~\cite{Bernhard.2016,Everett.2020,Monnai.2019,Parotto.2018} and also for understanding possible phase transitions~\cite{Aarts:2023LatticeQCD}, apart from a vivid conclusive exploration of the QCD phase diagram, constructed in the $T-\muB$ plane. Although Lattice
QCD, which is the preferred method of calculating observables in the non-perturbative regime of QCD, performs calculations and determines the chiral crossover temperature upto sufficient accuracy and precision separating the hadronic and quark-gluon plasma (QGP) phases at $\muB = 0$, it suffers a breakdown when the baryon chemical potential $\mu_B$ is 
non-zero. This is the well-known infamous sign problem of Lattice QCD~\cite{Alford.2001.61,Fodor.2002,deForcrand.2009,Nagata.2022,Barbour.1986,Kogut.1995}. Despite recent progress~\cite{Cristoforetti.2012, Sexty.2013, Fukuma.2019, Aarts.2009,
Aarts.2013, Fodor.2015}, the current state-of-the-art results for the QCD EoS have been obtained by using 
either analytical continuation from imaginary to real $\mu_B$~\cite{Borsanyi.2018,Ratti.2018,Gunther.2016,Takaishi.2010,D'Elia.2009}, or by expanding the Equation of state (EoS) in a Taylor series in the chemical potential $\mu_B$ and calculating the first $N$ coefficients~\cite{Allton.2005,Bazavov.2017,Bollweg.2021,Choe.2002,Gavai.2003,Allton.2003,Gavai..2005,Allton.2002}. 
In the latter case, a knowledge of the first several coefficients is necessary, not only to obtain the EoS for a fairly wide range of chemical
potentials but also to determine the radius of convergence of the Taylor series beyond which the Taylor expansion must break down~\cite{Gavai.2008, Dimopoulos.2021, Giordano.2019slo}. 

Unfortunately, the calculation of the higher order Taylor coefficients is computationally very challenging and it is natural to ask whether something can be learned about them from a knowledge of the first few Taylor coefficients. It turns out that this is indeed possible because 
the first $N$ derivatives $D_1,\dots,D_N$ of $\ln\det \M(\mu)$, where $\M(\mu)$ is the fermion matrix, also contribute to the higher order Taylor coefficients through products such as $D_N^2$, $D_ND_1$, etc. In fact, the contribution of the $n$th derivative $D_n$ to all higher orders can be shown to take the form of an exponential $\exp(D_n\muB^n/n!)$~\cite{Mondal.2022}. Thus, if $D_n$ is known exactly, then its contribution to the Taylor series can be resummed to all orders through exponentiation. Exponential resummation can be shown to have several advantages compared to the original Taylor series: First, the resummed EoS converges faster than the Taylor series. Moreover, since the odd derivatives $D_1,D_3,\dots$ are purely imaginary, the resummed expression directly gives us a phase factor whose expectation value approaches zero as $\muB$ is increased, leading to a breakdown of the calculation. This breakdown is physical and related to the presence of poles or branch cut singularities of the QCD partition function in the complex $\muB$ plane. The resummed expression for the partition function also makes it possible to calculate these singularities directly. Some of these advantages have been recently demonstrated through analytical calculations in a low-energy model of QCD~\cite{Mukherjee.2021}.

Despite its advantages, a technical drawback of exponential resummation is that the derivatives $D_1,\dots,D_N$ are not known exactly in an actual Lattice calculation. As is easily seen from the identity $\ln\det \M = \Tr\ln \M$, the $D_n$ can be expressed in terms of traces of various operators, all of which involve the inverse of the fermion matrix $\M$. Since $\M$ is typically of size $10^8$ or greater, its exact inverse is too expensive to calculate. Instead the various traces, and hence the derivatives $D_n$, are estimated stochastically using $\mathcal{O}(10^2$-$10^3)$ random volume sources per gauge configuration. Now, products of such stochastically estimated quantities e.g. $D_N^2$, need to be evaluated in an unbiased manner i.e. estimates coming from the same random vector must not be multiplied together. If $D_N^{(i)}$, $i=1,2,\dots,\NR$ are the $\NR$ stochastic estimates of the trace $D_N$, then the Unbiased Estimate (UE) of $D_N^2$ is given by
\begin{equation}
    \text{UE}\left[D_N^2\right] = \frac{2}{\NR(\NR-1)} \sum_{i=1}^{\NR}\sum_{j=i+1}^{\NR} D_N^{(i)}D_N^{(j)}.
\label{eq:unbiased_square}
\end{equation}
By contrast, the naive Biased Estimate (BE) is given by
\begin{equation}
    \text{BE}\left[D_N^2\right] = \left[\frac{1}{\NR}\sum_{i=1}^{\NR}D_N^{(i)}\right]^2.
\label{eq:biased_square}
\end{equation}
~\autoref{eq:unbiased_square} and \autoref{eq:biased_square} can both be readily generalized to any finite power or to the product of a finite number of traces.
In Ref.~\cite{Mitra.2022.prd} of this thesis, we present formulas for evaluating the unbiased estimate of such finite products in an efficient manner.
However we do not know of any corresponding formula to calculate the unbiased estimate of an infinite series such as an exponential. 
In this chapter, we will present a new way of calculating the QCD EoS based on the well-known cumulant expansion from statistics. 

The cumulant expansion method is intermediate between a strict Taylor series expansion and exponential resummation in the sense that the contribution of $D_1,\dots,D_N$ are resummed 
only up to a maximum order $M$. However, since the order is finite it is possible to evaluate the terms of the expansion in an unbiased manner. 
The cumulant expansion agrees exactly with the Taylor series expansion to $\mathcal{O}(\mu^N)$ provided that $M \ge N$. However, it contains additional
contributions at $\mathcal{O}(\mu^{N+2},\dots,\mu^{MN})$ which are exactly the contributions of $D_1,\dots,D_N$ to the higher order Taylor coefficients $\chi_{N+2},\dots,\chi_{MN}$. As we have seen before, the CP symmetry of QCD ensures that $N$ is even, implying even powers of $\mu$ only. 

Although the cumulant expansion method also works for $\muB\ne0$, in this chapter, we will present the formalism for the simpler case of finite isospin chemical 
potential $\muI$ instead. For $\muI\ne0$, the fermion determinant is real and one has no sign problem. Thus one only works with real quantities which in turn
simplifies the presentation. Moreover, the absence of the sign problem allows us to calculate observables for much larger values of
$\muI$ than would be possible for the $\muB$ case, and it is precisely for these large values that bias can become significant. And in the abscence of sign problem, it would be easy to identify the bias and the difference caused by its prescence in subsequent calculations. Lastly, the QCD
phase diagram in the $T$-$\muI$ plane is known from several studies to be interesting in its own right~\cite{Son.2000,Brandt.2017,Adhikari.2020}, and
we hope that we would be able to apply our formalism for a possible study in the future.

 In this chapter, we discuss the basic formalism of cumulant expansion in \autoref{sec:cumu form}. The setup and features of the lattice, including the associated Line of Constant Physics (LCP) is enlightened in \autoref{sec:cumu setup}. After discussing the lattice and its characteristic features, we present the results of cumulant expansion formalism in \autoref{sec:cumu results} for isospin chemical potential $\mu_I$ at $T=135$ MeV. This is to curtail the sign problem as already highlighted before, so as to get the idea of working of this formalism and understand what difference it does make in the larger picture after implementing this formalism. We then highlight the shortcomings and loopholes of this formalism in \autoref{sec:cumu drawbacks}, which will eventually motivate our new work of unbiased exponential resummation, as we will see later. We first foray into a discussion of the basic formalism of cumulant expansion in the next section as follows.

\section{Formalism and Discussion}
\label{sec:cumu form}
We consider Lattice QCD with $2+1$ flavors of rooted staggered quarks. The partition function at non-zero isospin chemical potential $\muI$ is given by 
\begin{equation}
    \Z(T,\muI) = \int \mathcal{D}U e^{-S_G(T)} \,\det \M(T,\muI),
\label{eq:partition_function}
\end{equation}
where $\det \M(T,\muI)$ is shorthand for
\begin{equation}
\det \M(T,\muI) = \prod_{f=u,d,s}\big[\det \M_f(m_f,T,\mu_f)\big]^{1/4},
\label{eq:determinant}
\end{equation}
with $m_u=m_d$, $\mu_u=-\mu_d=\muI$ and $\mu_s=0$. The excess pressure $\Delta P(T,\muI) \equiv P(T,\muI) - P(T,0)$ is given by
\begin{equation}
    \frac{\Delta P(T,\muI)}{T^4} = \frac{1}{VT^3} \, \ln \left[\frac{\Z(T,\muI)}{\Z(T,0)}\right]
\label{eq:excess_pressure}
\end{equation}
where $V$ is the spatial volume and $T$ is the temperature of the system, considered as a grand canonical ensemble for the present thermodynamic analysis. By employing the same arguments as in Ref.~\cite{Mondal.2022}, we can write
\begin{equation}
    \frac{\Z(T,\muI)}{\Z(T,0)} = \Bigg \langle
    \exp\left[\sum_{n=1}^\infty \frac{D_{2n}^I(T)}{(2n)!}\mT^{2n}\right] \Bigg \rangle,
    \label{eq:allorders}
\end{equation}
where the angular brackets $ \langle \mathcal{O} \rangle $ represent the expectation value of observable $\mathcal{O}$, where the expectation value is taken over a gauge field ensemble generated at $\mu_u=\mu_d=\mu_s=0$, and
\begin{equation}
    D^I_{n}(T) = \frac{\partial^n \left[\ln\det M(\muI)\right]}{\partial(\mu_I/T)^n}\Bigg\rvert_{\muI=0}.
\label{eq:Dn}
\end{equation}
The presence of only even powers is because the odd $\muI$ derivatives vanish identically. Since even derivatives of the quark determinant are purely real, we see that $\det \M(\muI)$ is purely real and hence there is no sign problem. It must be noted that this is true even when $\muI$ is purely imaginary.

The $D_n^I$ can be expressed as traces of various operators~\cite{Allton.2005,Gavai.2004}. In Lattice calculations, the first $N$ derivatives $D^I_1,\dots,D_N^I$ are calculated stochastically using $\NR$ random vectors, where $\NR$ is typically of order $\sim \mathcal{O}(10^2 - 10^3)$. Then $\dP(T,\muI)/T^4$ is
approximately equal to
\begin{equation}
    \frac{\Delta P_N^R(T,\muI)}{T^4} = \frac{\Nt^3}{\Ns^3} \ln \Bigg\langle \exp\left[\sum_{n=1}^{N/2} \frac{\D_{2n}^I(T)}{(2n)!}\mT^{2n}\right]\Bigg\rangle 
\label{eq:resummation}
\end{equation}

\begin{equation}
    \D_{2n}^I = \frac{1}{\NR} \sum_{k=1}^{\NR} \left[D_{2n}^{I(k)}\right] 
    \label{eq:Bar DnI}
\end{equation}

Here $\Ns$ and $\Nt$ are the number of lattice sites in the spatial and temporal directions respectively, while $\D^I_{2n}$ is the average of the $\NR$ stochastic
estimates of $D^I_{2n}$. Eq.~\autoref{eq:resummation} is the $N^{\text{th}}$ order exponential resummation formula for $\dP(T,\muI)/T^4$. In the limit
$\NR\to\infty$, it accurately resums the contribution of the first $N$ derivatives $D_1^I,\dots,D_N^I$ to all orders in $\muI$~\cite{Mondal.2022}. For finite $\NR$, i.e for 
$\NR<\infty$ however, the formula contains bias. This is easily seen if one writes the exponential as an infinite series in $\muI$. The series expansion leads to terms
such as $(\D^I_{2m})^p (\D^I_{2n})^q\cdots$, and we have already seen that such products are biased due to multiplication of estimates coming from
the same random vector, out of a sample of stochastically generated mutually independent random volume sources. 

The well-known cumulant expansion formula from statistics states that
\begin{equation}
    \ln \big\langle e^{tX} \big\rangle = \sum_{k=1}^\infty \frac{t^k}{k!}\,\Ca_k(X).
\label{eq:cumulant_expansion}
\end{equation}
The coefficients $\Ca_k(X)$ are known as the cumulants of $X$~\cite{Kubo.1962, Endres.2011}. The first four cumulants are given by
\begin{align}
    \Ca_1(X) &= \langle X \rangle, \notag \\
    \Ca_2(X) &= \langle X^2 \rangle - \langle X \rangle^2, \notag \\
    \Ca_3(X) &= \langle X^3 \rangle - 3 \langle X^2 \rangle \langle X \rangle + 2 \langle X \rangle^3, \notag \\
    \Ca_4(X) &= \langle X^4 \rangle - 4 \langle X^3 \rangle \langle X \rangle - 3 \langle X^2 \rangle^2 
            + 12 \langle X^2 \rangle \langle X \rangle^2 - 6 \langle X \rangle^4.
\label{eq:first_four_cumulants}
\end{align}
In our case $t=1$, which we assume lies within the radius of convergence of the cumulant expansion, and $X\equiv X_N(T,\muI)$, where
\begin{equation}
    X_N(T,\muI) = \sum_{n=1}^{N/2} \frac{\D_{2n}^I(T)}{(2n)!}\mT^{2n}
\label{eq:exp_arg}
\end{equation}
$\Bar{D}_{2n}^I(T)$ is already defined in Eq.~\autoref{eq:Bar DnI}. Truncating Eq.~\autoref{eq:cumulant_expansion} at $k=M \ge N/2$ gives us yet another way to estimate $\dP/T^4$, namely
\begin{equation}
    \frac{\dP^C_{N,M}(T,\muI)}{T^4} = \frac{\Nt^3}{\Ns^3}\; \sum_{k=1}^{M \ge N/2} \frac{1}{k!}\, \mathcal{C}_k\big(X_N(\muI)\big), \quad \big(\muI \equiv \muI/T\big)
\label{eq:cumulant_pressure}
\end{equation}
Here, we have $M\ge N/2$ instead of $M \ge N$ since the first non-vanishing isospin derivative is $D_2^I$ rather than $D_1^I$. Eq.~\autoref{eq:cumulant_pressure} may be compared to the familiar Taylor series expansion of $\dP/T^4$, which in our case is given by
\begin{equation}
    \frac{\Delta P^T_N(T,\muI)}{T^4} = \sum_{n=1}^{N/2} \frac{\chi_{2n}^I(T)}{(2n)!}\mT^{2n}
\label{eq:QNS_pressure}
\end{equation}
The restriction $M\ge N/2$ in Eq.~\autoref{eq:cumulant_pressure} ensures that the cumulant and Taylor expansions of the pressure agree term-by-term up to $\mathcal{O}(\muI^N)$. However, the cumulant expansion also contains additional terms proportional to $\muI^{N+2},\dots,\muI^{MN}$. These extra terms are the same terms that appear in the calculation of the higher order Taylor coefficients $\chi_{N+2}^I,\dots,\chi_{MN}^I$. The cumulant expansion thus manages to capture some of the higher order contributions to $\dP/T^4$, even though it is not a resummation to all orders in $\mu$ like Eq.~\autoref{eq:resummation}.
Unlike Eq.~\autoref{eq:resummation} however, only finite products of traces appear in Eq.~\autoref{eq:cumulant_pressure}. 
Thus, the cumulant expansion is free of the bias that can affect exponential resummation, upto a finite order in $\mu$, here $\muI$.

Finally, we will also present results for the net isospin density $\dN(T,\muI)$ which is given by
\begin{equation}
    \frac{\dN(T,\muI)}{T^3} = \frac{\partial}{\partial \left(\mu_I/T\right)}\left[\frac{\dP(T,\mu_I)}{T^4}\right].
\end{equation}
The Taylor series expression $\dN^T_N(T,\muI)$ for the same is straightforward. The resummed and cumulant expansion expressions $\dN^R_N(T,\muI)$ and 
$\dN^C_{N,M}(T,\muI)$ can be obtained by differentiating ~\autoref{eq:resummation} and \autoref{eq:cumulant_pressure} respectively. We do not write down the explicit expressions here. It must be noted however, that the resummed formula for number density $\dN_N^R(T,\muI)$, unlike the cumulant expansion expression and Taylor expansion, involves a ratio of expectation values.

\section{Computational Setup}
\label{sec:cumu setup}

To verify our formalism, we made use of the data generated by the HotQCD collaboration for their calculations of the finite-density EoS, finite-density chiral
crossover temperature and conserved charge cumulants using Taylor series expansions~\cite{Bazavov.2017,Bazavov.2018,Bollweg.2022}. The data
consists of $2+1$-flavor gauge configurations with $\Nt=8, 12$ or $16$ and $\Ns=4\Nt$ in the temperature range $125$~MeV $\lesssim T\lesssim 178$ MeV. The 
configurations were generated using a Symanzik-improved gauge action~\cite{Symanzik.1983.187,Symanzik.1983.205,Weisz.1985} and the Highly Improved Staggered Quark (HISQ)
action~\cite{Follana.2007,Bazavov.2010,Bazavov:2011nk,Bazavov:2014pvz,Bollweg:2021vqf} for fermions. The lattice spacing was determined using both the Sommer parameter $r_1$ as well as the
decay constant $f_K$. The temperature values quoted in this paper were obtained using the $f_K$ scale. For each lattice spacing, the light and strange quark bare 
masses were tuned so that the pseudo-Goldstone meson masses reproduced the physical pion and kaon masses. A description of the gauge ensembles, along with scale setting and Line of constant physics (LCP) fixation 
can be found in Ref.~\cite{Bollweg.2021} and \autoref{Appendix 2}.

The results presented here were obtained with around $20,000$ configurations for $T=135$~MeV, generated with $\Nt=8$ and $N_{\sigma}=4N_{\tau}$. On each gauge configuration, the first eight 
derivatives $D_1^f,\dots,D_8^f$ for each quark flavor were estimated stochastically using around $2000$ Gaussian random volume sources for $D^f_1$ and around $500$
sources for the rest. We used the exponential-$\mu$ formalism~\cite{Hasenfratz.1983} to calculate the first four derivatives, while the
linear-$\mu$ formalism~\cite{Gavai.2012linear,GAVAI.2015} was used in calculating all higher derivatives, the details of which have been depicted in \autoref{Appendix 3}.

\section{Results and important aspects}
\label{sec:cumu results}

\begin{figure}
 \centering
      \includegraphics[width=0.49\textwidth]{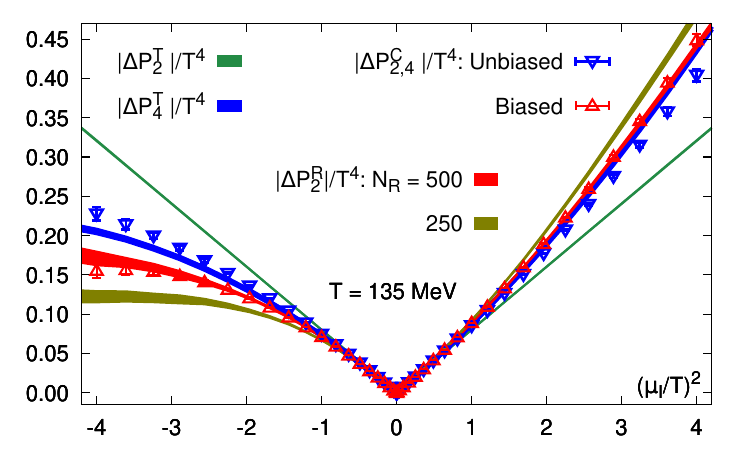}
      \includegraphics[width=0.49\textwidth]{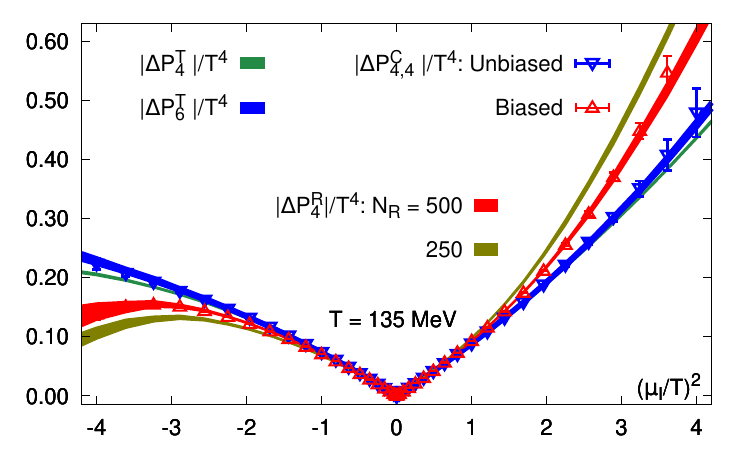} 
      \caption{\small \hspace{2mm}Comparison of the results for the excess pressure $\dP(T,\muI)$ obtained using Taylor series expansion, exponential resummation and
cumulant expansion. Upper~(lower) plots show the results for $2^{\text{nd}}$ ($4^{\text{th}})$ order exponential resummation and  $(N,M)=(2,4)$ ($(N,M)=(4,4)$) cumulant
expansion. Results for real and imaginary $\muI$ are plotted on the positive and negative $(\muI/T)^2$ axis respectively.
Biased and unbiased cumulant expansion results are the upright and inverted triangles respectively. Red and yellow bands depict the resummed results
calculated using $\NR=500$ and $250$ Gaussian random sources respectively. Finally, green and blue bands depict the $2^{\text{nd}}$ and $4^{\text{th}}$ ($4^{\text{th}}$ and
$6^{\text{th}}$) order Taylor results respectively.}
      \label{fig:PT4_second_fourth}
  \end{figure}

\begin{figure}
      \includegraphics[width=0.49\textwidth]{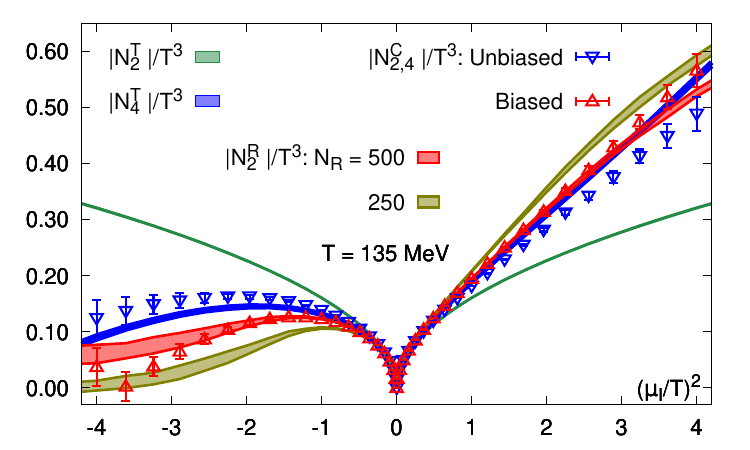}
      \includegraphics[width=0.49\textwidth]{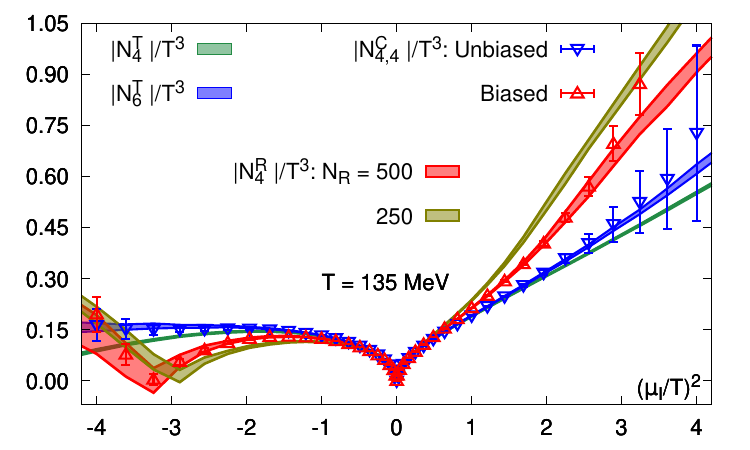}
      
      \caption{Comparison of the results for the net isospin density $\dN(T,\muI)$ obtained using Taylor series expansion, exponential resummation and
cumulant expansion. Upper~(lower) plots show the results for $2^{\text{nd}}$ and $4^{\text{th}}$ ($4^{\text{th}}$ and $6^{\text{th}}$) order Taylor, $2^{\text{nd}}$ ($4^{\text{th}})$ order
exponential resummation and $(N,M)=(2,4)$ ($(N,M)=(4,4)$) cumulant expansion. All colors and symbols are the same as in Fig.~\ref{fig:PT4_second_fourth}.}

      \label{fig:NT3_second_fourth}
\end{figure}

We present our results for the excess pressure $\dP(T,\muI)$ and the net isospin density $\dN(T,\muI)$ in Figs.~\ref{fig:PT4_second_fourth} and
\ref{fig:NT3_second_fourth} respectively. These observables were calculated using a $2^{\text{nd}}$, $4^{\text{th}}$ and $6^{\text{th}}$ order Taylor series expansion, $2^{\text{nd}}$
and $4^{\text{th}}$ order exponential resummation and cumulant expansion with $(N,M) = (2,4)$ and $(4,4)$. The results were obtained for both real and imaginary
$\muI$, in the range $0 \leqslant \lvert \muI/T \rvert \leqslant 2$.
The upper plots in each figure compare the results of a $2^{\text{nd}}$ order resummation and a $(2,4)$ cumulant expansion to $2^{\text{nd}}$ and $4^{\text{th}}$ order Taylor
expansions, while the lower plots compare $4^{\text{th}}$ order resummation and a $(4,4)$ cumulant expansion to $4^{\text{th}}$ and $6^{\text{th}}$ order
Taylor expansions respectively.

Focusing first on the upper plots, we see that the $2^{\text{nd}}$ and $4^{\text{th}}$ order Taylor results 
start to differ significantly around $\lvert\mu_I/T\rvert=1$. For real $\muI$, this difference is seemingly captured by the resummed result, which
almost agrees with the $4^{\text{th}}$ order Taylor results both for $\dP(T,\muI)$ as well as for $\dN(T,\muI)$. For imaginary $\muI$ however, the resummed result
for both observables lies even lower than the $4^{\text{th}}$ order Taylor result. By contrast, the cumulant expansion result is in good 
agreement with the $4^{\text{th}}$ order Taylor result both for real as well as imaginary $\muI$. Note that the cumulant result too contains higher order contributions, in this case the contribution of $D_2^I$ to the Taylor coefficients $\chi^I_4$, $\chi_6^I$ and $\chi^I_8$. Despite this, the cumulant result always lies between
the two Taylor results, both for real as well as imaginary $\muI$.

One explanation for the difference between the resummed and cumulant results is the higher order contributions that are present in the former but not in the
latter. Another possibility is the bias that is present in the resummed but not in the cumulant result. To distinguish between the two possibilities, we
recalculated the cumulant result using the biased formulas for the trace products e.g. Eq.~\autoref{eq:biased_square}, rather than the unbiased
ones. We find that the biased result agrees very well with the resummed result, thus suggesting that bias, rather than the contribution from higher orders, is
responsible for the difference.

To further confirm that this is the case, we recalculated the resummed result using only $250$ random vectors instead of $500$.
The HotQCD data contains roughly $\Ob(2000)$ random estimates of $D_1$ and $\Ob(500)$ random estimates of $D_n$ for $n \geq 2$. For the isospin case however, all $n$-point correlation functions for odd $n$ are zero, implying naturally
$D_1^I$ is identically equal to zero. Therefore, we had a maximum of $500$ random volume sources available to us for the cross-check. Since bias
vanishes in the limit $\NR\to\infty$, we should expect the bias to increase when we use fewer random vectors. From Figs.~\ref{fig:PT4_second_fourth} and
\ref{fig:NT3_second_fourth}, we see that the $\NR=250$ result lies further from the Taylor and unbiased cumulant results than the $\NR=500$ result, starting from around $\lvert\muI/T\rvert=1$. Thus we see that the resummed result is indeed affected by bias for large values of the chemical potential ($\muI$ in this case).

The presence of bias must especially be accounted for when

\begin{itemize}
    \item we compare higher order results like the $4^{th}$ order results for excess pressure and number density in 
Figs.~\ref{fig:PT4_second_fourth} and
\ref{fig:NT3_second_fourth}. 
    \item we study behaviour of observables which are higher order $\mu$ derivatives of the free energy, like $\mathcal{N}/T^3$ over $\Delta P/T^4$ in this case.
\end{itemize}

This is purely because, the more fluctuating and non-monotonic higher order correlation functions come into the picture with increasing orders of cumulant expansion. And with successive higher-order $\mu$ derivatives, the multiplicative factors associated with these terms, increase multiplicatively, thereby augmenting the contributions of the individual fluctuations in the calculations. These observations are clearly reflected in the relative magnitude of errorbars among different orders of pressure and number density calculations as well as between pressure and number density for a given order of cumulant expansion. 

We see that the sixth order Taylor correction to the fourth order result is small over the entire 
range of $\muI$ considered here, and in fact it has the opposite sign for imaginary $\muI$. The unbiased cumulant calculation reproduces both these features, even though it does not contain the trace $D_6^I$. By contrast, the resummed correction is both 
large and of the same sign as the second order calculation for imaginary $\muI$. The biased cumulant calculation again reproduces the resummed calculation, and the $\NR=250$ calculation mostly increases rather than decreases this discrepancy.

\section{Drawbacks}
\label{sec:cumu drawbacks}

 Despite all these advantages, the major drawback of cumulant expansion is the truncation of the original exponential series, as shown in \autoref{eq:cumulant_expansion}. We have already seen the partition function being given as the gauge ensemble average of the various reweighting factors obtained from the ensemble of gauge field configurations stochastically generated at $\mu=0$. 
 
 Strictly speaking, although, we have seen from our previous discussion on cumulant expansion that the $6^{th}$ order correction is small as compared to the $4^{th}$ order Taylor expansion in $\muI$, that does not however imply that successive higher order corrections get smaller and can be safely ignored, while considering the complete series in $\mu$. In fact, there is no comprehensive generic theory or sufficient piece of evidence till date, to prove this fact. Consequently, the series truncation to some finite order in $\mu$ also does not therefore guarantee that the behaviour of the full series can be encapsulated through this truncation. 
 
 So, not only we are deprived of a proper reweighting factor and an associated phasefactor, it is logically sensible for the time being to comment that the knowledge of partition function gets also lost through this process of cumulant expansion, and so is the full QCD thermodynamics. And it is this very setback of cumulant expansion, that motivates us towards the new formalism of unbiased exponential resummation, which forms the second part of my PhD work. 

\vspace{5cm}

 	\cleardoublepage

\chapter{Unbiased Exponential Resummation}
\label{Chapter 8}

\graphicspath{{Figures/Chapter-8figs/PDF/}{Figures/Chapter-8figs/}}

\vspace{.8cm}

 \hspace{3mm}In this chapter, I have extensively discussed about the second part of my doctoral work, which aims to find a new formalism of unbiased exponential resummation at a finite chemical potential $\mu$. The work has been comprehensively discussed and has been elaborated substantially with necessary plots. This chapter gives an elaborate discussion of two of my recent papers~\cite{Mitra.2022.pos,Mitra.2022.arx}.
 
\section{Motivation and Introduction}
\label{sec:unb exp mot}
\hspace{3mm}The cumulant expansion method~\cite{Mitra.2022.prd}, as outlined in the last chapter, proves to be a very useful and fruitful method, when it comes to controlling the emergence of biased estimates order-by-order in $\mu$. It is basically the drawbacks of this approach that eventually motivates and leads us to the new and novel approach of an unbiased exponential resummation. As already mentioned, the biased estimates of different correlation functions $D_n$ do appear in the method of exponential resummation, entirely because of using exponential function in estimating the partition function. The cumulant expansion truncates the series surely and provides a way to replace these biased estimates with unbiased counterparts.

But due to this truncation, there is no reweighting factor and phasefactor, which have already been outlined in \autoref{Chapter 2}. As already mentioned, the exponentiation of $D_n$ series $\sim \exp{\left(\mu^n D_n\right)}$ is the reweighting factor and the imaginary part of this exponentiation gives the phaseangle, the cosine of which gives the measure of phasefactor. These quantities are invaluable from the perspective of identifying the resummation breakdown and the singularities of $\Z$ in the complex $\mu$ plane. Also, the truncation poses another problem. Since there is no well-defined knowledge of higher order correlation functions appearing along with higher powers of $\mu$ in the series expansion of $\Z$, it is difficult to comment on the nature and degree of convergence or divergence of the series. Hence, it is safe to conclude that the truncated version of $\Z$ provides an inadequate estimate of $\Z$ and so, retaining the exponential form of $\Z$ is paramount for a sufficient and adequate estimate of $\Z$. And at the end of the day, there is no thermodynamics without a proper partition function $\Z$.

All these setbacks are plugged in by the new formalism of unbiased exponential resummation. This formalism preserves the anatomy of the original exponential resummation approach, by expressing the partition function $\Z$ as the gauge ensemble average of the exponential of some argument $\left(\Z \sim \LA \exp{\left[f\left(D_n,\mu\right)\right]}\RA\right)$, involving chemical potential $\mu$ and $n$ point correlation functions $D_n$. This also ensures that we have a new well-defined reweighting factor and phasefactor, enabling us to re-calculate roots of $\Z$ in complex $\mu$ plane. Most importantly, all these are obtained preserving unbiased estimates to a finite order in $\mu$, thereby ensuring greater statistical reliability of our results at least within the breakdown domain of resummation. As mentioned before, this is  analogous to the radius of convergence of the corresponding Taylor series, being considered for resummation method. 

 In this chapter, we present the basic formalism in \autoref{sec:Unb exp formalism} and discuss the working of the unbiased exponential resummation in the two bases, namely the chemical potential basis in \autoref{Chem pot basis} and cumulant basis in \autoref{Cumu basis}. We then conduct a profound comparative study between these two bases and substantiate the advantages and disadvantages with supportive plots in \autoref{subsec:chem and cumu}. The setup of the lattice and details for the Line of Constant Physics (L$CP$) is enlightened in \autoref{sec:unb exp setup}, before an explanation of exponential and linear $\mu$ formalism in \autoref{Appendix 3} to illustrate the difference in calculation between correlation functions $D_n$ upto $n \leq 4$ and $D_n$ for $n \geq 5$. Finally, we present the results in \autoref{sec:unb exp results} and do so for $\muI$ in \autoref{subsec:unb exp results isospin} and $\muB$ in \autoref{subsec:unb exp results baryon}, before enlightening the computationally beneficial aspects of this new formalism in \autoref{sec:comp benefits}.        

The mathematical formalism is discussed in two bases as follows:

\section{Formalism and Discussion}
  \label{sec:Unb exp formalism}
  In this formalism, we consider lattice QCD with $2+1$ flavors of rooted staggered quarks defined on an $\Ns^3~\times~\Nt$ lattice. As before, we consider $u,d,s$ quarks with $m_u=m_d=m_l$ and $m_l=m_s/27$, where $m_u,m_d$ and $m_s$ are the masses of the up, down and strange quarks respectively. The partition function $\Z(T,\mu)$ at temperature $T$ and baryochemical potential $\muB$ is given by 
\begin{equation}
    \Z(T,\mu) = \int \mathcal{D}U e^{-S_G(T,U)} \,\det \M(T,\mu,U)
    \label{eq:Partition_function}
\end{equation}
where $S_G(T,U)$ is the gauge action and with $\mu \in (\muB, \mu_S, \muI)$, $\det \M(T,\mu, U)$ is the fermion determinant given by
\begin{equation}
    \det \M(T,\mu) = \prod_{f=u,d,s}\big[\det \M_f(m_f,T,\mu_f)\big]^{1/4}
    \label{eq:fermion_determinant}
\end{equation}
with $m_u=m_d$. 
From $\Z(T,\mu)$, the excess pressure $\dP(T,\mu) \equiv P(T,\mu)-P(T,0)$ can be calculated as
\begin{equation}
    \frac{\dP(T,\mu)}{T^4} = \frac{1}{VT^3} \, \ln \left[\frac{\Z(T,\mu)}{\Z(T,0)}\right]
    \label{eq:Excess_pressure}
\end{equation}
where $V$ is the volume of the system. Owing to the sign problem of lattice QCD, it is not possible to evaluate Eqn. \eqref{eq:excess_pressure} directly. An alternative approach is to instead expand the right hand side in a Taylor series in $\mu$ up to some (even) order $N$ viz.
\begin{equation}
    \frac{\dP_N^T(T,\mu)}{T^4} = \sum_{n=1}^{N/2} \frac{\chi_{2n}(T)}{(2n)!} \left(\frac{\mu}{T}\right)^{2n}
    \label{eq:taylor_pressure}
\end{equation}
This is the $N$th order Taylor estimate of $\dP(T,\mu)$. Due to the particle-antiparticle symmetry or the $CP$ symmetry of the system, only the even powers of $\mu$ appear in the expansion, making Eqn.~\eqref{eq:taylor_pressure} even in $\mu$. The calculation of the $2n^{th}$ coefficient $\chi_{2n}$ requires calculating terms such as $\langle D_1^a D_2^b \cdots D_N^k\rangle$ where
\begin{equation}
    D_n(T) = \frac{\partial^n \ln \det\M(T,\mu)}{\partial (\mu/T)^n}\, \bigg\vert_{\mu=0}
    \label{eq:DN}
\end{equation}
 $1\cdot a + 2\cdot b + \dots + N\cdot k = 2n$, and the angular brackets $\langle\Ob\rangle$ denote the expectation value of an observable $\Ob$ with respect to an ensemble of gauge configurations generated at the same temperature $T$ and at $\mu=0$~\cite{Allton.2005,Gavai.2004}, as follows:
\begin{equation}
    \big\langle \mathcal{O}(T,\mu)\big\rangle = \frac{\int \mathcal{D} U \, e^{-S_G(T)} \, \Ob(T,\mu)\det \M(T,\mu=0)}{\int \mathcal{D} U\,e^{-S_G(T)} \det \M(T,\mu=0)}.
    \label{eq:angular_brackets}
\end{equation}
The derivatives or the $n$-point correlation functions $D_1,\dots,D_N$ also contribute to higher-order Taylor coefficients through products such as $D_ND_1$, $D_N^2$, etc. As already mentioned in \autoref{sec:Exp resummation}, the contribution of $D_1,\dots,D_N$ to all orders in $\mu$ takes the form of an exponential as in \autoref{eq:pr}. Hence the resummed estimate of $\dP(T,\mu)$ can be written as
\begin{equation}
    \frac{\dP_N^R(T,\mu)}{T^4}=\frac{\Nt^3}{\Ns^3} \ln \left\langle \exp \left[ \sum_{n=1}^N \frac{\overline{D_n}(T)}{n!}\left(\frac{\mu}{T}\right)^n \right] 
    \right\rangle.
    \label{eq:resummed_pressure}
\end{equation}

In a typical Lattice QCD calculation, the $D_n$ are not known exactly but rather estimated stochastically, using $\NR \sim \Ob(10^2$ - $10^3)$ random volume sources per gauge configuration. Hence in \autoref{eq:resummed_pressure} we have replaced $D_n$ by $\overline{D_n}$, where the overline denotes the average of the $\NR$ stochastic estimates of $D_n$. As $\NR\to\infty$, the average $\overline{D_n}$ approaches the true value $D_n$ and \autoref{eq:resummed_pressure} becomes exact. This is outlined in Ref.~\cite{Mitra.2022.prd} and also briefly in \autoref{Chapter 6}. For finite $\NR$ however, the exponential factor in \autoref{eq:resummed_pressure} contains stochastic bias, which can be seen as follows: If we expand the exponential in a Taylor series, then we get terms such as $(\overline{D_m})^p (\overline{D_n})^q\cdots$ which contain products of estimates coming from the same random vector and are hence not truly independent estimates. The contribution coming from such products is the stochastic bias; although it is suppressed by powers of $\NR$, it can still be significant depending upon the observable and the value of the chemical potential. It therefore needs to be subtracted in order to obtain a better estimate of the exponential.

Stochastic bias is not an issue in the calculation of the Taylor coefficients, although such products also appear there, because there exist formulas to efficiently evaluate the unbiased estimate of finite products of the derivatives~\cite{Mitra.2022.prd}. Taking advantage of this, one way of avoiding stochastic bias is through the cumulant expansion of the excess pressure as discussed in \autoref{sec:cumu form} in \autoref{Chapter 5}, which follows from
\begin{align}
&\ln\left\langle\exp\left[\sum_{n=1}^N \frac{D_n(T)}{n!}\left(\frac{\mu}{T}\right)^n \right]\right\rangle
= \sum_{m=1}^\infty\frac{\Ka_m\big(X_N(T,\mu)\big)}{m!}, \notag \\
&X_N \equiv \sum_{n=1}^N \frac{D_n(T)}{n!}\left(\frac{\mu}{T}\right)^n, \notag \\
&\Ka_1 = \left\langle X_N \right\rangle,\; \Ka_2 = \left\langle X_N^2 \right\rangle - \left\langle X_N \right\rangle^2,\;\text{etc.}
\end{align}
However as already noted, all-orders resummation is lost in this approach, as is knowledge of the phase factor. Therefore, instead of the above approach, 
we define an improved stochastic estimate of the exponential operator, in which the bias is subtracted up to a certain order in $\mu$, by replacing the argument of the exponential in \autoref{eq:resummed_pressure} as follows:
\begin{equation}
    \sum_{n=1}^N \frac{\overline{D_n}(T)}{n!}\left(\frac{\mu}{T}\right)^n \longrightarrow \sum_{m=1}^M \frac{\mathcal{C}_m(T)}{m!}\left(\frac{\mu}{T}\right)^m,
    \label{eq:XN_to_CM}
\end{equation}
with the $\mathcal{C}_m$ chosen so that the Taylor expansion of the new exponential is unbiased to $\Ob(\mu^M)$ viz.
\begin{align}
    &\mathcal{C}_1 = \overline{D_1}, \notag \\
    &\mathcal{C}_2 = \overline{D_2} + \left(\overline{D_1^2} - \overline{D_1}^2\right), \notag \\
    &\mathcal{C}_3 = \overline{D_3} + 3\left(\overline{D_2D_1} - \overline{D_2}\;\overline{D_1}\right) + 
           \left(\overline{D_1^3} - 3\,\overline{D_1^2}\;\overline{D_1} + 2\,\overline{D_1}^3\right), \notag \\
    &\mathcal{C}_4 = \overline{D_4} + 3\left(\overline{D_2^2} - \overline{D_2}^2\right)+ 4\left(\overline{D_3D_1} - \overline{D_3}\;\overline{D_1}\right) \notag \\ &+
      6\left( \overline{D_2D_1^2} - \overline{D_2}\;\overline{D_1^2}\right)
    - 12 \left(\overline{D_2D_1}\;\overline{D_1} - \overline{D_2}\;\overline{D_1}^2\right) \notag \\ &+ 
     \left(\overline{D_1^4} - 4\,\overline{D_1^3}\;\overline{D_1} + 12\,\overline{D_1^2}\;\overline{D_1}^2 - 6\,\overline{D_1}^4 - 3\,(\overline{D_1^2})^2\right),
\label{eq:Cm}
\end{align}
and so on. The overlines in the above equations represent unbiased estimates e.g. $\overline{D_m^pD_n^q}$ is the unbiased estimate of $D_m^pD_n^q$. When this exponential is substituted in \autoref{eq:resummed_pressure}, we obtain as follows:
\begin{equation}
    \frac{\dP^{R(\text{unb})}_M(T,\mu)}{T^4} = \frac{\Nt^3}{\Ns^3} \, \ln \left\langle \exp \left[ \sum_{m=1}^M \frac{\mathcal{C}_m(T)}{m!}\left(\frac{\mu}{T}\right)^m \right]\right\rangle,
    \label{eq:unbiased_resummed-1}
\end{equation}
 and $\dP^{R(\text{unb})}(T,\mu)/T^4$ expanded in a Taylor series in $\mu$, the resulting expression is also unbiased up to the same order. We note the following points:
\begin{enumerate}
\item Since the Taylor expansion is unbiased to $\Ob(\mu^M)$, and since we work with the first $N$ derivatives, the Taylor expansion of \autoref{eq:unbiased_resummed-1} will be identical to the QCD Taylor series \autoref{eq:taylor_pressure} up to $\Ob(\mu^L)$, where $L = \min\,\{M,N\}$. For the rest of this paper, we will set $M=N$ in evaluating \autoref{eq:unbiased_resummed-1} and compare our results with the $N$th order Taylor series.

\item From \autoref{eq:Cm}, we see that the first term in each $\mathcal{C}_m$ is simply $\overline{D_m}$. In the limit $\NR \to \infty$, this term approaches the correct value of $D_m$. The rest of the terms for each $\mathcal{C}_m$ also cancel each other out as $\NR\to\infty$, since in that limit the distinction between biased and unbiased products vanishes. Thus $\mathcal{C}_m\to D_m$ as $\NR\to\infty$ and hence \autoref{eq:unbiased_resummed-1} too represents an all-orders resummation of the derivatives $D_1,\dots,D_N$, the only difference this time being that the stochastic bias is eliminated to $\Ob(\mu^N)$.
\end{enumerate}

Although \autoref{eq:unbiased_resummed-1} is an improvement over \autoref{eq:resummed_pressure}, it is possible to do still better. In a typical Lattice QCD calculation, each stochastic estimate of the $D_1,\dots,D_N$ is constructed using the same random source. Therefore, the different stochastic estimates can be actually thought of as different estimates of the operator $X_N(T,\muB) \equiv \sum_{n=1}^N D_n(T)(\muB/T)^n / n!$. It is possible to write a version of \autoref{eq:resummed_pressure} in which the bias is eliminated up to a certain power of $X_N$ itself, by making the replacement
\begin{equation*}
    \sum_{n=1}^N \frac{\overline{D_n}(T)}{n!}\left(\frac{\mu}{T}\right)^n \longrightarrow \sum_{m=1}^M \frac{\mathcal{W}_m\big(X_N(T,\muB)\big)}{m!}\mT^m
    \label{eq:XN_to_WM}
\end{equation*}
where
\begin{align}
   \Wc_1 &= \overline{X_N}, \notag \\
   \Wc_2 &= \overline{X_N^2} - \big(\overline{X_N}\big)^2, \notag \\
   \Wc_3 &= \overline{X_N^3} - 3\,\big(\overline{X_N}\big)\;\big(\overline{X_N^2}\big) + 2\,\big(\overline{X_N}\big)^3, \notag \\
   \Wc_4 &= \overline{X_N^4}- 4\,\big(\overline{X_N^3}\big)\;\big(\overline{X_N}\big) - 3\,\big(\overline{X_N^2}\big)^2 \notag \\ 
         &+ 12\,\big(\overline{X_N}\big)^2\;\big(\overline{X_N^2}\big) - 6\,\big(\overline{X_N}\big)^4,
    \label{eq:Lm}
\end{align}
and so on. The resulting expression for $\dP(T,\mB)$ viz.
\begin{equation}
    \frac{\dP^{R(\text{unb})}_{N,M}(T,\muB)}{T^4} = \frac{\Nt^3}{\Ns^3} \, \ln \left\langle \exp \left[ \sum_{m=1}^M \frac{\Wc_m(X_N(T,\muB))}{m!}\right]\right\rangle,
    \label{eq:unbiased_resummed-2}
\end{equation}
reproduces the unbiased cumulant expansion of the resummed pressure~\cite{Mitra.2022.prd} to order $m=M$ viz.
\begin{align}
\frac{\dP^{C}_{N,M}(T,\mB)}{T^4} &= \frac{\Nt^3}{\Ns^3} \, \ln \Big \langle e^{X_N(T,\,\mB)} \Big \rangle, \notag \\
&= \frac{\Nt^3}{\Ns^3} \, \sum_{m=1}^M \frac{\Ka_m\big(X_N(T,\mB)\big)}{m!} \text{ + higher orders.}
\end{align}
In fact, Eqs.~\eqref{eq:Lm} are similar to those for the cumulant expansion of the pressure with two differences: (i) the powers $X_N^p$ are replaced by their respective unbiased estimates $\overline{X_N^p}$, and (ii) the expansion is in the space of all random estimates for a single gauge configuration, rather than in the space of all gauge configurations. In the limit $\NR\to\infty$, the difference between biased and unbiased estimates vanishes, and the $\Wc_m$ become the cumulants of $X_N$ over the set of all random estimates for a single gauge configuration. In the double limit $\NR\to\infty$ and $M\to\infty$ therefore, \autoref{eq:XN_to_WM} is just the cumulant expansion of $\overline{\;e^{X_N}\,}$.  This observation helps to understand the construction of the unbiased exponential: It is the systematic (order-by-order) replacement of the incorrect (biased) estimate $e^{\;\overline{X_N}}$ by the correct estimate $\overline{\;e^{X_N}\,}$ of the exponential factor.

\hspace{1mm}The two bases in which we have implemented this formalism and carried out our calculations are as follows
\begin{itemize}
    \item Chemical potential basis ($\mu$ basis)
    \item Cumulant basis ($X$ basis)
\end{itemize}

Before going into the discussions of these bases separately, it is important to address the question that what is the uniqueness of this method, or how this method is different from the old exponential resummation. A naive and quick overview tells us that the argument of the exponential is different; the real deal or novelty of this formalism lies in designing a suitable argument, which after exponentiation and consideration of a natural logarithm, yields a thermodynamic pressure $\Delta P$ which exactly is identical to the Taylor counterpart upto the desired order in $\mu$. This ensures the prescence of unbiased estimates upto that order in $\mu$, since the Taylor coefficients or the scaled Quark number susceptibilities are constructed using unbiased powers of appropriate correlation functions $D_n$. 

Once the pressure $\Delta P$ gets set or tuned properly as an offset, the number density $\mathcal{N}$ and higher order susceptibilities which are subsequent higher order $\mu$ derivatives of pressure follow suit. Because the issue of biased and unbiased estimates is confined only within the different correlation functions; it does not propagate to the associated $\mu$ raised to appropriate powers. For example, if $\mathcal{O}(\mu^N)$ is unbiased in $\Delta P$, then automatically $\mathcal{O}(\mu^{N-1})$ is unbiased in $\mathcal{N}$ and so on for higher order $\mu$ derivatives. 

  \subsection{Chemical potential basis}
  \label{Chem pot basis}
  \hspace{2mm}In chemical potential ($\mu$) basis, we define the excess pressure from a newly defined partition function, following the usual prescription of the exponential resummation, but differing conspicuously in exponential argument as follows:
        \begin{align}   
     \frac{\Delta P_{ub}^{N}(\mu)}{T^4} &= \frac{1}{VT^3} \hspace{1mm}\ln \hspace{1mm} \mathcal{Z}_{ub}^{N}(\mu) , \notag \\ \mathcal{Z}_{ub}^{N}(\mu) &= \left \langle \text{Re} \bigg[\exp \Big(A_N(\mu)\Big)\bigg] \right \rangle , \notag \\
    A_N(\mu) &= \sum_{n=1}^{N} \left(\frac{\mu}{T}\right)^n \hspace{.5mm}\frac{\mathcal{C}_{n}}{n!}
     \label{eq:mu basis}
      \end{align}  
      where the first four $\mathcal{C}_n$ are given as follows:

 
\begin{align}
    \mathcal{C}_1 &= \overline{D_1}, \notag \\
    \mathcal{C}_2 &= \overline{D_2} + \left(\overline{D_1^2} - \overline{D_1}^2\right), \notag \\
    \mathcal{C}_3 &= \overline{D_3} + 3\left(\overline{D_2D_1} - \overline{D_2}\;\overline{D_1}\right) 
    + \left(\overline{D_1^3} - 3\,\overline{D_1^2}\;\overline{D_1} + 2\,\overline{D_1}^3\right), \notag \\
    \mathcal{C}_4 &= \overline{D_4} + 3\left(\overline{D_2^2} - \overline{D_2}^2\right)+ 4\left(\overline{D_3D_1} - \overline{D_3}\;\overline{D_1}\right) \notag \\
    &\hspace{3mm}+ 6\left( \overline{D_2D_1^2} - \overline{D_2}\;\overline{D_1^2}\right) - 12 \left(\overline{D_2D_1}\;\overline{D_1} - \overline{D_2}\;\overline{D_1}^2\right) \notag \\
   &\hspace{3mm}+ \left(\overline{D_1^4} - 4\,\overline{D_1^3}\;\overline{D_1} + 12\,\overline{D_1^2}\;\overline{D_1}^2 - 6\,\overline{D_1}^4 - 3\,(\overline{D_1^2})^2\right) 
\label{eq:mu coefficients}
\end{align}

 Here, in the list of $C_n$ as given in above eqn.\eqref{eq:mu coefficients}, $\overline{D_n^pD_m^q}$ indicates unbiased $\left(p+q\right)^{th}$ power of $D_n$ and $D_m$. The mechanism of calculating unbiased powers from random volume source estimates of $D_n$, available in every gauge field configuration is discussed extensively in \autoref{Chapter 6}. The $CP$ symmetry of QCD is palpable from eqn.\eqref{eq:mu basis}, in the sense that the real part of the exponential is considered while estimating the partition function $\Z_{ub}(\mu)$ to $\mathcal{O}(\mu^N)$.
 
 The utility of considering this basis is that it is exceedingly simple in this basis, to understand the extent or order of unbiased estimates in the series. That is done by observing the degree of the polynomial $A(\mu)$ in eqn.\eqref{eq:mu basis}. In our work, we considered all correlation functions $D_n$ satisfying $n \leq N=4$ implying the degree of $A$ in \autoref{eq:mu basis} is $4$. This indicates that unbiased estimates are present upto $\mathcal{O}(\mu_B^4)$ and $\mathcal{O}(\mu_I^4)$, with $\mu_B$ and $\mu_I$ having the conventional meanings. An explicit proof how the above \autoref{eq:mu basis} reproduces $4^{th}$ order Taylor series in $\mu$ is presented in \autoref{Appendix 12}.

  \subsection{Cumulant basis}
\label{Cumu basis}
  
  \hspace{2mm}In cumulant basis, a new variable $X_N$ is defined, where 
  \begin{equation*}
  X_N =\sum_{n=1}^N \frac{\hat{\mu}^n}{n!}\hspace{.5mm}D_n  \hspace{2cm}   \left(\hat{\mu} \equiv \mu/T\right)
  \end{equation*}
 and we subsequently define excess pressure as follows
      \begin{align}   
     \frac{\Delta P_{ub}^{M}(X_N)}{T^4} &= \frac{1}{VT^3} \hspace{1mm}\ln \hspace{1mm} \mathcal{Z}_{ub}^{M}(X_N) , \notag \\ \mathcal{Z}_{ub}^{M}(X_N) &= \LA \text{Re} \bigg[\exp \Big(Y_M(X_N)\Big)\bigg] \RA , \notag \\
    Y_M(X_N) &= \sum_{n=1}^{M} \frac{\mathcal{L}_{n}(X_N)}{n!}
      \label{eq:cumulant basis}
      \end{align}  
    which would reproduce exactly the first $M$ cumulants in unbiased cumulant expansion of excess pressure as highlighted in Ref.~\cite{Mitra.2022.prd}. The $\mathcal{L}_n (X_N)$ of eqn.~\eqref{eq:cumulant basis} upto $M = 4$ are as follows:

 \begin{align}
    \mathcal{L}_1(X_N) &= \overline{X_N} \notag \\                                                 
    \mathcal{L}_2(X_N) &= \left(\overline{X_N^2}\right) - \left(\overline{X_N}\right)^2 \notag \\  
    \mathcal{L}_3(X_N) &= \left(\overline{X_N^3}\right) - 3 \hspace{.5mm}\left(\overline{X_N^2}\right) \left(\overline{X_N}\right) + 2 \hspace{.5mm}\left(\overline{X_N}\right)^3 \notag \\
    \mathcal{L}_4(X_N) &= \left(\overline{X_N^4}\right) - 4 \hspace{.5mm}\left(\overline{X_N^3}\right) \left(\overline{X_N}\right) + 12 \hspace{.5mm}\left(\overline{X_N^2}\right) \left(\overline{X_N}\right)^2 \notag \\
    &\hspace{7mm}- 6 \hspace{.5mm}\left(\overline{X_N}\right)^4 - 3\hspace{.5mm} \left(\overline{X_N^2}\right)^2
    \label{eq:Cumulant coefficients}
 \end{align}

 Just like in \autoref{Chem pot basis}, $\overline{X_N^p}$ denotes the unbiased $p^{th}$ power of $X_N$ and the method of calculating unbiased powers is exactly similar as mentioned in \autoref{Chem pot basis}.
 All the above calculations can be performed for dimensionless number density $\mathcal{N}/T^3$ and further higher order susceptibilities as well, taking the different order $\hat{\mu}$ derivatives of the excess pressure $\Delta P$ in both cumulant and $\mu_B$ bases, where $\hat{\mu} \equiv \mu/T$. One can always evaluate more number of cumulants in cumulant basis or equivalently $C_n$ for higher values of $n$ in $\mu$ basis to obtain unbiased estimates for even further higher orders in $\hat{\mu}$.
 
\section{A brief comparative discussion}
\label{subsec:chem and cumu}
\hspace{2mm} As discussed before, in $\mu$ basis, the argument of the exponential resembling the reweighting factor is expressed as a power series in terms of $\hat{\mu}$, where $\hat{\mu} \equiv \mu/T$ is the lattice compatible dimensionless chemical potential, scaled in terms of finite temperature $T$. Whereas in cumulant basis, the corresponding exponential argument is the sum of the first cumulants in $X_N$, similar to the structure of cumulant expansion formalism outlined in \autoref{Chapter 5}, with $X_N$ defined previously in \autoref{Cumu basis}. The expansion as a power series in $\mu$ is not so apparent here, although, implicitly, the same thing is performed here, mainly because $X_N$ is a function of $\mu/T$.

The advantage of using cumulant basis is that it is therefore, possible to capture more number of higher order contribution terms in this basis as compared to $\mu$ basis. Let us illustrate this with a brief simple example. Let us consider the number of cumulants $M=2$ as in \autoref{eq:cumulant basis}, only the first two coefficients $\mathcal{C}_n, n=2$ as in \autoref{eq:mu coefficients} and only the first two non-zero correlation functions $D_n$ so that $D_n=0$, for $n \geq 3$. Then, in chemical potential $(\mu)$ basis, the argument of the exponential is upto $\Ob(\mu^2)$ whereas because of the second cumulant containing an $X^2$, the argument contains terms upto $\Ob(\mu^4)$ in cumulant basis. This difference surely increases with more number of cumulants alongside increasing more number of coefficients in chemical potential $(\mu)$ basis i.e. increasing $n$ of $\mathcal{C}_n$. This gulf among the bases enhances even more with increasing number of non-zero $n$-point correlation functions $D_n$. 

As evident from the above discussion and argument, in our present calculations with $i)\hs \hs \hs \hs M=4$, $ii)\hs \hs \hs \hs \mathcal{C}_n, \hs \hs 1 \leq n \leq 4$ and $iii)\hs \hs \hs \hs D_n, \hs \hs 1 \leq n \leq 4$, the cumulant basis captures terms upto $\Ob(\mu^{16})$ in the argument, whereas the highest term spanned in the $\mu$ basis is of $\Ob(\mu^4)$. In the cumulant basis, this happens because in the $4^{th}$ cumulant, there is a term $X^4$ where $X \sim \mu\,D_1 + \cdots \mu^4\,D_4$. Although in both the bases, the completely unbiased contributions are captured only upto $\Ob(\mu^4)$ and the higher order terms in $\mu$ do contain bias estimates partially (in this case for $\mu^n$ where $5 \leq n \leq 16$), the utility and essence of using the cumulant basis is that it encapsulates a greater spectrum of the true infinite Taylor series of thermodynamic observables in terms of $\mu$ and hence, on any day, is more preferable, since we do not know anything about the nature or trend of higher order correlation functions $D_n$ and so, correspondingly the series. This is already discussed in the drawback of cumulant expansion in \autoref{Chapter 5}. Hence, one observes in this context is that if one considers $n$-point correlation functions $D_n$ where $A \leq n \leq B$ with total cumulants $M$, then unbiased powers will appear for all $\Ob(\mu^n)$, where $A \leq n \leq BM$. Further observations are summarised in a point form as follows : 

\begin{enumerate}
  \item Considering $AM$ and $BM$ are even, and $B$ is even
   \begin{itemize}
    \item If $AM < B$, then complete unbiased contributions will appear upto $\Ob(\mu^{AM})$. This means that in Taylor series, the Taylor coefficients $c_n$ are completely reproduced with $n$ being even satisfying $2 \leq n \leq AM$, and the remaining Taylor coefficients $c_n$ for $(AM+2) \leq n \leq BM$ possess unbiased contributions in $D_n$ to some partial extent. This extent decreases with increasing $n$ i.e. while increasing $n$ from $AM+2$ to $BM$.
    
    \item If $AM \geq B$, then complete unbiased contributions will appear upto $\Ob(\mu^{B})$. Correspondingly, $c_n$ with $2 \leq n \leq B$ are produced completely and partial unbiased contributions of $D_n$ remain in $c_n$ for $(B+2) \leq n \leq BM$.
    \end{itemize}

    \item Considering $M$ is even, $B$ is odd, so that $AM$ and $BM$ are even
   \begin{itemize}
    \item If $AM < B$, then complete unbiased contributions will appear upto $\Ob(\mu^{AM})$. This means that in Taylor series, the Taylor coefficients $c_n$ are completely reproduced where $2 \leq n \leq AM$, and the remaining Taylor coefficients $c_n$ for $(AM+2) \leq n \leq BM$ possess unbiased contributions in $D_n$ to some partial extent. This extent decreases with increasing $n$ i.e. while increasing $n$ from $(AM+2)$ to $BM$.
    
    \item If $AM \geq B$, then complete unbiased contributions will appear upto $\Ob(\mu^{B})$. Since the Taylor series as we have seen is even series in $\mu$, correspondingly we will have $c_n$ with $2 \leq n \leq (B-1)$ are produced completely and partial unbiased contributions of $D_n$ remain in $c_n$ for $(B+1) \leq n \leq BM$. Clearly, this means $B \geq 3$, or else we won't observe that it reproduces a valid Taylor series.
    \end{itemize}

    \item Considering $M$ is odd, $B$ is odd, $A$ is even so that $AM$ is even but $BM$ is odd
   \begin{itemize}
    \item If $AM < B$, then complete unbiased contributions will appear upto $\Ob(\mu^{AM})$. This means that in Taylor series, the Taylor coefficients $c_n$ are completely reproduced where $2 \leq n \leq AM$, and the remaining Taylor coefficients $c_n$ for $(AM+2) \leq n \leq (BM-1)$ possess unbiased contributions in $D_n$ to some partial extent. This extent decreases with increasing $n$ i.e. while increasing $n$ from $(AM+2)$ to $(BM-1)$.
    
    \item If $AM \geq B$, then complete unbiased contributions will appear upto $\Ob(\mu^{B})$. Since the Taylor series as we have seen is even series in $\mu$, correspondingly we will have $c_n$ with $2 \leq n \leq (B-1)$ are produced completely and partial unbiased contributions of $D_n$ remain in $c_n$ for $(B+1) \leq n \leq (BM-1)$. Clearly, this means $B \geq 3$ and $M > 1$, because then only one has $BM - 1 > B - 1$.
    \end{itemize}

     \item Considering $M$ is odd, $B$ is odd, $A$ is odd so that $AM$ and $BM$ are odd
   \begin{itemize}
    \item If $AM < B$, then complete unbiased contributions will appear upto $\Ob(\mu^{AM})$. This means that in Taylor series, the Taylor coefficients $c_n$ are completely reproduced where $2 \leq n \leq (AM-1)$, and the remaining Taylor coefficients $c_n$ for $(AM+1) \leq n \leq (BM-1)$ possess unbiased contributions in $D_n$ to some partial extent. This extent decreases with increasing $n$ i.e. while increasing $n$ from $(AM+1)$ to $(BM-1)$.
    
    \item If $AM \geq B$, then complete unbiased contributions will appear upto $\Ob(\mu^{B})$. Since the Taylor series as we have seen is even series in $\mu$, correspondingly we will have $c_n$ with $2 \leq n \leq (B-1)$ are produced completely and partial unbiased contributions of $D_n$ remain in $c_n$ for $(B+1) \leq n \leq (BM-1)$. Clearly, this means $B \geq 3$ and $M > 1$, because then only one has $BM - 1 > B - 1$.
    \end{itemize}
    
\end{enumerate}

  \begin{figure}[H]
   \centering
      \includegraphics[width=0.47\textwidth]{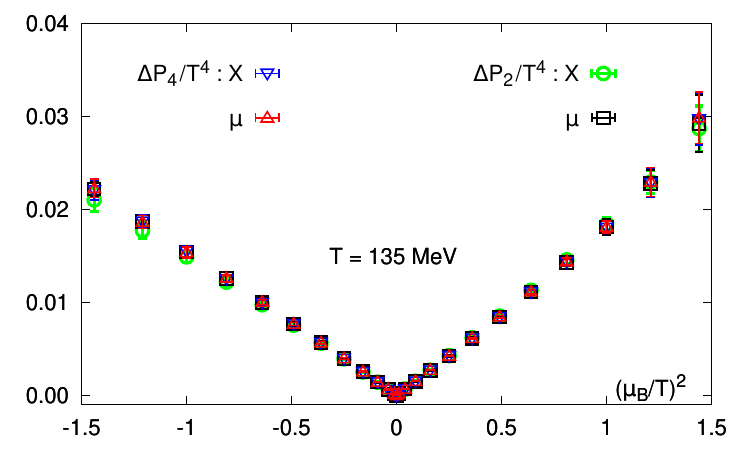}                         
      \includegraphics[width=0.47\textwidth]{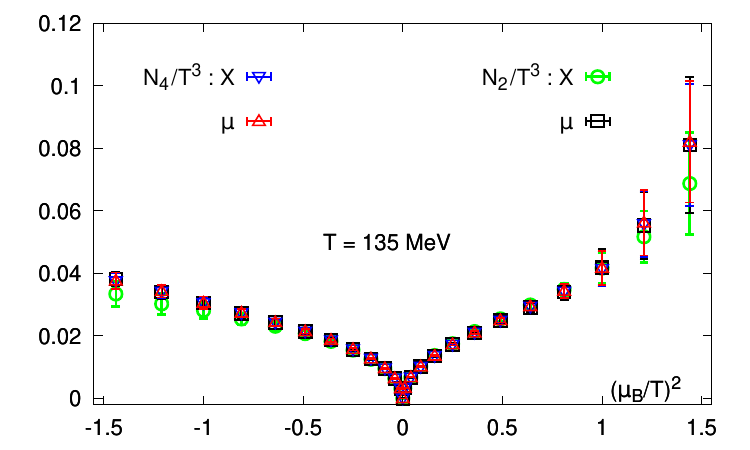}  \\             
      \includegraphics[width=0.52\textwidth]{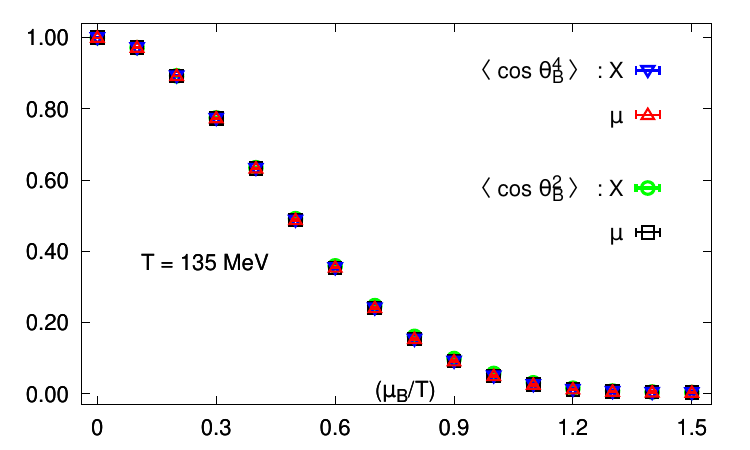}                    
       \caption{\hspace{1mm} $\Delta P/T^4$ (top left), $\mathcal{N}/T^3$ (top right) and  $\LA \cos{\theta}\RA$ (bottom) in $X$ and $\mu$ bases in $2^{nd}$ and $4^{th}$ orders in $\muB/T$ at $T=135$ MeV
       }
      \label{fig:135_X_mu}
    \end{figure}

\begin{figure}
    \centering
           \includegraphics[width=0.49\textwidth]{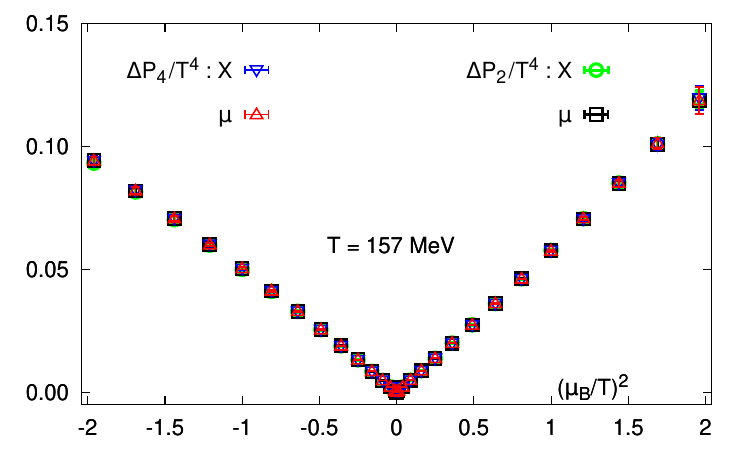} 
      \includegraphics[width=0.49\textwidth]{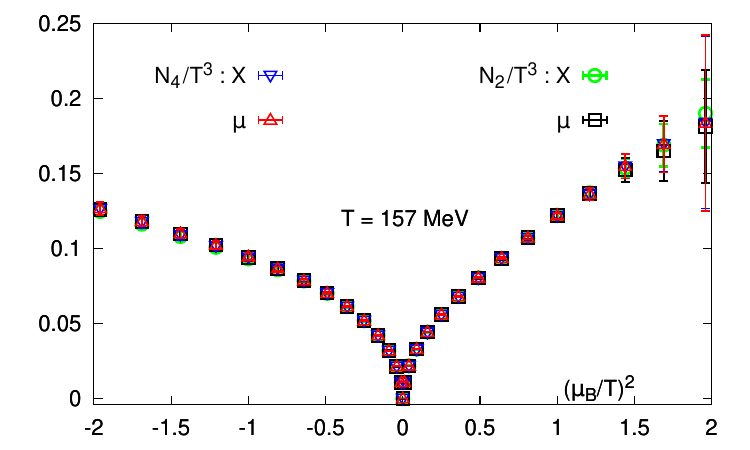} \\
      \includegraphics[width=0.55\textwidth]{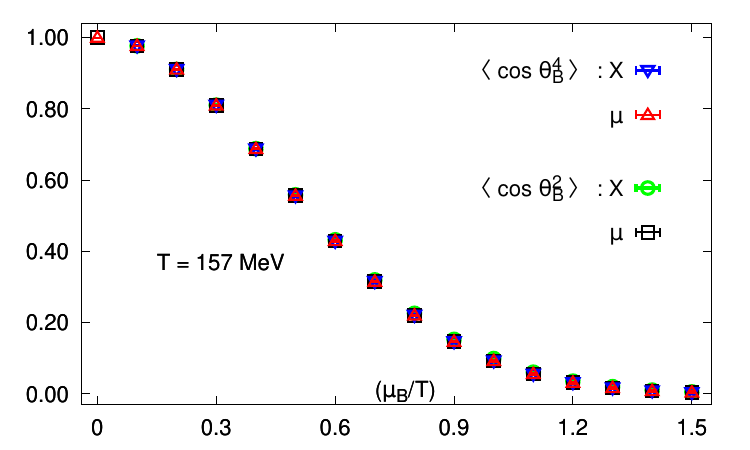}   
    \caption{\hspace{1mm}$\Delta P/T^4$ (top left), $\mathcal{N}/T^3$ (top right) and  $\LA \cos{\theta}\RA$ (bottom) in $X$ and $\mu$ bases in $2^{nd}$ and $4^{th}$ orders in $\muB/T$ at $T=157$ MeV
    }
    \label{fig:157_X_mu}
\end{figure}

\begin{figure}
    \centering
                \includegraphics[width=0.49\textwidth]{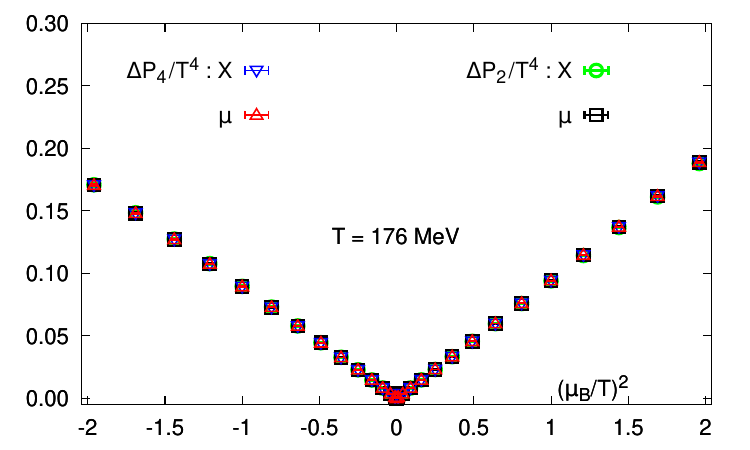} 
      \includegraphics[width=0.49\textwidth]{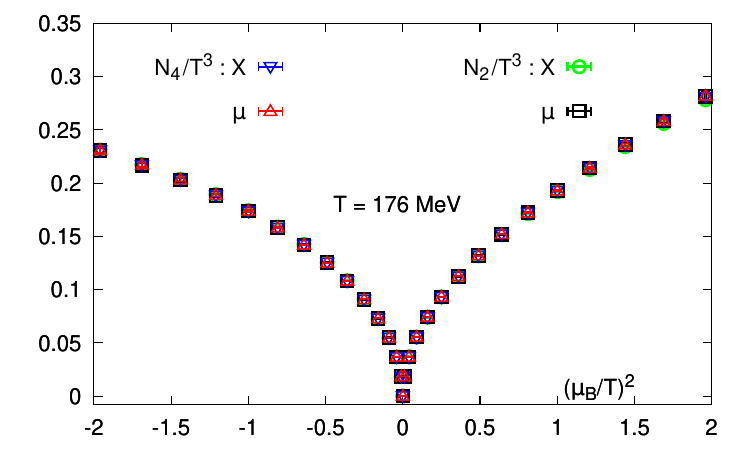} \\
      \includegraphics[width=0.55\textwidth]{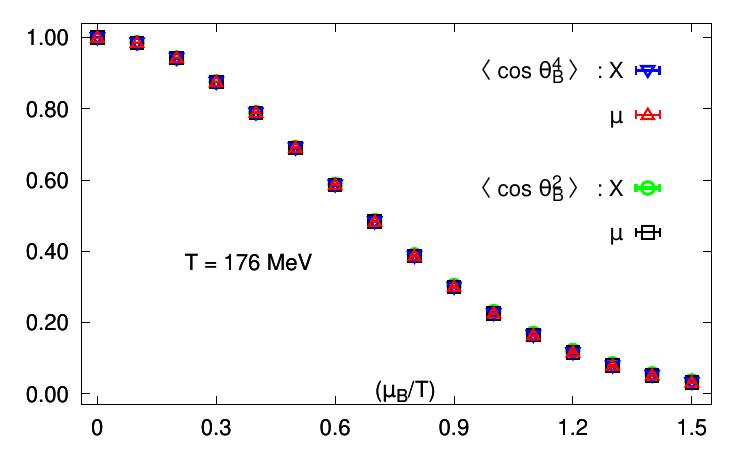} 
    \caption{\hspace{2mm}$\Delta P/T^4$ (top left), $\mathcal{N}/T^3$ (top right) and  $\LA \cos{\theta}\RA$ (bottom) in $X$ and $\mu$ bases in $2^{nd}$ and $4^{th}$ orders in $\muB/T$ at $T=176$ MeV
    }
    \label{fig:176_X_mu}
\end{figure}

In the above point-wise discussion, the Taylor series being even in $\mu$ due to the $CP$ symmetry of QCD, the index $n$ is even for all the corresponding Taylor coefficients $c_n$.

 Despite spanning a larger regime of the full series, we find excellent agreement between the unbiased results in the two bases order-by-order in pressure $\Delta P/T^4$ and number density $\mathcal{N}/T^3$ in $\mu_B$ for all the three temperatures viz. $135$ MeV in~\autoref{fig:135_X_mu}, $157$ MeV in ~\autoref{fig:157_X_mu} and $176$ MeV in~\autoref{fig:176_X_mu}, upto $4^{th}$ order. The $\Delta P/T^4$ and $\mathcal{N}/T^3$ plots are shown for both real and imaginary values of $\muB/T$, whereas the average phasefactor $\LA \cos{\theta} \RA$ is shown only for real part of $\muB$ which is the non-trivial regime of interest in this case. This is because, $\LA \cos{\theta} \RA = 1$  for imaginary regime, suggesting the abscence of sign problem for purely imaginary $\muB$. We find the fluctuations of correlation functions decrease with increasing temperature. We also find the average phasefactor $\LA \cos{\theta} \RA$ showing good agreement order-by-order in $\muB$ for both the orders, suggesting that the breakdown of calculations happen amost at the same value of $\muB$ for both the bases for both $2^{nd}$ and $4^{th}$ order calculations. This is also demonstrated in the subsequent upcoming sections about the results for $\mu_B$ and $\mu_I$. 
 
 This evidently implies that, at least, with the finite number of higher order terms considered, there is no appreciable contribution coming from the higher order terms containing higher order correlation functions glued with larger powers of $\mu$. The series in $\mu$ as formed in cumulant basis, therefore bears the blueprint of a genuine series expansion, where the higher order terms contribute progressively lesser as compared to the leading order terms. We discuss the results in the subsequent sections.

\section{Setup of lattice and calculation}

\label{sec:unb exp setup}
\hspace{2mm}We must unambiguously mention the source of our data for the gauge configurations and all the relevant related actions, like the gauge and the fermion actions that we have used in our analysis. In order to verify our formalism, we used the data generated by the HotQCD collaboration for its ongoing Taylor expansion calculations of the finite density Equation of State (EoS), finite $\mu_B$ chiral crossover temperature and conserved charge cumulants at finite density~\cite{Bazavov.2017,Bazavov.2018,Bollweg.2022}. For these calculations, $2+1$-flavor gauge configurations of $\ord(10^4$ - $10^6)$ were generated in the temperature range 125~MeV~$\lesssim~T~\lesssim$~178~MeV using a Symanzik-improved gauge action~\cite{Symanzik.1983.187,Symanzik.1983.205,Weisz.1985} and the Highly Improved Staggered Quark (HISQ) fermion action~\cite{Follana.2007,Bazavov.2010,Bazavov:2011nk,Bazavov:2014pvz} with $\Nt=8$, $12$ and $16$ and $\Ns=4\Nt$. The temperature for each $\Nt$ was varied by varying the lattice spacing $a$ through the appropriate tuning of gauge coupling $\beta$. The HotQCD collaboration used both the Sommer parameter $r_1$ as well as the kaon decay constant $f_K$ to determine the function $a(\beta)$, where $a$ is the lattice spacing and $\beta$ is the inverse coupling. The temperature values quoted in this thesis were obtained using the $f_K$ scale. In addition to the gauge coupling, the bare light and strange quark masses $m_l(a)$ and $m_s(a)$, with the masses being functions of lattice spacing $a$ were also tuned so that the pseudo-Goldstone pion and kaon masses were respectively equal to the physical pion and kaon masses for each value of $a$. A complete description of the gauge ensembles and scale setting, which eventually sets up the Line of Constant Physics ($LCP$) for the respective lattice calculations, can be found in Ref.~\cite{Bollweg.2021}.

To calculate the Taylor coefficients, on each gauge configuration the first eight derivatives $D_1^f,\dots,D_8^f$ for each quark flavor were estimated stochastically using $2000$ Gaussian random volume sources for $D^f_1$ and $500$ sources for the higher derivatives. The exponential-$\mu$ formalism~\cite{Hasenfratz.1983} was used to calculate the first four derivatives i.e. $D_n^f$, for $1 \leq n \leq 4$ while the linear-$\mu$ formalism~\cite{Gavai.2012linear,GAVAI.2015} was used to calculate the higher derivatives i.e. for $D_n^f$, with $n > 4$. A quick and short note on these two formalisms is given in the following section has been vividly outlined in \autoref{Appendix 3}.

\section{Results}
\label{sec:unb exp results}
Using this data, we calculated the excess pressure and number density for both real and imaginary baryon as well as isospin chemical potentials $\muB$ and $\muI$, in the range $0 \leqslant \lvert \mu_{B,I}/T \rvert \leqslant 2$. This is done using $100K$ gauge field configurations per temperature for the baryon potential $\muB$ and $20K$ configurations per temperature for the isospin potential $\muI$. Our results were obtained on $\Nt=8$ lattices for three temperatures viz. $T \sim 135$, $157$ and $176$ MeV. These temperatures were chosen as being approximately equal to $\Tpc$ and $\Tpc\pm20$~MeV, where $\Tpc=156.5 \pm 1.5$~MeV is the chiral crossover temperature at $\muB=0$~\cite{Bazavov.2018}, around which the crossover chiral phase transition from the chiral symmetry broken hadronic phase to the chiral symmetry restored quark gluon plasma phase happens, which is well-established from zero density lattice QCD.

  \subsection{For isospin chemical potential}
 \label{subsec:unb exp results isospin}
   Before considering finite $\muB$, let us consider the simpler case of finite isospin chemical potential $\muI$ which is obtained by setting $\mu_u=-\mu_d=\mu_I$, $\mu_s=0$ instead. The $\muI\ne0$ case has the advantage that the fermion determinant is real and there is no Sign Problem. Hence it is possible to calculate observables for much larger values of the chemical potential compared to the $\muB$ case, and it is precisely for these value that bias can become significant. The QCD phase diagram in the $T$-$\muI$ plane is also a topic of interest in its own right~\cite{Son.2000, Brandt.2017, Adhikari.2020}, and our formalism could prove useful in the study and probing of future Taylor series based Lattice QCD approaches.

\vspace{3mm}

 \begin{figure}[H]
 \centering
\includegraphics[width=0.49\textwidth]{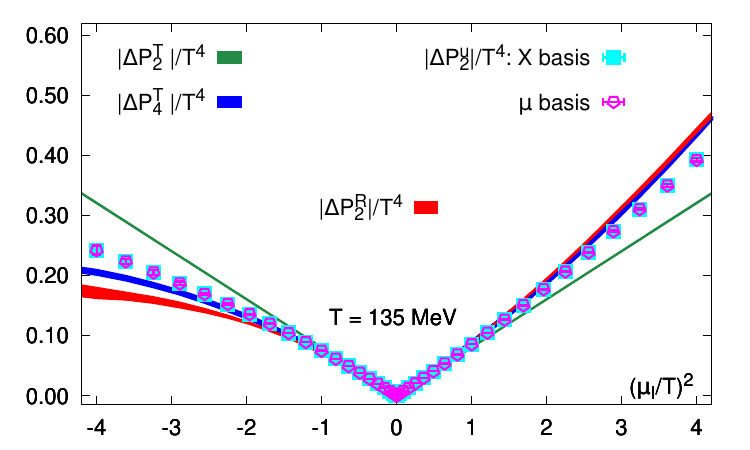}
\includegraphics[width=0.49\textwidth]{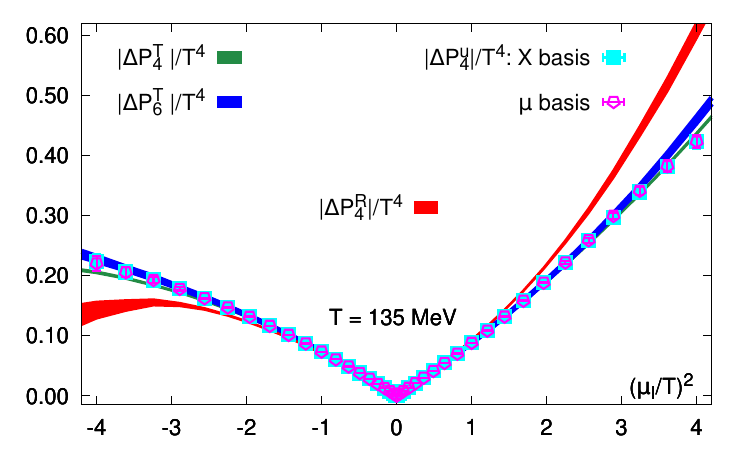}
\includegraphics[width=0.49\textwidth]{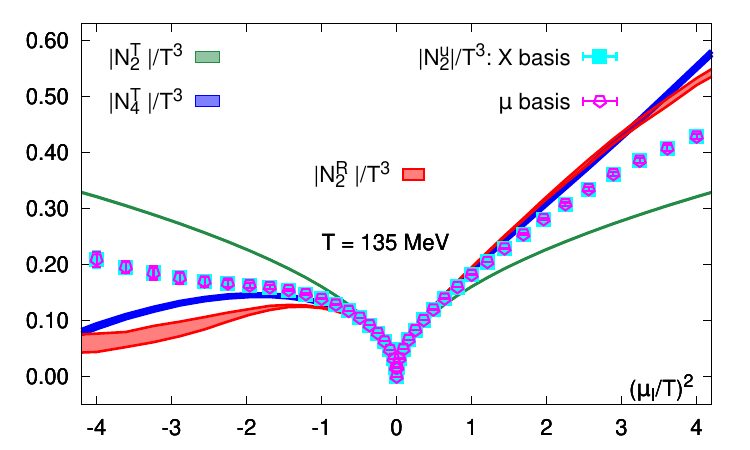}
\includegraphics[width=0.49\textwidth]{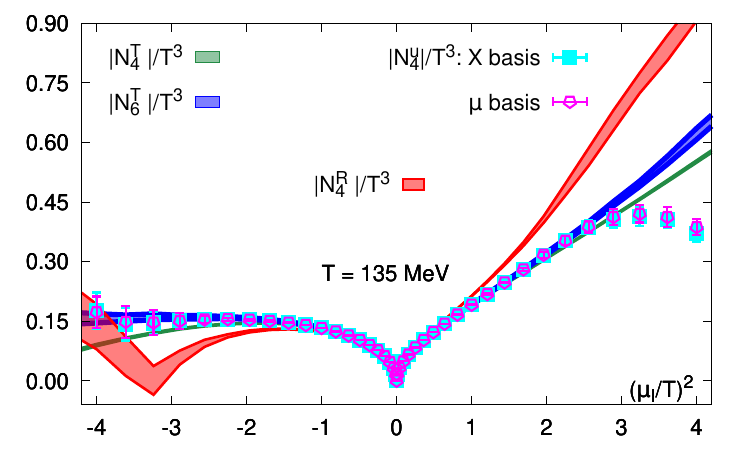}
\caption{$\Delta P_2/T^4$ and $\Delta P_4/T^4$ (in top row), $\mathcal{N}_2/T^3$ and $\mathcal{N}_4/T^3$ (in bottom row) plotted in $(\muI/T)^2$ at $T=135$ MeV with Taylor results $\Delta \text{P}^{\text{T}}$ , $\text{N}^{\text{T}}$, old biased resummed results $\Delta \text{P}^{\text{R}}$ , $\text{N}^{\text{R}}$ and unbiased results $\Delta \text{P}^{\text{u}}$ , $\text{N}^{\text{u}}$} 
\label{fig:135_mu_I}
\end{figure}

\vspace{1mm}

\begin{figure}[H]
  \centering
     \includegraphics[width=0.49\textwidth]{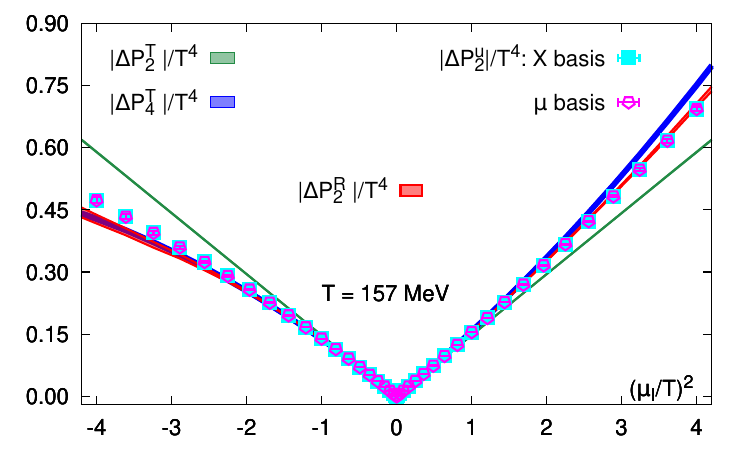} 
     \includegraphics[width=0.49\textwidth]{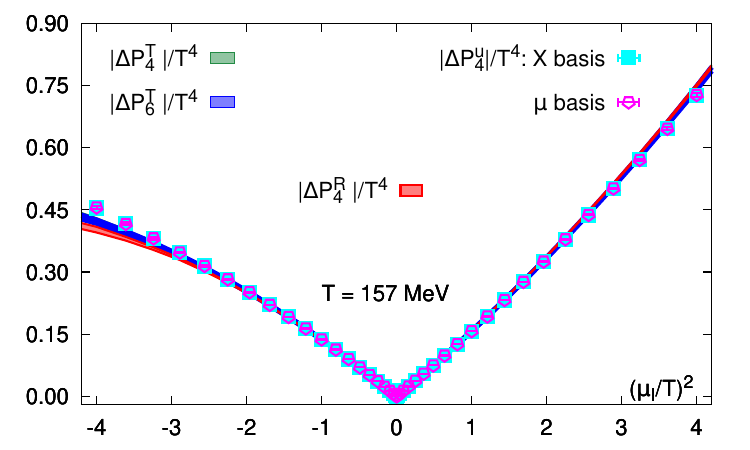} 
     \includegraphics[width=0.49\textwidth]{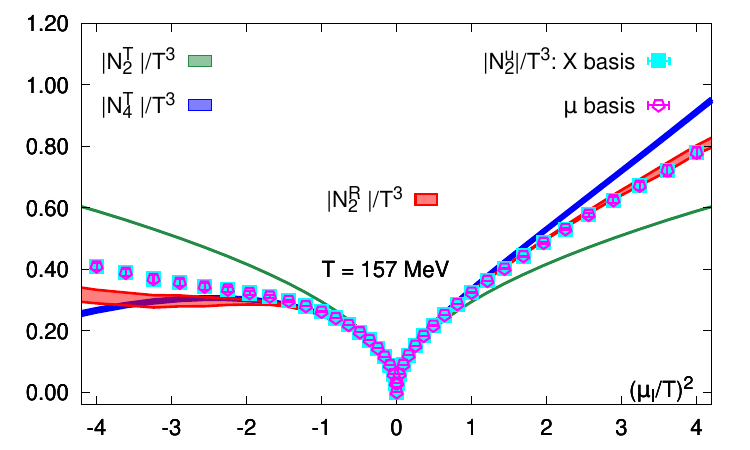} 
     \includegraphics[width=0.49\textwidth]{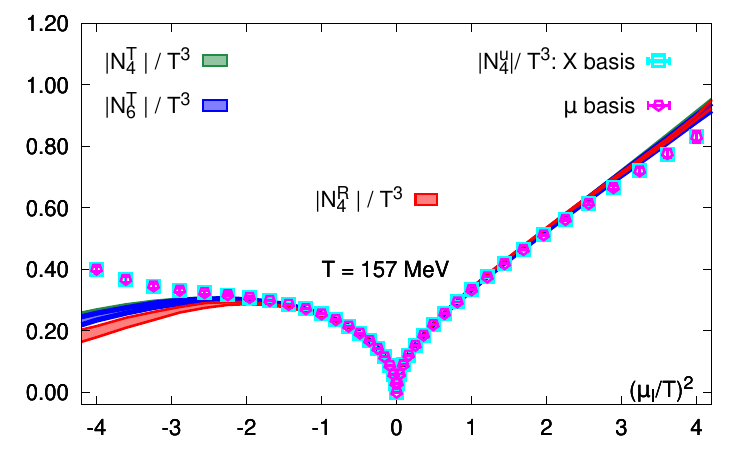} 
     \caption{$\Delta P_2/T^4$ and $\Delta P_4/T^4$ (in top row), $\mathcal{N}_2/T^3$ and $\mathcal{N}_4/T^3$ (in bottom row) plotted in $(\muI/T)^2$ at $T=157$ MeV with Taylor results $\Delta \text{P}^{\text{T}}$ , $\text{N}^{\text{T}}$, old biased resummed results $\Delta \text{P}^{\text{R}}$ , $\text{N}^{\text{R}}$ and unbiased results $\Delta \text{P}^{\text{u}}$ , $\text{N}^{\text{u}}$}  
     \label{fig:157_mu_I}
    \end{figure}

\begin{figure}[H]
   \centering
     \includegraphics[width=0.49\textwidth]{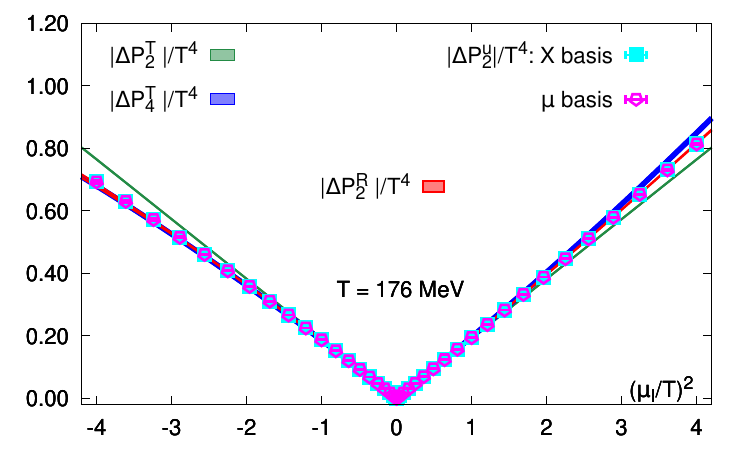} 
          \includegraphics[width=0.49\textwidth]{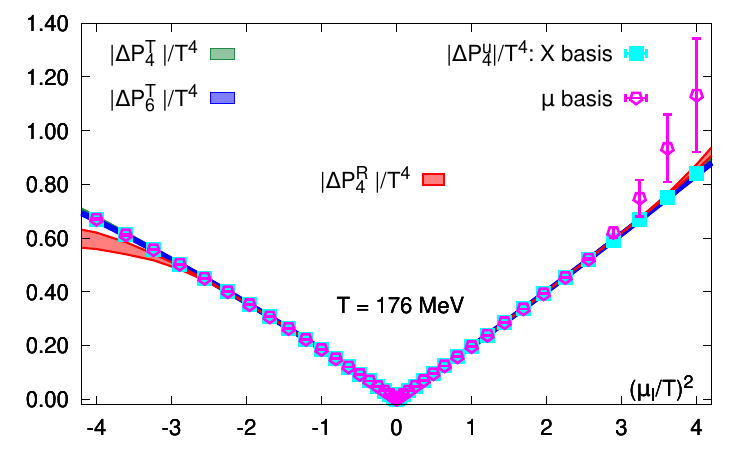} 
     \includegraphics[width=0.49\textwidth]{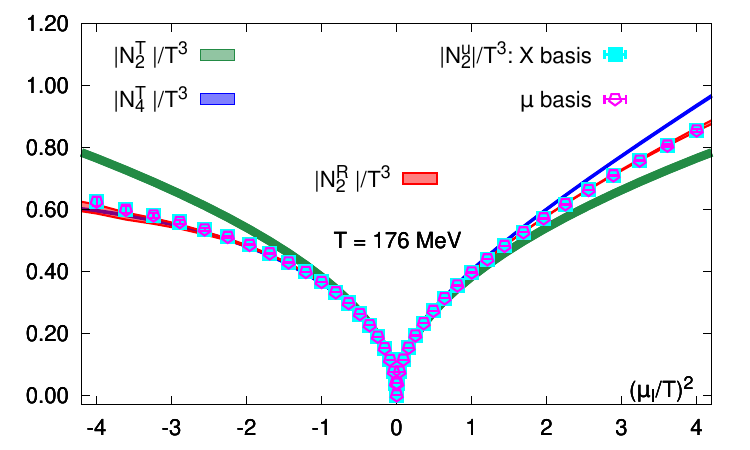} 
     \includegraphics[width=0.49\textwidth]{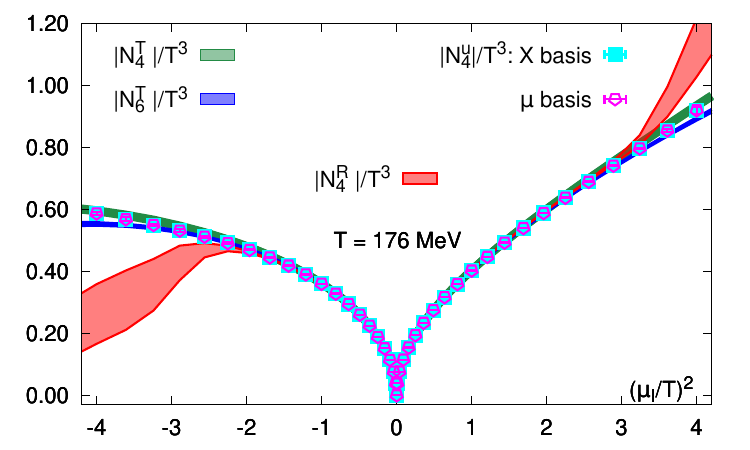} 
     \caption{$\Delta P_2/T^4$ and $\Delta P_4/T^4$ (in top row), $\mathcal{N}_2/T^3$ and $\mathcal{N}_4/T^3$ (in bottom row) plotted in $(\muI/T)^2$ at $T=176$ MeV with Taylor results $\Delta \text{P}^{\text{T}}$ , $\text{N}^{\text{T}}$, old biased resummed results $\Delta \text{P}^{\text{R}}$ , $\text{N}^{\text{R}}$ and unbiased results $\Delta \text{P}^{\text{u}}$ , $\text{N}^{\text{u}}$}
     \label{fig:176_mu_I}
\end{figure}

  We plot our second and fourth order results for the excess pressure in the top row and net isospin density in the bottom row for $T=135$~MeV, in \autoref{fig:135_mu_I}, for $157$ MeV, in \autoref{fig:157_mu_I} and for $176$ MeV, in \autoref{fig:176_mu_I} respectively. The resummed results were obtained using both the biased (\autoref{eq:resummed_pressure}) as well as unbiased exponentials (\autoref{eq:unbiased_resummed-1} with $(N,M)=(2,2)$ and \autoref{eq:unbiased_resummed-2} with $(N,M)=(2,4)$). The nomenclatures for the different symbols used have already been mentioned in \autoref{Chapter 5}. In both figures, we also plot both the second and fourth order Taylor expansion results (\autoref{eq:taylor_pressure} with $N=2$ and 4) for comparison. In all these figures, the Taylor results are plotted in the form of green and blue bands, the old biased resummed results in red bands and the new unbiased results are represented in the form of points.

\begin{figure}[H]
  \centering
     \includegraphics[width=0.49\textwidth]{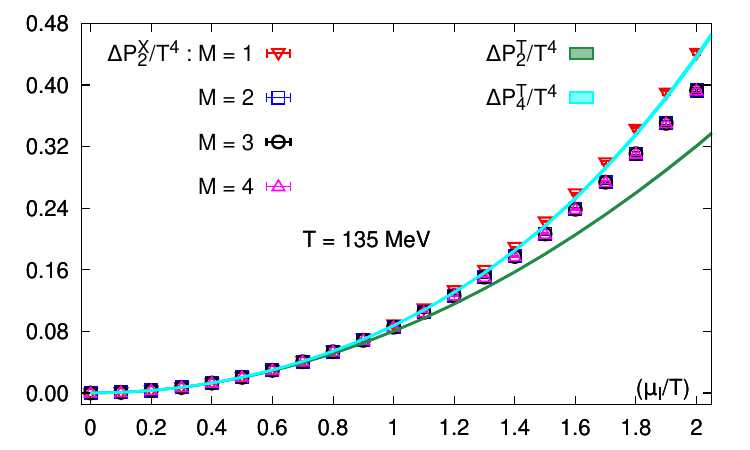} 
     \includegraphics[width=0.49\textwidth]{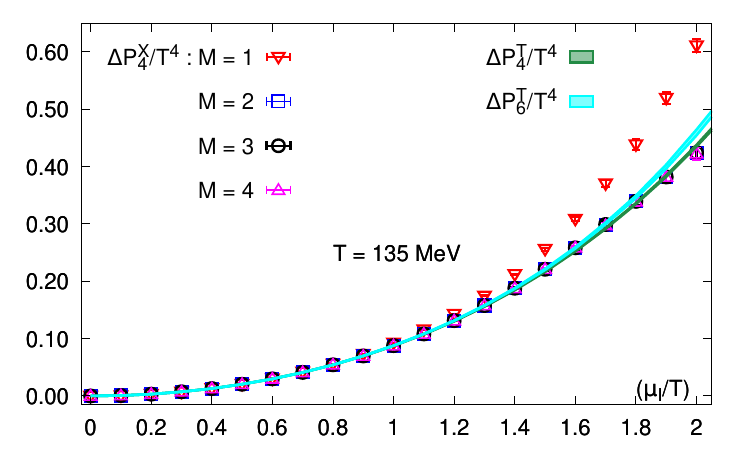} 
     \includegraphics[width=0.49\textwidth]{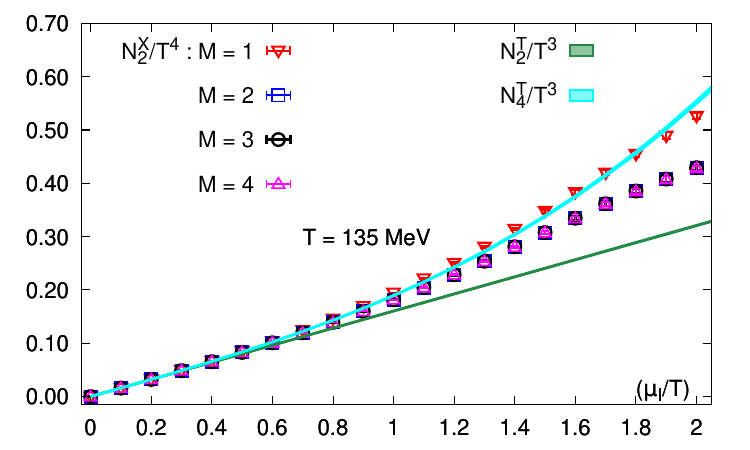} 
     \includegraphics[width=0.49\textwidth]{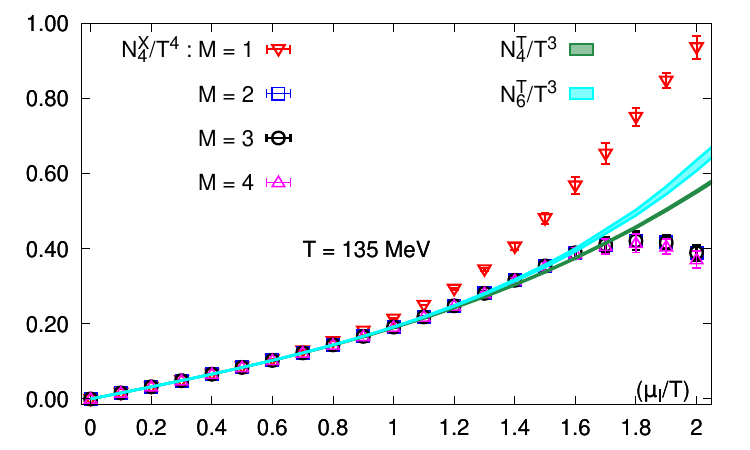} 
     \caption{$\Delta P_2^X/T^4$ and $\Delta P_4^X/T^4$ (in top row), $\mathcal{N}_2^X/T^3$ and $\mathcal{N}_4^X/T^3$ (in bottom row) plotted in $\muI/T$ at $T=135$ MeV with Taylor results $\Delta \text{P}^{\text{T}}$ , $\text{N}^{\text{T}}$ in bands and the unbiased results for different number of cumulants in points  }  
     \label{fig:diff cumu mu_I}
    \end{figure}

We find that the second and fourth order Taylor expansions start to differ for $\big\lvert\muI/T\big\rvert\approx1$ both for real and imaginary $\mu_I$. With regard to the resummed results, while all three schemes predict significant corrections to the second order Taylor results for $\big\lvert\muI/T\big\rvert\gtrsim1$, the biased resummation corrections also exceed the fourth order corrections while the unbiased resummation results always lie between the second and fourth order Taylor results. This suggests that the large corrections suggested by \autoref{eq:resummed_pressure} are at least in part due to bias. To verify this, we also compared the results of \autoref{eq:unbiased_resummed-2} for different $M$. Below in \autoref{fig:diff cumu mu_I}, we have plotted $\Delta P/T^4$ and $\mathcal{N}/T^3$ for different numbers of cumulants namely from $M=1$ to $M=4$ for $T=135$ MeV. We also plot the corresponding Taylor series results for both the observables. Note that \autoref{eq:resummed_pressure} corresponds to the simplest case $M=1$ of \autoref{eq:unbiased_resummed-2}. We find that the biased results smoothly approach the unbiased results as $M$ is increased, thus proving that bias needs to be subtracted from the results of \autoref{eq:resummed_pressure}. It must be noted that all the unbiased results here in \autoref{fig:diff cumu mu_I} have been obtaineed using the cumulant or $X$ basis.

Another important observation is the non-monotonic behaviour of number density $\mathcal{N}/T^3$ for $4^{th}$ order from $(\muI/T)^2 \geq 3$ in \autoref{fig:135_mu_I} and equivalently in \autoref{fig:diff cumu mu_I}. The unbiased results for $M=1$ are monotonic, whereas results from $M=2$ onwards show non-monotonicity. This raises eyebrows, since the excess pressure $\Delta P/T^4$ for the same order and value of $\muI$, exhibits monotonic quadratic behaviour and hence, being the first order $\mu$ derivative of $\Delta P$, should be linear, at least monotonic. The $M=1$ results are the old biased results and hence, highly unreliable, as we understand from our long discussion on stochastic bias. The unbiased powers start popping in the calculations from $M=2$ and this strongly sends out the message that this is a pure genuine manifestation of stochastic bias. 

Had we not diagnosed this stochastic bias, we never would be in a position to identify the non-monotonicity in the number density, which is the real genuine picture. This non-monotonicity is attributable to a genuine breakdown in case of $\muI$~\cite{Mitra:2023isospin}. Although there is no sign problem in $\muI$, the exponential resummation (biased and unbiased) must exhibit signs of breakdown since, there exists a genuine phase diagram in $T-\muI$ plane, which indicates the formation of a pion ($\pi$) condensate at some finite value of $\muI$ even for a finite temperature $T$. This is one of the several future works put forward.

  \subsection{For baryon chemical potential}
\label{subsec:unb exp results baryon}
The resummed results for the QCD EoS at finite baryon chemical potential $\muB$, obtained using the biased equation \autoref{eq:resummed_pressure}, were presented in Ref.~\cite{Mondal.2022}. In obtaining those results, the full set of 2000 independent random estimates for $D_1$ was made use of as the $D_1$ derivative does not vanish for the $\muB$ case. Note that at any given order $n$, the operator $D_1$ not only appears with the highest power viz. $D_1^n$, but also multiplies the highest power $(\muB/T)^n$ of the chemical potential. Hence one would expect the effects of bias to be the greatest for this particular operator. As we will see, the use of $2000$ random vectors over $500$ random sources for this operator greatly reduces this bias. At the same time, a much greater reduction in the bias can be achieved by using the unbiased exponential formalism with only $500$ random vectors, without the need for a separate calculation of $2000$ independent estimates of $D_1$. Note also that the unbiased exponential simultaneously decreases the bias due to all the derivatives $D_1,\dots,D_N$ up to the same order in $\muB$, since all the correlation functions are estimated using equal number of random volume sources.  

  \begin{figure}
  \centering
 \includegraphics[width=0.49\textwidth]{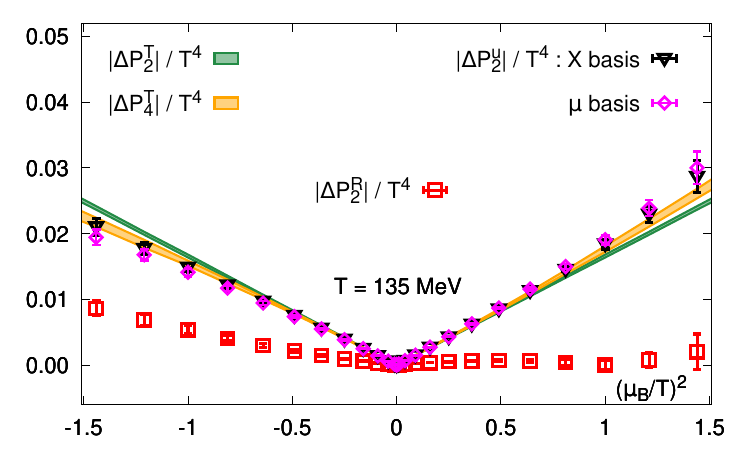} 
  \includegraphics[width=0.49\textwidth]{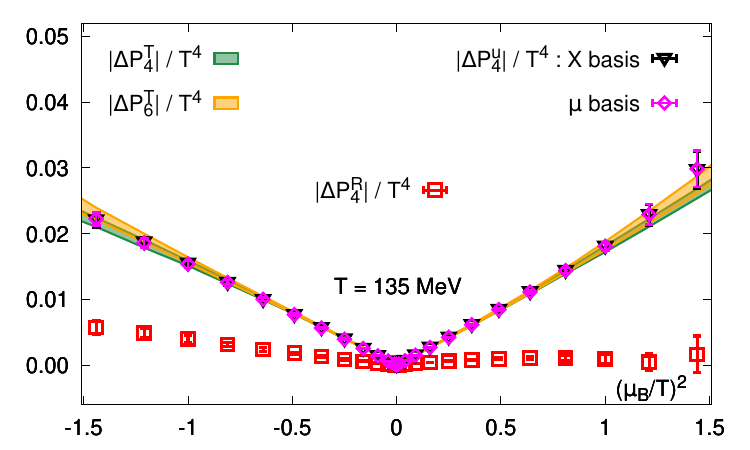} 
    \includegraphics[width=0.49\textwidth]{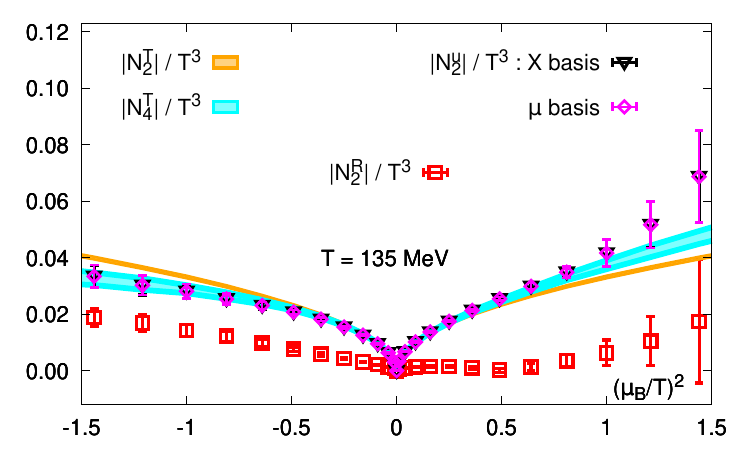} 
    \includegraphics[width=0.49\textwidth]{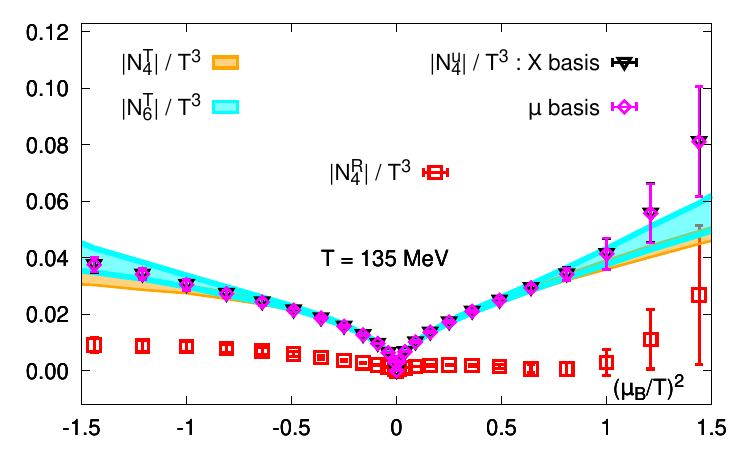} 
    \caption{$\Delta P/T^4$ (top row) and $\mathcal{N}/T^3$ (bottom row) plots for $(\muB/T)^2$ at $T=135$ MeV}
    \label{fig:muB 135 MeV}
\end{figure}

\begin{figure}
  \centering
    \includegraphics[width=0.49\textwidth]{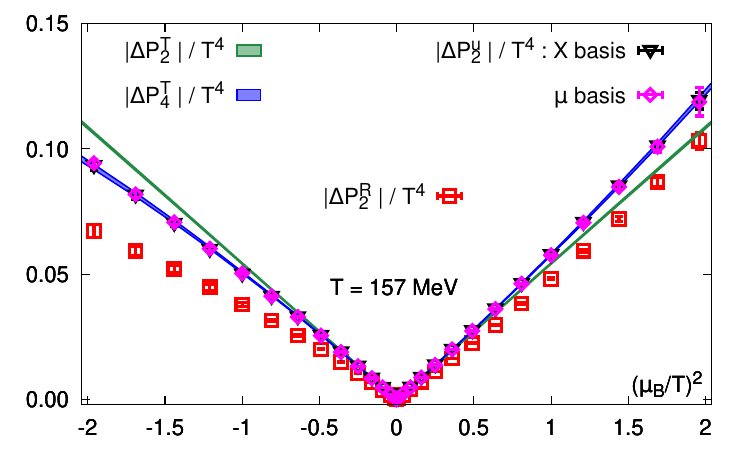} 
    \includegraphics[width=0.49\textwidth]{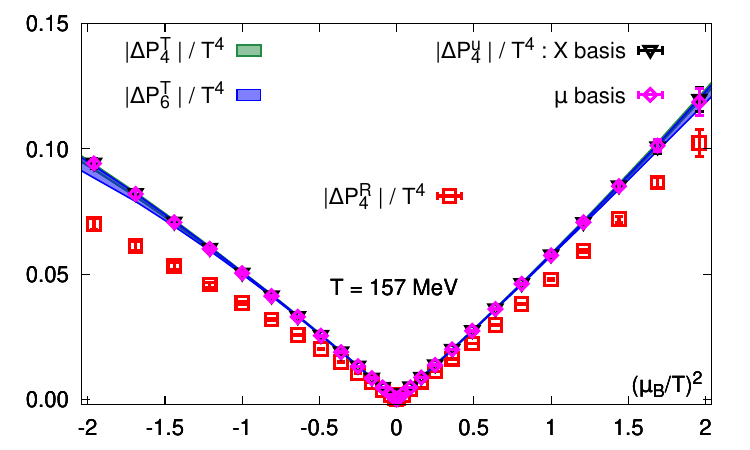}
    \includegraphics[width=0.49\textwidth]{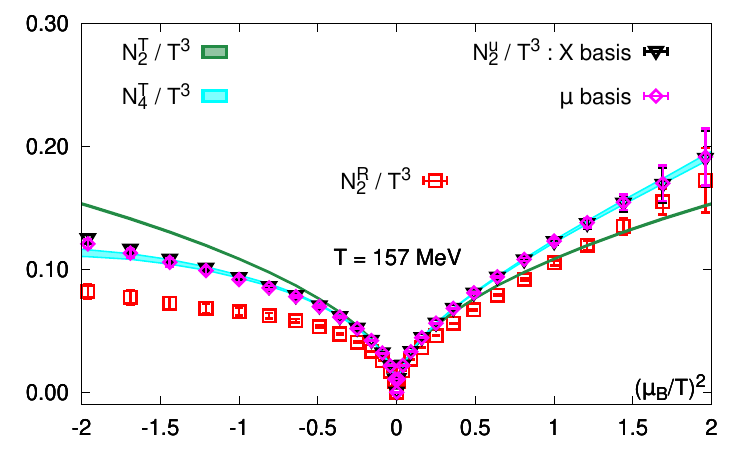} 
     \includegraphics[width=0.49\textwidth]{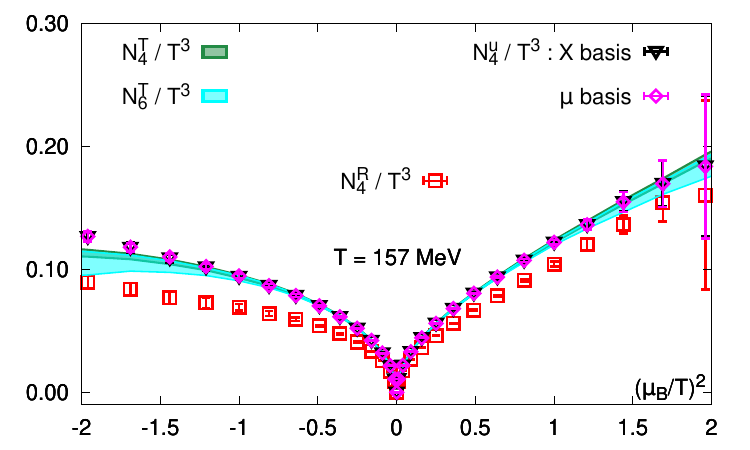} 
     \caption{$\Delta P/T^4$ (top row) and $\mathcal{N}/T^3$ (bottom row) plots for $(\muB/T)^2$ at $T=157$ MeV}
     \label{fig:muB 157 MeV}
\end{figure}

\begin{figure}
\centering
    \includegraphics[width=0.49\textwidth]{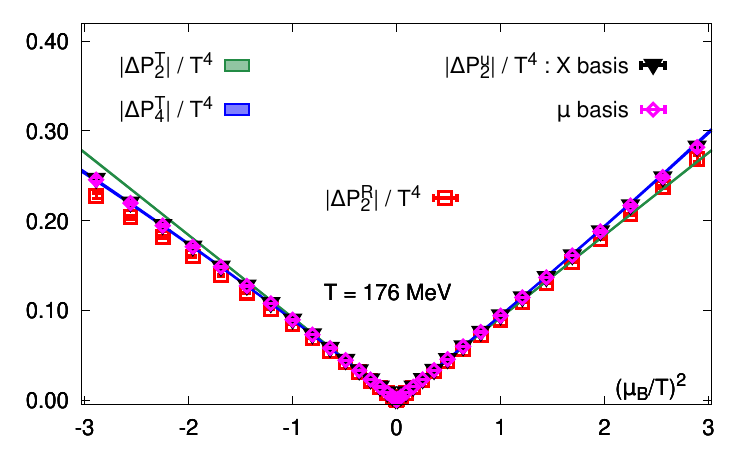}  
    \includegraphics[width=0.49\textwidth]{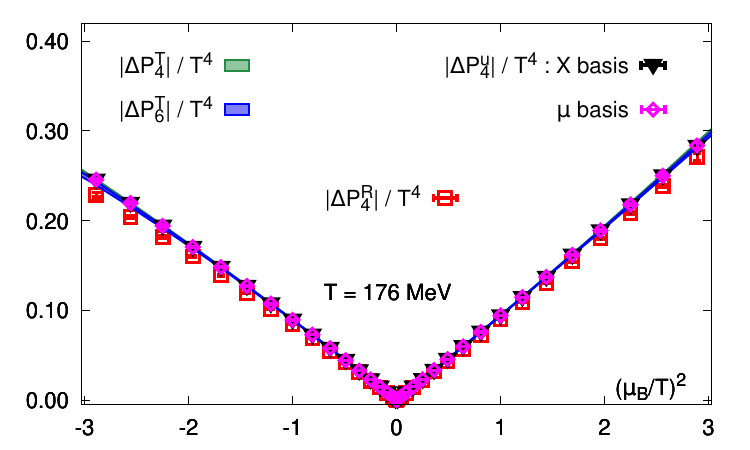} 
    \includegraphics[width=0.49\textwidth]{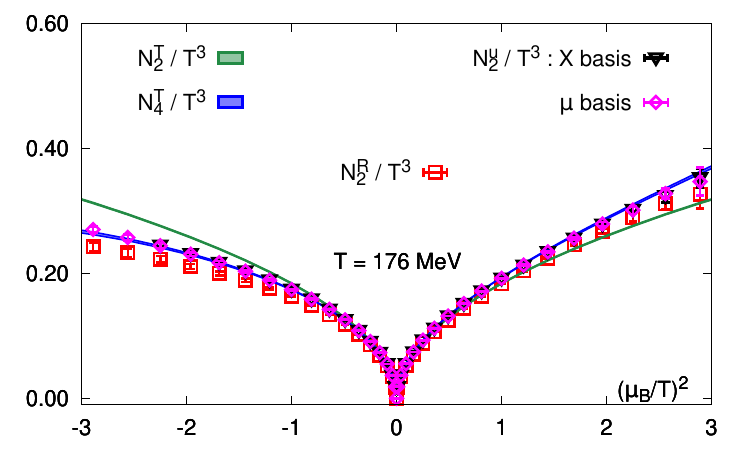}
    \includegraphics[width=0.49\textwidth]{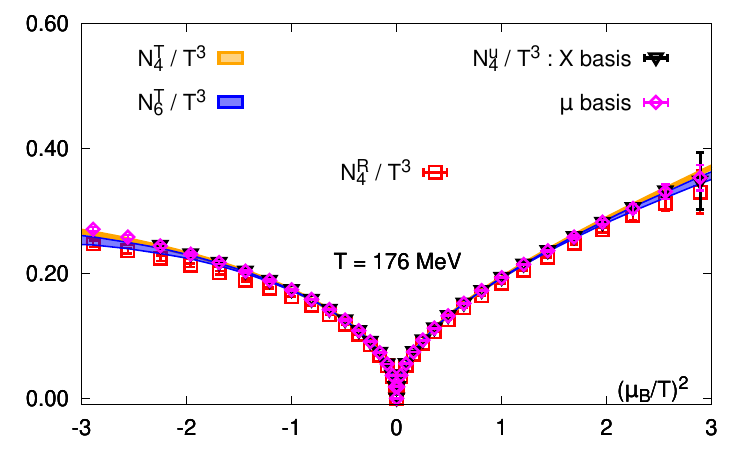} 
    \caption{$\Delta P/T^4$ (top row) and $\mathcal{N}/T^3$ (bottom row) plots for $(\muB/T)^2$ at $176$ MeV}
    \label{fig:muB 176 MeV}
\end{figure}

Here, we have demonstrated the comparative results among the Taylor series, the old biased exponential resummation and the unbiased exponential resummation for real and imaginary $\muB$  for $T=135,157$ and $176$ MeV, namely in \autoref{fig:muB 135 MeV}, \autoref{fig:muB 157 MeV} and \autoref{fig:muB 176 MeV} respectively. As before, the Taylor results are plotted in bands and since, we expect $n^{th}$ order exponential resummation to capture Taylor series of order at least $n+2$, hence we have plotted $2^{nd}$ and $4^{th}$ order Taylor series for $2^{nd}$ order biased and unbiased exponential resummation and similarly $4^{th}$ and $6^{th}$ order for $4^{th}$ order exponential resummation. For these figures, we have estimated the correlation functions $D_n, 1 \leq n \leq 4$ using $\Ob(500)$ random vectors and have used $100K$ gauge configurations.

Clearly, we find that the bias effect is most dominant at $T = 135$ MeV, for which the biased resummed results, shown in red points in \autoref{fig:muB 135 MeV} are far away from the Taylor series results, which we know, are constructed purely using unbiased powers (\autoref{Chapter 4}) of $D_n$. This effect decreases with increasing temperature, as we see the red points approaching Taylor bands in successive \autoref{fig:muB 157 MeV} and \autoref{fig:muB 176 MeV}. This signifies that the estimates of $D_n, 1 \leq n \leq 4$, stochastically generated in the sample, show less stochastic fluctuations with reducing variance $\sigma^2$ and standard deviation $\sigma$ of the statistical sample with increasing temperature.

This can become highly misleading, specially at lower temperatures as calculations with the specified aforementioned conditions yield signed excess pressure to become negative at $T=135$ MeV. One can therefore easily conclude that the system or ensemble of particles is under an attractive force field, causing particles to interact more and hence, pressure at finite $\muB$ is less than $\muB=0$ pressure. The issue of stochastic bias simply goes under the radar, unidentified and we would have stepped onto wrong physics. \autoref{fig:muB 135 MeV} demonstrates that even with the same specified number of random sources and gauge configurations, the unbiased results exude very good agreement with the Taylor results and even capture an order higher Taylor results, justifying the true utility of exponential resummation.

\begin{figure}[H]
\centering
      \includegraphics[width=0.49\textwidth]{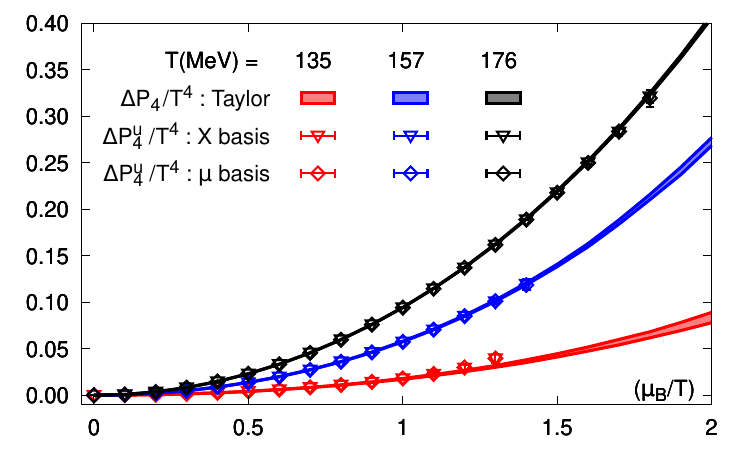}
      \includegraphics[width=0.49\textwidth]{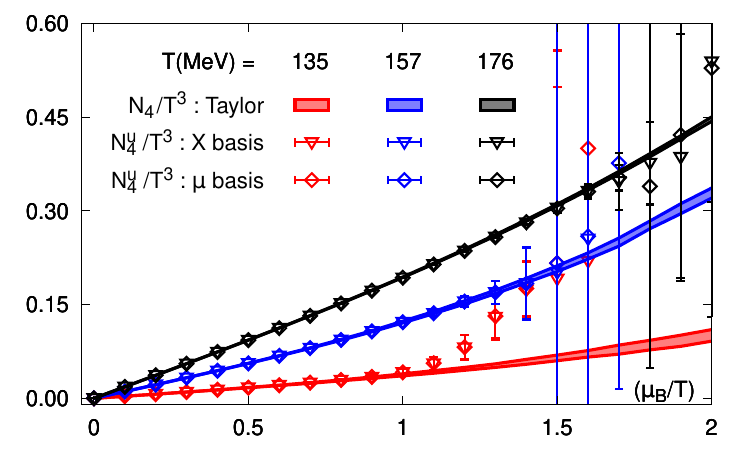}
      \caption{Comparisons between results obtained from Taylor and unbiased exponential resummation using cumulant $(X)$ and chemical potential $(\mu)$ basis at all working temperatures $T = 135$, $157$ and $176$ MeV}
  \label{fig:observables for all temp}
  \end{figure}

Similar story holds for $T=157$ and $176$ MeV also, where the distinction between the biased and Taylor series is not so tangible. A similar degree of agreement in both real and imaginary regimes of $\muB$ between unbiased and Taylor series results clearly proves the thermal stability and temperature-independent efficiency of the formalism. On a greater note, the efficiency is phase-independent to some extent also, as all these three temperatures do bear the signature of the hadronic, crossover and QGP phases respectively.

\begin{figure}[H]
\centering
      \includegraphics[width=0.49\textwidth]{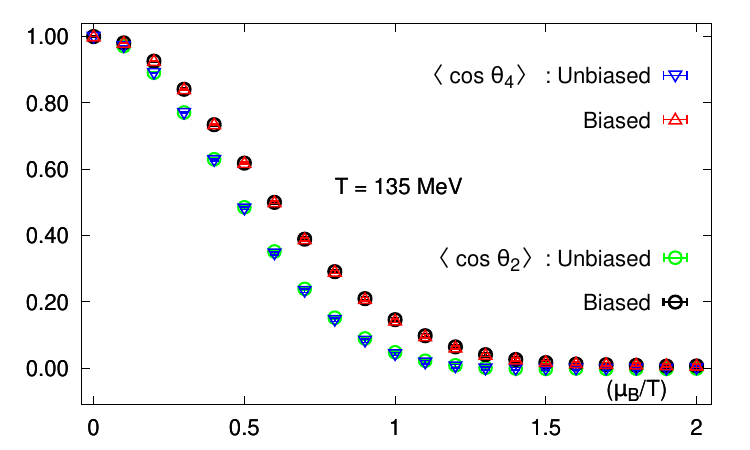}
      \includegraphics[width=0.49\textwidth]{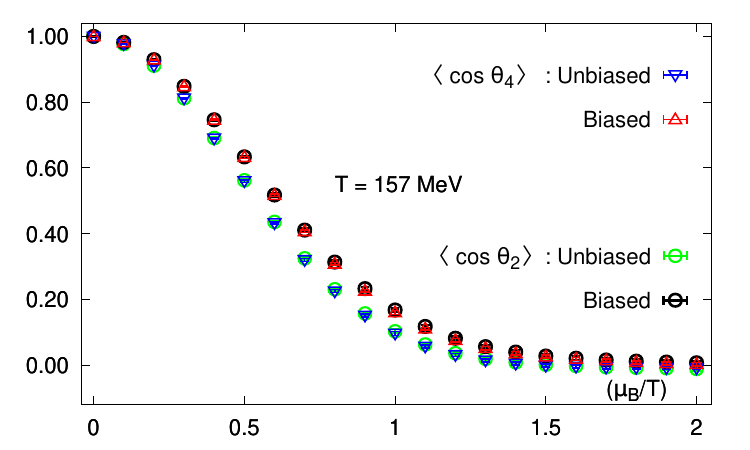}
      \includegraphics[width=0.49\textwidth]{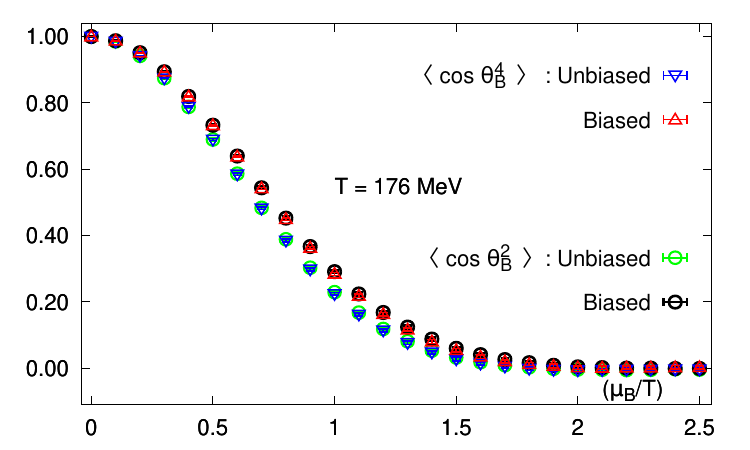} \\
      \caption{Phasefactor plots with $X$ basis at $T = 135$, $157$ and $176$ MeV}
  \label{fig:phasefactor for all temp}
  \end{figure}

However, as the above \autoref{fig:phasefactor for all temp} suggests, the breakdown of the resummation procedure happens almost at the same value of $\muB$ for both the biased and unbiased cases. We find that the phasefactor obtained from unbiased calculations plummet to zero faster than the biased counterparts and with increasing temperature, the unbiased results seem to approach the biased results. The $2^{nd}$ and $4^{th}$ order results agree well for both biased and unbiased approaches. Although, the unbiased results provide greater reliability of the results within the domain of validity of resummation, the job still remains to search for methods or approaches to enhance the regime of validity.

\section{Associated Computational benefits}
\label{sec:comp benefits}
So far, we have discussed that how the new formalism promises to eliminate stochastic bias upto a finite order in $\mu_X$, $X \equiv (B,I)$ irrespective of the working temperature. The particular choice of our working temperatures and its utility and significance has already been elaborated in the previous section. In this section, we will vividly give a discourse on how the usage of this formalism saves sizeable computational time and precious storage space, which are invaluable for users and people actively involved in computational calculations and evaluations. 
 
 As outlined before, all the Taylor coefficients or the scaled quark number susceptibilities constitute linear combination of different $n$ point correlation functions $D_n$, raised to appropriate powers. It is well-known that $D_1$ is the noisiest and also, for any order in $\mu_B$ in Exponential Resummation, this derivative is raised to the highest integral powers among all $D_n$, evident from series expansions of eqns.~\eqref{eq:pr} and ~\eqref{eq:Z}. This therefore makes it imperative to reduce the estimate bias at least in $D_1$ as much as possible, which implies calculating $D_1$ to the highest number of random vector estimates for every single gauge field configuration, while performing the exponential resummation. This vividly explains the computation of $D_1$ using around $2000$ Gaussian random volume sources per configuration as highlighted in Ref.~\cite{Mondal.2022}. The remaining derivatives $D_n$ for $1 < n \leq N$ are calculated using approximately $500$ volume sources per configuration. In principle, in the limit of infinite random volume sources, the estimate of each $D_n$ for every $n$ is unbiased, since the random vectors follow the following familiar orthonormal relation

 \begin{equation}
 \lim_{\NR\to\infty} \frac{1}{\NR} \sum_{n=1}^{\NR} \eta_i^{(n)}\eta_j^{(n)\dagger} = \delta_{ij}    
 \label{eq:RVS}
 \\\\\\*\
 \end{equation}

 This has been discussed comprehensively in \autoref{Chapter 4}. The above eqn.~\eqref{eq:RVS} allows to calculate exact unbiased estimate of the individual traces of operators~\cite{Allton.2005, Gavai.2004}, which unfortunately does not hold for finite number of random volume sources. One can therefore, always argue to estimate a given derivative using more and more number of random sources, which eventually makes this process of obviating bias by regulating number of random sources, a never-ending and an impractical one in the long run. This has eventually motivated the quest of finding a new formalism, which will eliminate the formalism bias (see \autoref{Chapter 4}) to finite order in $\mu$ ($\muB$ in this case), without focusing too much on the number of random vectors used per gauge configuration constituting the gauge ensemble.

\begin{figure}[H]
\centering
 \includegraphics[width=0.49\textwidth]{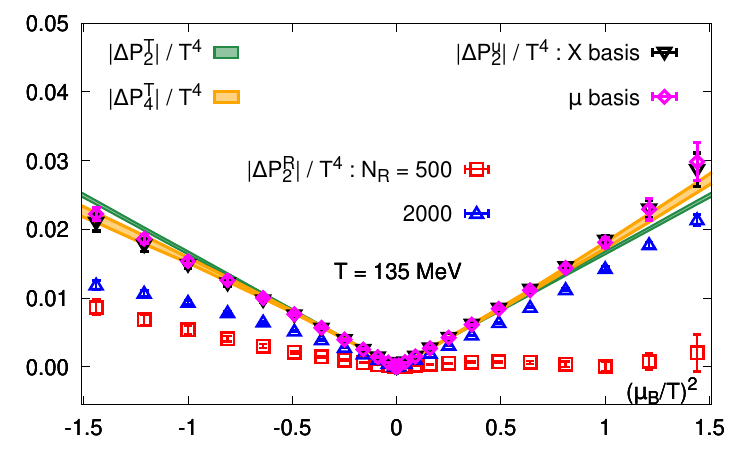} 
    \includegraphics[width=0.49\textwidth]{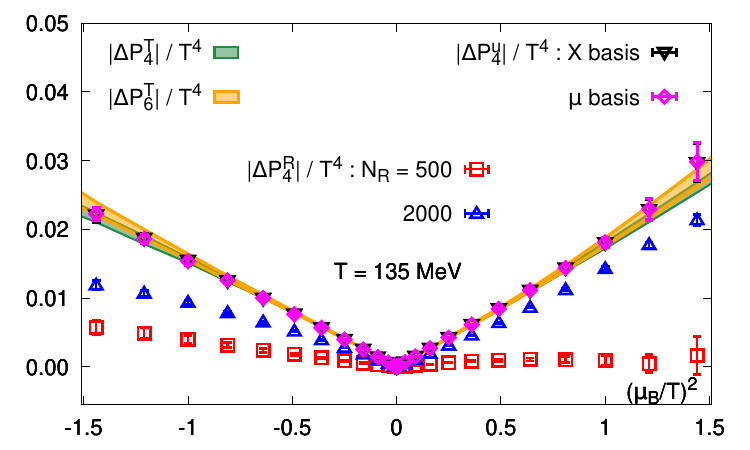} 
    \includegraphics[width=0.49\textwidth]{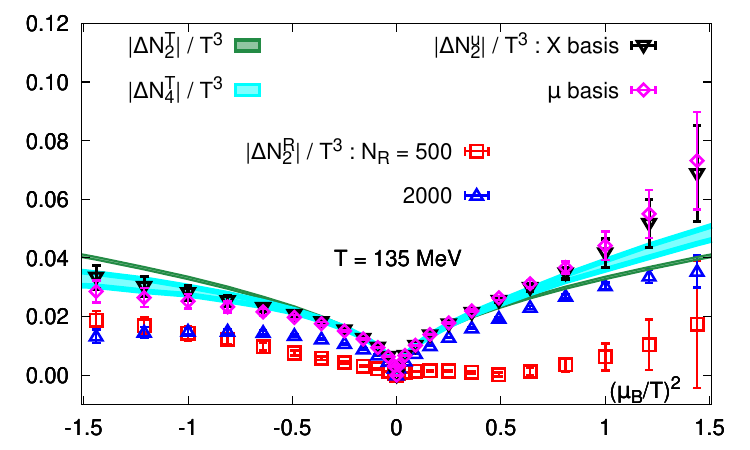} 
     \includegraphics[width=0.49\textwidth]{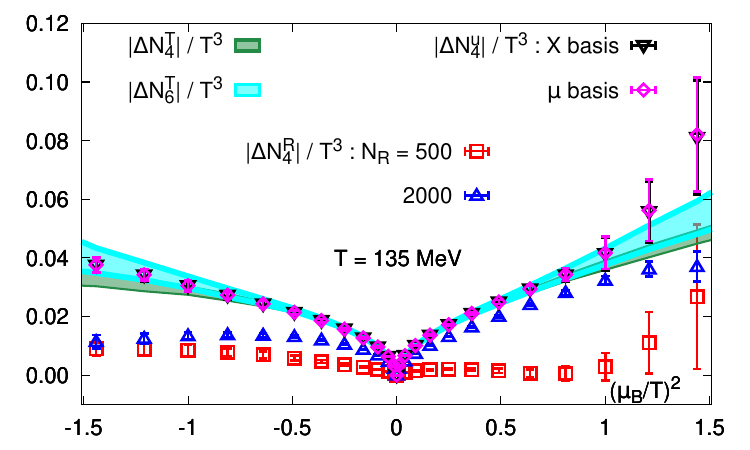} 
     \caption{Comparison of Biased and Unbiased results at $T=135$ MeV. The old biased results are calculated using $500$ and $2000$ random sources respectively. The Taylor results are shown in bands and the unbiased results are plotted in $X$ and $\mu$ bases.}
     \label{fig:2000 135 MeV}
\end{figure}

Here, like before, we have plotted excess pressure and number density as a function of $\muB$ for $T=135,157$ and $176$ MeV, but this time, we have also included the old biased resummed results using $2000$ random sources for $D_1$ and $500$ sources for remaining $D_n$ where $2 \leq n \leq 4$. This is purely to reduce the estimate bias (see \autoref{Chapter 4}) of $D_1$ fourfold, which as we have mentioned before, appears with the highest power in the computation of any Taylor coefficient. 

\begin{figure}[H]
\centering
 \includegraphics[width=0.49\textwidth]{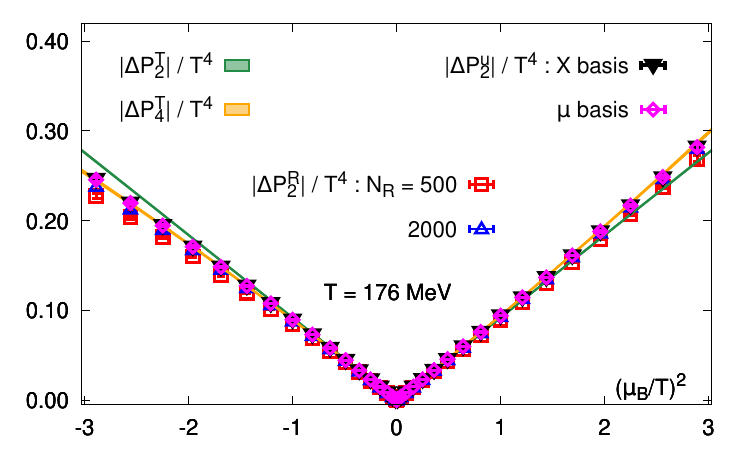} 
    \includegraphics[width=0.49\textwidth]{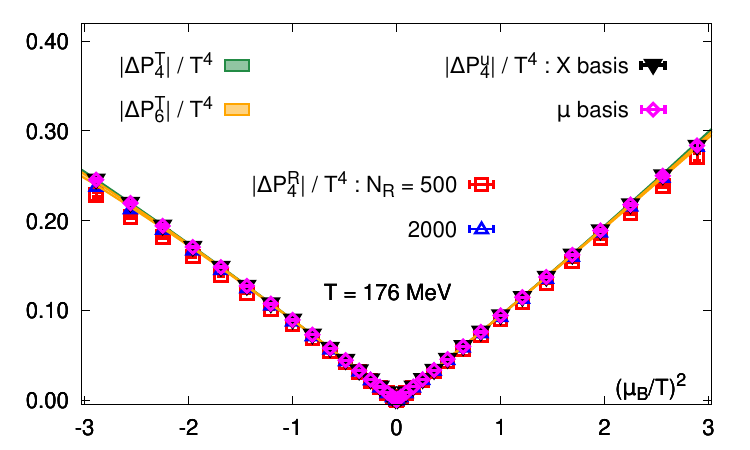} 
    \includegraphics[width=0.49\textwidth]{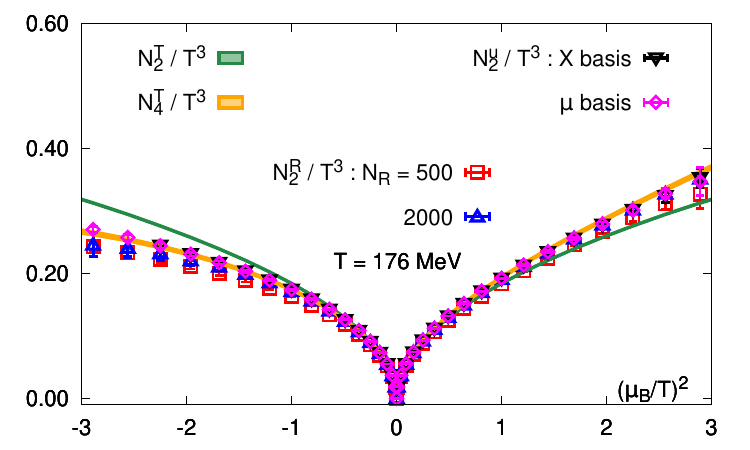} 
     \includegraphics[width=0.49\textwidth]{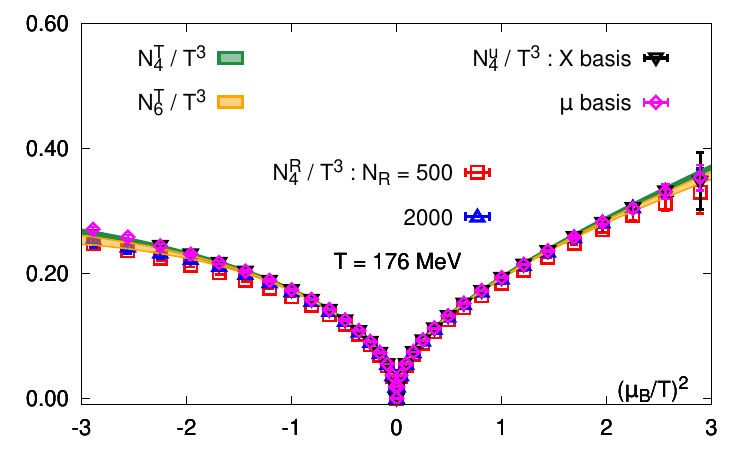} 
     \caption{Comparison of Biased and Unbiased results at $T=176$ MeV. The old biased results are calculated using $500$ and $2000$ random sources respectively. The Taylor results are shown in bands and the unbiased results are plotted in $X$ and $\mu$ bases.}
     \label{fig:2000 176 MeV}
\end{figure}

We find a remarkable agreement between the results obtained using the new formalism and Taylor series results up to similar orders in $\mu_B$, despite using only $500$ random volume sources for all $D_n, 1 \leq n \leq 4$. The plots in the following figures, namely \autoref{fig:2000 135 MeV}, \autoref{fig:2000 157 MeV} and \autoref{fig:2000 176 MeV} vindicate the agreement, for both real and imaginary $\mu_B$. 

Let us analyse the results for the important, yet computationally difficult $T=135$ MeV. Clearly, we see the effect of reduced estimate bias of $D_1$, following which the biased results with $2000$ sources of $D_1$, shown by the blue points in \autoref{fig:2000 135 MeV} approach the Taylor results in both the real and imaginary domains of $\muB$. As demonstrated from the figure itself, the results from the unbiased formalism using $\Ob(500)$ random vectors for $D_1$ exhibit even better agreement with the Taylor series results and in fact, surpass the Taylor results for $\left|\muB/T\right| \geq 1.1$, thereby capturing higher order Taylor series results. 

\begin{figure}[H]
\centering
 \includegraphics[width=0.49\textwidth]{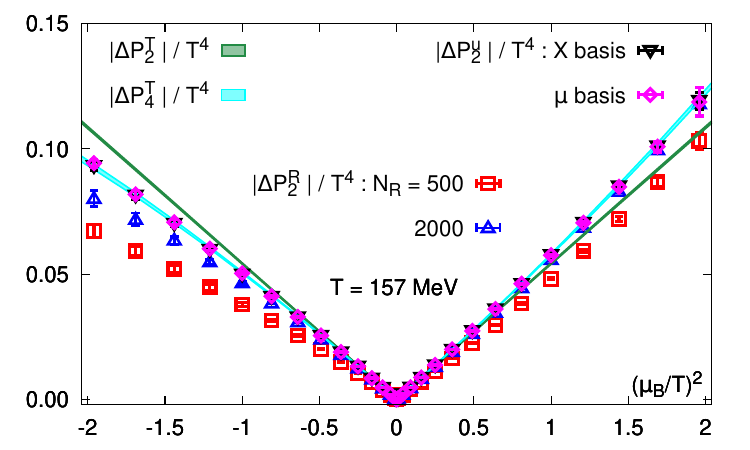} 
    \includegraphics[width=0.49\textwidth]{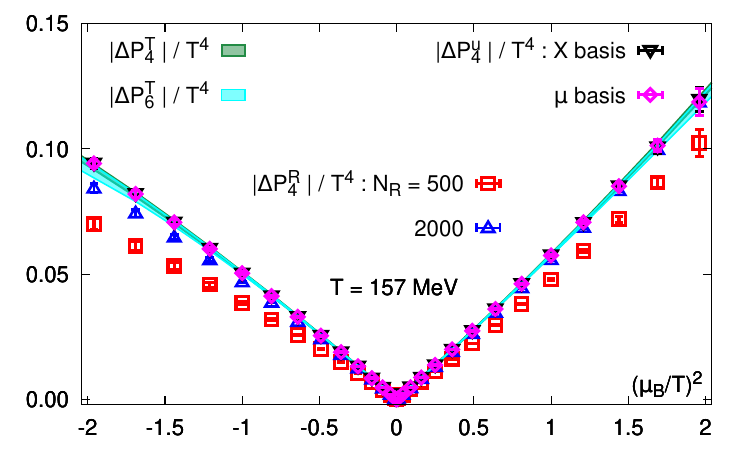} 
    \includegraphics[width=0.49\textwidth]{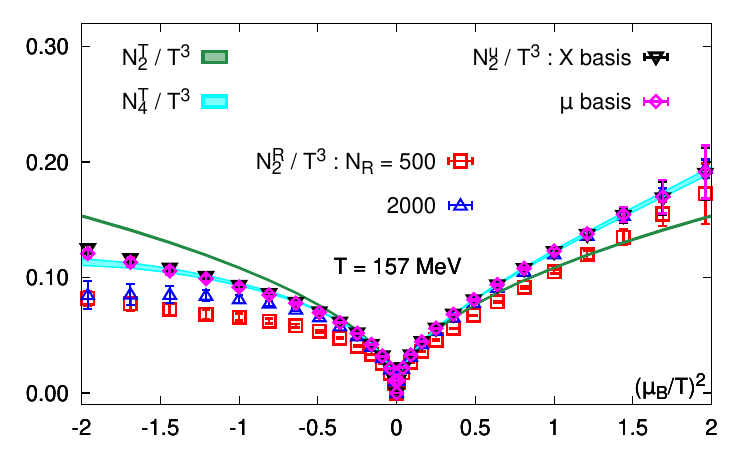} 
     \includegraphics[width=0.49\textwidth]{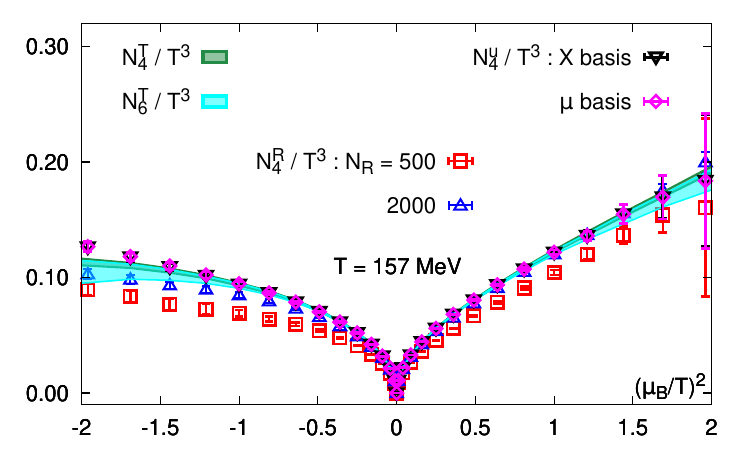} 
     \caption{Comparison of Biased and Unbiased results at $T=157$ MeV. The old biased results are calculated using $500$ and $2000$ random sources respectively. The Taylor results are shown in bands and the unbiased results are plotted in $X$ and $\mu$ bases.}
     \label{fig:2000 157 MeV}
\end{figure}

This difference is even more stark for imaginary $\muB$. Unlike the real regime, where there is a palpable trend of increasing values of observables ($\Delta P$ and $\mathcal{N}$) with increasing orders in $\muB$, the imaginary regime does not possess any such trend, at least upto $6^{th}$ order of calculation. We find that the $4^{th}$ order lies quite lower than the $2^{nd}$ order observables for imaginary $\muB$, but again the $6^{th}$ order lies just above $4^{th}$ order due to very small $6^{th}$ order corrections, as compared to the $4^{th}$ order. In this context therefore, the unbiased results are more reliable, as they provide better agreement with the Taylor series, specially in a situation, where we cannot confidently comment on its ability to capture higher order Taylor results.    

 This is promising, particularly from the perspective of analytic continuation approach from imaginary to real $\mu_B$~\cite{Borsanyi.2018, Gunther.2016}, determining the Equation of state (EoS) for real, finite $\mu_B$. It goes without saying that the old resummed results improved markedly by increasing the number of random volume sources fourfold, as evident from the above~\autoref{fig:2000 135 MeV}, \autoref{fig:2000 157 MeV} and \autoref{fig:2000 176 MeV} respectively. Another significant takeaway is the faster convergence of the new formalism, which is manifested by the excellent agreement between the $2^{nd}$ order unbiased Exponential Resummation results and the $4^{th}$ order Taylor series results. The agreement remains equally sustained for both real and imaginary $\mu_B$ regimes. The old resummed results, particularly using $2000$ random volume sources for $D_1$ appear to agree with the Taylor series results from $4^{th}$ order on-wards, where $D_n$ upto $D_4$ are taken into account. This agreement is exactly what we usually expect from a proper exponential resummation scheme. This also eliminates the need for separate, tedious calculations of $2000$ separate independent estimates of $D_1$ since our formalism moreover treats all the $D_n$ on an equal footing. As expected, all the three sets of results narrow in with increasing temperatures as evident from \autoref{fig:2000 157 MeV} and \autoref{fig:2000 176 MeV} respectively.
 
 \begin{figure}[H]
 \centering
\includegraphics[width=0.49\textwidth]{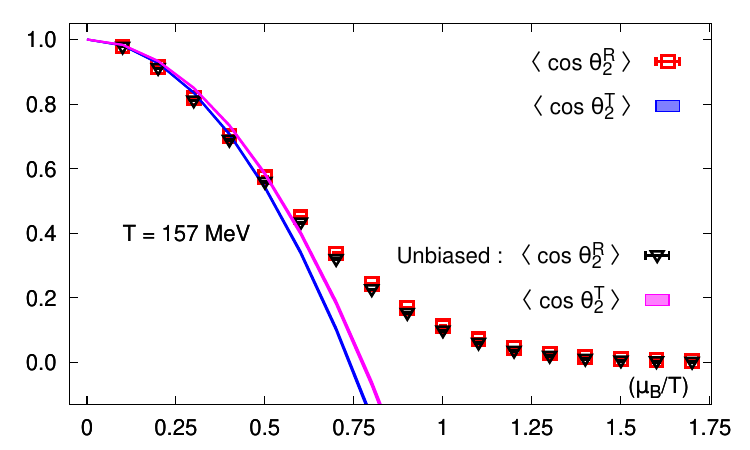}
\includegraphics[width=0.49\textwidth]{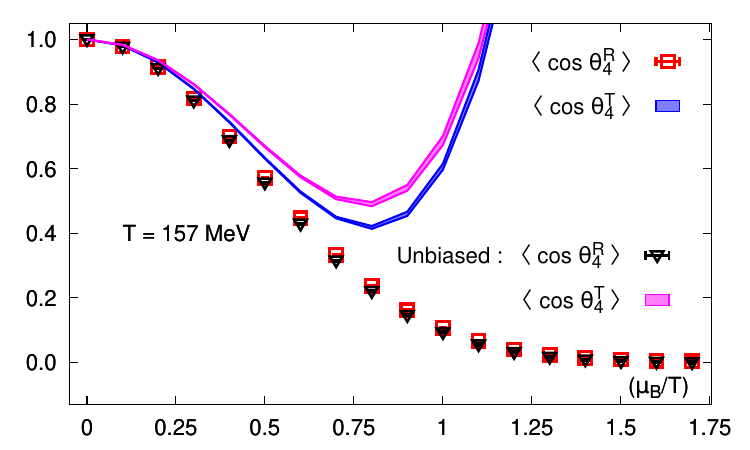}

\caption{Comparison of the Resummed and the Taylor expanded results for the average phasefactor $\LA \text{cos} \hspace{.7mm}\theta_2 \RA$ (top) and $\LA \text{cos} \hspace{.7mm}\theta_4 \RA$ (bottom). The bands illustrate the Taylor expanded results, whereas the usual points represent the usual resummed results for both biased and unbiased calculations.}
\label{fig:phasefactor Taylor and resummation}
\end{figure}

 The above \autoref{fig:phasefactor Taylor and resummation} reiterates that resummation is a far better approach than Taylor series in the sense that the latter has a slow rate of convergence, whereas the former indicates a proper physical breakdown from the zeroes of the phasefactor. The biased and unbiased Taylor estimates are not drastically different and both of them signify the slow convergence property of an usual Taylor series. As mentioned before, these zeroes of phasefactor in case of exponential resummation prove to be a reliable indicator for the singularities of partition function $\Z$ in complex $\mu$ plane as well as reflect the severity of the sign problem for finite real $\muB$.

 	\cleardoublepage
 	\chapter{Summary and conclusions}
\label{Chapter 9}

 In this final chapter of the thesis, I briefly summarise my work revolving mainly around this new formalism of unbiased exponential resummation at a finite chemical potential, which has been discussed vividly and comprehensively to the best of my efforts in \autoref{Chapter 8}. Post discussion of summary, I have also tried to enlighten possible future work in this direction.
 
\section{Summary of the thesis}

 In this thesis, a new formalism of unbiased exponential resummation has been presented which can completely reproduce the Taylor series expansion of thermodynamic observables upto a given finite order in chemical potential. As mentioned before, the conclusive exploration and evidence of QCD phase diagram is a significant problem in the regime of high-energy physics and also equally important for knowing the chronological evolution of Universe, paving way for the present form of the living world. Till date, the circle of our knowledge is very limited, the state-of-the-art results provide very little conclusive evidence and inadequate clarity to replace these conjectures. Nevertheless, we have come along a long way, despite knowing that one of our principal hurdles in this voyage, the notorious sign problem at a finite baryon chemical potential still remains existent and intact. 

 We have presented the origin and cause of this sign problem and despite knowing that till date, there is no existing general solution to sign problem in QCD formulated on lattice, we have come across different techniques like reweighting, analytic continuation which, instead of eliminating, can prolong the problem to some extent.

 One of these approaches adopted in an attempt to shrug off sign problem, is the Taylor series expansion of thermodynamic observables. We have considered excess pressure and number density as two thermodynamic observables for our study. The Taylor series definitely provides reliable results of observables in chemical potential, but that up to the radius of convergence of the series measured in units of chemical potential. All throughout the work and the thesis, we have considered dimensionless versions of physical quantities, since we are working on a lattice discretised spacetime. In spite of the ease of implementation and understanding, the Taylor series suffer from a very slow rate of convergence and highly non-monotonic behaviour for a wide range of working temperatures. This require calculations to sufficiently higher order in chemical potential, where we find the computations of higher order Taylor series, starting from $6^{th}$ order, to be extremely tedious and very much expensive computationally.

 This drawback of Taylor series, motivated the idea of resummation of lower order Taylor series. Unlike Taylor series which is finite, resummation series usually contains transcendental functions, which can capture finite ordered contributions of different correlation functions to all (infinite) orders in chemical potential. We have discussed exponential resummation in this thesis in detail, whereas just touched upon another form of resummation which is Pad{\'e} resummation. 

 While performing exponential resummation, we encounter another problem in form of biased estimates of different correlation functions, which would emerge uncontrollably, due to the exponential function and its series expansion in the resummation formulae. The main reason is rooted deep into the setup of our calculations. The different correlation functions are linear combinations of traces of various operators, each of which involves the computation of inverse of fermion matrix. This is impossible to calculate analytically, at least in practically relevant lattices with large volumes and fine spacings. Hence, we have to numerically estimate these traces and so the different correlation functions. This is done by using random volume sources per configuration and due to this stochastic averaging over random volume sources, we find biased estimates in exponential resummation. This stochastic bias is important to diagnose and treat properly at an early stage since, in due course of evaluation, it may mislead calculations and misdirect consequent inferences about the physics, specially in the regime of higher values, orders of chemical potential and also dealing with observables forming higher order chemical potential derivatives of free energy. We may mistake bias effect with a breakdown phenomenon.

 To deal with these biased estimates, we perform cumulant expansion of the series, which enable one to calculate biased estimates order-by-order in chemical potential in a controllable fashion. We performed this for isospin chemical potential, rather than the more relevant baryon chemical potential since, unlike the latter, the former has no sign problem. We have adequately supported our arguments with proper figures, where we have also demonstrated how stochastic bias increase with decreasing number of random volume sources per configuration and how this effect propagates into the calculation of observables. We also have showed how we are able to capture genuine large fluctuations of higher-order correlation functions appearing alongside higher ordered chemical potential, when we replace these biased estimates with unbiased counterparts order-by-order in chemical potential. These are manifested through large errorbars and also, these errorbars increase for number density over excess pressure since, the former is a first order derivative of free energy with respect to chemical potential. But amidst all these improvements and satisfaction, we dispossessed reweighting factor, phasefactor which constituted the exponential resummed estimate of partition function.

 This ultimately motivated us towards formulating a new formalism, which will produce unbiased estimates, but preserving the structure and anatomy of the original exponential resummed form. This meant, that one is only free to fiddle with the argument of the exponential in the expression of the partition function in the formula of excess pressure in such a manner, that when the series expansion of the exponential and the log is performed, it produces the original Taylor series order-by-order to a finite order in chemical potential. Because, we have seen that the Taylor coefficients constitute unbiased powers and estimates of appropriate correlation functions, the same which is used in exponential resummation too. Through figures, we have demonstrated the comparative effects between biased and unbiased results for our three working temperatures, which span all the hadronic, crossover and the QGP phases and have established the thermal stability and phase independent property of our formalism. We have demonstrated the computational benefits of our formalism, using different number of random volume sources for the $1$-point correlation function, which is expected to contribute the most in a calculation. Finally, we also provide the plots of phasefactor and roots of partition function in complex chemical potential plane since, we now have a newly defined partition function with a valid reweighting factor and phasefactor.

\section{Conclusions and Outlook}

This thesis concludes here introducing the new formalism of unbiased exponential resummation at finite chemical potential along with its possible advantages and disadvantages. 

The job is not over definitely. We need to investigate more observables, extrapolate upto higher orders, see for which number of random volume sources, the old results seem to tally with our results. Any alternative approach or any tweak, which can improve the radius of convergence, in forms of zeroes of the phasefactor and can push it further towards higher values of baryon chemical potential. This definitely augments our sphere of reliability on these approaches for higher and higher values of chemical potential, which is the fundamental key toward knowing finite density QCD and which is instrumental to conclusive and concrete understanding of the QCD phase diagram.

In the meantime, we have proposed a new way to understand and identify the breakdown of this unbiased exponential resummation at a finite isospin chemical potential $\mu_I$. Unlike baryon chemical potential $\mu_B$ or other forms of chemical potentials which suffer from sign problem, $\muI$ does not experience a sign problem and hence the usual phasefactor formalism, which used to signify the breakdown prominently for $\mu_B$ does not work for $\mu_I$. In this recent work, we have looked at the Newton-Raphson singularities of the QCD partition function estimated using unbiased exponential resummation, unbiased upto $\mathcal{O}(\mu_I^4)$. We have made $\mu_I$ complex and calculated a non-trivial phasefactor for these complex values of $\mu_I$. From the plots of phasefactor and radius of convergence, we observe that this phasefactor can capture the singularities efficiently. We also investigated the overlap problem and have found that this problem becomes extremely severe with large errorbars typically as one goes beyond the value of real $\mu_I$, from which these phasefactor values condense to zero. We are yet to explore the correspondence of these observations with the isospin QCD phase diagram in the $T-\mu_I$ plane and also, we need to check if these findings go well with the zero temperature calculations and observations. As mentioned before, a detailed explanation of this work is presently beyond the scope of the thesis. We refer interesting readers to Ref.~\cite{Mitra:2023isospin}.

A comparative study of exponential and Pad{\'e} resummation is always on the cards. At present, these two methods of resummation are the two frequently used approaches used for probing finite density QCD. 

 	\cleardoublepage
   \end{spacing}

\appendix
   \begin{spacing}{1.3}

\chapter{Line of Constant Physics (LCP)}
\label{Appendix 2}

The Lattice QCD action in $2+1$ flavor signature is given as follows

\begin{align}
    S_{QCD} &= \beta \sum_{n}\sum_{\mu=1}^4 \sum_{\nu>\mu} \text{Re} \Tr \left[\mathbb{I}-P_{\mu\nu}(n)\right] \notag\\
    &+ \,\sum_{n,m} \sum_{\mu=1}^{4} \Bar{\Psi}_u(n) \left[ U_{\mu}(n) \delta_{n+1,m} - U_{\mu}^{\dagger}(n) \delta_{n-1,m} + m_l \delta_{n,m}\right] \Psi_u(m)\notag \\
    &+ \,\sum_{n,m} \sum_{\mu=1}^{4} \Bar{\Psi}_d(n) \left[ U_{\mu}(n) \delta_{n+1,m} - U_{\mu}^{\dagger}(n) \delta_{n-1,m} + m_l \delta_{n,m}\right] \Psi_d(m)\notag \\
    &+ \sum_{n,m} \sum_{\mu=1}^{4} \Bar{\Psi}_s(n) \left[ U_{\mu}(n) \delta_{n+1,m} - U_{\mu}^{\dagger}(n) \delta_{n-1,m} + m_s \delta_{n,m}\right] \Psi_s(m)
    \label{eq:LCP action}
\end{align}
 The $2+1$ flavored QCD action as in \autoref{eq:LCP action}, contains three coupling constants namely the inverse coupling $\beta = 2N/g^{2}$, where $N$ is the number of colors of quarks considered, the light quark mass $m_l$ and the strange quark mass $m_s$. The factor of $2$ in the second term of this equation indicates inclusion of up and down quarks as they are equally massive with mass $m_l$ in $2+1$ flavor QCD. The $P_{\mu\nu}(n)$ is the standard plaquette which is constructed using four vertices, namely $n$, $n+a\hat{\mu}$, $n+a\hat{\mu}+a\hat{\nu}$ and  $n+a\hat{\nu}$ in anti-clockwise direction using the four gauge fields $U$, acting as four gauge links on the lattice. The plaquette is given by
 \begin{equation*}
     P_{\mu\nu}(n) = U_{\mu}(n) \hspace{1mm}U_{\nu} (n+a\hat{\mu}) \hspace{1mm}U_{\mu}^{\dagger} (n+a\hat{\nu})\hspace{1mm} U_{\nu}^{\dagger} (n)
 \end{equation*}
 where $a$ is the value of the lattice spacing and $\hat{\mu}$, $\hat{\nu}$ are the $4$-directions on lattice. We know the following relations from the gauge invariance of the fermion action, namely
 
 \begin{align*}
     U_{-\mu}(n+a\hat{\mu}+a\hat{\nu}) &= U_{\mu}^{\dagger} (n+a\hat{\nu}) \\
     U_{-\nu}(n+a\hat{\nu}) &= U_{\nu}^{\dagger} (n)
 \end{align*}
 The up and down quarks are considered mass degenerate and hence, are termed together as light quarks. Among these parameters, $\beta$ is used to set the scale or spacing of lattice, whereas the respective quark masses $m_l$ and $m_s$ are tuned accordingly so as to reproduce the known physical meson masses. 

 Leading order chiral perturbation theory $(\chi PT)$ dictates that the square of the meson mass is proportional to the masses of the quarks constituting the meson. This is mathematically expressed as

 \begin{equation}
     M_{\pi}^2 = B(\beta)m_l, \quad M_K^2 = B(\beta)\left(\frac{m_l+m_s}{2}\right), \quad M_S^2 = B(\beta)m_s
     \label{eq:LCP all mass}
 \end{equation}
 where in the above \autoref{eq:LCP all mass}, $B(\beta)$ is a constant of proportionality which is a function of $\beta$, and $M_{\pi}$ and $M_K$ are the masses of the pion and kaon which are roughly $140$ MeV and $490$ MeV respectively. $M_S$ is the mass of a hypothetical $\eta_{\bar{s}s}$ meson. This clearly implies that the mass of this hypothetical meson is related to pion and kaon masses though the mathematical relation
 \begin{equation}
     M_S^2 = 2M_K^2 - M_{\pi}^2
     \label{eq:LCP ms}
 \end{equation}

 After substituting the masses of pion ($M_{\pi}$) and kaon ($M_K$) in \autoref{eq:LCP ms}, we find $M_S \approx 686$ MeV.

\begin{figure}
\centering
 \includegraphics[width=0.68\textwidth]{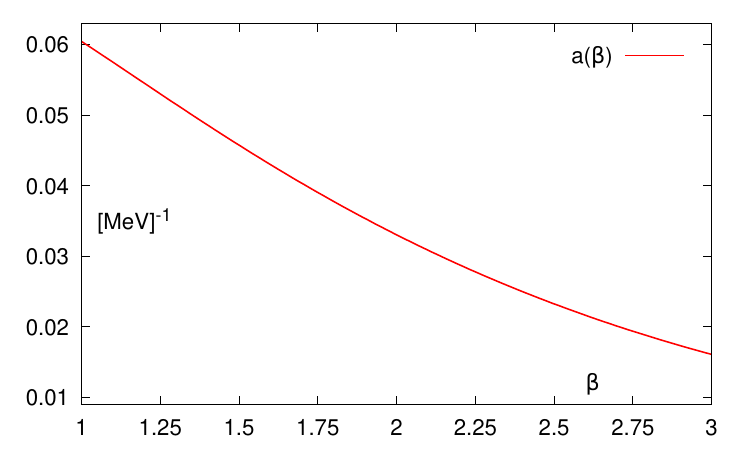}
\caption{Lattice spacing $a$ (in MeV$^{-1}$) as a function of inverse gauge coupling $\beta$}
\label{fig:Lat_beta}
\end{figure}

For the sake of completeness, it is useful to discuss a quick intuitive physics here. It is quite understandable from our basic knowledge of thermal agitation that with increase in temperature of a system of particles, the degree of thermal agitation increases, as a result of which, the coupling among the particles decrease and vice versa. This implies that the inverse coupling $\beta \propto g^{-2}$ increases. Since, we know that $T = (aN_{\tau})^{-1}$, where the symbols have the conventional usual meanings, for a fixed lattice with a fixed number of temporal sites $N_{\tau}$, the lattice spacing $a$ decreases with increase in $\beta$ and vice versa. This sets the function $a(\beta)$, which is what \autoref{fig:Lat_beta} implies exactly.

\begin{figure}
\centering
  \includegraphics[width=0.68\textwidth]{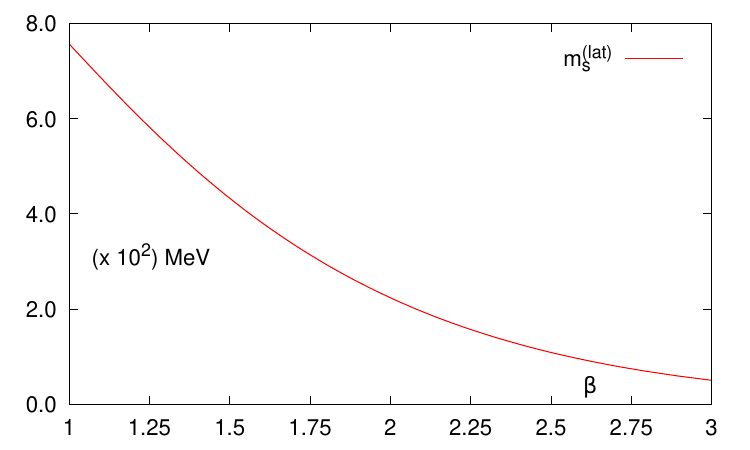}
\caption{Bare strange quark mass $m_s^{(lat)}$ (in MeV) on lattice as a function of $\beta$, setting the Line of constant physics}
\label{fig:m_s_beta}
\end{figure}

Also, with increase in $\beta$, since the particles in a system become less coupled and hence less interacting, the mass of the system can be considered as the unmixed sum of the masses of these constituent particles. This implies, as per the above Eqn.~\eqref{eq:LCP all mass}, that $B(\beta)$ increases and decreases with increase and decrease in $\beta$, in short, $B \propto \beta^{n}$, where $n > 0$ is positive. This means, that since, the physical masses of kaon and pion i.e. $M_{\pi}$ and $M_K$ remains fixed and so, is $M_S$, which is in the LHS of the eqn.~\eqref{eq:LCP all mass}, the masses $m_l$ and $m_s$ will decrease with increase in $\beta$ and vice-versa.  
 And this is precisely what is found while comparing Tables~\ref{table:32*8} , \ref{table:48*12} and \ref{table:64*16}, where the $N_{\tau}$ increases progressively. The \autoref{fig:m_s_beta}  describes the tuning of bare $m_s$ in units of MeV on lattice with $\beta$, the behaviour of which is expected and already explained as above. Both \autoref{fig:Lat_beta} and \autoref{fig:m_s_beta} describe the scale setting in HISQ discretisation scheme. 
 
 It must be noted that all these discussions in the present context, pertain to bare parameters of the theory $\beta$ and $a$, and hence must be taken in the spirit of evoking a feeling of motivation in the minds of the readers only. For a more correct discussion, these non-renormalised parameters require renormalisation, which involve incorporating QCD beta function defined as $\beta_{QCD} = \partial g/\partial (\ln \mu_E)$, where $g$ is the QCD running coupling constant and $\mu_E$ is the relevant energy scale. Although important, a detailed discussion on this and associated method of renormalisation is however beyond the scope of the present appendix and the thesis. The renormalised dependence of bare quark masses with $\beta$ in the context of HISQ action are presented in Ref.\,\cite{Bazavov:2014pvz,Bollweg:2021vqf}. 

 The HotQCD collaboration uses either Sommer parameter $r_1$ or kaon decay constant $f_K$ to set the scale, or in other words, the value of the lattice spacing $a$. Once the lattice spacing $a$ is known, then the dimensionless mass $a \cdot M_S$ of the hypothetical meson can be obtained from lattice calculations using the results for screening correlators on lattice. Dividing by the lattice spacing $a$, we obtain the physical value of the mass i.e. $M_S$. Depending upon whether the obtained value is less or greater than $686$ MeV, the quark mass $m_s$ or $m_l$ is adjusted or tuned accordingly. This is carried on, unless the correct physical value of $M_S$ is obtained i.e. $M_S = 686$ MeV.

 It is sufficient to tune any one of the bare quark masses, because, these bare quark masses are constrained by the relation $m_l^{(bare)} = m_s^{(bare)}/27$. This is the most realistic value of the ratio of light to strange quark masses, closest to the true physical value. In nature, the average up and down quark mass is $(2.1+4.79)/2 \approx 3.45$ MeV, while the strange quark mass is approximately equal to $95$ MeV. Thus $m_s/m_l \approx 27$. In this manner, the light $(m_l)$ and strange quark $(m_s)$ masses can be ascertained for every value of the lattice spacing $a$, which equivalently mean for each value of the inverse coupling $\beta$. Thus, we determine finally the functional form of quark mass as a function of $\beta$, like $m_k(\beta), k \in l,s$ and this is what is referred as the familiar and important Line of Constant Physics (LCP). The most recent determination of LCP, conducted by HotQCD collaboration is given in the Ref.~\cite{Thakkar.2019}

\vspace{3mm}

 Following are \autoref{table:32*8}, \autoref{table:48*12} and \autoref{table:64*16}, illustrating how light $m_l$ and strange $m_s$ quark masses get adjusted on lattices of different volumes, namely $32^3 \times 8$, $48^3 \times 12$ and $64^3 \times 16$, with a fixed aspect ratio $N_{\sigma}/N_{\tau}$. All these are shown at different temperatures $T$ in MeV scales, which correspond to different inverse gauge couplings $\beta$. The mass ratio i.e. $m_l/m_s$ however is kept fixed at its physical value of $27$.

\vspace{6mm}

\renewcommand{\arraystretch}{1.5}

\begin{table}[H]
\centering 
\begin{tabular}{|P{3cm}|P{3cm}|P{3cm}|P{3cm}|} 
\hline
$\beta$ & $T$[MeV] & $m_l$ & $m_s$    \\ \hline 
$6.245$ & $134.84$ & $0.00307$ & $0.0830$  \\ \hline
$6.285$ & $140.62$ & $0.00293$ & $0.0790$   \\ \hline
$6.315$ & $145.11$ & $0.00281$ & $0.0759$   \\ \hline
$6.354$ & $151.14$ & $0.00270$ & $0.0728$   \\ \hline
$6.390$ & $156.92$ & $0.00257$ & $0.0694$   \\ \hline
$6.423$ & $162.39$ & $0.00248$ & $0.0670$   \\ \hline
$6.445$ & $166.14$ & $0.00241$ & $0.0652$   \\ \hline
$6.474$ & $171.19$ & $0.00234$ & $0.0632$   \\ \hline
$6.500$ & $175.84$ & $0.00228$ & $0.0614$ \\ \hline
\end{tabular}

\vspace{6mm}

\caption{\hspace{1mm}LCP for $m_l = m_s/27$ on a $N_{\tau}=8$ lattice, with $N_{\sigma}=4N_{\tau}$}
\label{table:32*8} 
\end{table}

\begin{table}[H]
\centering
\begin{tabular}{|P{3cm}|P{3cm}|P{3cm}|P{3cm}|}
    \hline
$\beta$ & $T$[MeV] & $m_l$ & $m_s$    \\ \hline 
$6.640$ & $135.24$ & $0.00196$ & $0.0523$  \\ \hline
$6.680$ & $140.80$ & $0.00187$ & $0.0505$   \\ \hline
$6.712$ & $145.40$ & $0.00181$ & $0.0489$   \\ \hline
$6.754$ & $151.62$ & $0.00173$ & $0.0467$   \\ \hline
$6.794$ & $157.75$ & $0.00167$ & $0.0451$   \\ \hline
$6.825$ & $162.65$ & $0.00161$ & $0.0435$   \\ \hline
$6.850$ & $166.69$ & $0.00157$ & $0.0424$   \\ \hline
$6.880$ & $171.65$ & $0.00153$ & $0.0413$   \\ \hline
$6.910$ & $176.73$ & $0.00148$ & $0.0400$ \\ \hline
\end{tabular}

\vspace{6mm}

\caption{\hspace{1mm}LCP for $m_l = m_s/27$ on a $N_{\tau}=12$ lattice, with $N_{\sigma}=4N_{\tau}$}
\label{table:48*12}
\end{table}

\vspace{6cm}

\begin{table}[ht]
\centering
\begin{tabular}{|P{3cm}|P{3cm}|P{3cm}|P{3cm}|}
    \hline
$\beta$ & $T$[MeV] & $m_l$ & $m_s$    \\ \hline 
$6.935$ & $135.80$ & $0.00145$ & $0.0392$  \\ \hline
$6.973$ & $140.86$ & $0.00139$ & $0.0375$   \\ \hline
$7.010$ & $145.95$ & $0.00132$ & $0.0356$   \\ \hline
$7.054$ & $152.19$ & $0.00129$ & $0.0348$   \\ \hline
$7.095$ & $158.21$ & $0.00124$ & $0.0335$   \\ \hline
$7.130$ & $163.50$ & $0.00119$ & $0.0321$   \\ \hline
$7.156$ & $167.53$ & $0.00116$ & $0.0313$   \\ \hline
$7.188$ & $172.60$ & $0.00113$ & $0.0305$   \\ \hline
$7.220$ & $177.80$ & $0.00110$ & $0.0297$ \\ \hline
\end{tabular}

\vspace{6mm}

\caption{\hspace{1mm}LCP for $m_l = m_s/27$ on a $N_{\tau}=16$ lattice, with $N_{\sigma}=4N_{\tau}$}
\label{table:64*16}
\end{table}

\vspace{3mm}

\renewcommand{\arraystretch}{1}

	\cleardoublepage
	\newcommand{\1}{\M^{-1}\frac{\partial \M}{\partial \mu}}
\newcommand{\2}{\M^{-1}\frac{\partial^2 \M}{\partial \mu^2}}
\newcommand{\3}{\M^{-1}\frac{\partial^3 \M}{\partial \mu^3}}
\newcommand{\4}{\M^{-1}\frac{\partial^4 \M}{\partial \mu^4}}
\newcommand{\5}{\M^{-1}\frac{\partial^5 \M}{\partial \mu^5}}
\newcommand{\6}{\M^{-1}\frac{\partial^6 \M}{\partial \mu^6}}
\newcommand{\7}{\M^{-1}\frac{\partial^7 \M}{\partial \mu^7}}
\newcommand{\8}{\M^{-1}\frac{\partial^8 \M}{\partial \mu^8}}

\chapter{Taylor expansion coefficients}
\label{Appendix 3}

In this appendix, we derive some equations which are used in the 
calculation of the various thermodynamic quantities and expansion
coefficients of the Taylor series presented in this thesis. 
The partition function $\Z$ is given by 
\begin{eqnarray}
{\cal Z} = \int 
{D}U (\det \M)^{n_{\rm f}/4} e^{-S_g} \quad , 
\label{eq:partition} 
\end{eqnarray}

with $U\in SU(3)$ are gauge field configurations of QCD, with $n_f$ flavors of quarks.  The expectation value of a physical quantity, 
$\left\langle {\cal O} \right\rangle$ is then obtained as 

\begin{eqnarray}
\left\langle {\cal O} \right\rangle 
=
\frac{1}{\cal Z} \int {D}U {\cal O} 
(\det \M)^{n_{\rm f}/4} e^{-S_g} \quad 
\end{eqnarray}

and its derivatives with respect to dimensionless quark chemical potential and dimensionless quark
mass are given as follows

\begin{eqnarray}
\frac{\partial \langle {\cal O} \rangle}{\partial \mu} 
\hspace{-2mm} &=& \hspace{-2mm} 
\left\langle 
\frac{\partial {\cal O}}{\partial \mu} \right\rangle 
+ \frac{n_{\rm f}}{4} \left(
\left\langle {\cal O} \; 
\frac{\partial (\ln \det \M)}{\partial \mu} \right\rangle 
-\left\langle {\cal O} \right\rangle 
\left\langle  
\frac{\partial (\ln \det \M)}{\partial \mu} \right\rangle \right) \quad , 
\\ \nonumber \\
\frac{\partial \langle {\cal O} \rangle}{\partial m} 
\hspace{-2mm} &=& \hspace{-2mm} 
\left\langle 
\frac{\partial {\cal O}}{\partial m} \right\rangle 
+ \frac{n_{\rm f}}{4} \left(
\left\langle {\cal O}  \;
\frac{\partial (\ln \det \M)}{\partial m} \right\rangle 
-\left\langle {\cal O} \right\rangle 
\left\langle  
\frac{\partial (\ln \det \M)}{\partial m} \right\rangle \right) \nonumber \\
\hspace{-2mm} &=& \hspace{-2mm} 
\left\langle 
\frac{\partial {\cal O}}{\partial m} \right\rangle 
+ \frac{n_{\rm f}}{4} \left(
\left\langle {\cal O}  \;
{\rm tr} \left(\M^{-1}\right) \right\rangle 
-\left\langle {\cal O} \right\rangle 
\left\langle {\rm tr} \left(\M^{-1}\right) \right\rangle \right) \quad 
\end{eqnarray}

Here we define $m \equiv m_qa$ as the dimensionless quark mass, with dimensional mass $m_q$ 
and also $\mu \equiv \mu_q a$ as the dimensionless quark chemical potential, with dimensional chemical potential $\mu_q$. Here, $a$ is the lattice spacing of the lattice used.

The temperature is $T=(N_{\tau} a)^{-1}$ and the volume is 
$V=(N_{\sigma} a)^3$, where $\Nt$ and $\Ns$ are the number of lattice sites in temporal and spatial directions respectively.
Moreover, for simplification, we introduce 

\begin{eqnarray}
D_n = \frac{n_{\rm f}}{4} \frac{\partial^n \ln \det \M} 
{\partial \mu^n} \quad 
\label{eq:basic}
\end{eqnarray}

All Taylor expansion coefficients used in this thesis can be expressed
in terms of expectation values of certain combinations of different powers and orders of correlation functions
$D_n$. The required derivatives of $\ln \det \M$ are explicitly given in the following.

\begin{align}
\frac{\partial \ln \det \M}{\partial \mu} 
 &=
{\rm tr} \left( \M^{-1} \frac{\partial \M}{\partial \mu} \right) 
\label{eq:dermu1} \\
\nonumber\\
\frac{\partial^2 \ln \det \M}{\partial \mu^2} 
&= 
{\rm tr} \left( \M^{-1} \frac{\partial^2 \M}{\partial \mu^2} \right)
 - {\rm tr} \left( \M^{-1} \frac{\partial \M}{\partial \mu}
                   \M^{-1} \frac{\partial \M}{\partial \mu} \right) 
\label{eq:dermu2} \\
\nonumber\\
\frac{\partial^3 \ln \det \M}{\partial \mu^3} 
&= 
{\rm tr} \left( \M^{-1} \frac{\partial^3 \M}{\partial \mu^3} \right)
 -3 \hspace{1mm}{\rm tr} \left( \M^{-1} \frac{\partial \M}{\partial \mu}
              \M^{-1} \frac{\partial^2 \M}{\partial \mu^2} \right) 
\nonumber \\ & 
+2 \hspace{1mm} {\rm tr} \left( \M^{-1} \frac{\partial \M}{\partial \mu}
        \M^{-1} \frac{\partial \M}{\partial \mu}
        \M^{-1} \frac{\partial \M}{\partial \mu} \right) 
\label{eq:dermu3} \\
\nonumber\\
\frac{\partial^4 \ln \det \M}{\partial \mu^4} 
&=
{\rm tr} \left( \M^{-1} \frac{\partial^4 \M}{\partial \mu^4} \right)
 -4 \hspace{1mm} {\rm tr} \left( \M^{-1} \frac{\partial \M}{\partial \mu}
              \M^{-1} \frac{\partial^3 \M}{\partial \mu^3} \right) \nonumber \\
&
-3 \hspace{1mm}{\rm tr} \left( \M^{-1} \frac{\partial^2 \M}{\partial \mu^2}
        \M^{-1} \frac{\partial^2 \M}{\partial \mu^2} \right) \nonumber \\
 &+12 \hspace{1mm} {\rm tr} \left( \M^{-1} \frac{\partial \M}{\partial \mu}
        \M^{-1} \frac{\partial \M}{\partial \mu}
        \M^{-1} \frac{\partial^2 \M}{\partial \mu^2} \right) \nonumber \\
&
-6 \hspace{1mm} {\rm tr} \left( \M^{-1} \frac{\partial \M}{\partial \mu}
        \M^{-1} \frac{\partial \M}{\partial \mu}
        \M^{-1} \frac{\partial \M}{\partial \mu}
        \M^{-1} \frac{\partial \M}{\partial \mu} \right) 
\label{eq:dermu4} 
\end{align}

\vspace{8mm}

 Since the UV divergences remain upto $\Ob(\textbf{a}^4)$, where $\textbf{a}$ is the inverse lattice spacing, equivalent to the mass dimension of $\mu$, all the $n$ derivatives of $\M$ w.r.t $\mu$ must be considered for $n \leq 4$, in order to cancel out the divergences. This implies that all the traces in the above equations from Eqn.~\eqref{eq:dermu1} to Eqn.~\eqref{eq:dermu4} must be considered while calculating $D_n$ upto $n=4$. However, one can always consider sticking to exponential formalism to evaluate $D_n$, for all $n$. It has been already observed that the results from the exponential and linear formalisms are almost similar for $n \geq 5$ and hence, one usually recommend using linear $\mu$ formalism, while working for $n \geq 5$, since, it saves substantial computational time as well as avoid violating the underlying physics.
 
\begin{eqnarray}
\frac{\partial^5 \ln \det \M}{\partial \mu^5} 
\hspace{-2mm} &=& \hspace{-2mm} 
{\rm tr} \left( \M^{-1} \frac{\partial^5 M}{\partial \mu^5} \right)
 -5 {\rm tr} \left( \M^{-1} \frac{\partial \M}{\partial \mu}
              \M^{-1} \frac{\partial^4 M}{\partial \mu^4} \right) \nonumber \\
&& \hspace{-25mm}
 -10 {\rm tr} \left( \M^{-1} \frac{\partial^2 M}{\partial \mu^2}
        \M^{-1} \frac{\partial^3 M}{\partial \mu^3} \right) 
 +20 {\rm tr} \left( \M^{-1} \frac{\partial \M}{\partial \mu}
        \M^{-1} \frac{\partial \M}{\partial \mu} 
        \M^{-1} \frac{\partial^3 M}{\partial \mu^3} \right) \nonumber \\
&& \hspace{-25mm}
 +30 {\rm tr} \left( \M^{-1} \frac{\partial \M}{\partial \mu}
        \M^{-1} \frac{\partial^2 M}{\partial \mu^2}
        \M^{-1} \frac{\partial^2 M}{\partial \mu^2} \right) \nonumber \\
&& \hspace{-25mm}
 -60 {\rm tr} \left( \M^{-1} \frac{\partial \M}{\partial \mu}
        \M^{-1} \frac{\partial \M}{\partial \mu}
        \M^{-1} \frac{\partial \M}{\partial \mu}
        \M^{-1} \frac{\partial^2 M}{\partial \mu^2} \right) \nonumber \\
&& \hspace{-25mm}
 +24 {\rm tr} \left( \M^{-1} \frac{\partial \M}{\partial \mu}
        \M^{-1} \frac{\partial \M}{\partial \mu}
        \M^{-1} \frac{\partial \M}{\partial \mu}
        \M^{-1} \frac{\partial \M}{\partial \mu}
        \M^{-1} \frac{\partial \M}{\partial \mu} \right) ,
\label{eq:dermu5} 
\end{eqnarray}

\begin{eqnarray}
\frac{\partial^6 \ln \det \M}{\partial \mu^6} 
\hspace{-2mm} &=& \hspace{-2mm} 
{\rm tr} \left( \M^{-1} \frac{\partial^6 M}{\partial \mu^6} \right)
 -6 {\rm tr} \left( \M^{-1} \frac{\partial \M}{\partial \mu}
              \M^{-1} \frac{\partial^5 M}{\partial \mu^5} \right) \nonumber \\
&& \hspace{-25mm}
 -15 {\rm tr} \left( \M^{-1} \frac{\partial^2 M}{\partial \mu^2}
        \M^{-1} \frac{\partial^4 M}{\partial \mu^4} \right) 
 -10 {\rm tr} \left( \M^{-1} \frac{\partial^3 M}{\partial \mu^3} 
        \M^{-1} \frac{\partial^3 M}{\partial \mu^3} \right) \nonumber \\
&& \hspace{-25mm}
 +30 {\rm tr} \left( \M^{-1} \frac{\partial \M}{\partial \mu}
        \M^{-1} \frac{\partial \M}{\partial \mu} 
        \M^{-1} \frac{\partial^4 M}{\partial \mu^4} \right)
 +60 {\rm tr} \left( \M^{-1} \frac{\partial \M}{\partial \mu}
        \M^{-1} \frac{\partial^2 M}{\partial \mu^2}
        \M^{-1} \frac{\partial^3 M}{\partial \mu^3} \right) \nonumber \\
&& \hspace{-25mm}
 +60 {\rm tr} \left( \M^{-1} \frac{\partial^2 M}{\partial \mu^2}
        \M^{-1} \frac{\partial \M}{\partial \mu}
        \M^{-1} \frac{\partial^3 M}{\partial \mu^3} \right)
 +30 {\rm tr} \left( \M^{-1} \frac{\partial^2 M}{\partial \mu^2}
        \M^{-1} \frac{\partial^2 M}{\partial \mu^2}
        \M^{-1} \frac{\partial^2 M}{\partial \mu^2} \right) \nonumber \\
&& \hspace{-25mm}
 -120 {\rm tr} \left( \M^{-1} \frac{\partial \M}{\partial \mu}
        \M^{-1} \frac{\partial \M}{\partial \mu}
        \M^{-1} \frac{\partial \M}{\partial \mu}
        \M^{-1} \frac{\partial^3 M}{\partial \mu^3} \right) \nonumber \\
&& \hspace{-25mm}
 -180 {\rm tr} \left( \M^{-1} \frac{\partial \M}{\partial \mu}
        \M^{-1} \frac{\partial \M}{\partial \mu}
        \M^{-1} \frac{\partial^2 M}{\partial \mu^2}
        \M^{-1} \frac{\partial^2 M}{\partial \mu^2} \right) \nonumber \\
&& \hspace{-25mm}
 -90 {\rm tr} \left( \M^{-1} \frac{\partial \M}{\partial \mu}
        \M^{-1} \frac{\partial^2 M}{\partial \mu^2}
        \M^{-1} \frac{\partial \M}{\partial \mu}
        \M^{-1} \frac{\partial^2 M}{\partial \mu^2} \right) \nonumber \\
&& \hspace{-25mm}
 +360 {\rm tr} \left( \M^{-1} \frac{\partial \M}{\partial \mu}
        \M^{-1} \frac{\partial \M}{\partial \mu}
        \M^{-1} \frac{\partial \M}{\partial \mu}
        \M^{-1} \frac{\partial \M}{\partial \mu}
        \M^{-1} \frac{\partial^2 M}{\partial \mu^2} \right) \nonumber \\
&& \hspace{-25mm}
 -120 {\rm tr} \left( \M^{-1} \frac{\partial \M}{\partial \mu}
        \M^{-1} \frac{\partial \M}{\partial \mu}
        \M^{-1} \frac{\partial \M}{\partial \mu}
        \M^{-1} \frac{\partial \M}{\partial \mu}
        \M^{-1} \frac{\partial \M}{\partial \mu}
        \M^{-1} \frac{\partial \M}{\partial \mu} \right) .
\label{eq:dermu6}
\end{eqnarray}

\begin{align}
       \frac{\partial^7 \ln \det \M}{\partial \mu^7} &= {\rm tr} \left( \7 \right) - 7\hspace{1mm}{\rm tr} \left( \1 \6 \right) \notag\\&\hspace{-30mm}- 21\hspace{1mm}{\rm tr} \left( \2 \5 \right) - 35\hspace{1mm}{\rm tr} \left( \3 \4 \right) \notag\\&\hspace{-30mm}+ 42\hspace{1mm}{\rm tr} \left( \1 \1 \5 \right) + 105\hspace{1mm}{\rm tr} \left( \1 \2 \4 \right) \notag\\&\hspace{-30mm}+ 105\hspace{1mm}{\rm tr} \left( \2 \1 \4 \right) + 140\hspace{1mm}{\rm tr} \left( \1 \3 \3 \right) \notag\\&\hspace{-30mm}+ 210\hspace{1mm}{\rm tr} \left( \2 \2 \3 \right) \notag\\&\hspace{-30mm}- 210\hspace{1mm}{\rm tr} \left( \1 \1 \1 \4 \right) \notag\\&\hspace{-30mm}- 420\hspace{1mm}{\rm tr} \left( \1 \1 \2 \3 \right) \notag\\&\hspace{-30mm}- 420\hspace{1mm}{\rm tr} \left( \1 \2 \1 \3 \right) \notag\\&\hspace{-30mm}- 420\hspace{1mm}{\rm tr} \left( \1 \1 \3 \2 \right) \notag\\&\hspace{-30mm}- 630\hspace{1mm}{\rm tr} \left( \1 \2 \2 \2 \right) \notag\\&\hspace{-30mm}+ 840\hspace{1mm}{\rm tr} \left( \1 \1 \1 \1 \3 \right) \notag\\&\hspace{-30mm}+ 1260\hspace{1mm}{\rm tr} \left( \1 \1 \1 \2 \2 \right) \notag\\&\hspace{-30mm}+ 1260\hspace{1mm}{\rm tr} \left( \1 \1 \2 \1 \2 \right) \notag\\&\hspace{-30mm}- 2520\hspace{1mm}{\rm tr} \left( \1 \1 \1 \1 \1 \2 \right) \notag\\&\hspace{-30mm}+ 720\hspace{1mm}{\rm tr} \left( \1 \1 \1 \1 \1 \1 \1\right)
    \label{eq:dermu7}
\end{align}

\begin{align}
   \frac{\partial^8 \ln \det \M}{\partial \mu^8} &= 
    \hspace{1mm}{\rm tr} \big( \M_8 \big)  \notag\\
    &- 8\hspace{1mm}{\rm tr} \big( \M_1 \M_7 \big)-28\hspace{1mm}{\rm tr} \big( \M_2 \M_6 \big) -56\hspace{1mm}{\rm tr} \big( \M_3 \M_5 \big) -35\hspace{1mm}{\rm tr} \big( \M_4 \M_4 \big) \notag\\
    &+ 56\hspace{1mm}{\rm tr} \big( \M_1 \M_1 \M_6 \big) + 168\hspace{1mm}{\rm tr} \big( \M_1 \M_2 \M_5 \big) + 168\hspace{1mm}{\rm tr} \big( \M_2 \M_1 \M_5 \big) \notag\\
    &+ 280\hspace{1mm}{\rm tr} \big( \M_1 \M_3 \M_4 \big) + 280\hspace{1mm}{\rm tr} \big( \M_1 \M_4 \M_3 \big) + 420\hspace{1mm}{\rm tr} \big( \M_2 \M_2 \M_4 \big) \notag\\
    &+ 560\hspace{1mm}{\rm tr} \big( \M_2 \M_3 \M_3 \big) \notag\\
    &- 336\hspace{1mm}{\rm tr} \big( \M_1 \M_1 \M_1 \M_5 \big) -840\hspace{1mm}{\rm tr} \big( \M_1 \M_1 \M_2 \M_4 \big) \notag\\
    &- 840\hspace{1mm}{\rm tr} \big( \M_1 \M_2 \M_1 \M_4 \big)- 840\hspace{1mm}{\rm tr} \big( \M_1 \M_1 \M_4 \M_2 \big) \notag\\
    &- 1120\hspace{1mm}{\rm tr} \big( \M_1 \M_1 \M_3 \M_3 \big)- 560\hspace{1mm}{\rm tr} \big( \M_1 \M_3 \M_1 \M_3  \big) \notag\\
    &- 1680\hspace{1mm}{\rm tr} \big( \M_1 \M_2 \M_2 \M_3 \big) - 1680\hspace{1mm}{\rm tr} \big( \M_1 \M_2 \M_3 \M_2 \big) \notag\\
    &- 1680\hspace{1mm}{\rm tr} \big( \M_1 \M_3 \M_2 \M_2 \big) - 630\hspace{1mm}{\rm tr} \big( \M_2 \M_2 \M_2 \M_2 \big) 
    \notag\\&+ 1680\hspace{1mm}{\rm tr} \big( \M_1 \M_1 \M_1 \M_1 \M_4 \big) + 3360\hspace{1mm}{\rm tr} \big( \M_1 \M_1 \M_1 \M_2 \M_3 \big) \notag\\&+ 3360\hspace{1mm}{\rm tr} \big(\M_1 \M_1 \M_2 \M_1 \M_3 \big) + 3360\hspace{1mm}{\rm tr} \big( \M_1 \M_1 \M_3 \M_1 \M_3 \big) \notag\\&+ 3360\hspace{1mm}{\rm tr} \big( \M_1 \M_1 \M_1 \M_3 \M_2 \big) + 5040\hspace{1mm}{\rm tr} \big( \M_1 \M_1 \M_2 \M_2 \M_2 \big) \notag\\&+ 5040\hspace{1mm}{\rm tr} \big( \M_1 \M_2 \M_1 \M_2 \M_2 \big) 
    \notag\\&- 6720\hspace{1mm}{\rm tr} \big( \M_1 \M_1 \M_1 \M_1 \M_1 \M_3 \big) \notag\\&-10080\hspace{1mm}{\rm tr} \big( \M_1 \M_1 \M_1 \M_1 \M_2 \M_2 \big) \notag\\&- 10080\hspace{1mm}{\rm tr} \big( \M_1 \M_1 \M_1 \M_2 \M_1 \M_2 \big) \notag\\&- 5040\hspace{1mm}{\rm tr} \big( \M_1 \M_1 \M_2 \M_1 \M_1 \M_2 \big) 
    \notag\\&+ 20160\hspace{1mm}{\rm tr} \big( \M_1 \M_1 \M_1 \M_1 \M_1 \M_1 \M_2 \big) 
    \notag\\&- 5040\hspace{1mm}{\rm tr} \big( \M_1 \M_1 \M_1 \M_1 \M_1 \M_1 \M_1 \M_1 \big)
    \label{eq:dermu8}
\end{align}

where in the above Eqn.~\eqref{eq:dermu8}, we define
\begin{equation*}
  \M_k \equiv \M^{-1} \frac{\partial ^k \M}{\partial \mu^k}  \quad , \quad  k \in \mathbb{Z}^{+} \quad,\quad 1 \leq k \leq 8 
\end{equation*}
In linear $\mu$ formalism, the trace containing only the linear $\mu$ derivatives are considered and hence, Eqns. from \eqref{eq:dermu5} to \eqref{eq:dermu8} can be written as

\begin{align*}
    &\hspace{-.2cm}\frac{\partial^5 \ln \det \M}{\partial \mu^5} = 24 \hspace{1mm}{\rm tr} \left( \M^{-1} \frac{\partial \M}{\partial \mu}
        \M^{-1} \frac{\partial \M}{\partial \mu}
        \M^{-1} \frac{\partial \M}{\partial \mu}
        \M^{-1} \frac{\partial \M}{\partial \mu}
        \M^{-1} \frac{\partial \M}{\partial \mu} \right) \notag \\
    &\hspace{-.2cm}\frac{\partial^6 \ln \det \M}{\partial \mu^6} = -120 \hspace{1mm} {\rm tr} \left( \M^{-1} \frac{\partial \M}{\partial \mu}
        \M^{-1} \frac{\partial \M}{\partial \mu}
        \M^{-1} \frac{\partial \M}{\partial \mu}
        \M^{-1} \frac{\partial \M}{\partial \mu}
        \M^{-1} \frac{\partial \M}{\partial \mu} 
        \M^{-1} \frac{\partial \M}{\partial \mu} \right) \notag \\
    &\hspace{-.2cm}\frac{\partial^7 \ln \det \M}{\partial \mu^7} = 720 \hspace{1mm}{\rm tr} \left( \M^{-1} \frac{\partial \M}{\partial \mu}
        \M^{-1} \frac{\partial \M}{\partial \mu}
        \M^{-1} \frac{\partial \M}{\partial \mu}
        \M^{-1} \frac{\partial \M}{\partial \mu}
        \M^{-1} \frac{\partial \M}{\partial \mu} 
        \M^{-1} \frac{\partial \M}{\partial \mu} 
        \M^{-1} \frac{\partial \M}{\partial \mu} \right) \notag \\
    &\hspace{-.2cm}\frac{\partial^8 \ln \det \M}{\partial \mu^8} = -5040 \hspace{1mm} {\rm tr} \, \bigg( \M^{-1} \frac{\partial \M}{\partial \mu}
        \M^{-1} \frac{\partial \M}{\partial \mu}
        \M^{-1} \frac{\partial \M}{\partial \mu}
        \M^{-1} \frac{\partial \M}{\partial \mu}
        \M^{-1} \frac{\partial \M}{\partial \mu} 
        \M^{-1} \frac{\partial \M}{\partial \mu} \notag \\
        &\hspace{11cm}\M^{-1} \frac{\partial \M}{\partial \mu} 
        \M^{-1} \frac{\partial \M}{\partial \mu} \bigg) \notag \\
        &\hspace{8cm}\vdots \\
        &\hspace{8cm}\vdots \\
    &\hspace{3cm}\frac{\partial^p \ln \det \M}{\partial \mu^p} = \left(-1\right)^{p+1} \left(p-1\right)! \hspace{1mm} {\rm tr} \Bigg(\prod_{i=1}^p \left[\M^{-1} \frac{\partial \M}{\partial \mu}\right] \Bigg)  \hspace{2mm} \text{for} \hspace{2mm} p \geq 5  
\end{align*}

\noindent
Having defined the explicit representation of ${D}_n$
we now can proceed to define the expansion coefficients for various
thermodynamic quantities discussed in this paper.

\paragraph{Pressure \boldmath $(p)$ \unboldmath :}
The pressure is obtained from the logarithm of the QCD partition function. The leading
expansion coefficient $c_0$ is given by the pressure calculated at $\mu_q=0$.
All higher order expansion coefficients are given in terms of 
derivatives of $\ln {\cal Z}$.
\begin{eqnarray}
\frac{p}{T^4} \equiv \Omega = 
\frac{1}{VT^3} \ln {\cal Z} = \sum_{n=0}^{\infty} c_n 
\left( \frac{\mu_q}{T} \right)^n \quad ,
\end{eqnarray}
with 
\begin{equation}
c_n = \frac{1}{n! VT^3} \frac{\partial^n \ln {\cal Z}}{\partial (\mu_q/T)^n} 
\biggr|_{\mu=0}
\end{equation}
To generate the expansion we first consider derivatives of $\ln {\cal Z}$
for $\mu \ne 0$. For the first derivative we find
\begin{equation}
\frac{\partial \ln {\cal Z}}{\partial \mu} \equiv {\cal A}_1  = 
\left\langle {D}_1 \right\rangle \quad .
\end{equation}
Higher order derivatives are generated using the relation
\begin{eqnarray}
\frac{\partial {\cal A}_n}{\partial \mu} 
= {\cal A}_{n+1} - {\cal A}_n {\cal A}_1 \quad ,
\label{eq:An1}
\end{eqnarray}
where ${\cal A}_n$ is defined as
\begin{equation}
{\cal A}_n \equiv \left\langle 
\exp\{ -{D}_0 \} \frac{\partial^n \exp\{ {D}_0 \} }{\partial \mu^n}
 \right\rangle \quad .
\label{eq:an}
\end{equation}
With this we can generate higher order derivatives of $\ln {\cal Z}$ 
iteratively using 
\begin{equation}
\partial^{n+1} \ln {\cal Z}/\partial \mu^{n+1} =
\partial^{n} {\cal A}_1 /\partial \mu^{n} \quad .
\label{eq:Zn1}
\end{equation}
Explicitly we find from Eq.~(\ref{eq:an})
\begin{eqnarray}
{\cal A}_2 \hspace{-2mm} &=& \hspace{-2mm} 
\left\langle {D}_2 \right\rangle 
+\left\langle {D}_1^2 \right\rangle, 
\\
{\cal A}_3 \hspace{-2mm} &=& \hspace{-2mm} 
\left\langle {D}_3 \right\rangle 
+3\left\langle {D}_2 {D}_1 \right\rangle 
+\left\langle {D}_1^3 \right\rangle, 
\\
{\cal A}_4 \hspace{-2mm} &=& \hspace{-2mm} 
\left\langle {D}_4 \right\rangle 
+4\left\langle {D}_3 {D}_1 \right\rangle 
+3\left\langle {D}_2^2 \right\rangle 
+6\left\langle {D}_2 {D}_1^2 \right\rangle 
+\left\langle {D}_1^4 \right\rangle, 
\\
{\cal A}_5 \hspace{-2mm} &=& \hspace{-2mm} 
\left\langle {D}_5 \right\rangle 
+5\left\langle {D}_4 {D}_1 \right\rangle 
+10\left\langle {D}_3 {D}_2 \right\rangle 
+10\left\langle {D}_3 {D}_1^2 \right\rangle 
+15\left\langle {D}_2^2 {D}_1 \right\rangle 
\nonumber \\ &&\hspace{-3mm}
+10\left\langle {D}_2 {D}_1^3 \right\rangle 
+\left\langle {D}_1^5 \right\rangle, 
\\
{\cal A}_6 \hspace{-2mm} &=& \hspace{-2mm} 
\left\langle {D}_6 \right\rangle 
+6\left\langle {D}_5 {D}_1 \right\rangle 
+15\left\langle {D}_4 {D}_2 \right\rangle 
+10\left\langle {D}_3^2 \right\rangle 
+15\left\langle {D}_4 {D}_1^2 \right\rangle 
\nonumber \\ && \hspace{-3mm}
+60\left\langle {D}_3 {D}_2 {D}_1 \right\rangle 
+15\left\langle {D}_2^3 \right\rangle 
+20\left\langle {D}_3 {D}_1^3 \right\rangle 
+45\left\langle {D}_2^2 {D}_1^2 \right\rangle
\nonumber \\ && \hspace{-3mm}
+15\left\langle {D}_2 {D}_1^4 \right\rangle 
+\left\langle {D}_1^6 \right\rangle  \quad .
\\
{\cal A}_7 \hspace{-2mm} &=& \hspace{-2mm} 
\left\langle {D}_7 \right\rangle 
+7\left\langle {D}_6 {D}_1 \right\rangle 
+21\left\langle {D}_5 {D}_2 \right\rangle 
+35\left\langle {D}_4 {D}_3 \right\rangle
+21\left\langle {D}_5 {D}_1^2 \right\rangle
\nonumber \\ && \hspace{-3mm}
+105\left\langle {D}_4 {D}_2 {D}_1 \right\rangle
+70\left\langle {D}_3^2 {D}_1 \right\rangle
+105\left\langle {D}_3 {D}_2^2 \right\rangle
+35\left\langle {D}_4 {D}_1^3 \right\rangle
\nonumber \\ && \hspace{-3mm}
+210\left\langle {D}_3 {D}_2 {D}_1^2 \right\rangle
+105\left\langle {D}_2^3 {D}_1 \right\rangle
+35\left\langle {D}_3 {D}_1^4 \right\rangle
+105\left\langle {D}_2^2 {D}_1^3 \right\rangle
\nonumber \\ && \hspace{-3mm}
+21\left\langle {D}_2 {D}_1^5 \right\rangle
+\left\langle {D}_1^7 \right\rangle \quad .
\\
{\cal A}_8 \hspace{-2mm} &=& \hspace{-2mm} 
\left\langle {D}_8 \right\rangle 
+8 \left\langle {D}_7 {D}_1 \right\rangle
+28 \left\langle {D}_6 {D}_2 \right\rangle
+56 \left\langle {D}_5 {D}_3 \right\rangle
+35 \left\langle {D}_4^2 \right\rangle
\nonumber \\ && \hspace{-3mm}
+28 \left\langle {D}_6 {D}_1^2 \right\rangle
+168 \left\langle {D}_5 {D}_2 {D}_1  \right\rangle
+280 \left\langle {D}_4 {D}_3 {D}_1 \right\rangle
+210 \left\langle {D}_4 {D}_2^2 \right\rangle
\nonumber \\ && \hspace{-3mm}
+280 \left\langle {D}_3^2 {D}_2 \right\rangle
+56 \left\langle {D}_5 {D}_1^3 \right\rangle
+420 \left\langle {D}_4 {D}_2 {D}_1^2 \right\rangle
+280 \left\langle {D}_3^2 {D}_1^2 \right\rangle
\nonumber \\ && \hspace{-3mm}
+840 \left\langle {D}_3 {D}_2^2 {D}_1 \right\rangle
+105 \left\langle {D}_2^4 \right\rangle
+70 \left\langle {D}_4 {D}_1^4 \right\rangle
+560 \left\langle {D}_3 {D}_2 {D}_1^3 \right\rangle
\nonumber \\ && \hspace{-3mm}
+\hspace{1mm}420 \left\langle {D}_2^3 {D}_1^2 \right\rangle
+56 \left\langle {D}_3 {D}_1^5 \right\rangle
+210 \left\langle {D}_2^2 {D}_1^4 \right\rangle
+28 \left\langle {D}_2 {D}_1^6 \right\rangle
\nonumber \\ && \hspace{-3mm}
+ \left\langle {D}_1^8 \right\rangle
\quad .
\end{eqnarray}
From Eq.~(\ref{eq:Zn1}) we then obtain through repeated 
application of Eq.~(\ref{eq:An1}),
\begin{eqnarray}
\frac{\partial \ln {\cal Z}}{\partial \mu} 
\hspace{-2mm} &=& \hspace{-2mm} 
{\cal A}_1 ,
\\
\frac{\partial^2 \ln {\cal Z}}{\partial \mu^2} 
\hspace{-2mm} &=& \hspace{-2mm} 
{\cal A}_2 - {\cal A}_1^2 ,
\\
\frac{\partial^3 \ln {\cal Z}}{\partial \mu^3} 
\hspace{-2mm} &=& \hspace{-2mm} 
{\cal A}_3 -3 {\cal A}_2 {\cal A}_1 +2 {\cal A}_1^3 ,
\\
\frac{\partial^4 \ln {\cal Z}}{\partial \mu^4} 
\hspace{-2mm} &=& \hspace{-2mm} 
{\cal A}_4 -4 {\cal A}_3 {\cal A}_1 -3 {\cal A}_2^2 
+12 {\cal A}_2 {\cal A}_1^2 -6 {\cal A}_1^4 ,
\\
\frac{\partial^5 \ln {\cal Z}}{\partial \mu^5} 
\hspace{-2mm} &=& \hspace{-2mm} 
{\cal A}_5 -5 {\cal A}_4 {\cal A}_1 -10 {\cal A}_3 {\cal A}_2 
+20 {\cal A}_3 {\cal A}_1^2 +30 {\cal A}_2^2 {\cal A}_1 \nonumber\\
&&\hspace{-3mm}-60 {\cal A}_2 {\cal A}_1^3 +24 {\cal A}_1^5 ,
\\
\frac{\partial^6 \ln {\cal Z}}{\partial \mu^6} 
\hspace{-2mm} &=& \hspace{-2mm} 
{\cal A}_6 -6 {\cal A}_5 {\cal A}_1 -15 {\cal A}_4 {\cal A}_2 
-10 {\cal A}_3^2 +30 {\cal A}_4 {\cal A}_1^2 
\nonumber \\ &&\hspace{-3mm} 
+120 {\cal A}_3 {\cal A}_2 {\cal A}_1 
+30 {\cal A}_2^3 -120 {\cal A}_3 {\cal A}_1^3 
-270 {\cal A}_2^2 {\cal A}_1^2 \nonumber\\
&&\hspace{-3mm}+360 {\cal A}_2 {\cal A}_1^4 -120 {\cal A}_1^6,
\\
\frac{\partial^7 \ln {\cal Z}}{\partial \mu^7} 
\hspace{-2mm} &=& \hspace{-2mm} 
{\cal A}_7 -7 {\cal A}_6 {\cal A}_1 -21 {\cal A}_5 {\cal A}_2 
-35 {\cal A}_4 {\cal A}_3 + 42 {\cal A}_5
{\cal A}_1^2 \nonumber\\&&\hspace{-3mm}+ 210 {\cal A}_4{\cal A}_2{\cal A}_1
+ 140 {\cal A}_3^2 {\cal A}_1 + 210 {\cal A}_3 {\cal A}_2^2
-210 {\cal A}_4 {\cal A}_1^3 \nonumber\\
&&\hspace{-3mm}- 1260 {\cal A}_3 {\cal A}_2 {\cal A}_1^2 - 630 {\cal A}_2^3 {\cal A}_1
+ 840 {\cal A}_3 {\cal A}_1^4 + 2520 {\cal A}_2^2 {\cal A}_1^3 \nonumber\\&&\hspace{-3mm}- 2520 {\cal A}_2 {\cal A}_1^5 + 720 {\cal A}_1^7,
\\
\frac{\partial^8 \ln {\cal Z}}{\partial \mu^8} 
\hspace{-2mm} &=& \hspace{-2mm} 
{\cal A}_8 -8 {\cal A}_7 {\cal A}_1 -28 {\cal A}_6 {\cal A}_2 -56 {\cal A}_5 {\cal A}_3 - 35 {\cal A}_4^2 + 56 {\cal A}_6 {\cal A}_1^2 \nonumber\\&&\hspace{-3mm}+ 336 {\cal A}_5 {\cal A}_2 {\cal A}_1 
+ 560 {\cal A}_4 {\cal A}_3 {\cal A}_1 + 420 {\cal A}_4 {\cal A}_2^2 + 560 {\cal A}_2 {\cal A}_2 - 336 {\cal A}_5 {\cal A}_1^3 \nonumber\\&&\hspace{-3mm}- 2520 {\cal A}_4 {\cal A}_2 {\cal A}_1^2 
- 1680 {\cal A}_3^2 {\cal A}_1^2 - 5040 {\cal A}_3 {\cal A}_2^2 {\cal A}_1 - 630 {\cal A}_2^4 + 1680 {\cal A}_4 {\cal A}_1^4 \nonumber\\&&\hspace{-3mm} + 13440 {\cal A}_3 {\cal A}_2 {\cal A}_1^3 
+ 10080 {\cal A}_2^3 {\cal A}_1^2 - 6720 {\cal A}_3 {\cal A}_1^5 - 25200 {\cal A}_2^2 {\cal A}_1^4 \nonumber\\&&\hspace{-3mm}+ 20160 {\cal A}_2 {\cal A}_1^6 - 5040 {\cal A}_1^8
\nonumber \\
\end{eqnarray}
These relations simplify considerably for $\mu=0$ as all odd
expectation values vanish, {\it i.e.} ${\cal A}_n =0$ for $n$ odd. In fact, 
$\partial^n (\ln \det \M)/\partial \mu^n$ 
is strictly real for $n$ even and pure imaginary for $n$ odd. 
Using this property, the odd derivatives of the pressure vanish 
and also the even derivatives become rather simple. This defines
the expansion coefficients $c_n$ introduced in \autoref{eq:Taylor series of excess pressure} and 
\autoref{eq:Taylor series of number density},
\begin{eqnarray}
c_2 &\equiv& \frac{1}{2}\frac{\partial^2 (p/T^4)}{\partial (\mu_{q}/T)^2}
\biggr|_{\mu_q=0} = 
\frac{1}{2} \frac{N_{\tau}}{N_{\sigma}^3}\; {\cal A}_2 \quad , \nonumber \\
c_4 &\equiv& \frac{1}{4!} \frac{\partial^4 (p/T^4)}{\partial (\mu_{q}/T)^4}
\biggr|_{\mu_q=0} = 
\frac{1}{4!} \frac{1}{N_{\sigma}^3 N_{\tau}} \left({\cal A}_4 -3 {\cal A}_2^2\right) \quad , 
\nonumber \\  
c_6 &\equiv& \frac{1}{6!} \frac{\partial^6 (p/T^4)}{\partial (\mu_{q}/T)^6} 
\biggr|_{\mu_q=0} = 
\frac{1}{6!} \frac{1}{N_{\sigma}^3 N_{\tau}^3} 
\left({\cal A}_6 -15 {\cal A}_4 {\cal A}_2 +30 {\cal A}_2^3\right) \quad ,
\nonumber \\
c_8 &\equiv& \frac{1}{8!} \frac{\partial^8 (p/T^4)}{\partial (\mu_{q}/T)^8} 
\biggr|_{\mu_q=0} = 
\frac{1}{8!} \frac{1}{N_{\sigma}^3 N_{\tau}^5} 
\big({\cal A}_8 -28 {\cal A}_6 {\cal A}_2 - 35{\cal A}_4^2 \nonumber\\&&\hspace{5cm}+ 420 {\cal A}_4{\cal A}_2^2 - 630{\cal A}_2^4\big) \quad ,
\label{eq:dpmu0}
\end{eqnarray}
Here all expectation values ${\cal A}_n$ are now meant to be evaluated at
$\mu =0$. In general, the $n^{th}$ order coefficient is given as

\begin{equation}
    \frac{\partial^n(\Delta P/T^4)}{\partial(\mu_q/T)^n} = \frac{1}{VT^3} \frac{\partial^n \ln \Z}{\partial(\mu_q/T)^n}
    = \frac{N_{\tau}^{3-n}}{N_{\sigma}^3} \frac{\partial \ln \Z}{\partial \mu^n}
    \label{eq:gen}
\end{equation}

\vspace{4mm}

where $\mu$ in above Eqn.~\eqref{eq:gen} is defined as $\mu = a\mu_q$ , with lattice spacing $a$.

	\cleardoublepage

\chapter{Generic formulae for Phasefactor and Phase-quenched reweighting factor}
\label{Appendix 4}

The reweighting factor $\mathcal{R} (\mu,T,U)$ upto order $\mathcal{O}(\mu^{N})$ is given by
  \begin{equation}
      \mathcal{R}(\mu,T,U)  =  \exp {\left[\sum_{k=1}^{N} \frac{\mu^{k}}{k!} D_{k} (T,U)\right]}
      \label{eq:D.1}
  \end{equation}
 
  where 
  \begin{equation*}
    D_{k} (T,U) = \frac{\partial^{k}}{\partial \mu^{k}} \ln {\big[ \det \M (\mu, T, U)\big]}\Bigg\vert_{\mu=0} 
  \end{equation*} 
  
  \vspace{7mm}
  where $\M$ is the familiar fermion matrix as defined throughout the thesis.

  By $CP$ symmetry of QCD, we know $D_{k}$ is purely real for all even $k$ , and purely imaginary for all odd $k$. 
  So, 
  
  \begin{align}
      D_{2k} &= \text{Re}\left(D_{2k}\right) \notag \\
      D_{2k-1} &= i \hspace{1mm} \text{Im}\left(D_{2k-1}\right)
      \label{eq:D_n}
  \end{align}

Rearranging Eqn.~\eqref{eq:D.1}, we find 

\begin{equation}
    \mathcal{R}(\mu,T,U)  =  \exp {\left[\sum_{k=1}^{N/2} \frac{\mu^{2k}}{(2k)!} D_{2k} (T,U)\right]} \exp{\left[\sum_{k=1}^{N/2} \frac{\mu^{2k-1}}{(2k-1)!} D_{2k-1} (T,U)\right]}
\end{equation}

  which as per Eqn.~\eqref{eq:D_n} becomes, 

\begin{equation}
    \mathcal{R}(\mu,T,U)  =  \exp {\left[\sum_{k=1}^{N/2} \frac{\mu^{2k}}{(2k)!} \text{Re}\left(D_{2k}\right)\right]} \exp{\left[\sum_{k=1}^{N/2} \frac{\mu^{2k-1}}{(2k-1)!} \text{Im}\left(D_{2k-1}\right)\right]}
    \label{eq:final D_n}
\end{equation}

  Now, the reweighting factor as given in Eqn.~\eqref{eq:D.1} is a complex number. From our basic knowledge of complex numbers, we know that the reweighting factor in eqn.~\eqref{eq:D.1} can be written in the form of 
  
  \begin{equation}
   \mathcal{R}(\mu,T,U) = R(\mu, T, U) \exp{\left[i\theta(\mu , T, U)\right]}   
   \label{eq:polar amp form}
  \end{equation}
  
  where $R(\mu, T, U)$ and  $\theta(\mu , T, U) \in \mathbb{R}$ are the amplitude and the phase of the reweighting factor $\mathcal{R}$ respectively. 
  In generic sense, $\mu$ is complex implying, $\mu^{k}$ is also complex and hence can be written as 
  
  \begin{equation}
   \mu^{k} = \text{Re}\left[\mu^{k}\right] + i \hspace{1mm}\text{Im}\left[\mu^{k}\right]   
  \end{equation}
  
  Using this in eqn.~\eqref{eq:final D_n}, we find, 
  \begin{equation*}
      \mathcal{R} = R e^{i\theta} = \exp\left[\sum_{k=1}^{N/2} \frac{\mu^{2k}}{(2k)!} D_{2k}\right]
         \exp{\left[\sum_{k=1}^{N/2} \frac{\mu^{2k-1}}{(2k-1)!} D_{2k-1}\right]}
  \end{equation*}
  
  \begin{equation*}
    =  \exp{\left[\sum_{k=1}^{N/2} \frac{\text{Re}\left(\mu^{2k}\right)+i \hspace{1mm}\text{Im}\left(\mu^{2k}\right)}{(2k)!} D_{2k}\right]}
         \exp{\left[i \left(\sum_{k=1}^{N/2} \frac{\text{Re}\left(\mu^{2k-1}\right)+i \hspace{1mm}\text{Im}\left(\mu^{2k-1}\right)}{(2k-1)!} \text{Im}\left(D_{2k-1}\right)\right)\right]}
  \end{equation*}
  
  \vspace{5mm}
  
  Equating the amplitude and phase parts in LHS and RHS and also from Eqn.~\eqref{eq:polar amp form}, we find, 
 
  \begin{align}
      R(\mu, T, U) &= \exp{\left[\mathlarger{\sum}_{k=1}^{N/2} \left( 
       \frac{\text{Re}\left(\mu^{2k}\right)}{(2k)!} D_{2k}(T,U) - 
       \frac{\text{Im}\left(\mu^{2k-1}\right)}{(2k-1)!} \text{Im}\Big[D_{2k-1}(T,U)\Big] \right) \right]} 
       \label{eq:phase-quenched} \\
    \notag \\
      \theta(\mu, T, U) &= \mathlarger{\mathlarger{\sum}}_{k=1}^{N/2} \left[ 
       \frac{\text{Im}\left(\mu^{2k}\right)}{(2k)!} D_{2k}(T,U) + 
       \frac{\text{Re}\left(\mu^{2k-1}\right)}{(2k-1)!} \text{Im}\Big[D_{2k-1}(T,U)\Big] \right] 
       \label{eq:phase}
  \end{align}
 
 \vspace{8mm}
  From Eqns.~\eqref{eq:phase-quenched} and \eqref{eq:phase}, we note the following observations : 

  \begin{itemize}
  \item It is trivial to show that $\mathcal{R} = 1$ and $\theta = 0$ at $\mu=0$.
  \item For purely imaginary $\mu$, we find $\text{Im}\left(\mu^{2n}\right) = 0$ and $\text{Re}\left(\mu^{2n-1}\right) = 0$, for which $\theta = 0$. 
  
  \item Due to $\theta=0$, one has no fermion sign problem for zero or purely imaginary $\mu$. 
  \item In fact, $\theta = 0$ for all $k$, which means $\cos {\theta} = 1$, term by term, order-by-order in $\mu$. 
  \item In case of $\muI$, $D_k$ vanishes identically for all odd $k$. Which makes $\theta=0$ and hence, we have no fermion sign problem for $\muI$. But this is not the same for $\muB$, where $D_k \neq 0$ for odd $k$.
  \item The severity of sign problem is measured by $\left\langle cos( \theta (\mu, T, U )) \right\rangle$. For $\mu$, where $\left\langle cos( \theta (\mu, T, U )) \right\rangle$ $\approx$ 1, the sign problem is very less severe, whereas for $\mu$, where $\left\langle cos( \theta (\mu, T, U )) \right\rangle$ $\approx$ 0 and $\ll$ 1, the FSP is highly severe. In such situations, one cannot rely on $\mu=0$ ensemble to generate observables for such $\mu$ values, that give $\left\langle cos( \theta (\mu, T, U )) \right\rangle$ $\ll$ 1.
 \end{itemize}
	\cleardoublepage
	\chapter{Fourier transformation on lattice}
\label{Appendix 6}

The goal of this appendix is to discuss the Fourier transform $\Tilde{f}(p)$ of functions
$f(n)$ defined on the lattice $\Lambda$. The lattice is given by

\be
\Lambda = \left \{ n = \left(n_1 , n_2 , n_3 , n_4 \right) \hspace{1mm}|\hspace{1mm} n_{\mu} = 0, 1, \cdots, N_{\mu}-1 \right \}
\label{eq:Lattice}
\ee
and in all of our applications regarding lattice in this thesis, we have $N_1 = N_2 = N_3 = N_{\sigma}$ , $N_4 = N_{\tau}$. 
The total number of spacetime points on lattice $\Lambda$ is therefore given by
\be
|\Lambda| = N_1 N_2 N_3 N_4 
\label{eq:Total points}
\ee

We impose toroidal boundary conditions, which is a generalization of periodic and anti-periodic boundary conditions given as follows : 

\be
f(n + \hat{\mu}\hspace{.5mm}a\hspace{.5mm}N_{\mu} ) = \exp{\left(2\pi i \theta_{\mu}\right)}f(n)
\label{eq:Bound}
\ee

for each of the directions $\hat{\mu}$. Here $\hat{\mu}$ denotes the unit vector in $\mu$-direction and $a$ is the lattice spacing. Periodic boundary conditions have $\theta_{\mu} = 0$, whereas anti-periodic boundary conditions correspond to $\theta_{\mu} = 1/2$.

The momentum space $\Tilde{\Lambda}$, which corresponds to the lattice $\Lambda$ with the
boundary conditions given in \autoref{eq:Bound}, is deﬁned as

\be
\Tilde{\Lambda} = \left \{ p = \left(p_1 , p_2 , p_3 , p_4 \right) \hspace{1mm}|\hspace{1mm} p_{\mu} = \frac{2 \pi}{aN_{\mu}} \left( k_{\mu} + \theta_{\mu} \right), k_{\mu} = -\frac{N_{\mu}}{2},\cdots, \frac{N_{\mu}}{2}+1 \right \}
\label{eq:Momentum lattice}
\ee

Like before, $a$ is the lattice spacing as used in \autoref{eq:Momentum lattice}. In \autoref{eq:Lattice} and \autoref{eq:Total points}, all are measured in units of $a$. The boundary phases $\theta_{\mu}$ have to be included in the deﬁnition of the momenta
$p_{\mu}$ such that the plane waves $\exp{\left( i p \cdot na\right)}$, satisfying the following condition

\begin{equation*}
    p \cdot n = \sum_{\beta=1}^4 p_{\beta} n_{\beta}
\end{equation*}

follows the boundary conditions as given in \autoref{eq:Bound}.

The basic formula, governing Fourier transformation on the lattice, is

\be
\frac{1}{N} \mathlarger{\sum}_{k=-N/2}^{N/2+1} \exp{\left( i \frac{2 \pi l k}{N} \right)} = 
\frac{1}{N} \mathlarger{\sum}_{k=0}^{N-1} \exp{\left( i \frac{2 \pi l k}{N} \right)} = \delta_{l0}
\label{eq:Fourier basic}
\ee
Here $l$ is an integer with $0 \leq l \leq N-1$.

For $l = 0$, \autoref{eq:Fourier basic} is trivial. For $l \neq 0$, \autoref{eq:Fourier basic} follows from applying the well-known algebraic identity 

\be
\sum_{k=0}^{N-1} x^k = \frac{1-x^N}{1-x} \quad \text{with} \quad x = \exp{\left(i \frac{2 \pi l}{N} \right)}
\label{eq:GP}
\ee

We can combine four of the 1D sums in \autoref{eq:Fourier basic} to obtain the following identities:

\begin{align}
    \frac{1}{|\Tilde{\Lambda}|} \hspace{1mm} \mathlarger{\sum}_{p \in \Tilde{\Lambda}} \exp{\left[ i p \cdot \left(n-n^{'}\right)a\right]} &= 
    \delta \left(n-n^{'}\right) = \delta_{n_1 n_1^{'}} \delta_{n_2 n_2^{'}} \delta_{n_3 n_3^{'}} \delta_{n_4 n_4^{'}} \label{eq:mom to pos}\\
    \frac{1}{|\Lambda|} \hspace{1mm} \mathlarger{\sum}_{n \in \Lambda} \exp{\left[ i \left(p-p^{'}\right)\cdot na\right]} &= 
    \delta \left(p-p^{'}\right) = \delta_{p_1 p_1^{'}} \delta_{p_2 p_2^{'}} \delta_{p_3 p_3^{'}} \delta_{p_4 p_4^{'}}
    \label{eq:pos to mom}
\end{align}

We stress that the right-hand sides of \autoref{eq:mom to pos} and \autoref{eq:pos to mom} comprise a product of four Kronecker
deltas for the integers $n_{\mu}$ and $p_{\mu}$,  which label the position and momentum components $p_{\mu}$ respectively. 

If we now deﬁne the Fourier transform

\be
\Tilde{f}(p) = \frac{1}{\sqrt{|\Tilde{\Lambda}|}} \sum_{n \in \Lambda} f(n) \exp{\left( -ip \cdot na\right)}
\label{eq:FT}
\ee

 the inverse Fourier transformation gets automatically defined as follows

\be
f(n) = \frac{1}{\sqrt{|\Lambda|}} \sum_{p \in \Tilde{\Lambda}} \Tilde{f}(p) \exp{\left( ip \cdot na\right)}
\label{eq:inverse FT}
\ee

The last equation follows immediately from inserting \autoref{eq:FT} in \autoref{eq:inverse FT} and
using \autoref{eq:mom to pos}.
	\cleardoublepage
	\chapter{Basis transformation}
\label{Appendix 7}

In this Appendix, we present the basis transformation formulae from $\left(u,d,s\right)$ basis to $\left(B,S,I\right)$ basis, where the symbols have their usual conventional meanings. This basis transformation is important to study the more practical and relevant experimental regime of heavy ion collisions, where we usually find matter in some bound states of quarks in form of baryons, strange particles or isospin asymmetric matter, rather than free, isolated quarks themselves.

We use the usual quantum number conservation formulae as follows

\begin{align}
    B &= \frac{1}{3} \hspace{1mm}\bigg[N_u+N_d+N_s\bigg] \notag\\
    I &= \frac{1}{2} \hspace{1mm}\bigg[N_u-N_d\bigg] \notag\\
    S &= -N_s
    \label{eq:quant conservation}
\end{align}

where $N_u$, $N_d$, $N_s$ are the number of up, down and strange quarks respectively. Similarly, $B$, $S$, $I$ represent baryon, strangeness and isospin quantum numbers respectively. 

Using the fact that fugacity $= \sum_k \mu_k N_k$ is basis independent, where $k$ characterises basis, we have 

\begin{equation}
    \mu_B B + \mu_S S + \mu_I I = \mu_u N_u + \mu_d N_d + \mu_s N_s
    \label{eq:fuga conser}
\end{equation}

Applying relations of \autoref{eq:quant conservation} in the above \autoref{eq:fuga conser} and equating coefficients of $N_u$, $N_d$ and $N_s$ which are all independent terms in $\left(u,d,s\right)$ basis, we obtain

\begin{align}
    \mu_u &= \bigg[\frac{\mu_B}{3} + \frac{\mu_I}{2}\bigg] \notag \\
    \mu_d &= \bigg[\frac{\mu_B}{3} - \frac{\mu_I}{2}\bigg] \notag \\
    \mu_s &= \bigg[\frac{\mu_B}{3} - \mu_S\bigg]
    \label{eq:B to u}
\end{align}

Using the inverse relations of \autoref{eq:B to u}, we find the relevant chemical potentials like baryon $(\mu_B)$, strangeness $(\mu_S)$ and isospin $(\mu_I)$ chemical potentials in terms of up $(\mu_u)$, down $(\mu_d)$ and strange $(\mu_s)$ quark chemical potentials as follows: 

\begin{align}
    \mu_B &= \frac{3}{2} \bigg[\mu_u+\mu_d\bigg] \notag \\
    \mu_S &= \frac{1}{2} \bigg[\mu_u+\mu_d\bigg] - \mu_s \notag \\
    \mu_I &=   \bigg[\mu_u-\mu_d\bigg]
    \label{eq:u to B}
\end{align}

	\cleardoublepage

\chapter{Real fermion determinant at zero and purely imaginary chemical potentials}
\label{Appendix 9}

In prescence of a non-zero finite chemical potential $\mu$, the $\gamma^5$ hermiticity of the fermion matrix $\M$ assumes the following form :

\begin{equation}
    \M^{\dagger}(-\mu) = \gamma^5 \hs \hs \M(\mu) \hs \hs \gamma^5
    \label{eq:gamma5 mu}
\end{equation}

The above \autoref{eq:gamma5 mu} implies that 

\begin{equation}
    [\det \M(-\mu)]^{\dagger} = \det \M(\mu)
    \label{eq:det}
\end{equation}

since $(\gamma_5)^2 = I_4$.
\autoref{eq:det} implies that if $\det \M(\mu) = A(\mu) + iB(\mu)$, then 
\begin{equation}
\det \M(-\mu) = A(\mu) - iB(\mu)    
\label{eq:det form}
\end{equation}
where $A,B \in \mathbb{R}$. Also since, $\det \M(\mu) = A(\mu) + iB(\mu)$, we therefore have from our knowledge of functions, the following 

\begin{equation}
\det \M(-\mu) = A(-\mu) + iB(-\mu)    
\label{eq:func indu}
\end{equation}

Combining \autoref{eq:det form} and \autoref{eq:func indu}, we find 

\begin{equation}
    A(-\mu) = A(\mu) \quad,\quad B(-\mu) = -B(\mu)
    \label{eq:decision}
\end{equation}

From this \autoref{eq:decision}, it is clear that $A$ and $B$ are real-valued even and odd functions of complex $\mu$. For any generic $\mu$, the functions $A$ and $B$ can be written as 

\begin{equation}
    A(\mu) = a_0 + \sum_{k=1}^{\infty} a_{2k} \hs \mu^{2k} \quad,\quad B(\mu) = \sum_{k=0}^{\infty} b_{2k+1} \hs \mu^{2k+1}
\end{equation}

where the Taylor coefficients $a_{2k},b_{2k+1} \in \mathbb{R}$. 

For $\boldsymbol{\mu=0}$, $A = a_0, B = 0$, implying $\det \M$ is real.

For $\boldsymbol{\mu = i\mu_I}$, where $\mu_I \in \mathbb{R}$, we have $A \in \mathbb{R}$, but a purely imaginary $B$, since the Taylor coefficients $a_{2k},b_{2k+1} \in \mathbb{R}$. Hence, $\det \M = A + iB$ is again real. 

The determinant is therefore complex for complex chemical potentials and finite non-zero real chemical potentials.

	\cleardoublepage

 	\chapter{Properties of n point correlation functions}
\label{Appendix 11}

\vspace{1cm}

In this appendix, we present a naive and brief, somewhat less descriptive proof of the fact that all $n$-point correlation functions are purely real for even $n$, and purely imaginary for odd $n$.

All the following equations have symbols with conventional meanings. On reweighting with respect to $\mu=0$, we can express the partition function as follows : 



\begin{align}
\mathcal{Z}(\mu) &= 
\int \mathcal{D}U\ e^{-S_{G}}\ \exp{\left[\frac{n_{f}}{4}\ \nsum[1.2]_{k=1}^{\infty}\ 
\frac{\mu^{k}}{k!}\, D_{k} \right]} \notag \\ 
 &= \int \mathcal{D}U\ e^{-S_{G}}\ \prod_{k=0}^{\infty}\ \exp \ \Big[ f_{k}(\mu)\ D_{k} \Big] 
 \label{eq:equation 1}
\end{align}

where  

\begin{align}
D_{k} &= \left[\frac{\partial^{ k}}{\partial \mu^{k}} \ln{det M(\mu)}\right]\Bigg|_{\mu = 0} \notag \\ 
f_k(\mu) &= \frac{n_{f} \mu^k}{4k!}
\label{eq:equation 2}
 \end{align}

  Now, from \autoref{eq:equation 2}, we understand that $f_{k}(\mu)$ is a monotonic increasing function in $\mu$ and $f_{k}(0) = 0 $ for all $k>0$. Also, $D_{k}$ is complex and hence, can be written as $D_{k} = D_{k}^{R} + iD_{k}^{I}$.  Also, we know that 

  \begin{equation*}
      \nsum[1.2]_{k=0}^{\infty} D_k = \nsum[1.2]_{k=0}^{\infty} D_{2k} + \nsum[1.2]_{k=1}^{\infty} D_{2k-1}
  \end{equation*}

  \vspace{6mm}
  
  Hence, the above \autoref{eq:equation 1} can be rewritten as 

\begin{align}
    \mathcal{Z}(\mu) = \int \mathcal{D}U\ e^{-S_{G}}\  &\prod_{k=0}^{\infty}\ \exp \ \left[ f_{2k}(\mu)\ \Big( D_{2k}^{R}+iD_{2k}^{I}\Big)\right] \notag \\
   &\exp \ \left[f_{2k+1}(\mu) \Big( D_{2k+1}^{R}+iD_{2k+1}^{I}\Big)\right]
   \label{eq:equation 3}
\end{align}

 Now, from physical arguments mentioned before, $\mathcal{Z}(\mu)$ = $\mathcal{Z}(-\mu)$ 
 Also  from \autoref{eq:equation 2}, we understand $f_{k}(-\mu)$ = $f_{k}(\mu)$ for all even $k$ and $f_{k}(-\mu)$ = $f_{k}(-\mu)$ for all odd $k$. Replacing $\mu$ with $-\mu$ and using the property of $f$, from \autoref{eq:equation 3}, we have 
 
\begin{align}
    \mathcal{Z}(-\mu) = 
\int \mathcal{D}U\ e^{-S_{G}}\  &\prod_{k=0}^{\infty}\ \exp \ \left[ f_{2k}(\mu)\ \Big( D_{2k}^{R}+iD_{2k}^{I}\Big)\right] \notag \\
 &\exp \ \left[-f_{2k+1}(\mu) \Big( D_{2k+1}^{R}+iD_{2k+1}^{I}\Big)\right]
\label{eq:equation 4}
\end{align}

\vspace{5mm}

 Equating \autoref{eq:equation 3} and \autoref{eq:equation 4}, because $\mathcal{Z}(\mu)$ = $\mathcal{Z}(-\mu)$, we  find that the respective integrands are equal, since the equality holds for arbitrary gauge configurations. This is given as

\begin{equation*}
    \prod_{k=0}^{\infty}\ \exp \ \left[-f_{2k+1}(\mu) \Big( D_{2k+1}^{R}+iD_{2k+1}^{I}\Big)\right]
    =  \prod_{k=0}^{\infty}\ \exp \ \left[f_{2k+1}(\mu) \Big( D_{2k+1}^{R}+iD_{2k+1}^{I}\Big)\right]
\end{equation*}

which implies the following

\begin{align}
    &\prod_{k=0}^{\infty}\ \exp \ \left[f_{2k+1}(\mu) \big( D_{2k+1}^{R}\big)\right]\, \prod_{k=0}^{\infty}\ \exp \ \left[f_{2k+1}(\mu) \big( iD_{2k+1}^{I}\big)\right] \notag \\
    &= \prod_{k=0}^{\infty}\ \exp \ \left[-f_{2k+1}(\mu) \big( D_{2k+1}^{R}\big)\right]\, \prod_{k=0}^{\infty}\ \exp \ \left[-f_{2k+1}(\mu) \big( iD_{2k+1}^{I}\big)\right]
    \label{eq:equation 5}
\end{align}

 Now, we know that 
\begin{equation} 
 \prod_{k=0}^{\infty}\ \exp \ \left[if_{2k+1}(\mu)  D_{2k+1}^{I}\right] = \prod_{k=0}^{\infty} exp \ \left[-if_{2k+1}(\mu)  D_{2k+1}^{I}\right]
 \label{eq:equation 6}
\end{equation}

From therefore \autoref{eq:equation 5} and \autoref{eq:equation 6}, we find

\begin{equation}
    \sum_{k=0}^{\infty} f_{2k+1}(\mu) D_{2k+1}^{R} = 0
    \label{eq:equation 7}
\end{equation}

Now, each power of $\mu$ is independent, which means that $f_{k}$ are independent terms for all $k$. So, as per above \autoref{eq:equation 7}, we have from the property of linear combination and independence, $D_{2k+1}^{R} = 0$. Also, we know, that phase factor at finite $\mu$ is non-zero, which means $D_{2k+1}^{I} \neq 0$. \\
This proves that $\textbf{All odd point correlation functions are purely imaginary}$.

\vspace{1cm}

Using the above even odd property of $f$ and rearranging \autoref{eq:equation 3}, we find,

\begin{equation}
    \mathcal{Z}(\mu) = 
\int \mathcal{D}U\ e^{-S_{G}}\  \prod_{k=0}^{\infty}\ \exp \ \left[ f_{2k}(\mu)\ D_{2k}^{R}\right]\ \prod_{k=0}^{\infty}\ \exp \ \left[i\Big(f_{2k}(\mu)  D_{2k}^{I}+f_{2k+1}(\mu)D_{2k+1}^{I}\Big)\right]
\label{eq:equation 8}
\end{equation}

 From the even odd property of $f$, we find,
 
\begin{equation}
 \mathcal{Z}(-\mu) = 
\int \mathcal{D}U\ e^{-S_{G}}\  \prod_{k=0}^{\infty}\ \exp \ \left[ f_{2k}(\mu)\ D_{2k}^{R}\right]\ \prod_{k=0}^{\infty}\ \exp \ \left[i\Big(f_{2k}(\mu)  D_{2k}^{I}-f_{2k+1}(\mu)D_{2k+1}^{I}\Big)\right]
\label{eq:equation 9}
\end{equation}

\vspace{6mm}

Equating \autoref{eq:equation 8} and \autoref{eq:equation 9}, we get,

\begin{equation*}
\prod_{k=0}^{\infty}\ \exp \ \left[i\Big(f_{2k}(\mu)  D_{2k}^{I}+f_{2k+1}(\mu)D_{2k+1}^{I}\Big)\right] = \prod_{k=0}^{\infty}\ \exp \ \left[i\Big(f_{2k}(\mu)  D_{2k}^{I}-f_{2k+1}(\mu)D_{2k+1}^{I}\Big)\right]
\end{equation*}
\vspace{2mm}

which implies
\begin{equation*}
  \exp \ \Bigg[\sum_{k=0}^{\infty}\left[i\Big(f_{2k}(\mu)  D_{2k}^{I}+f_{2k+1}(\mu)D_{2k+1}^{I}\Big)\right]\Bigg] = 
   \exp \ \Bigg[\sum_{k=0}^{\infty}\left[i\Big(f_{2k}(\mu)  D_{2k}^{I}-f_{2k+1}(\mu)D_{2k+1}^{I}\Big)\right]\Bigg]
\end{equation*}
\vspace{2mm}

 which again implies 
 \vspace{1cm}
 
\begin{equation}
    \sum_{k=0}^{\infty} f_{2k}(\mu)D_{2k}^{I} = 0
    \label{eq:equation 10}
\end{equation}
\vspace{3mm}

 Again \autoref{eq:equation 10} suggests that each power of $\mu$ is independent, which means that $f_{k}$ are independent terms for all $k$. This again follows from the property of linear combination and indpendence in case of power series.

 This therefore proves that $\textbf{All even point correlation functions are purely real}$.

	\cleardoublepage
 
 	\chapter{Chemical potential basis : Explicit proof}
\label{Appendix 12}

\newcommand{\mX}{\mu_X}

\vspace{.5cm}

Here, in this appendix, we illustrate explicit proof of how \autoref{eq:mu basis} removes stochastic bias upto $\Ob(\mX^4)$ and therefore reproduces Taylor series expansion of excess pressure $\Delta P/T^4$ with exact Taylor coefficients upto $\Ob(\mX^4)$. $\mX$ represents the generic flavor of a chemical potential where $X \in (B,S,I)$, with $B,S,I$ having the usual meanings as mentioned in the material of the thesis. As per the terms given in \autoref{eq:mu coefficients}, upto $N=4$, we find the following expression as the argument $A_4(\mX)$ of the exponential, given by 

\vspace{.5cm}   

\begin{align}
    A_4(\mX) &= \left(\frac{\mX}{T}\right)
    \overline{D_1} + \left(\frac{\mX}{T}\right)^2 \,\frac{1}{2!} \bigg[\overline{D_2} + \left(\overline{D_1^2} - \left(\overline{D_1}\right)^2\right)\bigg] \notag \\
    &+ \left(\frac{\mX}{T}\right)^3\,\frac{1}{3!}\bigg[\overline{D_3} + 3\left(\overline{D_2D_1} - \overline{D_2}\;\overline{D_1}\right) + 
           \left(\overline{D_1^3} - 3\,\overline{D_1^2}\;\overline{D_1} + 
           2\,\left(\overline{D_1}\right)^3\right)\bigg] \notag \\
    &+ \left(\frac{\mX}{T}\right)^4\, \frac{1}{4!} \bigg[\overline{D_4} + 3\left(\overline{D_2^2} - \left(\overline{D_2}\right)^2\right)+ 4\left(\overline{D_3D_1} - \overline{D_3}\;\overline{D_1}\right) \notag \\
    &+ 6\left( \overline{D_2D_1^2} - \overline{D_2}\;\overline{D_1^2}\right) - 3\,(\overline{D_1^2})^2\bigg]
    \label{eq:exponential argument}
\end{align}

\vspace{.5cm}

All the symbols in the above \autoref{eq:mu basis} have the usual meanings and conventional interpretations as described in the appropriate sections of the above paper. In the subsequent discussion we denote $\mX/T$ as $\boldsymbol{\mu}_X$. Now, we know, 

\vspace{.8cm}

\begin{equation}
    e^A = 1+A+\frac{A^2}{2!}+\frac{A^3}{3!}+\frac{A^4}{4!}+\mathcal{O}(A^5)
    \label{eq:exponential formula}
\end{equation}

\vspace{.8cm}

On performing exponential of $A$ as given in the above \autoref{eq:exponential argument} and using the above exponential formula (\autoref{eq:exponential formula}), we find the following expression of $\exp{(A)}$ after some simple, yet time-consuming and careful cancellations of various correlation terms along with the associated factors

\vspace{.8cm}

\begin{equation}
    e^A = 1+ \sum_{k=1}^4 \boldsymbol{\mu}_X^k\, \Ob\left(\boldsymbol{\mu}_X^k\right) + \Ob\left(\boldsymbol{\mu}_X^5\right)
\end{equation}

\vspace{.8cm}

where $\Ob(\mX^k)$ for $1 \leq k \leq 4$ are given as follows : 

\vspace{.8cm}

\begin{align}
    \Ob\left(\boldsymbol{\mu}_X^1\right) &:\quad \overline{D_1}\notag \\
    \Ob\left(\boldsymbol{\mu}_X^2\right) &:\quad \frac{1}{2!} \left[\overline{D_2} + \overline{D_1^2}\right] \notag \\
    \Ob\left(\boldsymbol{\mu}_X^3\right) &:\quad \frac{1}{3!} \left[\overline{D_3} + 3\,\overline{D_2 D_1} + \overline{D_1^3}\right] \notag \\
    \Ob\left(\boldsymbol{\mu}_X^4\right) &:\quad \frac{1}{4!} \left[\overline{D_4} + 3\,\overline{D_2^2} + 4\,\overline{D_3 D_1} + 6\,\overline{D_2 D_1^2} + \overline{D_1^4}\right] 
\end{align}

\vspace{.8cm}

Now, as per the formula of \autoref{eq:mu basis}, we have to extract the real part of the exponential i.e $e^A$, which means 
that the above series becomes an even series in $\mX$, since the coefficients appearing in odd powers of $\mX$ are purely imaginary. This is because, as per the CP symmetry of QCD, all $D_n$ are purely real for even $n$ and purely imaginary for odd $n$. Hence, we therefore have the following : 

\vspace{.8cm}

\begin{equation}
    \text{Re}(e^A) = 1 + \sum_{k=1}^2 \boldsymbol{\mu}_X^{2k}\, \Ob\left(\boldsymbol{\mu}_X^{2k}\right) + \Ob\left(\boldsymbol{\mu}_X^6\right)
\end{equation}

\vspace{.8cm}

Hence, this implies 

\vspace{.8cm}

\begin{equation*}
    \bigg \langle \text{Re}(e^A) \bigg \rangle = 1 + \sum_{k=1}^2 \boldsymbol{\mu}_X^{2k}\, \bigg \langle \Ob\left(\boldsymbol{\mu}_X^{2k}\right) \bigg \rangle + \bigg \langle \Ob\left(\boldsymbol{\mu}_X^6\right) \bigg \rangle
\end{equation*}  

and so, 
\begin{align}
    \frac{\Delta P_{ub}}{T^4} &= \frac{1}{VT^3}\,\ln \, \bigg \langle \text{Re}(e^A) \bigg \rangle \notag \\
    &= \frac{1}{VT^3}\,\ln \,\left[1 + \sum_{k=1}^2 \boldsymbol{\mu}_X^{2k}\, \bigg \langle \Ob\left(\boldsymbol{\mu}_X^{2k}\right) \bigg \rangle + \bigg \langle \Ob\left(\boldsymbol{\mu}_X^6\right) \bigg \rangle \right]
    \label{eq:real part of exponential}
\end{align}

\vspace{.8cm}

Now, we know the following expansion 

\vspace{.8cm}

\begin{equation}
    \ln (1+x) = x - \frac{x^2}{2} + \Ob(x^3)
    \label{eq:logarithm formula}
\end{equation}

\vspace{.8cm}

On computing $\Delta P_{ub}/T^4$ as given in \autoref{eq:real part of exponential} using the usual logarithm expansion given in \autoref{eq:logarithm formula}, and collecting coefficients upto $\Ob(\boldsymbol{\mu}_X^4)$, we find the following :

\vspace{.8cm}

\begin{align}
    \Ob(\boldsymbol{\mu}_X^2) &:\quad \frac{1}{VT^3} \, \left[\frac{1}{2!}\,\mathcal{A}_2\right] \notag \\
    \Ob(\boldsymbol{\mu}_X^4) &:\quad \frac{1}{VT^3} \, \bigg[\frac{1}{4!}\, \Big(\mathcal{A}_4 - 3\,\mathcal{A}_2^2\Big) \bigg]
    \label{eq:produce Taylor coefficients}
\end{align}

\vspace{.8cm}

where we have  
\vspace{.8cm}

\begin{align}
    \mathcal{A}_2 &= \bigg[ \LA \overline{D_2} \RA +  \LA \overline{D_1^2} \RA \bigg ] \notag \\
    \mathcal{A}_4 &=  \bigg[ \LA \overline{D_4} \RA + 4\,\LA \overline{D_3 D_1} \RA + 3\,\LA \overline{D_2^2} \RA + 6\,\LA \overline{D_2 D_1^2} \RA +  \LA \overline{D_1^4} \RA \bigg]
\end{align}

\vspace{.8cm}

Thus we find that \autoref{eq:produce Taylor coefficients} gives the exact Taylor coefficients $c_2$ and $c_4$ at $\Ob(\mX^2)$ and $\Ob(\mX^4)$ respectively, which would otherwise appear in the usual Taylor series expansion of excess pressure $\Delta P/T^4$. This clearly proves that the \autoref{eq:mu basis} with $\mathcal{C}_n$ for $1 \leq n \leq 4$ given in \autoref{eq:mu coefficients} exactly produces $4^{th}$ order Taylor series of excess pressure given as 

\begin{equation*}
    \frac{\Delta P_4^T}{T^4} = \nsum[1.4]_{n=1}^2 c_{2n}\, \left(\frac{\mX}{T}\right)^{2n}
\end{equation*}

	\cleardoublepage
    \end{spacing}

\renewcommand{\bibname}{References}

\begin{spacing}{1.4}

\end{spacing}

\end{document}